\documentclass[12pt]{article}
\usepackage{appendix}
\usepackage[utf8]{inputenc}
\usepackage[numbers]{natbib}
\usepackage{authblk, changepage}
\usepackage{pdfpages}
\usepackage{multicol}
\usepackage{multirow}
\usepackage{tabularx}
\usepackage{longtable}
\usepackage{amsmath}
\usepackage{amssymb}
\usepackage{amsfonts}
\usepackage{booktabs}
\usepackage{comment}
\usepackage{color}
\usepackage{subcaption}
\usepackage{bm}
\usepackage{pdflscape}
\usepackage{pdfpages}
\usepackage{array}
\usepackage{float}
\usepackage{fancyhdr}
\usepackage{adjustbox}
\usepackage{threeparttable}
\usepackage{makecell}

\usepackage[table]{xcolor}
\usepackage{lineno}
\usepackage{hyperref}
\hypersetup{colorlinks=true, pdfstartview=FitV, linkcolor=blue, citecolor=blue, urlcolor=blue}
\usepackage{xurl}

\usepackage[
  paper  = letterpaper,
  left   = 1.0in,
  right  = 1.0in,
  top    = 1.0in,
  bottom = 1.0in,
  ]{geometry}

\title{A cross-cultural dataset on the structure of social networks: The ENDOW project\thanks{This material is based upon work supported by the National Science Foundation under Award Nos. 1743019, 2218860, and 2218861.}}

\author{%
\footnotesize
Eleanor~A.~Power$^{16,25,\dagger}$,
Daniel~Redhead$^{41,18,11}$,
Tom~Rutter$^{26}$,
Sahana~Subramanyam$^{26}$,
Jeremy~Koster$^{18}$,
Monique~Borgerhoff~Mulder$^{35,25,18,33}$,
Samuel~Bowles$^{25}$,
Matthew~O.~Jackson$^{26,25}$,
Nurul~Alam$^{13}$,
Sarah~Alami$^{30}$,
Alexandra~Alvergne$^{29,5,10}$,
Curtis~Atkisson$^{48}$,
Michele~Barnes$^{46}$,
Bret~Beheim$^{18}$,
Christine~M.~Beitl$^{42}$,
Madeline~Brown$^{43}$,
Mark~Caudell$^{50}$,
Wendy~Ch\'{a}vez-P\'{a}ez$^{8}$,
Komal~Chauhan$^{16}$,
Joshua~Cinner$^{46}$,
Siobh\'{a}n~Cully$^{24}$,
Augusto~Dalla~Ragione$^{18}$,
Angelina~L.~DeMarco$^{47}$,
Ivan~Deschenaux$^{16}$,
Federico~Fernandez$^{6,27}$,
Juan~Pablo~Ferreiro$^{6,27}$,
Drew~Gerkey$^{20}$,
Matthew~Gervais$^{4}$,
Christopher~Golden$^{9}$,
Gianluca~Grimalda$^{17,21,12,15}$,
Werner~Hertzog$^{49}$,
Karen~Kramer$^{47}$,
Geoff~Kushnick$^{1}$,
Banrida~Langstieh$^{19}$,
Rodrigo~Lazo$^{44,23}$,
Sheina~Lew-Levy$^{7}$,
Shane~Macfarlan$^{47}$,
Emmanuel~Maliti$^{14}$,
Karl~J.~Mertens$^{2}$,
Madalena~Monteban$^{6,27}$,
Rafael~Morais~Chiaravalloti$^{31}$,
Daniel~Murphy$^{39}$,
Kathryn~Oths$^{32}$,
Alejandro~P\'{e}rez~Velilla$^{37}$,
Emily~Post$^{47}$,
Sean~Prall$^{36}$,
Cody~Ross$^{18}$,
Brooke~Scelza$^{36}$,
Michael~Schnegg$^{28}$,
Edmond~Seabright$^{30}$,
Mary~K.~Shenk$^{22}$,
Kathrine~E.~Starkweather$^{38}$,
Chun-Yi~Sum$^{3}$,
Cole~Tobin$^{45}$,
Bram~Tucker$^{40}$,
Bapu~Vaitla$^{35}$,
Vivek~Venkataraman$^{34}$,
Justin~Weltz$^{25}$,
John~P.~Ziker$^{2}$%
}

\date{October 2026}

\begin{document}

\maketitle

\vspace{-1cm}


\begin{abstract}
\noindent
We present multilayer social network data from $50$ communities, located in $29$ countries around the world, gathered as part of the ``ENDOW'' project.
The communities show exceptional diversity, from fishers in Bangladesh and Vanuatu, to horticulturalists in Fiji and Bolivia, to pastoralists in Namibia and Argentina, to farmers and wage labourers in India and Colombia.
Data collection embraced this variation, while still aiming for comparability.
Respondents answered ten network prompts on behalf of their ``sharing unit'' (household), covering different types of material and immaterial support.
From these, we construct networks representing flows of support between sharing units in each community.
We compare nomination patterns and network summary statistics across the ENDOW communities and to other benchmark datasets.
The summaries exhibit considerable heterogeneity, suggesting a need to revisit existing ``rules of thumb'' about social network structure.
These measures contribute novel reference points for research on the diversity of human sociality and social structure.
\end{abstract}

\newgeometry{top = 2cm, bottom = 2cm, left = 1.5cm, right = 1.5cm}


Studies of social networks over the preceding decades have made substantial empirical and theoretical contributions to the social and behavioural sciences, highlighting, for example, the importance of social influence for opinion formation and behaviour \cite{christakis_spread_2007, aral_distinguishing_2009, centola_spread_2010, banerjee_diffusion_2013} and noting fundamental regularities in the structure of social connections \cite{watts_collective_1998, barabasi_emergence_1999, mcpherson_birds_2001}.
However, our understanding of the structure of human social networks remains limited to specific contexts and certain types of social relations \cite[Table 10.1, p. 305]{newman2018networks}. 
Logistical constraints remain a crucial barrier for collecting sociocentric social network data (\textit{i.e.} reports on social relations from near--all members of a population) at scale \cite{marsden2005recent}.
This has led to the proliferation of two types of social network data.
First, researchers have collected large-scale egocentric personal network data that capture the characteristics of an individual's relations, but cannot map these relations onto others within the population to capture network structure \cite{perry2018egocentric, perry2020introduction}.
Second, researchers have collected sociocentric social network data mapping network structure in specific settings---predominantly schools and classrooms \cite{cillessen2009sociometric, veenstra2026harnessing} or organisations (\textit{e.g.}, businesses, public service organisations, religious groups:\cite{burcher2018social, doreian2017creating, kilduff2010organizational, lazega2001collegial})---where they can delineate a relatively natural `network boundary' within a population (\textit{e.g.}, sampling only individuals and their social relations within a certain classroom).
More recently, researchers have leveraged privileged access to proprietary large-scale data from online social media platforms such as Twitter (now X) and Facebook to construct sociocentric network data \cite{chetty_social_2022, chetty_social_2022-1, harris2025social, baileyetal2018,levin_dynamics_2021, rajkumar_causal_2022}.

The difficulty of collecting network data (whether egocentric or sociocentric) means that data are drawn predominantly from WEIRD (Western, Educated, Industrialised, Rich and Democratic) settings \cite{henrich2010weirdest}.
For example, a 2013 meta-review looking at personal social network size across the lifespan relied on a sample of 277 studies, of which $85\%$ were from North America or Europe, and nearly half from the United States \cite{wrzus_social_2013}.
There are multiple important exceptions, but they entailed substantial logistical and financial undertakings and are (understandably) limited to particular regions (\textit{e.g.}, $51$ villages in Thailand \cite{entwisle_networks_2007}, $75$ villages in Karnataka, India \cite{banerjee_diffusion_2013}, and $176$ villages in Honduras \cite{airoldi_induction_2024}).

As a result, many prevailing assumptions about how human social networks are structured are based on empirical estimates from a limited set of social settings that are unlikely to generalise to other contexts \cite{henrich2010weirdest}.
The mechanisms guiding the formation of social ties---and the resulting aggregate network structure---likely vary across social, cultural, and ecological settings.
Consider, for example, the research on friendship networks.
Much is known about how children and adolescents within formal education systems form and maintain friendships, as well as their social \cite{haynie2002friendship, salmivalli1997peer}, academic \cite{stadtfeld2019integration}, and health-related consequences \cite{mercken2009dynamics, parker2015hope}.
Yet, the meaning and function of friendship varies dramatically both within \cite{burt_distinguishing_1983} and across cultures \cite{hruschka2010friendship}, and opportunities for social interaction change over the life course \cite{carstensen1992social}.
Such variation will likely yield networks of fundamentally different structure than those observed in school classrooms, and the relative lack of sociocentric network data among adults or in more culturally diverse settings means that the extent of this variation remains unknown and unquantified.

In the absence of such comprehensive data, more or less explicit ``rules of thumb'' are often presented that draw on the limited available data.\footnote{There is a limited literature pointing out that many network measures do not follow rules of thumb (\textit{e.g.}, \cite{newman2003structure,jackson_meeting_2007,broido2019scale}), but those do not draw on similar settings and coordinated data collection.}
For example, a paper drawing on data from 82 Dutch school friendship networks, states (\cite{koskinen_multilevel_2023}, page 388):  ``Average degrees are about $4$, average reciprocity is about $0.60$, and average transitivity is about $0.56$. \emph{These are quite usual figures for friendship networks}'' (emphasis added).
Another example is the assertion that positive degree assortativity (individuals having ties with others who have a similar number of ties themselves) is an empirical regularity for social networks---a point that is often used to emphasise the distinction between social and other types of networks \citep{newman_why_2003, newman_assortative_2002, jackson_meeting_2007}.

Such ``rules of thumb'' must also necessarily be derived from comparisons drawn across datasets that may differ substantially in context, measurement, and construction. Networks in different datasets will be constructed on the basis of different prompts, with different samples and sampling strategies, and defined in different ways. All of these choices make comparison across even seemingly similar domains challenging, let alone widely varied ones. And yet, without consistent approaches, researchers have little choice but to draw potentially specious comparisons across datasets and settings.
To better understand how and why human social networks are structured in particular ways---and to establish whether existing heuristics are warranted---the scope of empirical inquiry must be broadened, systematically measuring different types of social relations across contexts that better reflect the diversity of human culture and sociality.

\begin{figure}[h]
\begin{center}
\vspace{-10pt}
\includegraphics[width=\textwidth]{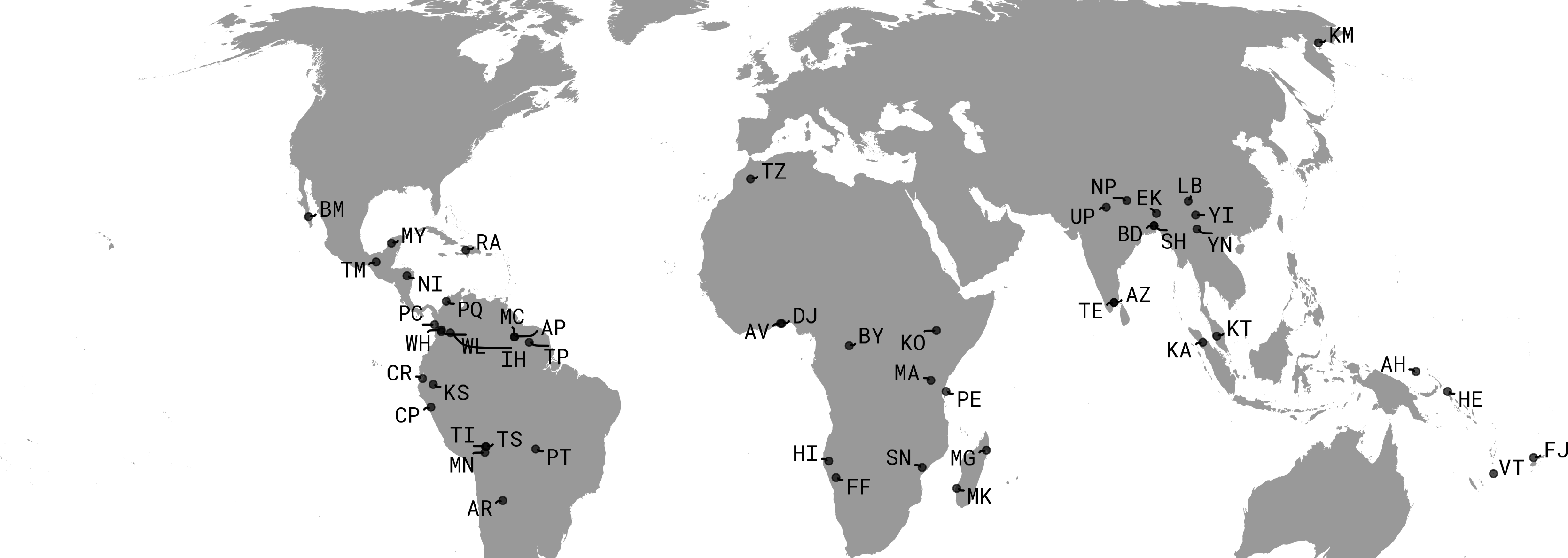}
\end{center}
\caption{The approximate location of ENDOW communities included in these analyses.}
\label{fig:endow-sites}
\end{figure}

Here, we present sociocentric network data from $50$ rural communities in $29$ countries around the world (see Figure~\ref{fig:endow-sites}).
These data were collected as part of the ongoing, longitudinal and comparative ENDOW (Economic Networks and the Dynamics Of Wealth inequality) project.
ENDOW began in 2017, with the aim of better understanding the social foundations of wealth inequality by assembling data to facilitate cross-cultural comparison \cite{power_social_2026}.
These communities are remarkably diverse in multiple dimensions.
While most of the communities could be considered rural or remote, some are in high population density areas.
Some of the communities lack substantial infrastructure and reliable electricity, while others are fully equipped with government services and facilities.
Although the communities can be roughly classified as horticulturalists, pastoralists, fishers, and similar discrete categories, almost all could also be effectively described as ``mixed economies,'' with families engaged in a range of productive activities \cite{burnsilver2016mixed, baird_pastoralist_2024}, increasingly including wage and salaried work.
Communities vary, too, in their post-marital residence patterns, in sharing norms, in how rights to land and resources are understood, in their division of labour, and in gender roles.
ENDOW communities range in size, with an average of a few hundred individuals divided into around $60-70$ households, which we here term ``sharing units'' so as to avoid any easy assumptions about their composition: while in some communities, sharing units equate to nuclear families centred on a monogamously married couple, in others they comprise larger polygynous compounds comprised of a man and his wives and their children, and in others they are multi-generational extended families.
The breadth of this variation among the ENDOW study communities is exceptional and provides a unique resource to advance our understanding of how human social networks are structured cross-culturally.

In what follows, we present descriptive measures of the ENDOW network data, starting first with respondent-level nomination patterns.
We then construct networks derived from these nominations (both distinct network layers based on each network prompt (\textit{i.e., name generator}) and one combined network based on the main layers), aggregating responses to the sharing unit level and therefore focusing on within-community ties.
With these networks in hand, we calculate a range of network summary statistics and look at the variation across networks and communities.
We additionally look at the overlap and correlation across network layers.
Throughout, comparisons are made to other datasets where similar measures have been presented, juxtaposing the data with alternative benchmarks.
These comparisons not only demonstrate the value of the ENDOW dataset as a resource for network scientists, giving new insights into the breadth of human social diversity and new reference points for future scholarship, but also demonstrate the caution that must otherwise be employed when comparing across datasets derived in such different ways. Across ENDOW networks, the substantial variation for numerous statistics indicates that prevailing ``rules of thumb'' about network features should be reconsidered while further suggesting that different social networks, which often serve fundamentally different purposes in human communities, will tend to have different structural properties.

\section*{Results}

\subsection*{Nomination patterns}

\begin{figure}[h]
\begin{center}
\includegraphics[width=\textwidth]{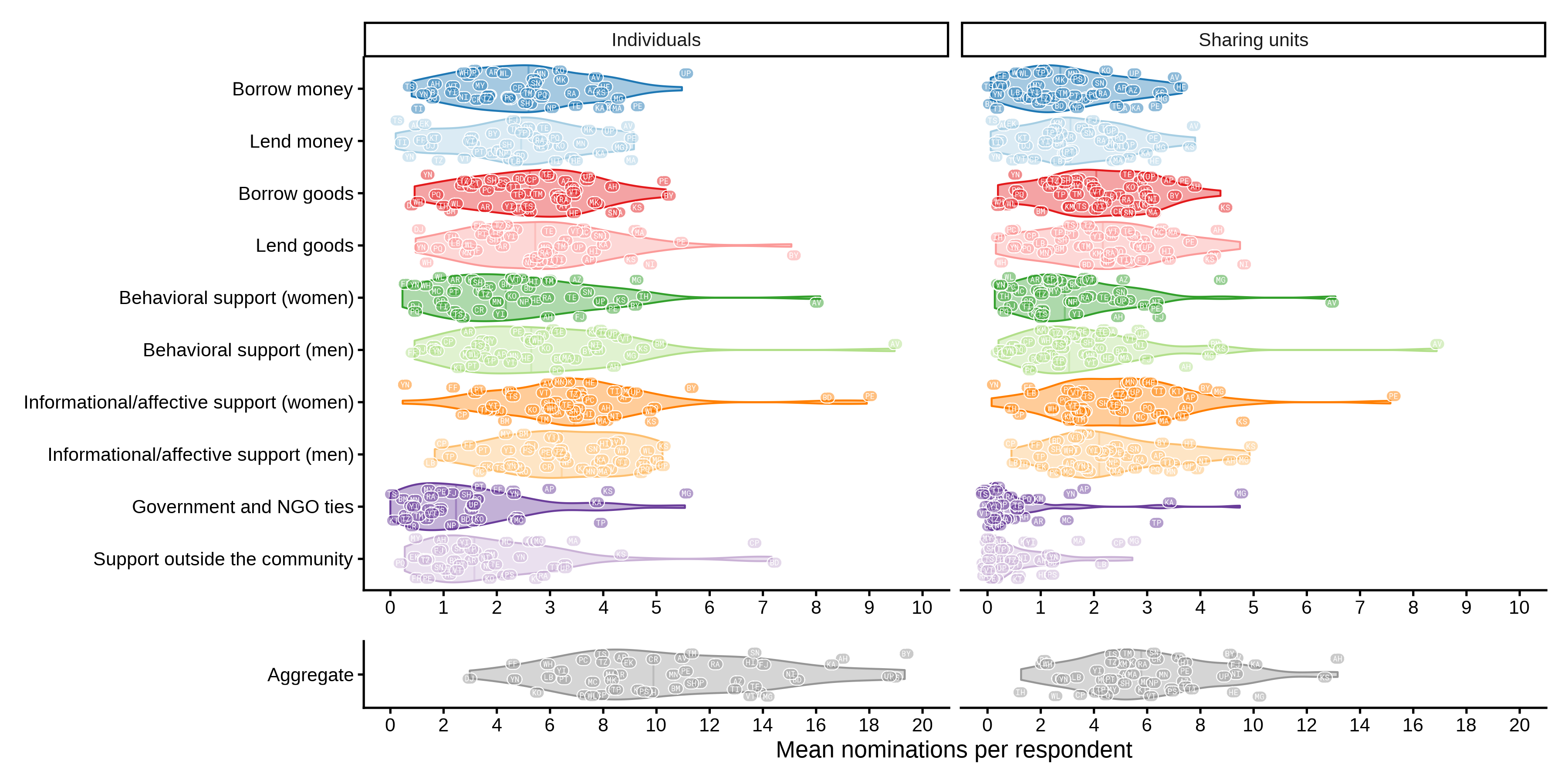}
\end{center}
\caption{Distribution of the mean number of nominations made by respondents across communities, showing (left) the number of \emph{individuals} and (right) the number of \emph{sharing units} named for each network prompt. The ``aggregate'' measure combines responses to give the total number of unique individuals or sharing units named across the network prompts that contribute to the aggregate network. Supplementary Figure~\ref{fig:nomination_counts_boxplot_each} shows the full distribution of nominations for each community and network prompt.}
\label{fig:nomination_counts_violin}
\end{figure}

Across the different network prompts posed, respondents could name whomever they liked, with no imposed minimum or maximum.
Figure~\ref{fig:nomination_counts_violin} shows the distribution of the mean number of nominations made by respondents in each community, showing (left) the number of \emph{individuals} and (right) the number of discrete \emph{sharing units} named for each network prompt.
Across communities, respondents named an average of two to three alters (\textit{i.e.}, individuals or partners) for each question, though there was substantial variation across communities.
Looking at the combined set of alters named for those prompts focused on within-community requests for support, we see across communities, respondents named an average of about $10$ individuals from about $6$ sharing units (see Table~\ref{tab:nominations-summary-global}).
Again, there is substantial variation across communities.

Existing datasets provide some ready comparisons for particular network prompts and the overall number of nominations given.
In Supplementary Section~\ref{si:nominations} the ENDOW network prompts are compared to existing data, revealing substantial variation across the data sources.
Regarding the aggregate number of nominations made, a useful reference is a meta-analysis looking at individual social network size, which found the number of alters to typically range between $10$ and $20$, and to decrease somewhat with age \cite{wrzus_social_2013}.
The authors note, however, that the majority of the studies they could include were based on samples from the United States, with only a handful from Africa or Asia. 

Supplementary Figure~\ref{fig:nomination_gender} illustrates the number of nominations made by men and women respondents across network prompts in study communities where there were a reasonable number of respondents of each gender.
Gender differences were slight, with the exception of the gender-specific questions.
As might be expected, men often named more alters for the two prompts focused on men, and equivalently so for women.
We also look at patterns across the age course and find generally negligible differences, with a few communities showing a suggestion of slightly fewer named alters in older age (see Supplementary Figure~\ref{fig:nomination_counts_scatter}).


While many of the nominations are to co-residents (\textit{i.e.}, members of other sharing units within the community), a substantial portion of nominations are to non-residents (\textit{i.e.}, those external to the community).
These ``external alters'' can vary substantially in how far removed they are from the community: some are a short walk away in the next village or neighbourhood, while others are separated by long international flights.
The reliance on ties outside the community varies substantially by community, with some, for example, being heavily reliant on remittances and seasonal or long-term migration and so with strong links beyond the community, and others being quite isolated and self-reliant.

\begin{figure}[h]
\begin{center}
\includegraphics[width=\textwidth]{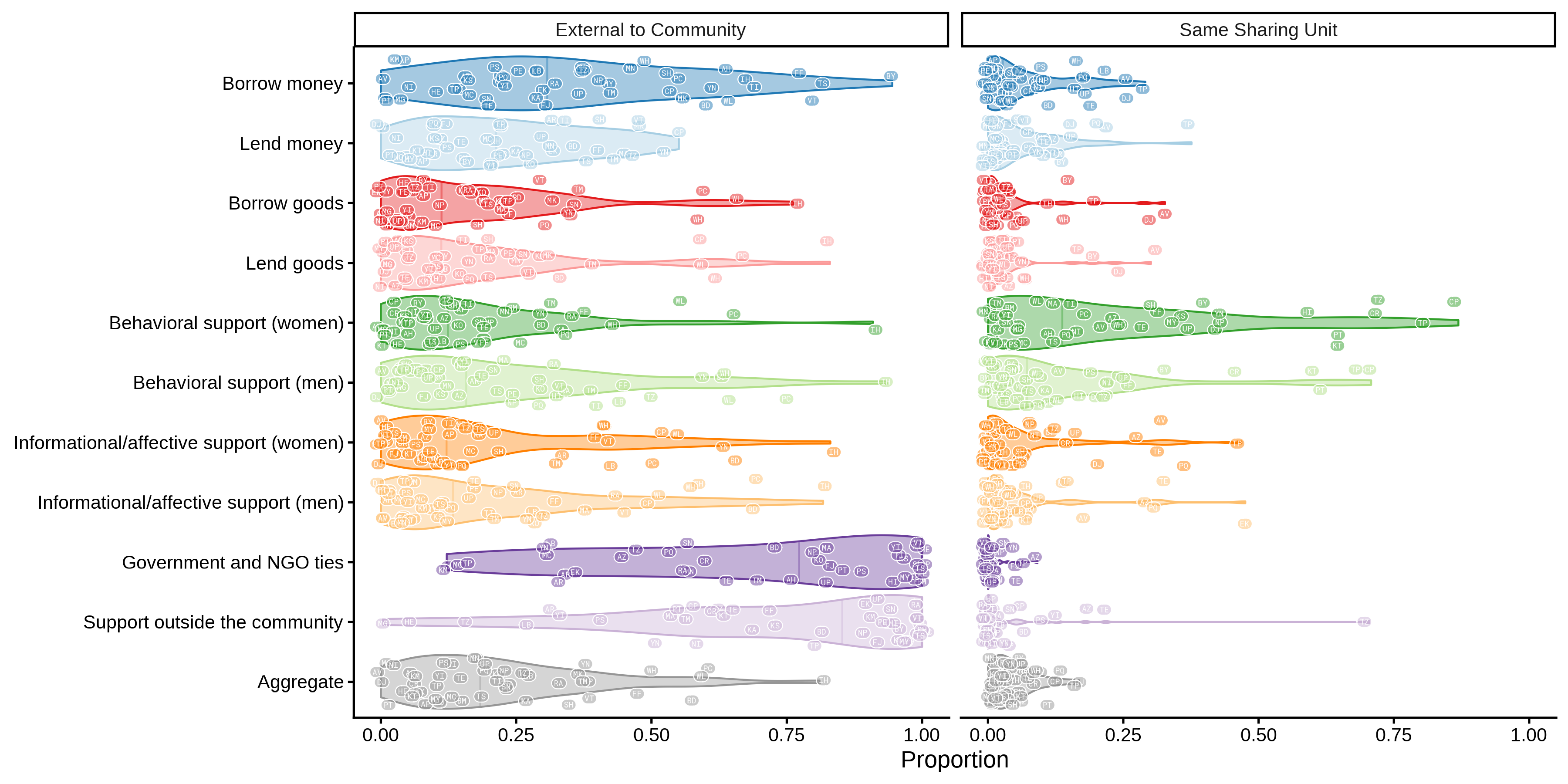}
\end{center}
\caption{Distribution of the mean proportion of alters named by each respondent who are external to the community (left) and who are members of their own sharing unit (right), for each network prompt. The ``aggregate'' measure considers those network prompts that contribute to our aggregate network. }
\label{fig:internal_external_violin}
\end{figure}

Across all communities, about one quarter of all named alters are non-residents, reinforcing the value of using an elicitation approach that allows for such nominations beyond the nominal boundary of the focal community \cite{laumann_boundary_1992}.
Figure~\ref{fig:internal_external_violin} shows the average proportion of external alters named by each respondent in each community, for each network prompt.
Even when focusing on the aggregated prompts that centre primarily on exchanges of support within the community, the average proportion of external alters in each community is around $20\%$ (see Table~\ref{tab:nominations-summary-global}).
As expected, the two network prompts geared towards eliciting external ties have a much greater proportion of nominations of non-residents across communities. (This aligns with the small number of sharing units named for these questions shown in Figure~\ref{fig:nomination_counts_violin}: most nominations for these questions are not to co-residents but rather those living elsewhere).
In general, the overall proportion of external alters broadly aligns with prior findings (see Supplementary Section~\ref{si:nominations} for direct comparisons).

Figure~\ref{fig:internal_external_violin}, also shows the proportion of nominations that are to members of respondents' own sharing units.
Across all communities, approximately $10\%$ of nominated alters are to people within the same sharing unit as the respondent.
Nominations to one's own sharing unit are notably high for behavioural support and low for external support and for borrowing and lending household goods (where we would generally expect people to need to venture beyond their own sharing unit for support).


With the sampling in many communities of multiple people from the same sharing unit, it is possible to examine the extent of concordance in responses among those who were effectively asked to report on the same relationships.
In this case, concordance entails naming the same individuals or, more broadly, naming the same sharing units.
As shown in Supplementary Section~\ref{si:concordance_su}, concordance varies between communities and is understandably much higher when considering nominations of sharing units (as opposed to individuals), where the median agreement between sharing unit respondents is typically well above $50\%$.

\subsection*{Network Structure}

\begin{figure}
\begin{center}
\includegraphics[width = \textwidth]{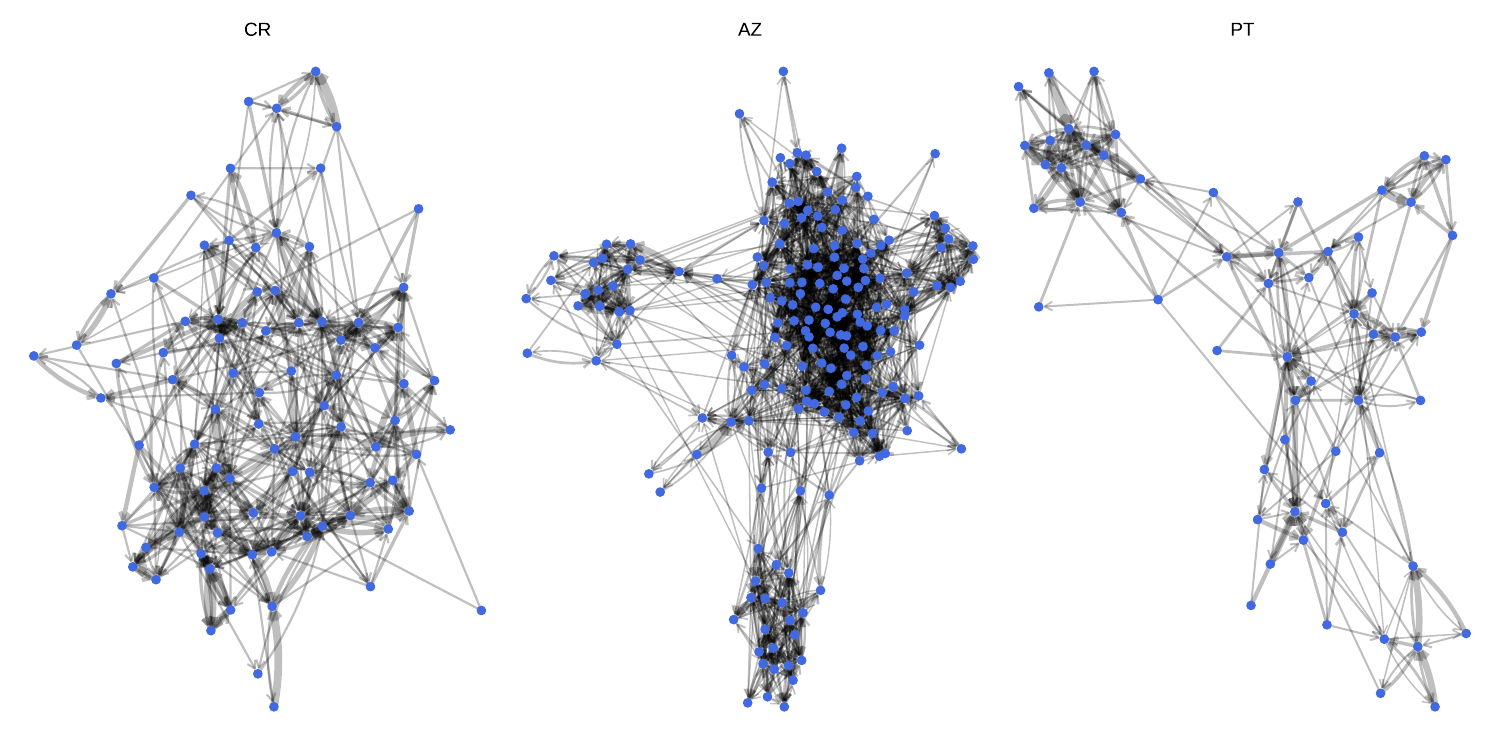}
\includegraphics[width = \textwidth]{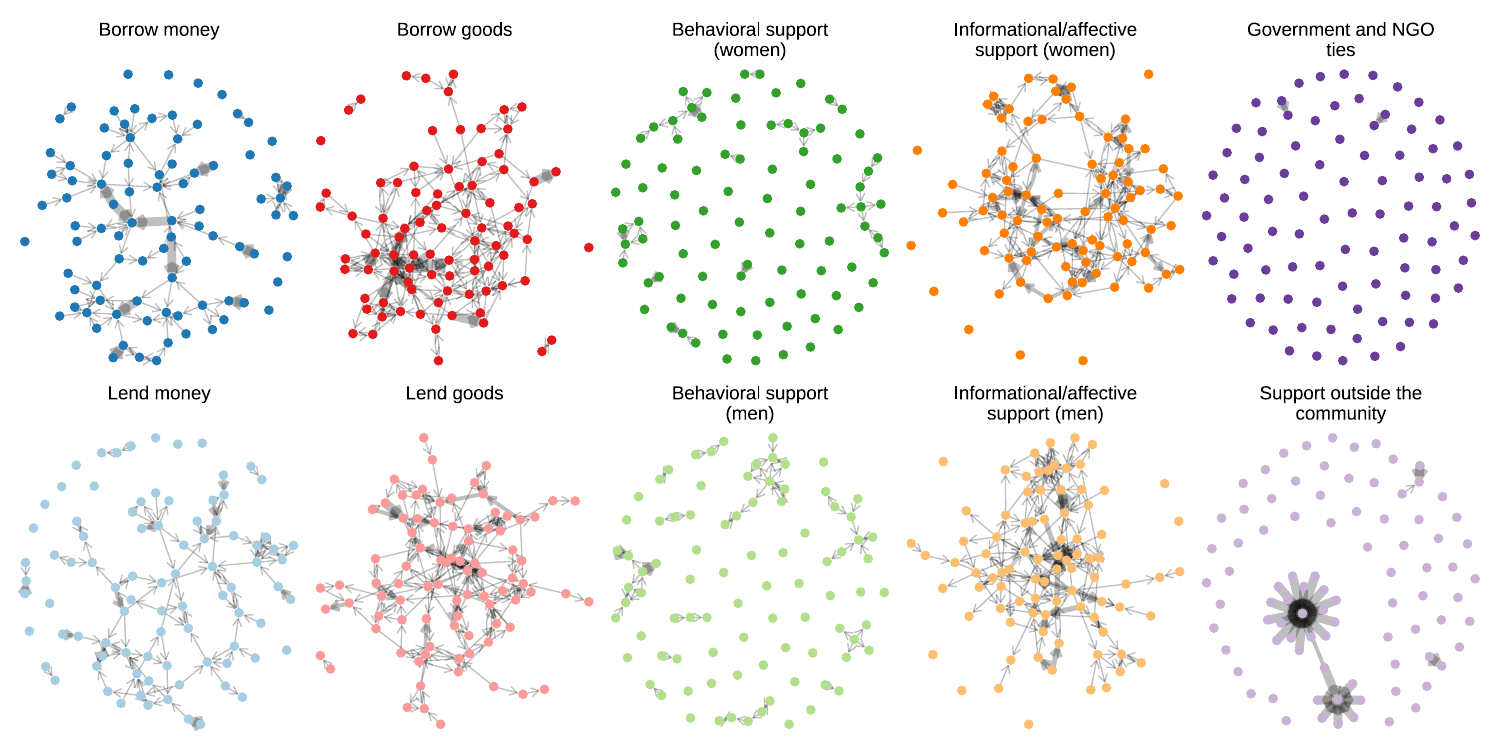}
\end{center}
\caption{(Top) Examples of aggregate networks (for CR: a coastal fishing community in Ecuador, AZ: a village of farmers and wage labourers in South India, and PT: a community in the Pantanal wetlands in Brazil). See full set in Supplementary Section~\ref{si:network_plots_agg}. (Bottom) Networks layers for CR. Note that edges are restricted to those between sharing units, explaining the sparsity of nominations in the two network layers focused on eliciting ties \emph{external} to the community (where very few community residents are nominated). See the full set of network diagrams in Supplementary Section~\ref{si:network_plots_each}. }
\label{fig:networks_plots}
\end{figure}

We construct distinct networks for each network prompt, representing sharing unit $i$'s nominations to another sharing unit, $j$, and one aggregate network, representing sharing unit $i$'s requests for support of another sharing unit, $j$. We remove nominations within the sharing unit (\emph{i.e.}, sel f-loops), so that attention is more cleanly on between-sharing unit relationships.
(As we consider only ties between sharing units, networks derived from the network prompts that focus on external relationships are quite sparse and do not merit strong interpretation, so this paper does not address their structural properties.) The resulting networks vary substantially in terms of their structure.
This holds both when comparing across layers within a community (as shown in the bottom panel of Figure~\ref{fig:networks_plots} for CR, a fishing community in Ecuador) and when comparing across communities (as shown in the top panel of Figure~\ref{fig:networks_plots}).

Supplementary Section~\ref{si:summary_stats} contains tables and figures presenting key summary statistics for individual network layers and the aggregate network, while Supplementary Section~\ref{si:network_plots} presents visualisations of each network.
Summary statistics are presented for
mean degree (\ref{si:mean_degree}),
in-degree centralisation (\ref{si:deg_centralization}),
degree assortativity (\ref{si:degree_assort}),
density (\ref{si:density}),
average path length (\ref{si:av_path_length}),
reciprocity (\ref{si:reciprocity}), and
transitivity (\ref{si:transitivity}).
Cross-site median, mean, and standard deviations for these measures in each layer are shown in Table~\ref{tab:mean-median-net-summary-stats}.
In Supplementary Section~\ref{si:sampling_impact}, we explore how differential sampling of sharing units within and across communities may shape some of these measures.

As one potential benchmark against which to compare the ENDOW data, Koskinen and Snijders (\cite{koskinen_multilevel_2023}, p. 388) describe a mean degree of about $4$, reciprocity of about $0.60$, and transitivity of $0.56$ as ``quite usual figures for friendship networks.'' Our closest corollary is the informational/affective support networks for men and women, respectively. We find a median mean degree of $2.5$ for the women's network and $2.2$ for the men's, quite a bit lower than what is deemed ``usual''.
Reciprocity is also much lower, with median values of $0.32$ (for the women's network) and $0.22$ (for the men's). Notably, reciprocity is much higher in the aggregate network, where the network construction is effectively allowing for reciprocity across different domains; Figure~\ref{fig:recip_across_layers} specifically illustrates reciprocity across all of the layers, showing that sharing units often have relationships that span multiple modalities, reciprocating not in kind but through other forms of support.
Transitivity values are again decidedly lower than the cited benchmark: our data exhibit a median global clustering coefficient of $0.26$ in the women's network and $0.23$ in the men's (and average local clustering coefficient medians of $0.31$ and $0.28$, respectively). Across all three measures, the ENDOW networks provide a very different candidate benchmark.

More general descriptors are often used in regards to the other measures we consider. Social networks are characterised as being ``small worlds,'' with short average path length combined with high clustering \cite{watts_collective_1998}.
While clustering in the ENDOW networks is lower than ``typical'' friendship networks, it is indeed substantially higher than what would be observed if connections happened at random, and average path length is generally short (median average path length of $2.82$).
Networks are often described as having ``heavy tailed'' degree distributions (with many nodes with relatively few connections and a few nodes with very many connections) \cite{newman2003structure}, or more strongly as following a scale-free power-law distribution \cite{barabasi_emergence_1999}; as seen in Supplementary Section~\ref{si:deg_distrib}, the ENDOW networks exhibit heavy-tailed degree distributions, but consistent with recent reviews \cite{broido2019scale} are mostly not likely to be strictly scale-free.
Social networks are also characterised as showing positive degree assortativity \cite{newman_assortative_2002, newman_why_2003, jackson_meeting_2007}.
While the cross-site median assortativity levels are generally positive, there are very many exceptions to this.
Indeed, the borrowing and lending money layers more often show \emph{negative} degree assortativity (\emph{i.e.} \emph{dis}assortativity) and a substantial number of communities have negative values in every layer.
This suggests that this overarching claim about social networks' degree assortativity needs some serious qualification, to at least consider the particular type of social support being represented in any social network.

Across these measures, our main takeaway is that there is substantial heterogeneity across networks. Across variables, coefficients of variation are often between $0.4-0.6$. In Supplementary Section~\ref{si:var_decomp} we quantify the sources of this heterogeneity by decomposing the variance in our measures into between-site and within-site (across-layer) components. As shown in Figure~\ref{fig:var_decomp}, a substantial share of the variation is associated with differences \textit{between} communities.  After accounting for systematic differences across network layers, community identity explains $61\%$ ($95\%$ CI: $49-70\%$) of the remaining variance in average degree and between $32-46\%$ of the remaining variance in reciprocity, global clustering, average local clustering, and in-degree centralization. The corresponding share for degree assortativity is lower, at $14\%$.  When we additionally control for some of the differences across communities that could impact these measures---including the number of sharing units, the sampling of sharing unit members, and reliance on alters outside the community---the between-site component of variation is attenuated but still prominent, suggesting that there are additional sources of cross-site heterogeneity in these measures.

\subsection*{Network Layer Overlap}

The rationale for asking multiple network prompts is typically that it will elicit meaningfully different alters and relationships \cite{marin_simplifying_2007, pustejovsky2009question}.
We therefore explore the extent to which different network prompts and network layers capture distinct, or aligned, alters and patterns of sociality.

\begin{figure}[h]
\begin{center}
\includegraphics[page=1, width = 0.49\linewidth]{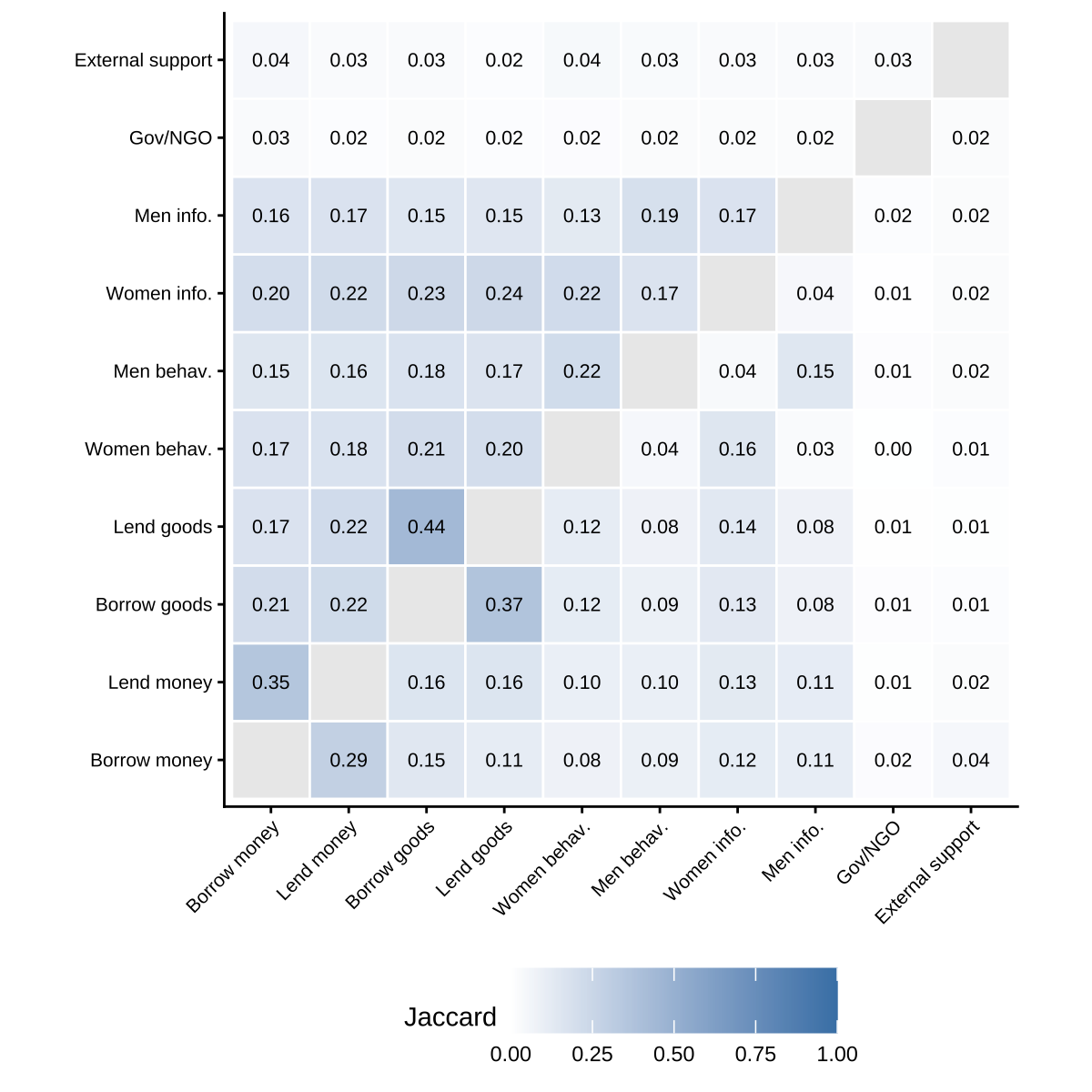}
\includegraphics[page=1, width = 0.49\linewidth]{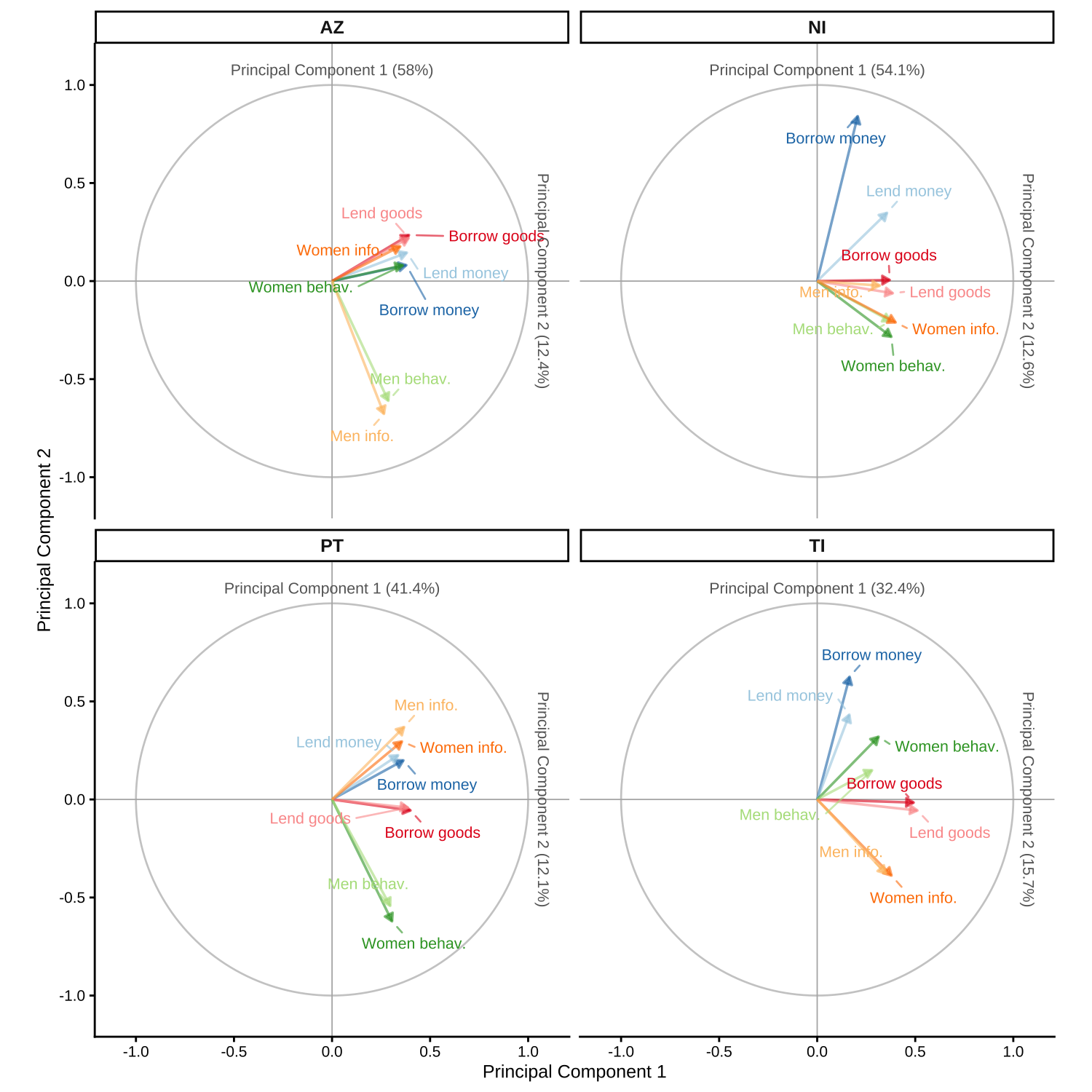}
\end{center}
\caption{(Left) Jaccard similarity of the sets of alters respondents nominate across network prompts, considering all \emph{individuals} named (lower triangle) and all \emph{sharing units} named (upper triangle). (Right) Examples of the community-specific loadings on the first and second principal components from AZ (a community in India), NI (a community in Nicaragua), PT (a community in Brazil), and TI (a community in Bolivia).}
\label{fig:alter_overlap_pooled_pca}
\end{figure}

In the left panel of Figure~\ref{fig:alter_overlap_pooled_pca}, we show the Jaccard similarity of the set of individuals and sharing units named by all respondents across all network prompts in all of the respective communities.
Supplementary Figure~\ref{fig:edge_overlap_pooled} reports similar measures using the derived sharing unit-level networks, calculating the Pearson's correlation coefficient of the $i \rightarrow j$ dyad in layer $k$ to the $i \rightarrow j$ dyad in layer $l$ across all pairs of network layers.
In Supplementary Sections~\ref{si:alter_overlap} and \ref{si:edge_overlap}, analogous plots are shown separately for each community.
There is no suggestion of negative correlation in the Pearson's correlations, where a connection to an alter for one relationship type makes it \emph{less} likely to be connected in another.
The Jaccard similarities show greater alignment of the sharing units named than of the particular individuals named, across effectively every relationship type.
Considering particular relationship types, it is notable that the set of individuals nominated for the two network prompts aimed at eliciting ties external to the community are quite distinct in the set of named alters.
There is also relatively low alignment in the sharing units named across the men- and women-specific relationship types.

Notably, the highest similarities are found for the two ``double-sampled'' question sets: many respondents report both borrowing \emph{and} lending household goods and (to a lesser extent) money with the same individuals and sharing units. This may be partially driven by a desire to be in balanced, reciprocal relationships.
However, that desire may not be borne out in the responses of their alters.
We can more formally assess the concordance of nominations across the two question sets: if a respondent in sharing unit $i$ says they borrow from sharing unit $j$, then respondents in $j$ should report that they \emph{lend} to $i$.
As shown in Supplementary Section~\ref{si:concordance_double} and Figure~\ref{fig:concordance_ds}, concordance on these questions is low, with an average across communities of $13\%$ for the monetary loan question and $21\%$ for the exchange of household items.
So, while there is relatively high overlap in the set of alters a sharing unit names across the two double-sampled questions, there is relatively low alignment across sharing units on who they would borrow/lend with, especially for monetary loans, suggesting that some of the alters named (especially for the lending networks) may be more aspirational than actual.
While these concordance levels are low if compared to those found for undirected relationships like ``best friends'' and ``closest intimates'' \cite{marsden_network_1990}, they are similar to levels found in a more comparable network dataset from 75 villages in Karnataka, India \cite{ready_measuring_2021}.

Cross-layer covariances beyond pairwise comparisons can also be elucidated by using principal component analysis.
The resulting loadings vary across communities, as seen in the right panel of Figure~\ref{fig:alter_overlap_pooled_pca} and in the full set of results in Supplementary Section~\ref{si:pca}.
Consistently across communities, the first principal component (capturing $42\%$ of the variance on average) captures basic connectedness: whether or not sharing units connected.
In many communities, the second principal component distinguishes between layers based on gender, separating the men--specific networks from the women--specific networks (as in AZ); a smaller number of communities (like TI) separate by support domains with the gender-specified questions being quite similar; and others (like NI) more strongly separate off particular support types, such as the monetary loan questions.
The double-sampled domains typically load closely together, consistent with the Jaccard similarity and concordance measures reported above.

The main inference for the ENDOW communities is that there are \emph{not} consistent overarching domains of support cross-culturally. Instead, there is substantial variation, with the particular amalgamations of support modalities differing across settings.
This diversity runs somewhat contrary to earlier findings showing greater alignment across settings (\textit{e.g.}, \cite{bernard_comparing_1990}).
For example, recent work looking at the Karnataka household networks \cite{chandrasekhar_multiplexing_2026} found that 
that network layers representing socialisation and borrowing kerosene or rice loaded consistently together. Across the seven ENDOW communities in the same region (India, Nepal, and Bangladesh), comparable alignment of the information/affective support and borrowing goods network layers is not evident. In other words, even among communities in the same region, the overarching finding is that of variability.

\subsection*{Kinship}

Kin relations often provide crucial support \cite{wellman_brothers_1989, hollinger_kinship_1990}.
Different marital and residency norms mean that the extent to which sharing units are linked by kinship relations varies substantially across the ENDOW communities, with some comprised primarily of unrelated families and others forming large extended kin groups (see Supplementary Section~\ref{si:kin} for plots, summary statistics, and further discussion of the close kin networks). Consequently, sharing units in different communities have highly variable access to co-residing kin: while in some communities (\textit{e.g.}, BD, FF, KO), the majority of sharing units have no kin among their neighbours, in others (\textit{e.g.}, MY, CR, TS) they have more than five sharing units (on average) with close kin and yet more with extended kin.
We focus here on close kin (parent-child and full sibling relationships) and on relationships between sharing units (we do not consider kin links outside the communities, nor links within a sharing unit).

\begin{figure}[h]
\begin{center}
\includegraphics[width=\textwidth]{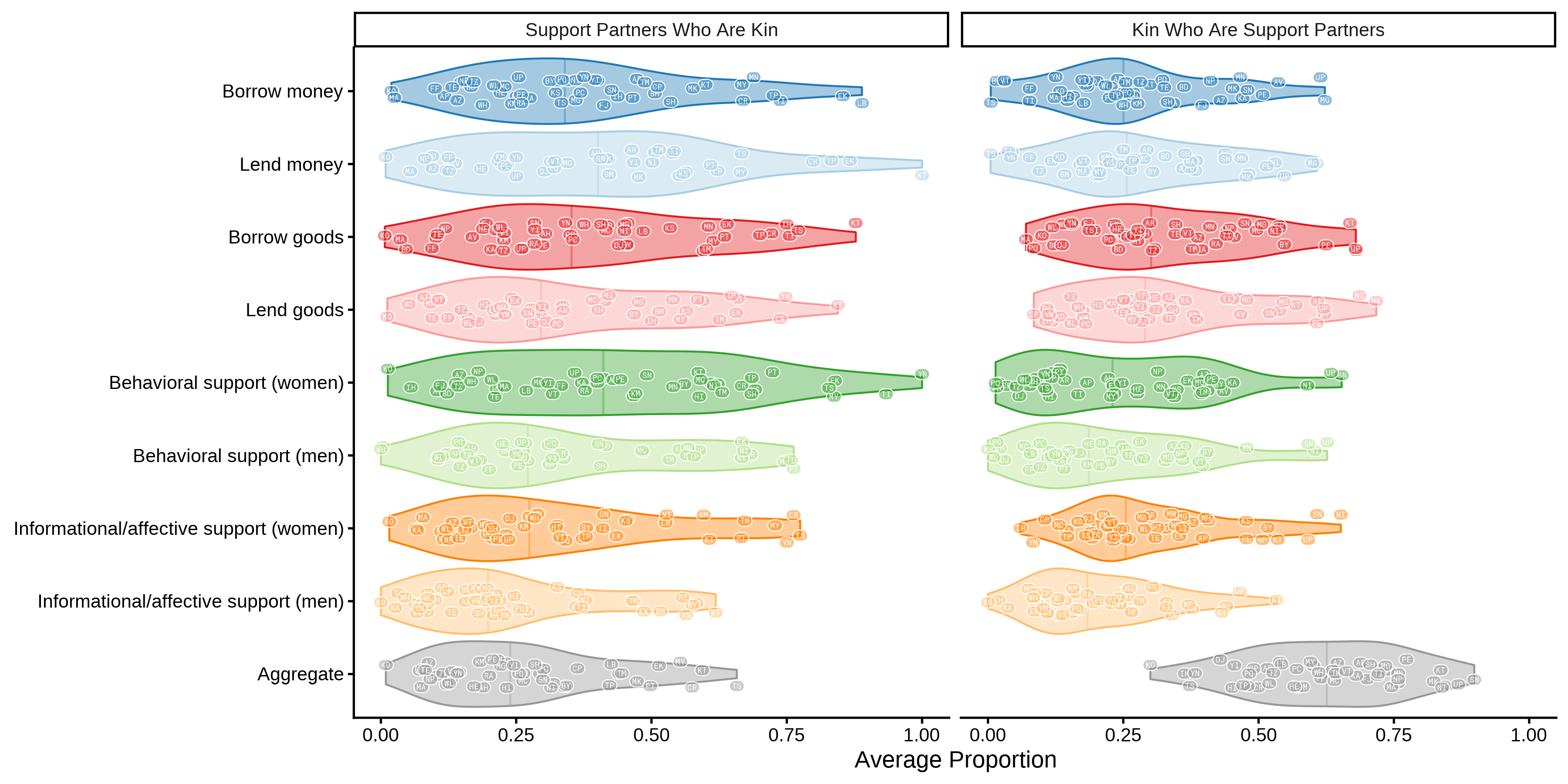}
\end{center}
\caption{Distribution of the overlap of close kin and social support alters. (Left) the average proportion of named sharing units which have close kin. (Right) the average proportion of sharing units with close kin that are named as support alters. }
\label{fig:kin_supp_violin}
\end{figure}

Figure~\ref{fig:kin_supp_violin} illustrates the extent to which kin are support partners in two ways: first, for each sharing unit, we look at the proportion of the sharing units named that are close kin; second, we look at the proportion of close kin sharing units that are named as support alters. The plots here show the average proportion for each community. Low values on the proportion of kin providing support could be the result either of low levels of relatedness between sharing units meaning that few kin are available, or because of active reliance on non-kin alters. For example, KO (a community in Ethiopia) has very few kinship links between sharing units, but $30\%$ of kin ties are named as providing some sort of support, on average.
Overall, reliance on kin is pervasive across communities: when considering the aggregate network, on average, about a quarter of the other sharing units named are ones with close kin, and over $60\%$ of other sharing units with close kin are named as providing some kind of support.

\section*{Discussion}

One of the main promises of (social) network analysis is that it brings a clearly defined structure (of nodes and edges) to the multifaceted complexities of social life.
This clarification (and necessary simplification) has the important potential to facilitate comparison across diverse contexts, through the shared vocabulary and structure of networks.

The common structure and metrics of network analysis, however, do not mean that comparison is always meaningful or warranted.
Researchers' decisions with regard to sampling strategies (\textit{e.g.}, sociocentric versus egocentric, self-report survey versus direct observation), elicitation techniques (\textit{e.g.}, roster versus network prompt), relationship definitions (\textit{e.g.}, network prompt phrasings), and subsequent network construction (\textit{e.g.}, taking the union versus intersection of nominations) all could undermine comparability.
While researchers may want to compare the structure of village networks of Karnataka with ones in Thailand or Honduras---or even compare them to school classrooms in the Netherlands or university cohorts in Copenhagen---such comparisons, while seemingly straightforward, would ignore the many divergences in measurement that would make such comparisons suspect.
Many of these risks are well appreciated, and substantial work has been done to establish what variation might be introduced by these different measurement choices (\textit{e.g.}, \cite{marsden_network_1990, bernard_comparing_1990, campbell_name_1991}).

And yet, clearly there is value and interest in such comparison.
Such comparison helps to establish potentially consistent patterns that arise across the breadth of social networks, even despite differences in measurement (\textit{e.g.}, small worlds).
Canonical textbooks (\textit{e.g.}, Newman's \textit{Networks} \cite
{newman2018networks}) reasonably place a range of social networks alongside those representing transport networks, food webs, and metabolic networks, to highlight some of those common features.

Our own efforts at comparison also draw on widely varied networks, using new data gathered in communities as distinct as wage labourers in Colombia and India, fishers in Brazil and Bangladesh, and pastoralists in Mexico and Namibia.
While these communities encompass much of the breadth of human social structural organization, there can be confidence that qualified comparisons are warranted because of the shared data collection protocols and common network construction. Further confidence comes from the ethnographic insights of the ENDOW team and the knowledge that the broad domains of everyday support that were queried are universally salient and materially consequential in these communities, as elsewhere. Given the reasonable desire to further place the variation observed in these communities alongside prior benchmarks, we have also sought to draw comparisons to other social network datasets while noting when cautious interpretations are needed.

Overall, reflecting the diversity of the ENDOW communities, substantial variation typifies nearly every measure that we examined, which in turn suggests that many of the tacit ``rules of thumb'' that researchers rely upon may merit reconsideration. Similar observations have been made for some specific network measures (\textit{e.g.}, for network degree distributions \cite{newman2003structure,jackson_meeting_2007,broido2019scale}), but here we show this for a much wider range of measures and across social networks that might have been expected to be more consistent in their attendant measures (due to their common elicitation and definition).

This variation is evident first in the number of nominations made by respondents. While in some communities people named on average over fifteen alters across the main support request network prompts, in others they named closer to five. For each particular network prompt, respondents typically named two or three alters, but here, too, the range across communities can be substantial. This generalization holds even for those network prompts that had more standardised phrasing across communities.

There is variation, too, in \emph{whom} people name as support alters.
The extent to which respondents named fellow sharing unit members or fellow community members differed substantially by community and by network prompt. While some domains, like connections to government and NGO officials, had people consistently seeking support outside not only their sharing unit but also their community, other domains were more varied: respondents' reliance on fellow sharing unit members for behavioural support is one notable case, with some communities largely getting support from within the sharing unit and others relying largely on neighbours.
Similar variation characterizes kin relations. Likely driven by differing marital and residency norms, ENDOW communities differ widely in the extent to which sharing units are linked by bonds of kinship, with some comprised of large extended kin groups and others of many unrelated nuclear family units. By extension, there is variation in the extent to which support relations are with close kin. Reliance on kin is, however, still pervasive: even in communities where sharing units have relatively few co-resident kin, the kin who are present in the community are frequently named as support alters.

Having elicited multiple domains of support, we can further assess whether there are consistent patterns in how different relationship domains overlap or remain distinct. Across communities, we find that different types of support elicit different alters, though there is certainly overlap in who is nominated, with people often turning to the same alters for multiple types of support, forming multiplexed relationships. The principal component analyses yet again highlight the variation behind this, however: while different support types may often prompt closely aligned sets of alters, \emph{which} types of support overlap varies across the ENDOW communities. This suggests different patterns of sociality, surely shaped by a complex combination of local social norms (such as those regarding gender) and varying modes of economic production and exchange, among other factors---a rich area for future study.

These findings also speak to fundamental questions of measurement and elicitation.
First, consider the reliance on those outside the community. External ties are clearly important across effectively all domains and communities. Had we used a roster method to elicit nominations (effectively limiting nominations to co-residents), about a quarter of all nominations across communities would have been missed.
Second, consider the overlap in nominations and edges with measurement in mind: the consistently tight alignment of the double-sampled questions may partially reflect our elicitation approach: these questions were asked back-to-back, making it much easier for respondents to name the same alters for the lending questions as for the borrowing, either due to those names already being front of mind, or due to a desire to be in balanced, reciprocal relationships. These patterns may then help account for the relatively low  concordance that is apparent across the double-sampled questions \cite[which is comparable to similar datasets:][]{ready_measuring_2021}.
Third, we can consider the concordance of reports given by members of the same sharing unit. Overall, there is reasonable agreement in respondents' overall set of nominations of sharing units, but there is decidedly less agreement (and, as always, substantial cross-site variation) for particular network prompts or particular individuals. This highlights that, while we may be relatively confident in the resulting aggregate network representation, who serves as representative(s) of a sharing unit will influence the resulting network.

These points of measurement matter further when the analysis turns to consider not individual nominations, but rather networks constructed from them. We took the common approach of constructing a sharing unit-level network by aggregating all responses made by each sharing unit's members. This was done in a bid for simplicity and comparability to prior work.
However, the different sampling of respondents across communities (ranging from communities with only one respondent per sharing unit to communities with all adult residents responding) coupled with the partial concordance in their responses means that some of the differences observed within and across communities may be due to their variable sampling in addition to ``true'' differences in local patterns of sociality.
Future work should build on network models that directly attempt to model such uncertainty and bias in self-reported network data \cite{butts_network_2003, de2023latent, redhead2024reliable}.

Alongside these caveats about comparability, it can be valuable to present common global summary statistics and compare them both within ENDOW communities and to other datasets. An overarching finding is that these networks differ in meaningful ways from the individual-level friendship networks that so often form the default benchmark for social network studies. The median values for reciprocity and transitivity, for example, are decidedly below what might be expected in friendship networks. Furthermore, there is substantial cross-layer and cross-site variation in these and effectively all other measures.
Beyond any particular ``rule of thumb'' point estimate, the ENDOW data even challenge more qualitative descriptors. Consider the common assertion that social networks are characterised by positive degree assortativity \cite{newman_assortative_2002, newman_why_2003, jackson_meeting_2007}: by contrast, a great deal of \emph{dis}assortativity across communities typifies this sample. The more comprehensive set of social support domains elicited (including ones where disassortativity may be more readily anticipated, such as borrowing/lending money) helps to account for some of this, but we find many communities with disassortativity even in the networks that are closest to ``friendship'' and about half of the communities display \emph{negative} degree assortativity in the aggregate network.
Clearly, insofar as the ENDOW communities provide a wide sampling of human sociocultural diversity, it should be clear that there is greater heterogeneity in the characteristics of human social networks than may have been appreciated.

Here, we offer summary measures from the ENDOW social support network data as a new addition to the collective understanding of the structure of human social relations. The project has documented substantial variation on effectively every measure considered, reflecting the wide sociocultural diversity of the ENDOW communities, and of human communities, writ large. These data move social and network scientists' reference set of social networks far beyond those that have served as the go-to benchmarks previously.
The variation that is evident in the data should facilitate future research into the axes upon which this variation rests. Some of the variation is surely due to differences in elicitation, sampling, and network construction, highlighting the need for further efforts to develop network models and measures that appropriately handle such measurement biases. Much of the variation, though, is due to meaningful differences in social relations across these widely varying communities.
We call for more empirical efforts to document the breadth of human social structural diversity, and, most importantly, for further theoretical work aimed at explaining the causes and consequences of such diversity for individuals, households, and societies. 

\section*{Methods}

The ENDOW project began in 2017 with the aim of collecting comparable demographic, economic, and social network data from a set of communities representing substantial cultural, ecological, and economic diversity.
The project is implemented by our interdisciplinary group of researchers who were already working in these communities, which provided the relevant knowledge and connections to facilitate the gathering of these data.
The data presented here consist of measures at both the level of the ``sharing unit'' and at the individual level.
The ``sharing unit'' is the fundamental unit of analysis, as it is the unit to which resources are (principally) pooled and shared.
This unit may often, but not always, align with the idea of a ``household''.
An overview of the communities is presented in Table~\ref{tab:su_summary}. ENDOW communities are typically small, with a median of about $60$ and a mean of about $90$ sharing units. The sharing units have, on average, five members.

\subsection*{Data Collection}

Researchers and research assistants conducted surveys during extended visits to their respective study communities, where they gathered comprehensive information through structured interviews.
The aim was to elicit information not just on sharing units and their social networks (\textit{e.g.}, material support), but also detailed individual-level demographic information (\textit{e.g.}, gender, age, occupation status), as well as measurements capturing an individual's ``embodied wealth,'' such as their locally valuable skills and knowledge \cite{borgerhoff_mulder_intergenerational_2009}.
The topics and exact wording of the questions included in the interviews for a study community were devised collaboratively among ENDOW researchers, with the aim of being as culturally- and community-specific as appropriate while also facilitating cross-cultural comparison.

The general sampling aim was to ensure as close to complete coverage of sharing units as possible for the social support network questions, and coverage is, on average, $96\%$ (see Supplementary Table~\ref{tab:respondents_by_site} for further details). Who, exactly, was the intended representative of a sharing unit varied from community to community, dependent on a number of factors, such as  who would likely be at home and available to respond and the time and resource constraints of the researchers. In $30\%$ of the communities, researchers aimed to survey all adult residents, in $22\%$ male and female household heads, and in the remaining $48\%$ either the male or female household head (which, specifically depending on the community). The intended respondents for each community are also shown in Supplementary Table~\ref{tab:respondents_by_site}, alongside the total number surveyed, by gender.

\subsection*{Social Networks}

To measure core aspects of social support between sharing units, we developed a series of ten network prompts.
Some of these questions used standardised phrasing, while it was necessary to tailor other questions to fit the context of the respective communities.
Two network prompts (one on monetary loans and one on exchange of basic household goods) were ``double sampled'', meaning that respondents were asked about both directions of a tie (i.e., they were asked who received from and also who they gave to).
This approach allows us to gather perspectives on the same directed ties from both individuals involved \cite{ready_measuring_2021}.
Respondents were first asked about whom they received from and were then asked about whom they gave to.
Given variation in survey respondents across communities, some questions were framed to ensure that respondents were thinking not only of themselves, but also of other sharing unit members. Specifically, two relationship domains had gender-specific network prompts, to make sure that, for example, a female head of household also considered the relationships of male sharing unit members, and vice versa.
Network prompts were typically posed in the same order across communities (and in the order presented here), with a few minor exceptions; within communities, every respondent was presented questions in the same set order.
Follow-up questions were asked for each named alter.
See Supplementary Section~\ref{si:name_gens} for tables showing the English-language phrasing of all network prompts asked in each community. As with all of the survey data, questions were posed in the relevant local language.

\paragraph{Financial Support}
The survey collected information about money lending using double-sampled network prompts. 
This was done by first asking respondents to report about who they would \emph{borrow} a week's wages and then about \emph{lending} a week's wages if there were a sudden and urgent need such as a medical emergency.
While the question was standardised to reflect the average wage an individual would earn for a week of work, the specific monetary amount provided to reflect a week's wages was based on local wage rates and presented in the local currency.

\paragraph{Material Support}
 We measured sharing unit-level exchanges of material goods, again, using double sampled questions (\textit{i.e.}, asking first about asking/borrowing and giving/lending).
 The particular phrasing was determined by the anthropologist and was thus specific to each community. 
 This meant that the specific goods that were the main focus of the questions necessarily differed between communities, with many questions focusing on food sharing and others focusing more on material items used for subsistence or other activities.

\paragraph{Behavioural Support}
 We measured gender-specific behavioural assistance and labour support between sharing units using two network prompts (which were not double-sampled), with one question focused on women's assistance and one on men's.
 This bifurcation by gender was done in recognition of the gendered division of labour in many communities, meaning that the types of behavioural assistance could plausibly differ for sharing unit members who were men to those who were women.
 This was further done to help respondents think holistically about the relationships of their entire sharing unit, not just their own personal connections.
 The particular prompts were specific to each community as the relevant types of behavioural assistance vary based on the set of livelihoods in the respective communities.
 That is, the women's behavioural assistance networks were generally focused on help during subsistence activities typically undertaken by women, while men's were primarily focused on subsistence activities typically conducted by men.

\paragraph{Informational and Affective Support}
The same approach was used for the two subsequent questions on informational and affective support (\textit{i.e.}, gender-specific questions, specific to each community).
Prompts were typically focused on who the men/women socialised, chatted, or spent their leisure time with.

\paragraph{External Support}
The survey further collected information about support from beyond the community, particularly about ties to organisations and people who can facilitate the navigation of bureaucracy and economic opportunities.
First, the survey included a question on government and NGO-employees who were well known and so could render assistance, which was a question with standardized phrasing across communities.
Second was a community-specific question on other important sources of support \emph{outside} of the community, with a particular focus on economic opportunities (\textit{e.g.}, accessing markets).

\paragraph{Kinship}
Researchers gathered information on kinship, allowing us to identify consanguineal and affinal relatedness between all community residents, including both adults and children.
Practically, how this was done varied from community to community (based on what genealogical data the researcher already had at hand), but typically entailed identifying the parents and grandparents of all community residents. We present here a network derived from this sampling, linking together sharing units if they share any close kin (i.e., parent-child or full sibling).

\subsubsection*{Network Construction}

With the sharing unit as our key unit for analysis and sampling, our interest is in constructing networks representing connections between sharing units. Here, we present networks derived from each network prompt, as well as one composite network that combines responses across some of the key network prompts. For each, individual responses are collapsed to the sharing unit level, such that the resulting network has nodes reflecting sharing units, rather than individual respondents. This further means that while nominations of people external to the community are discussed in the section on nomination patterns above, the networks presented here focus exclusively on ties between sharing units, \textit{i.e.}, those within the community.

For each network, an edge weight is assigned by dividing the number of individuals in sharing unit $j$ named as alters by sharing unit $i$ by the number of individuals in sharing unit $i$ who were respondents for that network prompt. Specifically, if two individuals in sharing unit $i$ are interviewed, and one names $3$ people in sharing unit $j$ for the borrowing money question, and another names $2$ people, then the borrow money layer gets a weight of $(3+2)/2=2.5$. This weighting helps to account for sharing units' varying opportunities to name support alters (based on different numbers of respondents) and varying extent to which they might rely on a sharing unit (if multiple people within a sharing unit are named).

The aggregate network is constructed from a selection of the network prompts: borrowing money, borrowing household items, and the men- and women-specified behavioural support and informational/affective support prompts. In other words, we do not consider the ``double sampled'' prompts or the two prompts focused on external relationships. To construct an edge weight for the aggregate network, the edge weights of each constituent layer are summed. Where possible, we present both binary and weighted measures drawing on these edge weights.
Supplementary Section~\ref{si:sampling_impact} examines how the use of these edge weights helps to account for differences in sampling of sharing units, both within and across communities.

\subsection*{Network Overlap}
We compare across the network layers in three ways.
First, we compute alter-level overlap with a Jaccard index: for each pair of layers, we count the individuals (or, alternatively, sharing units) that each respondent nominates in both layers (the intersection) and those they nominate in at least one of the two layers (the union), sum these counts across all respondents, and divide the resulting total intersection by the total union, either in aggregate or separately for each study community. Second, we compute dyad-level overlap by constructing a dyad-by-layer matrix of each dyad's (binary or weighted) edge values across layers; using this, we calculate the Pearson correlation between the $i \rightarrow j$ dyad in layer $k$ and the $i \rightarrow j$ dyad in layer $l$ for every pair of network layers, both in aggregate and also separately for each community.
Third, to characterise shared structure across all layers jointly rather than pairwise, we use the same dyad-by-layer matrix and fit a principal component analysis for each study community; the resulting loadings indicate which layers co-vary and how much of each layer's variation is captured by the leading components.

\subsection*{Variance Decomposition}

To understand the extent to which the large amount of variation in many of the measures of network structure is attributable to differences across communities, as opposed to differences in the type of support being provisioned, we decompose the variance in our site-layer level network statistics into a within-site across-layer component and an across-site component. Specifically, we estimate random-intercept models (where the intercept varies across communities) that include dummy variables for our network layers, with the relevant network statistic as the dependent variable. This facilitates partitioning of variation between differences in the communities in which these locations were formed versus differences in the type of relationship formed (regardless of the community in which the relationship was formed). We estimate models with and without additional site-level controls. We summarize the share of variance attributable to between-site differences, conditional on the layer indicators and additional site-level controls, using an intraclass correlation coefficient, obtaining confidence intervals via bootstrap.

\clearpage

\bibliography{bibliography.bib}

\clearpage

\subsection*{Ethics}

Ethical approval for each community was secured by the relevant contributing researcher following local, national, and institutional policies. An overarching ethics review was also conducted by the University of Cincinnati's Institutional Review Board (Study ID 2016-4691). Survey respondents gave informed consent before participating.

\subsection*{Data availability}

Given the many varied informed consent processes and associated assurances of privacy and access made by the team of researchers across the various communities, as well as the associated risks of de-anonymisation with such relational data, we cannot currently share the full network data, and instead provide the network-level summary statistics presented herein. Versions of these tables are included as CSVs in the public GitHub repository: \url{https://github.com/endowproject/endow-net-descrip}.

\subsection*{Code availability}

The code used to generate the findings shown in this paper is available at \url{https://github.com/endowproject/endow-net-descrip}.

\clearpage


\section*{Author Affiliations}
\footnotesize$^{1}$Australian National University, Canberra, Australia.\and
\footnotesize$^{2}$Boise State University, Boise, ID, USA.\and
\footnotesize$^{3}$Boston University, Boston, MA, USA.\and
\footnotesize$^{4}$Brunel University of London, London, UK.\and
\footnotesize$^{5}$Centre National de la Recherche Scientifique (CNRS), France.\and
\footnotesize$^{6}$Consejo Nacional de Investigaciones Cient\'{i}ficas y T\'{e}cnicas (CONICET), Argentina.\and
\footnotesize$^{7}$Durham University, Durham, UK.\and
\linebreak[0]
\footnotesize$^{8}$German Institute of Development and Sustainability (IDOS), Bonn, Germany.\and
\footnotesize$^{9}$Harvard T.H.\ Chan School of Public Health, Boston, MA, USA.\and
\footnotesize$^{10}$Institut de Recherche pour le D\'{e}veloppement (IRD), France.\and
\footnotesize$^{11}$Institut des Sciences Cognitives Marc Jeannerod, CNRS/Universit\'{e} Claude Bernard Lyon 1, France.\and
\footnotesize$^{12}$Institute for Globally Distributed Open Research and Education (IGDORE).\and
\footnotesize$^{13}$International Centre for Diarrhoeal Disease Research, Bangladesh (icddr,b), Dhaka, Bangladesh.\and
\footnotesize$^{14}$iReview Limited, Tanzania.\and
\footnotesize$^{15}$Kiel Institute for the World Economy, Kiel, Germany.\and
\footnotesize$^{16}$London School of Economics and Political Science, London, UK.\and
\footnotesize$^{17}$Masaryk University, Brno, Czech Republic.\and
\footnotesize$^{18}$Max Planck Institute for Evolutionary Anthropology, Leipzig, Germany.\and
\footnotesize$^{19}$North-Eastern Hill University, Shillong, Meghalaya, India.\and
\footnotesize$^{20}$Oregon State University, Corvallis, OR, USA.\and
\footnotesize$^{21}$Passau University, Passau, Germany.\and
\footnotesize$^{22}$The Pennsylvania State University, University Park, PA, USA.\and
\footnotesize$^{23}$Pontificia Universidad Cat\'{o}lica del Per\'{u}, Lima, Peru.\and
\footnotesize$^{24}$Rutgers University, New Brunswick, NJ, USA.\and
\footnotesize$^{25}$Santa Fe Institute, Santa Fe, NM, USA.\and
\footnotesize$^{26}$Stanford University, Stanford, CA, USA.\and
\footnotesize$^{27}$Universidad Nacional de Jujuy, San Salvador de Jujuy, Argentina.\and
\footnotesize$^{28}$Universit\"{a}t Hamburg, Hamburg, Germany.\and
\footnotesize$^{29}$Universit\'{e} de Montpellier, Montpellier, France.\and
\footnotesize$^{30}$Universit\'{e} Mohammed VI Polytechnique, Ben Guerir, Morocco.\and
\footnotesize$^{31}$University College London, London, UK.\and
\footnotesize$^{32}$University of Alabama, Tuscaloosa, AL, USA.\and
\footnotesize$^{33}$University of Bristol, Bristol, UK.\and
\footnotesize$^{34}$University of Calgary, Calgary, Canada.\and
\footnotesize$^{35}$University of California Davis, Davis, CA, USA.\and
\footnotesize$^{36}$University of California Los Angeles, Los Angeles, CA, USA.\and
\footnotesize$^{37}$University of California Merced, Merced, CA, USA.\and
\footnotesize$^{38}$University of California Santa Barbara, Santa Barbara, CA, USA.\and
\footnotesize$^{39}$University of Cincinnati, Cincinnati, OH, USA.\and
\footnotesize$^{40}$University of Georgia, Athens, GA, USA.\and
\footnotesize$^{41}$University of Groningen, Groningen, Netherlands.\and
\footnotesize$^{42}$University of Maine, Orono, ME, USA.\and
\footnotesize$^{43}$University of Maryland, College Park, MD, USA.\and
\footnotesize$^{44}$University of Massachusetts Amherst, Amherst, MA, USA.\and
\footnotesize$^{45}$University of Missouri, Columbia, MO, USA.\and
\footnotesize$^{46}$University of Sydney, Sydney, Australia.\and
\linebreak[0]
\footnotesize$^{47}$University of Utah, Salt Lake City, UT, USA.\and
\footnotesize$^{48}$University of Washington, Seattle, WA, USA.\and
\footnotesize$^{49}$University of Zurich, Zurich, Switzerland.\and
\footnotesize$^{50}$Washington State University, Pullman, WA, USA.
\and
\linebreak[0]
\footnotesize$^{\dagger}$Corresponding author. Email: \href{mailto:e.a.power@lse.ac.uk}{e.a.power@lse.ac.uk}

\section*{Author Contributions}
E.A.P. and D.R. conceptualized the study.
E.A.P. and D.R. curated the data.
Formal analysis, visualization, and validation was conducted by E.A.P. and T.R.
Methodology was developed by E.A.P., D.R., and T.R.
Funding was acquired by E.A.P., M.B.M., S.B., M.O.J., J.K., D.R., M.Ba., M.C., S.C., K.K., G.K., D.M., and M.S.
Investigation was carried out by E.A.P., J.K., N.A., S.A., A.A., C.A., M.Ba., B.B., C.M.B., M.Br., M.C., W.C., K.C., J.C., S.C., A.D.R., A.L.D., I.D., F.F., J.P.F., D.G., M.G., C.G., G.G., W.B.H., K.K., G.K., B.L., R.L., S.L., S.M., E.M., K.J.M., M.M., R.M.C., K.O., A.P.V., E.P., S.P., C.R., B.S., M.S., E.S., M.K.S., K.E.S., C.S., C.T., B.T., B.V., V.V., and J.P.Z.
Project administration was handled by E.A.P., M.O.J., J.K., N.A., S.A., A.A., C.A., M.Ba., B.B., C.M.B., M.Br., M.C., W.C., K.C., J.C., S.C., A.D.R., A.L.D., I.D., F.F., J.P.F., D.G., M.G., C.G., G.G., W.B.H., K.K., G.K., B.L., R.L., S.L., S.M., E.M., K.J.M., M.M., R.M.C., D.M., K.O., A.P.V., E.P., S.P., C.R., B.S., M.S., E.S., M.K.S., K.E.S., C.S., C.T., B.T., B.V., V.V., and J.P.Z.
Resources were provided by E.A.P., K.K., S.C., M.M., and C.R.
Software was developed by E.A.P., D.R., T.R., S.S., and J.W.
Supervision was provided by E.A.P., M.O.J., D.R., S.C., M.M., R.M.C., C.R., and M.K.S.
E.A.P., D.R., and T.R. wrote the original draft.
E.A.P., M.B.M., S.B., M.O.J., J.K., D.R., and T.R. reviewed and edited the manuscript.

\section*{Competing Interests}
The authors declare no competing interests.

\section*{Acknowledgments}
\footnotesize{
This paper is based on fieldwork supported by the National Science Foundation under Award Nos. 1743019, 2218860, and 2218861.
We are also grateful for financial support from the Santa Fe Institute and the Omidyar Network through its Emergent Political Economies project.

Michele Barnes' fieldwork was co-funded by the Australian Research Council through a Discovery Early Career Fellowship Grant (DE190101583).
Christine Beitl's fieldwork was also supported by the University of Maine Faculty Summer Research Award.
Siobh\'{a}n Cully's fieldwork was supported by National Science Foundation grant Nos. 1461514, 1920812, and 2222262.
Karen Kramer's fieldwork was partially supported by National Science Foundation grant Nos. BCS-2051264, BCS-1632338.
Fieldwork by Ivan Deschenaux and Komal Chauhan was funded by Leverhulme Research Leadership Award RL-2022-039.
Matthew Jackson was partially supported by NSF grant SES-1629446.
Joshua Cinner was partially supported by Australian Research Council grants FL230100201, FT160100047.
Geoff Kushnick was partially supported by a grant from the ANU College of Arts and Social Science.

We thank all community residents and participants for their support and engagement with our fieldwork.
We further thank the research assistants who helped with data collection, entry, and processing.
Thanks to colleagues in the Department of Human Behavior, Ecology and Culture at the Max Planck Institute for Evolutionary Anthropology, including the student assistants who helped to process the data: Lucy Betke, Anastassiya Bublikova, Nikkin Dev, Gargi Kodgirwar, Enzo Lima, John Newton, Anne Buechner, Leonie Ette, Maria Froehlich, Vincent Kiepsch, Kristina Kunze, Carlotta Rueck, and Constanze Wiedemann.

Sheina Lew Levy thanks Prof. Clobite Bouka-Biona from IRSEN, who facilitated the acquisition of research permits and infrastructure, Dzabatou Moise, who served as a community liaison, Mekouno Pau for translation, and Sarah Pope-Caldwell and Adam Boyette for field support.
Geoff Kushnick thanks his Indonesian counterpart, Dr Fikarwin Zuska (Universitas Sumatera Utara), who facilitated research permits through RISTEK (2017) and BRIN (2023-24).
Michele Barnes thanks Jacqueline Lau for her assistance in the field and Sarah Sutcliffe for her assistance with data management.
Karen Kramer thanks Joe Hackman for his concerted help in managing the databases.
Rafael Morais Chiaravalloti thanks Mira\'{i}ra Noal Manfroi.
Eleanor Power thanks the Chella Meenakshi Centre for their support in facilitating fieldwork.
Mary Shenk thanks the International Centre for Diarrhoeal Disease Research, Bangladesh (icddr,b) and Tanima Rashid.
}

\appendix

\clearpage

\renewcommand{\figurename}{Figure}
\renewcommand{\tablename}{Table}
\renewcommand{\thefigure}{A\arabic{figure}}
\renewcommand{\thetable}{A\arabic{table}}

\setcounter{figure}{0}
\setcounter{table}{0}

\section{Supplementary Information}

\subsection{ENDOW Communities}

In Table~\ref{tab:su_summary} we show basic information on each of the ENDOW communities, including details of the sharing units and our sampling coverage of them.

\begin{figure}
    \centering
    \includegraphics[width=0.9\linewidth]{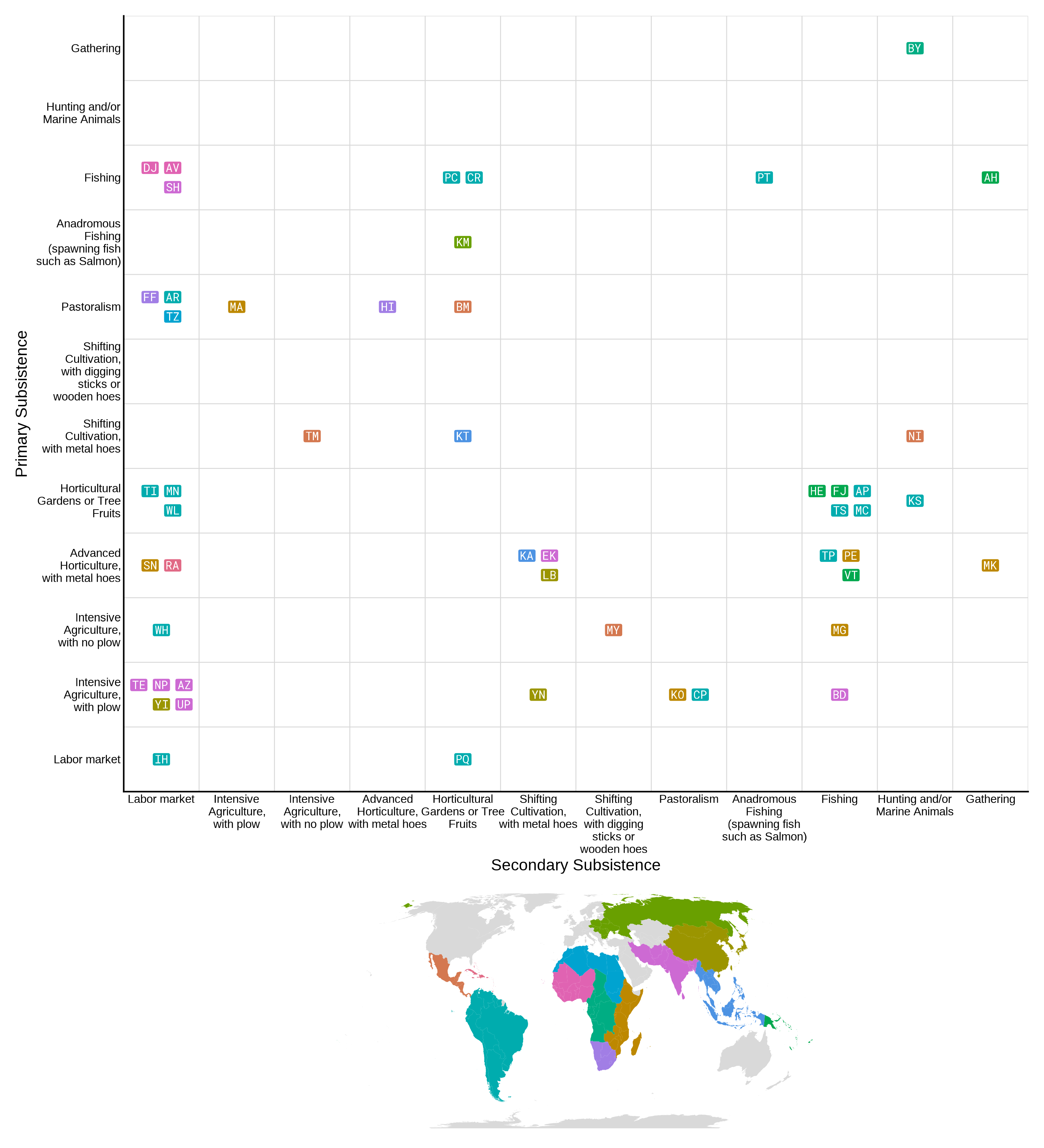}
    \caption{The primary and secondary mode of subsistence practised in each ENDOW community. Site codes are coloured by United Nations geoscheme regions, shown in the world map (with only represented regions coloured). Consult Figure~\ref{fig:endow-sites} for more exact locations.}
    \label{fig:subsistence}
\end{figure}

To give a sense of the diversity of the ENDOW communities, Figure~\ref{fig:subsistence} shows the primary and secondary mode of subsistence practised in each community, as reported by the contributing ENDOW researcher.

\begin{landscape}
\begingroup\fontsize{8}{10}\selectfont

\begin{longtable}[t]{l>{\raggedright\arraybackslash}p{3cm}>{\raggedright\arraybackslash}p{2cm}r>{\raggedright\arraybackslash}p{3cm}rrrrrr}
\caption{\label{tab:su_summary}Summary of each community and its sharing unit characteristics and sampling. The Members columns show the average, standard deviation, and maximum number of sharing unit members in each community. The Sampling columns show the average number of people surveyed in each sharing unit for the network questions, and subsequently the percent of sharing units with network data.}\\
\toprule
\multicolumn{6}{c}{ } & \multicolumn{3}{c}{Members} & \multicolumn{2}{c}{Sampling} \\
\cmidrule(l{3pt}r{3pt}){7-9} \cmidrule(l{3pt}r{3pt}){10-11}
Code & Participants & Country & Year & Lead(s) & \# SU & Mean & SD & Max & \# Surv & \% w/ Net\\
\midrule
\endfirsthead
\caption[]{Summary of each community and its sharing unit characteristics and sampling. The Members columns show the average, standard deviation, and maximum number of sharing unit members in each community. The Sampling columns show the average number of people surveyed in each sharing unit for the network questions, and subsequently the percent of sharing units with network data. \textit{(continued)}}\\
\toprule
Code & Participants & Country & Year & Lead(s) & \# SU & Mean & SD & Max & \# Surv & \% w/ Net\\
\midrule
\endhead

\endfoot
\bottomrule
\endlastfoot
AH & Ahus islanders & Papua New Guinea & 2018 & Michele Barnes and Joshua Cinner & 142 & 4.84 & 2.65 & 19 & 1.01 & 0.96\\
AP & Macushi & Guyana & 2017 & Curtis Atkisson & 55 & 4.60 & 2.23 & 10 & 1.62 & 1.00\\
AR & Argentinean & Argentina & 2024 & Madalena Monteban, Juan Pablo Ferreiro and Federico Fernandez & 176 & 3.10 & 1.81 & 10 & 1.00 & 0.90\\
AV & Xwla & Benin & 2023 & Augusto Dalla Ragione & 122 & 4.65 & 3.37 & 17 & 1.75 & 0.95\\
AZ & Tamil & India & 2017 & Eleanor Power & 191 & 3.23 & 1.54 & 9 & 2.32 & 0.99\\
BD & Bengalis & Bangladesh & 2017 & Mary Shenk and Nurul Alami & 79 & 3.78 & 1.58 & 8 & 1.00 & 1.00\\
BM & Rancheros or Choyeros & Mexico & 2017 & Shane Macfarlan & 32 & 3.53 & 1.52 & 8 & 1.90 & 0.97\\
BY & BaYaka & Congo-Brazzaville & 2018 & Sheina Lew-Levy & 43 & 6.58 & 3.39 & 14 & 2.73 & 0.95\\
CP & Chugurpampans & Peru & 2019 & Kathryn Oths & 90 & 3.22 & 1.81 & 8 & 1.11 & 1.00\\
CR & Ecuadorians & Ecuador & 2019 & Christine Beitl and Wendy Ch\'{a}vez-P\'{a}ez & 92 & 4.33 & 2.66 & 16 & 1.07 & 0.95\\
DJ & Xwla and Fon & Benin & 2023 & Augusto Dalla Ragione & 254 & 3.80 & 2.89 & 21 & 1.51 & 0.91\\
EK & Khasi Khynriam & India & 2020 & Alexandra Alvergne and Banrida Langstieh & 46 & 5.83 & 3.36 & 17 & 1.17 & 0.91\\
FF & Damara, Nama, Ovaherero, and Ovambo & Namibia & 2018 & Michael Schnegg & 44 & 3.68 & 2.66 & 12 & 1.00 & 1.00\\
FJ & Yasawans & Fiji & 2017 & Matthew Gervais & 44 & 3.98 & 2.20 & 9 & 1.00 & 0.95\\
HE & Solos & Papua New Guinea & 2019 & Gianluca Grimalda & 64 & 6.36 & 2.46 & 12 & 1.00 & 0.98\\
HI & Himba & Namibia & 2017 & Brooke Scelza and Sean Prall & 39 & 16.49 & 9.05 & 38 & 1.00 & 1.00\\
IH & Mestizo & Colombia & 2018 & Cody Ross & 78 & 3.41 & 1.92 & 10 & 1.99 & 0.99\\
KA & Karo & Indonesia & 2018 & Geoff Kushnick & 101 & 3.96 & 1.41 & 8 & 1.01 & 0.98\\
KM & Koryak & Russia & 2019 & Drew Gerkey & 95 & 2.66 & 1.83 & 12 & 1.00 & 0.76\\
KO & Kore & Ethiopia & 2017 & Mark Caudell & 67 & 7.28 & 3.76 & 20 & 1.84 & 1.00\\
KS & Jívaro & Peru & 2020 & Rodrigo Lazo & 55 & 5.62 & 2.59 & 11 & 1.00 & 0.96\\
KT & Semai & Malaysia & 2023 & Vivek Venkataraman & 18 & 5.00 & 1.81 & 10 & 1.00 & 1.00\\
LB & Mosuo & China & 2017 & Chun-Yi Sum and Siobh\'{a}n Cully & 30 & 3.83 & 1.42 & 7 & 1.00 & 1.00\\
MA & Maasai & Tanzania & 2017 & Mark Caudell & 42 & 6.98 & 3.76 & 19 & 1.98 & 1.00\\
MC & Macushi & Guyana & 2017 & Curtis Atkisson & 116 & 4.42 & 2.39 & 11 & 1.64 & 0.99\\
MG & Betsimisaraka & Madagascar & 2019 & Bapu Vaitla and Christopher Golden & 101 & 3.25 & 1.61 & 9 & 1.00 & 1.00\\
MK & Mikea & Madagascar & 2017 & Bram Tucker & 48 & 5.42 & 2.60 & 10 & 1.81 & 1.00\\
MN & Moseten & Bolivia & 2018 & Edmond Seabright & 31 & 4.58 & 1.88 & 9 & 1.27 & 0.97\\
MY & Yucatac Maya & Mexico & 2017 & Karen Kramer & 93 & 5.84 & 2.93 & 15 & 1.00 & 0.96\\
NI & Mayangna & Nicaragua & 2018 & Jeremy Koster & 42 & 8.60 & 4.42 & 20 & 2.39 & 0.98\\
NP & Nepali & Nepal & 2024 & Ivan Deschenaux & 163 & 2.71 & 1.37 & 8 & 2.04 & 0.99\\
PC & Afrocolombian, Embera, and Mestizo & Colombia & 2017 & Cody Ross & 61 & 4.49 & 2.21 & 10 & 2.12 & 0.97\\
PE & Pembans & Tanzania & 2018 & Emmanuel Maliti and Monique Borgerhoff Mulder & 44 & 6.95 & 2.88 & 13 & 1.93 & 1.00\\
PQ & Palenqueros & Colombia & 2026 & Alejandro P\'{e}rez Velilla and Cody Ross & 737 & 3.03 & 1.69 & 11 & 2.10 & 0.99\\
PT & Pantaneiros & Brazil & 2019 & Rafael Morais Chiaravalloti & 60 & 4.70 & 2.37 & 14 & 1.15 & 0.98\\
RA & Haitians & Haiti & 2018 & Angelina Demarco & 55 & 3.63 & 2.37 & 12 & 1.00 & 0.98\\
SH & Shodagor & Bangladesh & 2017 & Katherine Starkweather & 47 & 4.74 & 2.37 & 17 & 1.05 & 0.94\\
SN & Sena & Mozambique & 2019 & John Ziker and Karl Mertens & 34 & 9.12 & 5.53 & 22 & 1.85 & 1.00\\
TE & Tamil & India & 2017 & Eleanor Power & 154 & 3.19 & 1.48 & 7 & 2.28 & 0.99\\
TI & Tsimane' & Bolivia & 2019 & Bret Beheim & 43 & 5.23 & 2.39 & 10 & 1.95 & 0.88\\
TM & Tzotzil & Mexico & 2022 & Werner Hertzog & 75 & 5.45 & 2.59 & 16 & 1.79 & 0.96\\
TP & Tiriyó, Wayana, and Akuriyó & Suriname & 2017 & Cole Tobin & 56 & 6.38 & 4.77 & 26 & 1.74 & 0.96\\
TS & Tsimane & Bolivia & 2018 & Edmond Seabright & 65 & 5.18 & 2.34 & 10 & 1.06 & 0.72\\
TZ & High Atlas Amazighs & Morocco & 2023 & Sarah Alami & 77 & 5.47 & 2.84 & 14 & 2.74 & 0.91\\
UP & North Indian & India & 2025 & Komal Chauhan & 156 & 4.72 & 2.18 & 16 & 3.11 & 0.97\\
VT & Ni-Vanuatu & Vanuatu & 2018 & Siobh\'{a}n Cully & 34 & 4.76 & 2.58 & 9 & 1.03 & 0.97\\
WH & Mestizo & Colombia & 2019 & Cody Ross & 25 & 4.32 & 2.66 & 13 & 2.64 & 1.00\\
WL & Afrocolombian, Embera, and Mestizo & Colombia & 2017 & Cody Ross & 104 & 3.82 & 2.32 & 13 & 1.71 & 1.00\\
YI & Yi & China & 2018 & Madeline Brown & 74 & 5.38 & 1.42 & 9 & 1.00 & 0.93\\
YN & Mosuo & China & 2017 & Siobh\'{a}n Cully & 40 & 4.97 & 1.83 & 10 & 1.00 & 0.98\\*
\end{longtable}
\endgroup{}

\end{landscape}

\clearpage
\subsection{Survey Respondents}\label{si:survey_respondents}

Table~\ref{tab:respondents_by_site} includes details of our sampling coverage for the sharing units in each community. As this shows, the coverage of sharing units is very high -- over 96\% on average. Who was sampled within each sharing unit, however, varies substantially.

\begin{landscape}
\begingroup\fontsize{8}{10}\selectfont

\begin{longtable}[t]{lrrrrrrr>{\raggedright\arraybackslash}p{5cm}}
\caption{\label{tab:respondents_by_site}Summary of the sharing unit coverage and respondents for the social support network survey in each community.}\\
\toprule
Community & \# SU & Mean \# Members & \% Sampled & Mean Surveyed per SU & \# Surveyed & \# Women & \# Men & Intended Respondents\\
\midrule
\endfirsthead
\caption[]{Summary of the sharing unit coverage and respondents for the social support network survey in each community. \textit{(continued)}}\\
\toprule
Community & \# SU & Mean \# Members & \% Sampled & Mean Surveyed per SU & \# Surveyed & \# Women & \# Men & Intended Respondents\\
\midrule
\endhead

\endfoot
\bottomrule
\endlastfoot
AH & 142 & 4.84 & 96 & 1.01 & 138 & 38 & 100 & male head\\
AP & 55 & 4.60 & 100 & 1.62 & 89 & 49 & 40 & male and female heads\\
AR & 176 & 3.10 & 90 & 1.00 & 158 & 123 & 35 & female head\\
AV & 122 & 4.65 & 95 & 1.75 & 203 & 119 & 84 & all adult residents\\
AZ & 191 & 3.23 & 99 & 2.32 & 441 & 253 & 188 & all adult residents\\
BD & 79 & 3.78 & 100 & 1.00 & 79 & 79 & 0 & female head\\
BM & 32 & 3.53 & 97 & 1.90 & 59 & 28 & 31 & male and female heads\\
BY & 43 & 6.58 & 95 & 2.73 & 112 & 73 & 39 & all adult residents\\
CP & 90 & 3.22 & 100 & 1.11 & 100 & 77 & 23 & female head\\
CR & 92 & 4.33 & 95 & 1.07 & 93 & 88 & 5 & female head\\
DJ & 254 & 3.80 & 91 & 1.51 & 349 & 201 & 148 & all adult residents\\
EK & 46 & 5.83 & 91 & 1.17 & 49 & 40 & 9 & female head\\
FF & 44 & 3.68 & 100 & 1.00 & 44 & 24 & 20 & female head\\
FJ & 44 & 3.98 & 95 & 1.00 & 42 & 34 & 8 & female head\\
HE & 64 & 6.36 & 98 & 1.00 & 63 & 13 & 50 & male head\\
HI & 39 & 16.49 & 100 & 1.00 & 39 & 7 & 32 & male head\\
IH & 78 & 3.41 & 99 & 1.99 & 153 & 90 & 63 & all adult residents\\
KA & 101 & 3.96 & 98 & 1.01 & 100 & 98 & 2 & female head\\
KM & 95 & 2.66 & 76 & 1.00 & 72 & 45 & 27 & female head\\
KO & 67 & 7.28 & 100 & 1.84 & 123 & 64 & 59 & male and female heads (different prompts)\\
KS & 55 & 5.62 & 96 & 1.00 & 53 & 53 & 0 & female head\\
KT & 18 & 5.00 & 100 & 1.00 & 18 & 5 & 13 & male or female head\\
LB & 30 & 3.83 & 100 & 1.00 & 30 & 12 & 18 & male head\\
MA & 42 & 6.98 & 100 & 1.98 & 83 & 41 & 42 & male and female heads (different prompts)\\
MC & 116 & 4.42 & 99 & 1.64 & 189 & 98 & 91 & male and female heads\\
MG & 101 & 3.25 & 100 & 1.00 & 101 & 86 & 15 & female head\\
MK & 48 & 5.42 & 100 & 1.81 & 87 & 47 & 40 & male and female heads\\
MN & 31 & 4.58 & 97 & 1.27 & 38 & 23 & 15 & female head\\
MY & 93 & 5.84 & 96 & 1.00 & NA & NA & NA & male and female heads\\
NI & 42 & 8.60 & 98 & 2.39 & 98 & 52 & 46 & all adult residents\\
NP & 163 & 2.71 & 99 & 2.04 & 326 & 190 & 136 & all adult residents\\
PC & 61 & 4.49 & 97 & 2.12 & 125 & 66 & 59 & all adult residents\\
PE & 44 & 6.95 & 100 & 1.93 & 85 & 42 & 43 & male and female heads\\
PQ & 737 & 3.03 & 99 & 2.10 & 1533 & 857 & 676 & all adult residents\\
PT & 60 & 4.70 & 98 & 1.15 & 68 & 28 & 40 & male or female head\\
RA & 55 & 3.63 & 98 & 1.00 & 54 & 42 & 12 & female head\\
SH & 47 & 4.74 & 94 & 1.05 & 46 & 6 & 40 & male head\\
SN & 34 & 9.12 & 100 & 1.85 & 63 & 33 & 30 & male and female heads\\
TE & 154 & 3.19 & 99 & 2.28 & 346 & 191 & 155 & all adult residents\\
TI & 43 & 5.23 & 88 & 1.95 & 74 & 39 & 35 & all adult residents\\
TM & 75 & 5.45 & 96 & 1.79 & 129 & 69 & 60 & male and female heads\\
TP & 56 & 6.38 & 96 & 1.74 & 94 & 50 & 44 & male and female heads\\
TS & 65 & 5.18 & 72 & 1.06 & 50 & 47 & 3 & female head\\
TZ & 76 & 5.50 & 92 & 2.74 & 192 & 106 & 86 & all adult residents\\
UP & 156 & 4.72 & 97 & 3.11 & 470 & 224 & 246 & all adult residents\\
VT & 34 & 4.76 & 97 & 1.03 & 34 & 6 & 28 & male head\\
WH & 25 & 4.32 & 100 & 2.64 & 66 & 36 & 30 & all adult residents\\
WL & 104 & 3.82 & 100 & 1.71 & 178 & 113 & 65 & all adult residents\\
YI & 74 & 5.38 & 93 & 1.00 & 69 & 3 & 66 & male head\\
YN & 40 & 4.97 & 98 & 1.00 & 39 & 24 & 15 & female head\\*
\end{longtable}
\endgroup{}

\end{landscape}

\clearpage
\subsection{Nominations}\label{si:nominations}

In the main text Figure~\ref{fig:nomination_counts_violin} we show the mean number of nominations of individuals and sharing units for each network prompt and across the prompts that contribute to the aggregate network.
Table~\ref{tab:nominations-summary-global} shows global summaries of respondents' nominations, for each network prompt. Note that this is done having pooled all respondents across all communities, meaning that communities that are larger or have greater sampling of sharing unit members will make a greater contribution to these global summations.
Across all respondents, the greatest number of individuals and sharing units were named for the two questions on informational/affective support, and the greatest proportion of respondents not naming anyone was for the prompt on government and NGO ties. The highest proportion of nominations to others within the same sharing units were done for the behavioural support prompts. Borrowing/lending of household items appropriately has a low number of nominations within the sharing unit, and the two prompts meant to elicit relationships spanning beyond the community also appropriately have much higher proportions of nominations of individuals who are external to the community. Tables~\ref{tab:nomination-summary-loan_req}-\ref{tab:nomination-summary-external} show equivalent community-specific measures.

\vspace{0.5cm}

\begin{longtable}[t]{llllll}
\caption{\label{tab:nominations-summary-global}Mean (and standard deviation) values for nominations made across all respondents, for each name generator. `Prop. Zero' is the proportion of respondents nominating no alters. `N Indiv' is the number of individuals nominated. `N SU' is the number of sharing units. `Prop. Same SU' is the proportion of individuals nominated who are members of the same sharing unit as the respondent. `Prop. Ext' is the proportion of individuals nominated who are external to the community.}\\
\toprule
Name Generator & Prop. Zero & N Indiv. & N SU & Prop. Same SU & Prop. Ext.\\
\midrule
Borrow money & 0.13 & 2.79 (2.20) & 1.75 (1.64) & 0.12 (0.25) & 0.29 (0.37)\\
Lend money & 0.20 & 2.67 (2.19) & 1.94 (1.77) & 0.11 (0.24) & 0.17 (0.30)\\
Borrow goods & 0.25 & 2.30 (2.15) & 1.82 (1.82) & 0.05 (0.18) & 0.15 (0.30)\\
Lend goods & 0.24 & 2.43 (2.30) & 1.95 (1.93) & 0.05 (0.18) & 0.14 (0.29)\\
Women behav. & 0.28 & 2.38 (2.58) & 1.60 (1.91) & 0.26 (0.38) & 0.16 (0.31)\\
\addlinespace
Men behav. & 0.23 & 2.68 (2.59) & 1.86 (2.13) & 0.17 (0.32) & 0.20 (0.33)\\
Women info. & 0.10 & 3.73 (2.55) & 2.56 (1.99) & 0.15 (0.28) & 0.16 (0.28)\\
Men info. & 0.13 & 3.54 (2.63) & 2.53 (2.20) & 0.13 (0.26) & 0.18 (0.29)\\
Gov/NGO & 0.35 & 1.58 (1.75) & 0.68 (1.19) & 0.02 (0.11) & 0.58 (0.42)\\
External support & 0.41 & 1.61 (2.04) & 0.36 (0.88) & 0.07 (0.22) & 0.75 (0.39)\\
\addlinespace
Aggregate & 0.02 & 10.25 (6.55) & 5.85 (4.11) & 0.07 (0.11) & 0.20 (0.25)\\
\bottomrule
\end{longtable}

\vspace{0.5cm}

We can compare what we observe for particular network prompts with network data representing similar social support types in existing datasets.
Our informational/affective support prompts could be compared to the US General Social Survey question on ``important matters'' often glossed as representing people's ``discussion networks,'' which showed
a mean of $2$ and a mode of $0$ nominations in 2004 \cite{mcpherson_social_2006} (though see \cite{fischer_2004_2009}), or a European equivalent for older adults \cite{tomini_understanding_2016} with a mean of $2.46$ (ranging from $1.75$ in Slovenia to $2.89$ in Switzerland), both lower than what we observe in most ENDOW communities.
A closer corollary comes from $176$ villages in Honduras, where respondents across all villages reported on average $3.4$ or $4.1$ alters (across two waves of data collection) in response to three prompts (``\textit{Who do you spend your free time with}?'', ``\textit{Who is your closest friend}?'', and ``\textit{Who do you discuss personal matters with}?''). 
Decidedly more nominations were made by respondents to a personal networks study \cite{bernard_comparing_1990}: for a question on socialisation partners (which aligns with ours on information/affective support): respondents nominated $10.78$ alters in Jacksonville and $6.46$ in Mexico City.
This same study had prompts on household tasks (aligning with ours on behavioural support), where respondents nominated $3.40$ alters in Jacksonville and $0.86$ in Mexico City, and one on borrowing large sums of money (aligning with our monetary loan question), for which respondents nominated $2.48$ alters in Jacksonville and $1.47$ in Mexico City.
Here, we expand their documentation of the potential for substantial variation in the number of nominations across populations, and the relatively higher number of nominations for questions on socialisation.

\begin{figure}[h]
\begin{center}
\includegraphics[width=\textwidth]{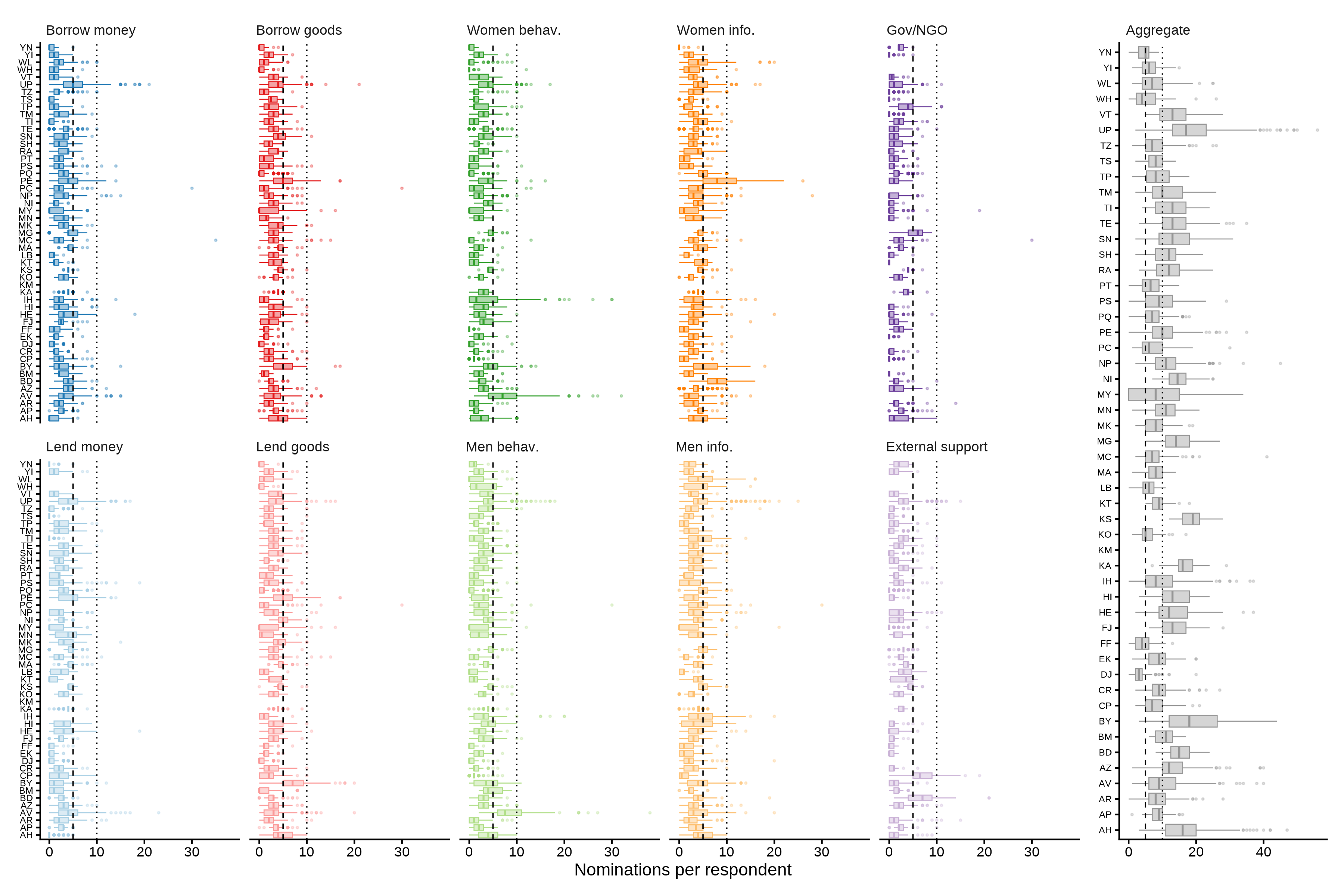}
\end{center}
\caption{\textbf{Distribution of the number of individuals named by respondents in each community for each network prompt}.The dashed lines demarcate $5$ and $10$, as references for hypothetical caps to the number of nominations respondents could have given. }
\label{fig:nomination_counts_boxplot_each}
\end{figure}

As a complement to Figure~\ref{fig:nomination_counts_violin} in the main text, Figure~\ref{fig:nomination_counts_boxplot_each} shows the distribution of the number of individuals named by respondents in each community for each network prompt. Showing the full distribution (as opposed to the means shown in Figure~\ref{fig:nomination_counts_violin}) highlights the range in the number of individuals named both across and within communities. For effectively every network prompt, there are some notable outliers of respondents who named very many individuals.

A common practice when gathering social network data is to place an upper bound on the number of nominations respondents can make.
For example, the General Social Survey ``important matters'' question limited respondents to five nominations \cite{marsden_core_1987},
in the Add Health studies, students could name up to five boys and five girls  \cite{fowler_model_2009};
in the Stanford dormitory study, students could name up to eight fellow dorm members \cite{morelli_empathy_2017};
in the Dutch school study, students could name up to twelve friends \cite{knecht_friendship_2010};
in the Karnataka microfinance study, respondents could name varying numbers of respondents for different network prompts with some uncapped \cite{banerjee_diffusion_2013}.
While authors of those studies often note that the cap is rarely reached, our unbounded network prompts provide some clarity across domains on the extent to which such caps may censor responses for different support domains.
The dotted lines in Figure~\ref{fig:nomination_counts_boxplot_each} demarcate $5$ and $10$, which we could take as hypothetical caps which could have been placed on the number of alters that could be named. While a cap at ten individuals would only rarely constrain people, a cap at five would have done so for a not insignificant number of respondents, particularly for the borrowing/lending goods and informational/affective support network prompts.

\begin{figure}[h]
\begin{center}
\includegraphics[width=\textwidth]{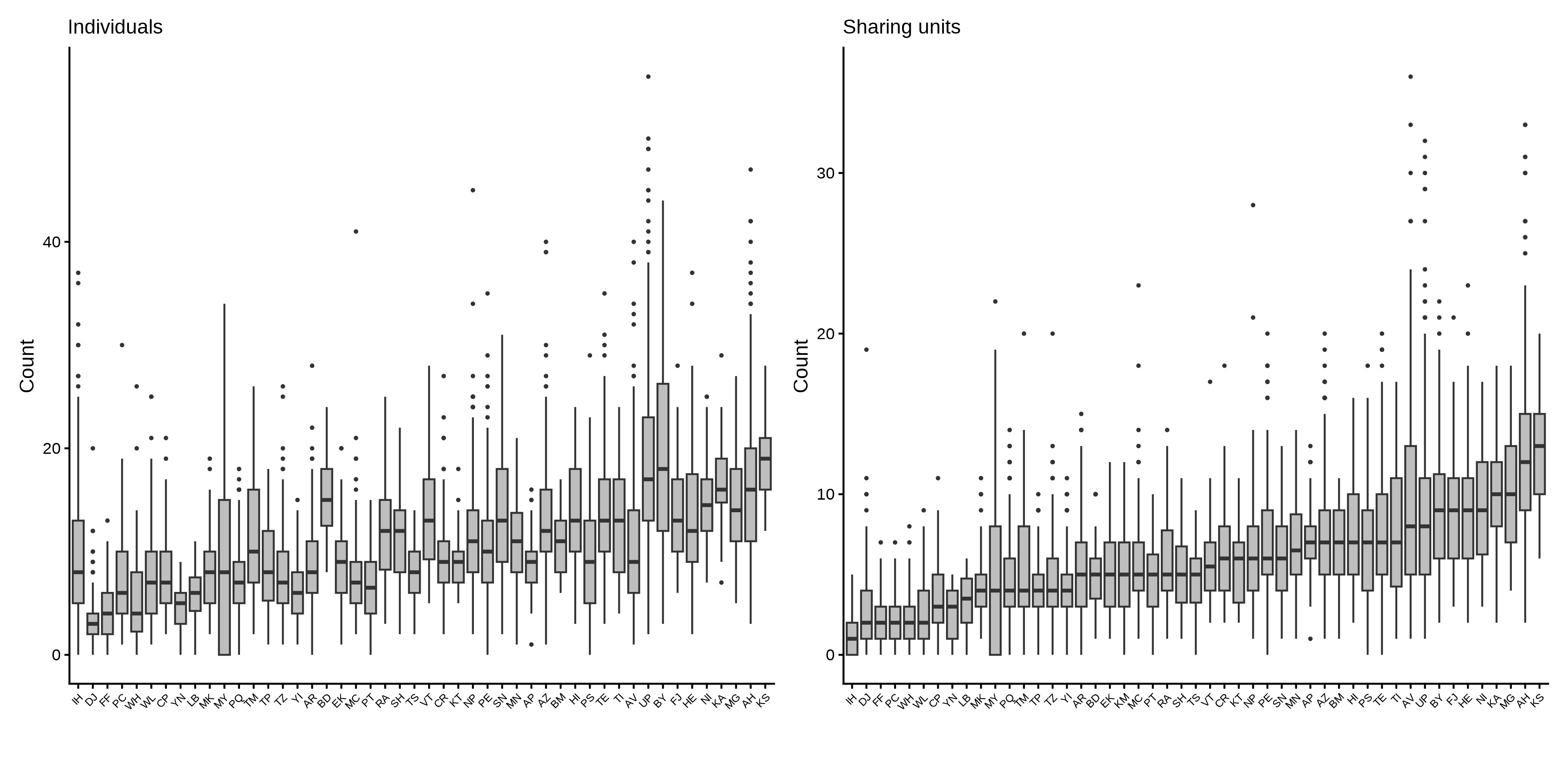}
\includegraphics[width=\textwidth]{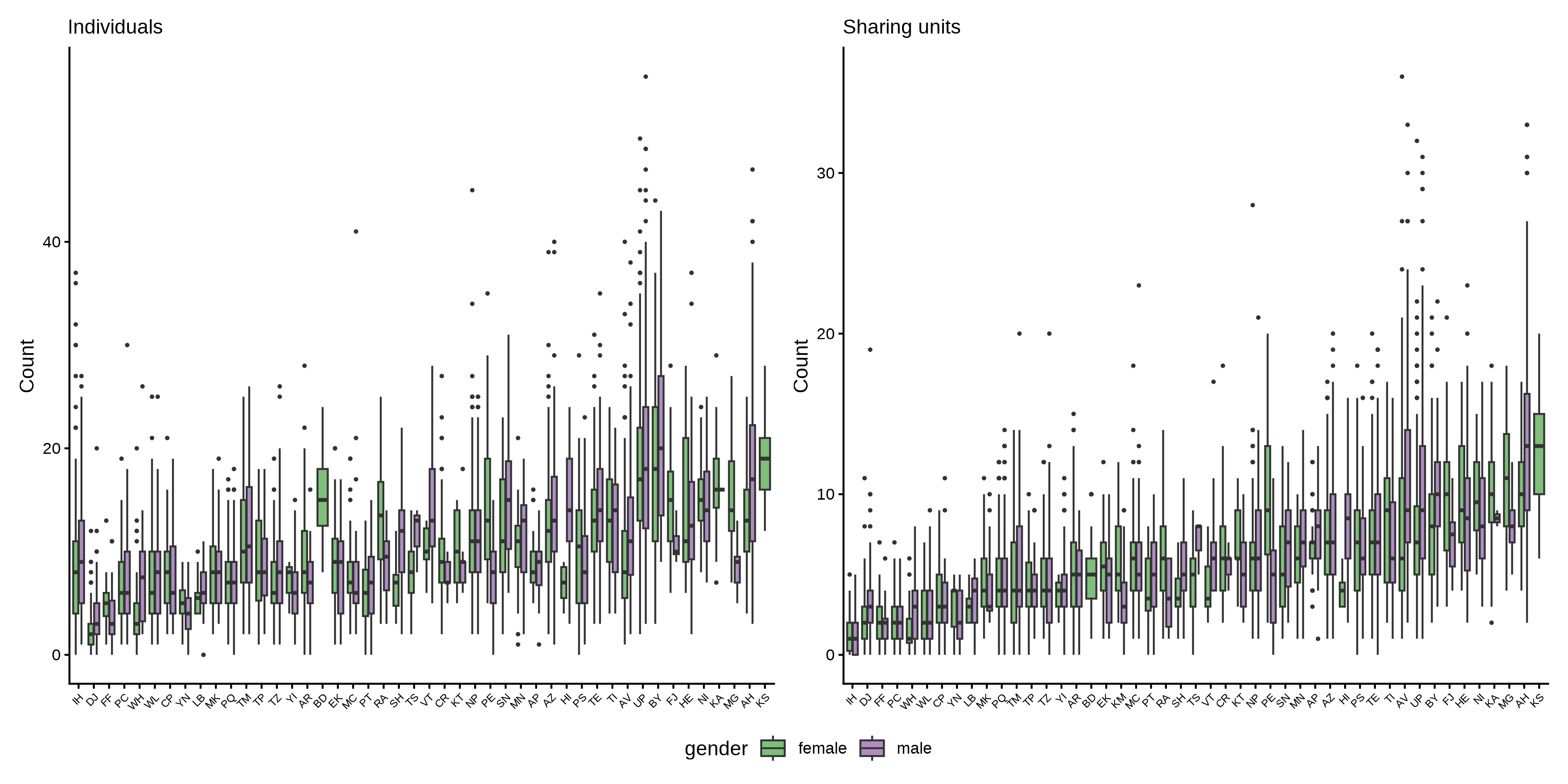}
\end{center}
\caption{\textbf{Distribution of the number of nominations made by respondents in each community jointly across the questions contributing to the aggregate network}, showing (left) the number of \emph{individuals} and (right) the number of \emph{sharing units} named for each network prompt, for all respondents (top) and broken out by gender (bottom).}
\label{fig:nomination_counts_boxplot_agg}
\end{figure}

In Figure~\ref{fig:nomination_counts_boxplot_agg} we again show the number of alters named, now including \emph{all} alters named in those network prompts that contribute to the aggregate network (``borrow money'', ``borrow goods'', ``behavioural support'' for men and women, and ``informational/affective support'' for men and women).
We can again observe substantial variation in the number of alters named.
Overall, respondents cumulatively named around $9$ or $10$ individuals across this set of network prompts, but this varies widely, from a median of $3$ in DJ (in Benin) to $18$ in BY (in Congo-Brazzaville). The most individuals named was 56, by a man in his late forties in UP in India; the most sharing units named was 36, by a woman in her seventies in AV in Benin.

When analysing nomination patterns by gender (in the lower panel of Figure~\ref{fig:nomination_counts_boxplot_agg}), there is little evidence of consistent differences; in some communities men named more alters and in some communities women, but in most settings the differences are very slight. Some of the differences (across communities and by gender) will be due to differences in the sets of network prompts posed: recall that in a number of communities the gender-specific questions were posed only to respondents of that gender, so while the overall number of prompts for each respondent would have been the same, the particular prompts posed would vary based on the gender of the respondent.
As a point of comparison for our findings, Bernard et. al. \cite{bernard_comparing_1990} compared the overall number of alters named across their different elicitation techniques, and found there to be no significant difference by gender in Jacksonville, and for women to nominate more overall alters in Mexico City.

\begin{figure}[h]
\begin{center}
\includegraphics[width=\textwidth]{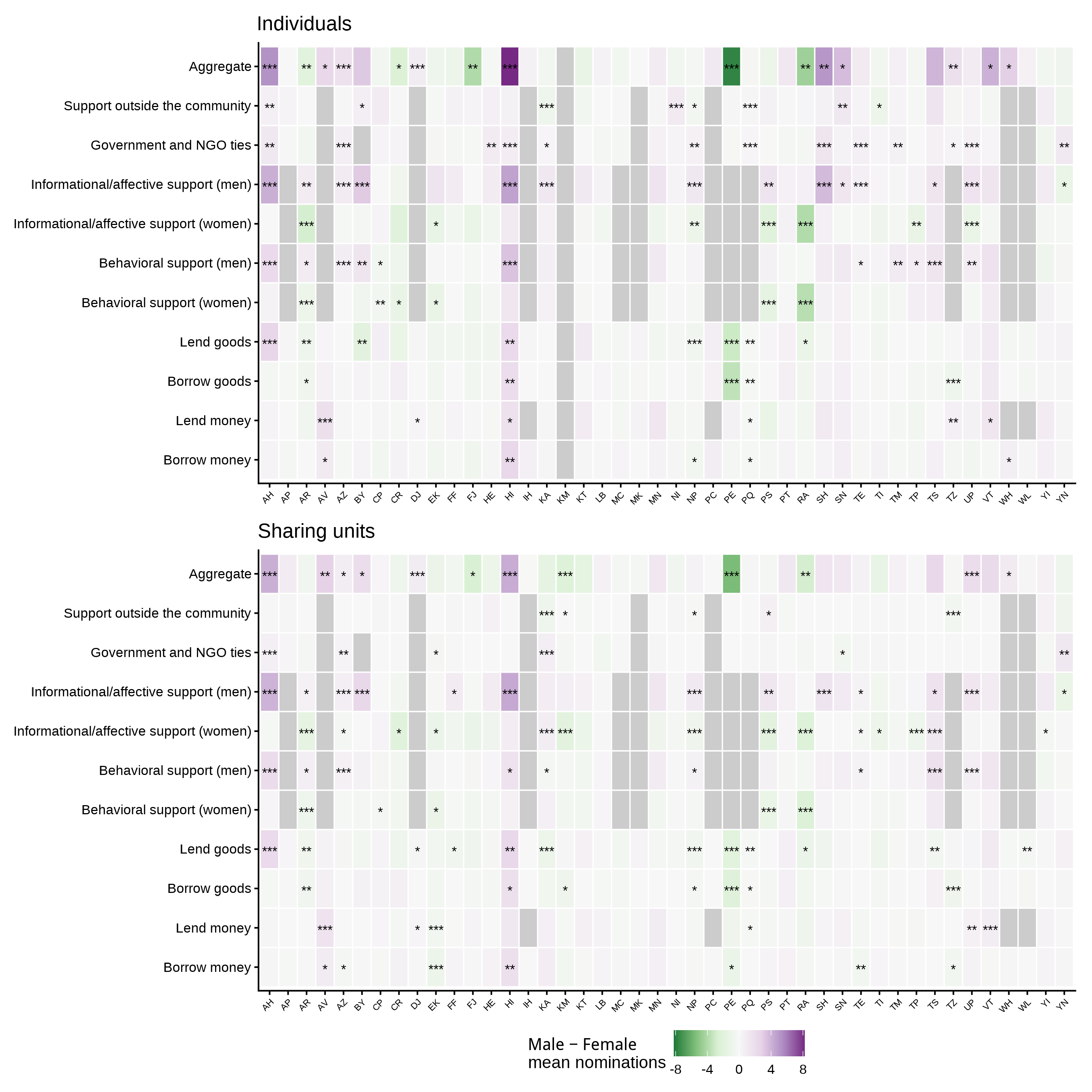}
\end{center}
\caption{\textbf{Differences in the mean number of alters named by gender, in each community, considering (top) all individuals mentioned and (bottom) all sharing units mentioned.} Cells are coloured by the difference in mean nominations between men and women; greater mean number of nominations by men would show as darker purple and greater mean number of nominations by women would show as darker green. The asterisks represent $p$-values of a simple $t$-test comparing the mean difference ($p < 0.05 / 0.01 / 0.001$). Grey cells represent cases where no comparison is possible, either because the question was not posed, or because it was posed only to members of one gender. }
\label{fig:nomination_gender}
\end{figure}

In Figure~\ref{fig:nomination_gender}, we more rigorously compare the alters named (either individuals or sharing units) by men and women in each community (where there were respondents of each gender across network prompts). Specifically, we compute simple $t$-tests of the differences in the number of alters named by gender, for each network prompt and community. Note that the number of respondents of each gender can be quite few in some cases (given a different intended set of respondents, see Table~\ref{tab:respondents_by_site}). 
AV, DJ, IH, MC, PE, TZ, WH, WL asked the gender-specific questions only of people of that gender.
AV, DJ, IH, MK, PC, WH, and WL did not ask the questions on government/NGO ties or support outside the community.
MK asked questions regarding behavioural support and informational/affective support, but did not specify the gender.
MG asked questions about a spouse's connections only to married women, meaning that men and women (and married versus unmarried women) were effectively asked a distinct number and set of questions, so this community is excluded from this figure.
The alters named for KM were recorded directly as sharing units for community nominations, so it is only shown for the sharing unit version.

Overall, it is evident that the differences in named alters are relatively slight, with some suggestion that men may have a slight tendency to name more alters across communities, but substantial variation. The clearest patterns are for the gender-specific questions: men often named more alters for the men-specific questions, and women often more for the women-specific questions, perhaps reflecting their greater direct knowledge or more ready and diligent reflection for those questions.

\begin{figure}[h]
\begin{center}
\includegraphics[width=\textwidth]{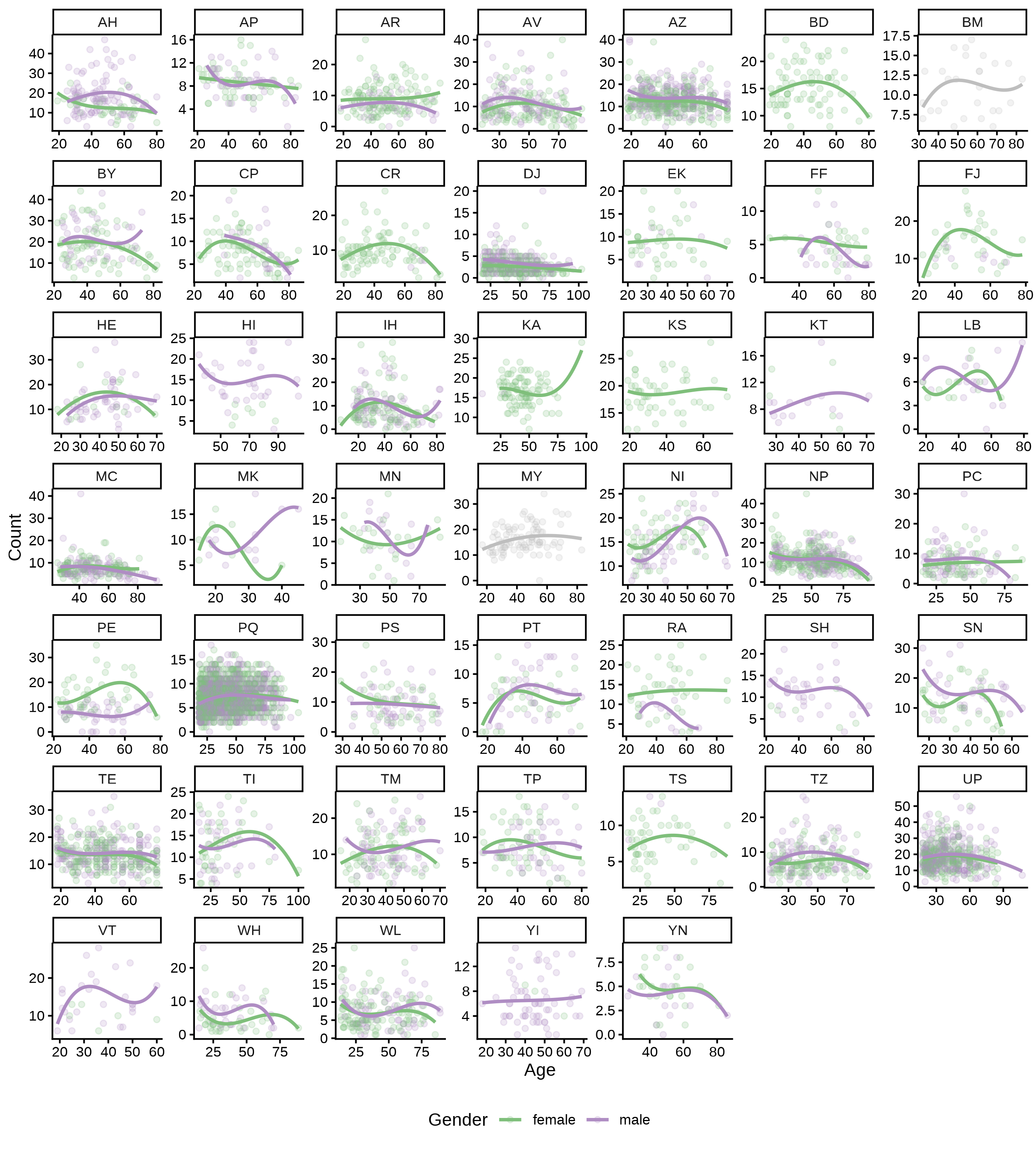}
\end{center}
\caption{\textbf{Distribution of the number of alters (individuals) named by respondents in each community for the questions contributing to the aggregate network, by age and gender.} Curves show cubic polynomial fits to the data, and are plotted only when there are at least 10 respondents of a gender. Communities BM and MY are shown in grey because couples were asked the question jointly; the relevant ages used in those plots are those of the female head of household. Note that axes vary across plots.}
\label{fig:nomination_counts_scatter}
\end{figure}

The effects of gender may be moderated by age. Figure~\ref{fig:nomination_counts_scatter} shows scatterplots (with fitted cubic polynomial curves) of the total number of alters (individuals) named by respondents across the questions contributing to the aggregate network. (Communities KO, MA, and MG are not included, as men and women were asked different questions, making their comparison difficult; KM is not included, as most alters named were recorded directly as sharing units, not individuals). The curves should be interpreted with some caution, especially when the number of respondents is relatively few (curves are only shown if there are at least ten respondents of that gender). They should also be interpreted with caution in those communities when the gender-specific questions were only asked of people of that gender (AP, AV, DJ, IH, MC, PC, PE, TZ, WH, and WL); while this should mean that respondents still answered the same overall number of questions, the set of questions would therefore have differed.

We can first note consistency with the prior plots in terms of the relative numbers of alters named by men and women (so, for example, the earlier figures show that men named more alters in AH, and here we can see that this holds consistently across age). In some communities (\textit{e.g.}, EK, KS, MY, YI) there is little suggestion of change in the number of alters across the age course. In others, there is some suggestion of the number of alters building into middle-age (though, where in middle age varies across communities) before declining again (\textit{e.g.}, BD, CR, FJ, NI, PT, TS), though the age at which alters peak varies across communities (from around age forty in FJ and PT to the mid-fifties in BD and CR and even later in NI). In some communities, men and women follow quite similar age trends (\textit{e.g.}, AZ, HE, NP, PT, TE, UP), whereas in others they seem to diverge (\textit{e.g.}, AH, MG),though in some cases this is surely due to relatively small numbers (as with MK or LB).
Studies in the US \cite{cornwell_social_2008} and South Africa \cite{harling_age_2020} have found that older adults have fewer social connections, but the gendering of these patterns differ; in South Africa, women showed a more substantial decrease with age. An older comparative study \cite{hollinger_kinship_1990} again found varied patterns in the number of friends reported across the age course across seven countries. Across the diversity of the ENDOW communities, there is a possible suggestion of lessening support with age, but plenty of heterogeneity overall and generally a small suggestion of any strong differences. Simple regression models predicting the number of nominations with covariates for age and age squared, for example, show only a very small handful with a significant correlation for either term.

In the main text (and particularly in Figure~\ref{fig:internal_external_violin}), we discuss the balance of internal versus external alters named. A few existing datasets similarly report the proportion of nominations made to those beyond the focal community. The classic ``East York'' study \cite{wellman_community_1979} reports about a quarter of named intimates as residing either in the same neighbourhood or elsewhere in East York, Toronto, Canada. 
Data on individual-level aggregate networks covering a range of social support domains in 32 communities in Kenya and Uganda \cite{chen_social_2021} showed the percent of non-community residents ranged from $8.9\%$ to $65.9\%$ of all nominations, with an average of $27.9\%$, again quite comparable to what we observe. 
A closer corollary specifically for the behavioural support questions is data on help with the rice harvest in 51 villages in Thailand \cite{entwisle_networks_2007}: they report that, on average, $46.7\%$ of household-level ties (SD $17.3\%$) extend beyond the village, quite a bit higher than what we observe.
As a point of comparison for the nominations of other sharing unit members, the network data from Kenya and Uganda shows an average of $21.2\%$ of ties being to other household members \cite{chen_social_2021}, substantially higher than what we observe in effectively all ENDOW communities. 

\begingroup\fontsize{8}{10}\selectfont

\begin{longtable}[t]{lrlllll}
\caption{\label{tab:nomination-summary-loan_req}Mean (and standard deviation) values for nominations made across all respondents, for the Borrow money name generator. `N Resp.' is the number of respondents. `Prop. Zero' is the proportion of respondents nominating no alters. `N Indiv' is the number of individuals nominated. `N SU' is the number of sharing units. `Prop. Same SU' is the proportion of individuals nominated who are members of the same sharing unit as the respondent. `Prop. Ext' is the proportion of individuals nominated who are external to the community.}\\
\toprule
Site & N Resp. & Prop. Zero & N Indiv. & N SU & Prop. Same SU & Prop. Ext.\\
\midrule
\endfirsthead
\caption[]{Mean (and standard deviation) values for nominations made across all respondents, for the Borrow money name generator. `N Resp.' is the number of respondents. `Prop. Zero' is the proportion of respondents nominating no alters. `N Indiv' is the number of individuals nominated. `N SU' is the number of sharing units. `Prop. Same SU' is the proportion of individuals nominated who are members of the same sharing unit as the respondent. `Prop. Ext' is the proportion of individuals nominated who are external to the community. \textit{(continued)}}\\
\toprule
Site & N Resp. & Prop. Zero & N Indiv. & N SU & Prop. Same SU & Prop. Ext.\\
\midrule
\endhead

\endfoot
\bottomrule
\endlastfoot
AH & 138 & 0.59 & 0.81 (1.17) & 0.26 (0.58) & 0.00 (0.00) & 0.64 (0.44)\\
AP & 89 & 0.01 & 2.74 (1.08) & 2.58 (1.04) & 0.00 (0.00) & 0.03 (0.10)\\
AR & 158 & 0.13 & 1.99 (1.34) & 1.15 (1.12) & 0.01 (0.06) & 0.37 (0.41)\\
AV & 203 & 0.06 & 3.83 (2.84) & 3.48 (2.57) & 0.25 (0.29) & 0.00 (0.00)\\
AZ & 441 & 0.02 & 3.78 (1.57) & 2.73 (1.47) & 0.17 (0.28) & 0.15 (0.25)\\
BD & 79 & 0.00 & 4.05 (1.87) & 1.27 (1.01) & 0.11 (0.24) & 0.60 (0.32)\\
BM & 59 & 0.00 & 2.69 (1.48) & 1.69 (1.21) & 0.00 (0.00) & 0.36 (0.37)\\
BY & 112 & 0.14 & 2.70 (2.46) & 0.12 (0.57) & 0.02 (0.14) & 0.94 (0.22)\\
CP & 100 & 0.10 & 2.39 (1.81) & 0.88 (0.90) & 0.08 (0.18) & 0.54 (0.42)\\
CR & 93 & 0.09 & 1.66 (1.17) & 1.17 (0.94) & 0.09 (0.26) & 0.22 (0.35)\\
DJ & 349 & 0.36 & 0.88 (0.81) & 0.84 (0.76) & 0.25 (0.41) & 0.00 (0.00)\\
EK & 49 & 0.12 & 1.53 (1.21) & 1.00 (1.04) & 0.02 (0.15) & 0.29 (0.40)\\
FF & 44 & 0.39 & 1.48 (1.61) & 0.30 (0.55) & 0.00 (0.00) & 0.77 (0.36)\\
FJ & 42 & 0.00 & 2.79 (1.52) & 1.81 (1.44) & 0.04 (0.12) & 0.31 (0.37)\\
HE & 63 & 0.03 & 4.05 (3.17) & 3.65 (3.01) & 0.00 (0.00) & 0.10 (0.27)\\
HI & 39 & 0.21 & 2.82 (2.56) & 2.05 (1.81) & 0.16 (0.27) & 0.15 (0.29)\\
IH & 153 & 0.22 & 2.20 (2.28) & 0.60 (0.81) & 0.09 (0.22) & 0.67 (0.37)\\
KA & 100 & 0.00 & 3.88 (0.88) & 2.74 (1.39) & 0.01 (0.07) & 0.29 (0.32)\\
KM & 72 & 0.26 & --- & 1.60 (1.33) & --- & 0.02 (0.09)\\
KO & 67 & 0.00 & 3.24 (1.18) & 2.28 (1.20) & 0.01 (0.04) & 0.28 (0.31)\\
KS & 53 & 0.00 & 4.04 (0.44) & 3.32 (1.01) & 0.02 (0.06) & 0.17 (0.23)\\
KT & 18 & 0.06 & 1.89 (1.45) & 1.11 (0.76) & 0.00 (0.00) & 0.21 (0.35)\\
LB & 30 & 0.47 & 0.73 (0.83) & 0.50 (0.73) & 0.21 (0.40) & 0.28 (0.45)\\
MA & 42 & 0.00 & 4.31 (1.16) & 2.60 (1.47) & 0.00 (0.00) & 0.38 (0.34)\\
MC & 189 & 0.02 & 2.21 (2.63) & 1.76 (1.64) & 0.00 (0.04) & 0.17 (0.31)\\
MG & 101 & 0.15 & 4.21 (2.13) & 3.21 (1.76) & 0.01 (0.05) & 0.03 (0.10)\\
MK & 87 & 0.03 & 3.21 (1.86) & 1.37 (1.37) & 0.01 (0.05) & 0.55 (0.38)\\
MN & 38 & 0.16 & 2.82 (2.02) & 1.58 (1.50) & 0.01 (0.06) & 0.47 (0.37)\\
MY & 177 & 0.51 & 1.67 (2.09) & 0.92 (1.37) & 0.03 (0.14) & 0.41 (0.36)\\
NI & 98 & 0.12 & 1.43 (0.93) & 1.29 (0.84) & 0.09 (0.26) & 0.05 (0.19)\\
NP & 326 & 0.10 & 2.96 (2.20) & 1.61 (1.61) & 0.09 (0.23) & 0.39 (0.37)\\
PC & 125 & 0.23 & 2.14 (3.10) & 0.71 (0.92) & 0.06 (0.20) & 0.55 (0.42)\\
PE & 85 & 0.15 & 4.58 (3.39) & 3.08 (2.34) & 0.00 (0.03) & 0.26 (0.32)\\
PQ & 1533 & 0.03 & 2.91 (1.32) & 2.04 (1.25) & 0.18 (0.27) & 0.23 (0.30)\\
PS & 91 & 0.10 & 2.63 (2.26) & 1.64 (1.50) & 0.11 (0.24) & 0.22 (0.34)\\
PT & 68 & 0.19 & 1.85 (1.45) & 1.65 (1.22) & 0.03 (0.10) & 0.01 (0.07)\\
RA & 54 & 0.02 & 3.37 (1.56) & 2.13 (1.37) & 0.00 (0.00) & 0.33 (0.34)\\
SH & 46 & 0.04 & 2.59 (1.67) & 1.02 (1.06) & 0.05 (0.12) & 0.53 (0.43)\\
SN & 63 & 0.14 & 2.75 (2.24) & 2.13 (1.77) & 0.00 (0.02) & 0.20 (0.34)\\
TE & 346 & 0.04 & 3.44 (1.48) & 2.38 (1.48) & 0.18 (0.31) & 0.19 (0.30)\\
TI & 74 & 0.72 & 0.53 (1.00) & 0.19 (0.57) & 0.05 (0.22) & 0.69 (0.39)\\
TM & 129 & 0.03 & 2.60 (1.56) & 1.43 (1.21) & 0.03 (0.15) & 0.42 (0.39)\\
TP & 94 & 0.31 & 1.49 (1.50) & 0.96 (1.13) & 0.29 (0.43) & 0.14 (0.32)\\
TS & 50 & 0.72 & 0.40 (0.70) & 0.06 (0.24) & 0.00 (0.00) & 0.82 (0.37)\\
TZ & 192 & 0.13 & 1.91 (1.65) & 1.15 (1.26) & 0.06 (0.21) & 0.37 (0.44)\\
UP & 470 & 0.01 & 5.48 (3.25) & 2.68 (2.18) & 0.18 (0.28) & 0.37 (0.30)\\
VT & 34 & 0.38 & 1.18 (1.40) & 0.24 (0.55) & 0.02 (0.11) & 0.79 (0.37)\\
WH & 66 & 0.30 & 1.47 (1.45) & 0.64 (0.83) & 0.16 (0.34) & 0.48 (0.44)\\
WL & 178 & 0.15 & 2.08 (1.88) & 0.62 (0.74) & 0.05 (0.18) & 0.65 (0.38)\\
YI & 69 & 0.42 & 1.25 (1.41) & 0.97 (1.26) & 0.03 (0.16) & 0.22 (0.36)\\
YN & 39 & 0.62 & 0.69 (1.08) & 0.26 (0.55) & 0.00 (0.00) & 0.61 (0.43)\\*
\end{longtable}
\endgroup{}

\begingroup\fontsize{8}{10}\selectfont

\begin{longtable}[t]{lrlllll}
\caption{\label{tab:nomination-summary-loan_giv}Mean (and standard deviation) values for nominations made across all respondents, for the Lend money name generator. `N Resp.' is the number of respondents. `Prop. Zero' is the proportion of respondents nominating no alters. `N Indiv' is the number of individuals nominated. `N SU' is the number of sharing units. `Prop. Same SU' is the proportion of individuals nominated who are members of the same sharing unit as the respondent. `Prop. Ext' is the proportion of individuals nominated who are external to the community.}\\
\toprule
Site & N Resp. & Prop. Zero & N Indiv. & N SU & Prop. Same SU & Prop. Ext.\\
\midrule
\endfirsthead
\caption[]{Mean (and standard deviation) values for nominations made across all respondents, for the Lend money name generator. `N Resp.' is the number of respondents. `Prop. Zero' is the proportion of respondents nominating no alters. `N Indiv' is the number of individuals nominated. `N SU' is the number of sharing units. `Prop. Same SU' is the proportion of individuals nominated who are members of the same sharing unit as the respondent. `Prop. Ext' is the proportion of individuals nominated who are external to the community. \textit{(continued)}}\\
\toprule
Site & N Resp. & Prop. Zero & N Indiv. & N SU & Prop. Same SU & Prop. Ext.\\
\midrule
\endhead

\endfoot
\bottomrule
\endlastfoot
AH & 138 & 0.77 & 0.54 (1.17) & 0.38 (0.89) & 0.03 (0.10) & 0.21 (0.36)\\
AP & 89 & 0.03 & 2.85 (1.05) & 2.62 (1.12) & 0.00 (0.00) & 0.08 (0.17)\\
AR & 158 & 0.11 & 2.75 (2.09) & 1.56 (1.38) & 0.01 (0.06) & 0.32 (0.36)\\
AV & 203 & 0.07 & 4.49 (3.45) & 3.90 (2.97) & 0.22 (0.28) & 0.00 (0.00)\\
AZ & 441 & 0.08 & 3.33 (1.63) & 2.59 (1.45) & 0.11 (0.23) & 0.10 (0.21)\\
BD & 79 & 0.06 & 2.48 (1.30) & 1.51 (1.02) & 0.03 (0.13) & 0.36 (0.33)\\
BM & 59 & 0.41 & 1.64 (1.88) & 1.44 (1.77) & 0.00 (0.00) & 0.15 (0.33)\\
BY & 112 & 0.47 & 1.98 (2.63) & 1.43 (1.99) & 0.13 (0.24) & 0.15 (0.27)\\
CP & 100 & 0.30 & 2.22 (2.32) & 0.85 (1.19) & 0.07 (0.16) & 0.55 (0.38)\\
CR & 93 & 0.10 & 1.98 (1.50) & 1.52 (1.15) & 0.12 (0.29) & 0.09 (0.24)\\
DJ & 349 & 0.50 & 0.80 (0.99) & 0.77 (0.92) & 0.16 (0.33) & 0.00 (0.00)\\
EK & 49 & 0.57 & 0.63 (0.86) & 0.45 (0.77) & 0.00 (0.00) & 0.31 (0.46)\\
FF & 44 & 0.68 & 0.61 (1.15) & 0.27 (0.54) & 0.05 (0.12) & 0.40 (0.45)\\
FJ & 42 & 0.07 & 2.31 (1.32) & 1.98 (1.35) & 0.05 (0.14) & 0.13 (0.30)\\
HE & 63 & 0.03 & 3.52 (3.08) & 3.17 (2.87) & 0.00 (0.00) & 0.04 (0.15)\\
HI & 39 & 0.23 & 3.13 (2.58) & 2.38 (1.87) & 0.12 (0.20) & 0.09 (0.17)\\
KA & 100 & 0.01 & 3.87 (1.01) & 2.90 (1.37) & 0.02 (0.08) & 0.24 (0.31)\\
KM & 72 & 0.31 & --- & 1.24 (1.26) & --- & 0.11 (0.29)\\
KO & 67 & 0.00 & 3.15 (1.25) & 2.27 (1.39) & 0.01 (0.06) & 0.28 (0.33)\\
KS & 53 & 0.00 & 4.38 (0.60) & 3.81 (1.23) & 0.01 (0.05) & 0.09 (0.19)\\
KT & 18 & 0.67 & 0.83 (1.42) & 0.67 (1.03) & 0.00 (0.00) & 0.07 (0.16)\\
LB & 30 & 0.27 & 2.43 (1.89) & 1.40 (1.28) & 0.02 (0.07) & 0.22 (0.32)\\
MA & 42 & 0.00 & 4.57 (1.77) & 2.50 (1.77) & 0.00 (0.00) & 0.44 (0.38)\\
MC & 189 & 0.06 & 2.07 (1.47) & 1.61 (1.38) & 0.01 (0.09) & 0.20 (0.31)\\
MG & 101 & 0.15 & 4.24 (2.01) & 3.19 (1.68) & 0.01 (0.07) & 0.02 (0.09)\\
MK & 87 & 0.06 & 3.70 (2.33) & 1.84 (1.73) & 0.01 (0.06) & 0.48 (0.35)\\
MN & 38 & 0.18 & 3.53 (2.54) & 2.29 (1.93) & 0.00 (0.00) & 0.32 (0.34)\\
MY & 177 & 0.55 & 1.44 (2.01) & 1.20 (1.68) & 0.02 (0.10) & 0.06 (0.19)\\
NI & 98 & 0.01 & 2.88 (0.96) & 2.56 (1.08) & 0.11 (0.25) & 0.04 (0.13)\\
NP & 326 & 0.26 & 2.07 (1.85) & 1.42 (1.50) & 0.04 (0.16) & 0.28 (0.35)\\
PE & 85 & 0.16 & 4.58 (3.39) & 3.16 (2.52) & 0.02 (0.12) & 0.22 (0.32)\\
PQ & 1533 & 0.03 & 3.10 (1.45) & 2.47 (1.37) & 0.21 (0.29) & 0.11 (0.22)\\
PS & 91 & 0.34 & 2.66 (3.49) & 1.84 (2.13) & 0.07 (0.16) & 0.11 (0.23)\\
PT & 68 & 0.28 & 1.69 (1.41) & 1.56 (1.34) & 0.04 (0.12) & 0.01 (0.07)\\
RA & 54 & 0.22 & 2.76 (1.89) & 2.09 (1.72) & 0.00 (0.00) & 0.20 (0.30)\\
SH & 46 & 0.11 & 2.24 (1.46) & 1.13 (1.05) & 0.05 (0.13) & 0.40 (0.42)\\
SN & 63 & 0.29 & 2.62 (2.24) & 1.97 (1.95) & 0.01 (0.04) & 0.21 (0.30)\\
TE & 346 & 0.12 & 2.85 (1.64) & 2.18 (1.52) & 0.11 (0.25) & 0.15 (0.30)\\
TI & 74 & 0.84 & 0.23 (0.59) & 0.16 (0.52) & 0.00 (0.00) & 0.33 (0.49)\\
TM & 129 & 0.25 & 2.33 (2.05) & 1.26 (1.52) & 0.00 (0.00) & 0.44 (0.36)\\
TP & 94 & 0.22 & 2.56 (2.27) & 1.26 (1.28) & 0.38 (0.41) & 0.23 (0.37)\\
TS & 50 & 0.92 & 0.10 (0.36) & 0.06 (0.24) & 0.00 (0.00) & 0.38 (0.48)\\
TZ & 192 & 0.62 & 0.84 (1.42) & 0.41 (0.93) & 0.10 (0.30) & 0.46 (0.47)\\
UP & 470 & 0.11 & 4.16 (2.94) & 2.37 (2.15) & 0.14 (0.27) & 0.29 (0.34)\\
VT & 34 & 0.41 & 1.41 (1.54) & 0.65 (0.81) & 0.06 (0.23) & 0.47 (0.40)\\
YI & 69 & 0.38 & 1.48 (1.71) & 1.07 (1.32) & 0.00 (0.00) & 0.21 (0.36)\\
YN & 39 & 0.85 & 0.36 (0.99) & 0.15 (0.43) & 0.08 (0.20) & 0.52 (0.32)\\*
\end{longtable}
\endgroup{}

\begingroup\fontsize{8}{10}\selectfont

\begin{longtable}[t]{lrlllll}
\caption{\label{tab:nomination-summary-exchange_req}Mean (and standard deviation) values for nominations made across all respondents, for the Borrow goods name generator. `N Resp.' is the number of respondents. `Prop. Zero' is the proportion of respondents nominating no alters. `N Indiv' is the number of individuals nominated. `N SU' is the number of sharing units. `Prop. Same SU' is the proportion of individuals nominated who are members of the same sharing unit as the respondent. `Prop. Ext' is the proportion of individuals nominated who are external to the community.}\\
\toprule
Site & N Resp. & Prop. Zero & N Indiv. & N SU & Prop. Same SU & Prop. Ext.\\
\midrule
\endfirsthead
\caption[]{Mean (and standard deviation) values for nominations made across all respondents, for the Borrow goods name generator. `N Resp.' is the number of respondents. `Prop. Zero' is the proportion of respondents nominating no alters. `N Indiv' is the number of individuals nominated. `N SU' is the number of sharing units. `Prop. Same SU' is the proportion of individuals nominated who are members of the same sharing unit as the respondent. `Prop. Ext' is the proportion of individuals nominated who are external to the community. \textit{(continued)}}\\
\toprule
Site & N Resp. & Prop. Zero & N Indiv. & N SU & Prop. Same SU & Prop. Ext.\\
\midrule
\endhead

\endfoot
\bottomrule
\endlastfoot
AH & 138 & 0.06 & 4.18 (2.61) & 3.91 (2.44) & 0.02 (0.11) & 0.01 (0.06)\\
AP & 89 & 0.02 & 3.63 (1.50) & 3.34 (1.50) & 0.00 (0.00) & 0.08 (0.14)\\
AR & 158 & 0.15 & 1.82 (1.48) & 1.57 (1.19) & 0.00 (0.02) & 0.06 (0.18)\\
AV & 203 & 0.05 & 3.33 (2.55) & 2.87 (2.24) & 0.33 (0.38) & 0.00 (0.00)\\
AZ & 441 & 0.05 & 3.30 (1.54) & 2.89 (1.46) & 0.04 (0.14) & 0.04 (0.14)\\
BD & 79 & 0.01 & 2.52 (1.18) & 1.80 (1.04) & 0.02 (0.12) & 0.25 (0.29)\\
BM & 59 & 0.25 & 1.19 (0.86) & 1.05 (0.80) & 0.00 (0.00) & 0.05 (0.21)\\
BY & 112 & 0.04 & 5.14 (2.86) & 3.43 (2.07) & 0.15 (0.23) & 0.08 (0.18)\\
CP & 100 & 0.08 & 2.60 (1.82) & 1.60 (1.58) & 0.03 (0.11) & 0.34 (0.38)\\
CR & 93 & 0.06 & 2.54 (1.86) & 2.38 (1.70) & 0.05 (0.19) & 0.03 (0.10)\\
DJ & 349 & 0.68 & 0.46 (0.79) & 0.42 (0.72) & 0.29 (0.43) & 0.00 (0.00)\\
EK & 49 & 0.18 & 1.39 (1.08) & 1.35 (1.03) & 0.00 (0.00) & 0.00 (0.00)\\
FF & 44 & 0.23 & 1.50 (1.36) & 1.11 (1.02) & 0.00 (0.00) & 0.23 (0.35)\\
FJ & 42 & 0.26 & 2.48 (2.19) & 2.02 (2.11) & 0.01 (0.07) & 0.19 (0.35)\\
HE & 63 & 0.03 & 3.38 (2.20) & 3.06 (1.99) & 0.00 (0.00) & 0.04 (0.17)\\
HI & 39 & 0.03 & 3.54 (2.21) & 3.03 (2.13) & 0.00 (0.00) & 0.16 (0.29)\\
IH & 153 & 0.49 & 0.99 (1.42) & 0.20 (0.46) & 0.10 (0.28) & 0.76 (0.37)\\
KA & 100 & 0.00 & 3.97 (0.73) & 3.07 (1.17) & 0.00 (0.02) & 0.21 (0.27)\\
KM & 72 & 0.21 & --- & 1.53 (1.33) & --- & 0.08 (0.20)\\
KO & 65 & 0.02 & 3.43 (1.29) & 2.71 (1.25) & 0.02 (0.08) & 0.20 (0.26)\\
KS & 53 & 0.00 & 4.55 (1.14) & 4.38 (1.26) & 0.00 (0.03) & 0.03 (0.09)\\
KT & 18 & 0.06 & 3.06 (1.63) & 2.39 (1.24) & 0.01 (0.05) & 0.04 (0.10)\\
LB & 30 & 0.03 & 3.20 (1.67) & 2.30 (1.24) & 0.03 (0.19) & 0.16 (0.27)\\
MA & 41 & 0.02 & 4.29 (1.50) & 3.12 (1.52) & 0.02 (0.07) & 0.23 (0.30)\\
MC & 189 & 0.01 & 3.52 (2.08) & 3.06 (1.92) & 0.00 (0.02) & 0.11 (0.20)\\
MG & 101 & 0.00 & 2.94 (1.06) & 2.91 (1.09) & 0.00 (0.00) & 0.01 (0.08)\\
MK & 87 & 0.07 & 3.87 (2.34) & 2.24 (1.65) & 0.01 (0.09) & 0.31 (0.33)\\
MN & 38 & 0.37 & 1.55 (1.57) & 1.21 (1.40) & 0.00 (0.00) & 0.20 (0.35)\\
MY & 177 & 0.53 & 2.10 (2.79) & 1.72 (2.23) & 0.01 (0.11) & 0.01 (0.06)\\
NI & 98 & 0.05 & 3.10 (1.56) & 3.05 (1.56) & 0.01 (0.05) & 0.01 (0.05)\\
NP & 326 & 0.09 & 2.43 (1.58) & 2.06 (1.42) & 0.01 (0.09) & 0.11 (0.24)\\
PC & 125 & 0.38 & 1.86 (3.18) & 0.59 (1.00) & 0.05 (0.16) & 0.60 (0.41)\\
PE & 85 & 0.08 & 5.18 (3.46) & 3.74 (2.79) & 0.01 (0.04) & 0.24 (0.28)\\
PQ & 1533 & 0.53 & 0.87 (1.20) & 0.61 (1.01) & 0.05 (0.19) & 0.31 (0.41)\\
PS & 91 & 0.26 & 2.22 (2.23) & 2.04 (2.02) & 0.01 (0.07) & 0.03 (0.14)\\
PT & 68 & 0.26 & 1.66 (1.48) & 1.51 (1.42) & 0.04 (0.16) & 0.00 (0.00)\\
RA & 54 & 0.04 & 3.19 (1.33) & 2.54 (1.25) & 0.01 (0.06) & 0.17 (0.26)\\
SH & 46 & 0.09 & 1.80 (1.17) & 1.37 (1.04) & 0.02 (0.07) & 0.18 (0.32)\\
SN & 63 & 0.06 & 4.19 (2.22) & 2.67 (1.92) & 0.00 (0.03) & 0.35 (0.32)\\
TE & 346 & 0.07 & 2.89 (1.41) & 2.58 (1.39) & 0.04 (0.15) & 0.04 (0.14)\\
TI & 74 & 0.14 & 2.41 (1.72) & 2.09 (1.66) & 0.03 (0.14) & 0.10 (0.25)\\
TM & 129 & 0.16 & 2.79 (1.77) & 1.70 (1.58) & 0.00 (0.03) & 0.37 (0.38)\\
TP & 94 & 0.19 & 2.35 (2.13) & 1.33 (1.39) & 0.20 (0.36) & 0.24 (0.39)\\
TS & 50 & 0.04 & 2.64 (1.26) & 1.82 (0.96) & 0.03 (0.15) & 0.20 (0.25)\\
TZ & 192 & 0.30 & 1.48 (1.37) & 1.35 (1.29) & 0.02 (0.12) & 0.05 (0.15)\\
UP & 470 & 0.08 & 3.66 (2.41) & 3.01 (1.91) & 0.06 (0.21) & 0.03 (0.11)\\
VT & 34 & 0.03 & 3.38 (1.92) & 2.00 (1.23) & 0.00 (0.00) & 0.30 (0.34)\\
WH & 66 & 0.65 & 0.52 (0.88) & 0.21 (0.48) & 0.13 (0.34) & 0.58 (0.45)\\
WL & 178 & 0.48 & 1.17 (1.50) & 0.38 (0.79) & 0.02 (0.09) & 0.66 (0.41)\\
YI & 69 & 0.23 & 2.14 (1.74) & 2.01 (1.70) & 0.00 (0.00) & 0.05 (0.17)\\
YN & 39 & 0.59 & 0.62 (0.94) & 0.44 (0.88) & 0.00 (0.00) & 0.34 (0.47)\\*
\end{longtable}
\endgroup{}

\begingroup\fontsize{8}{10}\selectfont

\begin{longtable}[t]{lrlllll}
\caption{\label{tab:nomination-summary-exchange_giv}Mean (and standard deviation) values for nominations made across all respondents, for the Lend goods name generator. `N Resp.' is the number of respondents. `Prop. Zero' is the proportion of respondents nominating no alters. `N Indiv' is the number of individuals nominated. `N SU' is the number of sharing units. `Prop. Same SU' is the proportion of individuals nominated who are members of the same sharing unit as the respondent. `Prop. Ext' is the proportion of individuals nominated who are external to the community.}\\
\toprule
Site & N Resp. & Prop. Zero & N Indiv. & N SU & Prop. Same SU & Prop. Ext.\\
\midrule
\endfirsthead
\caption[]{Mean (and standard deviation) values for nominations made across all respondents, for the Lend goods name generator. `N Resp.' is the number of respondents. `Prop. Zero' is the proportion of respondents nominating no alters. `N Indiv' is the number of individuals nominated. `N SU' is the number of sharing units. `Prop. Same SU' is the proportion of individuals nominated who are members of the same sharing unit as the respondent. `Prop. Ext' is the proportion of individuals nominated who are external to the community. \textit{(continued)}}\\
\toprule
Site & N Resp. & Prop. Zero & N Indiv. & N SU & Prop. Same SU & Prop. Ext.\\
\midrule
\endhead

\endfoot
\bottomrule
\endlastfoot
AH & 138 & 0.15 & 4.54 (3.09) & 4.25 (2.93) & 0.01 (0.07) & 0.03 (0.12)\\
AP & 89 & 0.01 & 3.83 (1.25) & 3.51 (1.31) & 0.00 (0.00) & 0.08 (0.14)\\
AR & 158 & 0.12 & 2.11 (1.65) & 1.75 (1.33) & 0.00 (0.03) & 0.10 (0.25)\\
AV & 203 & 0.14 & 3.13 (2.89) & 2.72 (2.51) & 0.30 (0.36) & 0.00 (0.00)\\
AZ & 441 & 0.06 & 3.28 (1.58) & 2.91 (1.47) & 0.05 (0.15) & 0.03 (0.11)\\
BD & 79 & 0.04 & 2.71 (1.48) & 1.82 (1.26) & 0.00 (0.01) & 0.33 (0.32)\\
BM & 59 & 0.53 & 1.42 (2.17) & 1.32 (1.91) & 0.00 (0.00) & 0.02 (0.06)\\
BY & 112 & 0.01 & 7.54 (3.78) & 4.21 (2.48) & 0.19 (0.24) & 0.12 (0.19)\\
CP & 100 & 0.35 & 1.99 (2.07) & 0.79 (1.39) & 0.02 (0.08) & 0.58 (0.41)\\
CR & 93 & 0.05 & 2.72 (2.00) & 2.38 (1.86) & 0.02 (0.09) & 0.05 (0.21)\\
DJ & 349 & 0.67 & 0.48 (0.80) & 0.44 (0.73) & 0.23 (0.40) & 0.00 (0.00)\\
EK & 49 & 0.24 & 1.61 (1.41) & 1.55 (1.32) & 0.00 (0.00) & 0.00 (0.00)\\
FF & 44 & 0.18 & 1.57 (1.32) & 1.09 (1.01) & 0.00 (0.00) & 0.27 (0.37)\\
FJ & 42 & 0.14 & 3.05 (2.06) & 2.95 (2.07) & 0.00 (0.00) & 0.04 (0.18)\\
HE & 63 & 0.03 & 3.63 (2.34) & 3.29 (2.07) & 0.01 (0.06) & 0.05 (0.16)\\
HI & 39 & 0.03 & 3.85 (2.18) & 3.38 (2.10) & 0.00 (0.00) & 0.11 (0.21)\\
IH & 153 & 0.45 & 1.12 (1.33) & 0.16 (0.43) & 0.05 (0.18) & 0.83 (0.34)\\
KA & 100 & 0.00 & 4.08 (0.85) & 2.88 (1.34) & 0.00 (0.02) & 0.29 (0.30)\\
KM & 72 & 0.26 & --- & 1.99 (2.02) & --- & 0.07 (0.24)\\
KO & 65 & 0.02 & 3.09 (1.20) & 2.54 (1.19) & 0.03 (0.11) & 0.16 (0.24)\\
KS & 53 & 0.02 & 4.53 (1.42) & 4.19 (1.46) & 0.00 (0.00) & 0.05 (0.12)\\
KT & 18 & 0.17 & 2.72 (1.90) & 2.17 (1.50) & 0.01 (0.05) & 0.05 (0.11)\\
LB & 30 & 0.40 & 1.13 (1.36) & 0.90 (1.06) & 0.00 (0.00) & 0.11 (0.32)\\
MA & 41 & 0.00 & 4.68 (1.60) & 3.49 (1.45) & 0.03 (0.08) & 0.20 (0.26)\\
MC & 189 & 0.01 & 3.56 (1.91) & 3.14 (1.80) & 0.00 (0.04) & 0.09 (0.19)\\
MG & 101 & 0.01 & 3.01 (1.09) & 2.98 (1.12) & 0.00 (0.00) & 0.01 (0.08)\\
MK & 87 & 0.05 & 4.03 (2.26) & 2.38 (1.75) & 0.01 (0.09) & 0.31 (0.35)\\
MN & 38 & 0.50 & 1.53 (2.02) & 1.16 (1.76) & 0.00 (0.00) & 0.25 (0.41)\\
MY & 177 & 0.51 & 2.28 (2.98) & 1.88 (2.41) & 0.02 (0.11) & 0.00 (0.02)\\
NI & 98 & 0.00 & 4.81 (1.54) & 4.74 (1.54) & 0.01 (0.06) & 0.01 (0.04)\\
NP & 326 & 0.09 & 2.61 (1.67) & 2.25 (1.53) & 0.01 (0.07) & 0.11 (0.22)\\
PC & 125 & 0.41 & 1.84 (3.30) & 0.50 (1.00) & 0.02 (0.13) & 0.67 (0.40)\\
PE & 85 & 0.07 & 5.51 (3.46) & 3.85 (2.72) & 0.01 (0.03) & 0.24 (0.28)\\
PQ & 1533 & 0.46 & 0.95 (1.18) & 0.80 (1.10) & 0.07 (0.23) & 0.16 (0.34)\\
PS & 91 & 0.20 & 2.66 (2.13) & 2.16 (1.74) & 0.00 (0.00) & 0.11 (0.23)\\
PT & 68 & 0.28 & 1.79 (1.64) & 1.65 (1.55) & 0.05 (0.18) & 0.00 (0.02)\\
RA & 54 & 0.17 & 2.85 (1.70) & 2.26 (1.63) & 0.00 (0.00) & 0.20 (0.30)\\
SH & 46 & 0.07 & 1.93 (1.36) & 1.39 (1.18) & 0.03 (0.16) & 0.20 (0.34)\\
SN & 63 & 0.05 & 3.97 (2.18) & 2.81 (1.87) & 0.00 (0.03) & 0.26 (0.30)\\
TE & 346 & 0.07 & 3.01 (1.41) & 2.67 (1.37) & 0.03 (0.13) & 0.04 (0.14)\\
TI & 74 & 0.11 & 3.15 (2.18) & 2.54 (2.03) & 0.04 (0.15) & 0.16 (0.29)\\
TM & 129 & 0.12 & 3.11 (1.88) & 1.75 (1.57) & 0.01 (0.05) & 0.39 (0.36)\\
TP & 94 & 0.19 & 1.93 (1.80) & 1.27 (1.17) & 0.16 (0.33) & 0.17 (0.34)\\
TS & 50 & 0.18 & 2.14 (1.76) & 1.52 (1.20) & 0.02 (0.11) & 0.20 (0.30)\\
TZ & 192 & 0.24 & 1.97 (1.81) & 1.82 (1.74) & 0.04 (0.18) & 0.04 (0.17)\\
UP & 470 & 0.10 & 3.60 (2.32) & 3.07 (1.93) & 0.04 (0.17) & 0.03 (0.12)\\
VT & 34 & 0.09 & 3.47 (2.05) & 2.35 (1.45) & 0.01 (0.06) & 0.27 (0.25)\\
WH & 66 & 0.55 & 0.67 (0.88) & 0.24 (0.56) & 0.07 (0.25) & 0.62 (0.47)\\
WL & 178 & 0.39 & 1.49 (1.65) & 0.53 (0.92) & 0.02 (0.14) & 0.59 (0.43)\\
YI & 69 & 0.22 & 2.28 (1.81) & 2.07 (1.79) & 0.00 (0.00) & 0.08 (0.22)\\
YN & 39 & 0.62 & 0.56 (0.88) & 0.49 (0.85) & 0.07 (0.26) & 0.17 (0.36)\\*
\end{longtable}
\endgroup{}

\begingroup\fontsize{8}{10}\selectfont

\begin{longtable}[t]{lrlllll}
\caption{\label{tab:nomination-summary-f_behav}Mean (and standard deviation) values for nominations made across all respondents, for the Behavioral support (women) name generator. `N Resp.' is the number of respondents. `Prop. Zero' is the proportion of respondents nominating no alters. `N Indiv' is the number of individuals nominated. `N SU' is the number of sharing units. `Prop. Same SU' is the proportion of individuals nominated who are members of the same sharing unit as the respondent. `Prop. Ext' is the proportion of individuals nominated who are external to the community.}\\
\toprule
Site & N Resp. & Prop. Zero & N Indiv. & N SU & Prop. Same SU & Prop. Ext.\\
\midrule
\endfirsthead
\caption[]{Mean (and standard deviation) values for nominations made across all respondents, for the Behavioral support (women) name generator. `N Resp.' is the number of respondents. `Prop. Zero' is the proportion of respondents nominating no alters. `N Indiv' is the number of individuals nominated. `N SU' is the number of sharing units. `Prop. Same SU' is the proportion of individuals nominated who are members of the same sharing unit as the respondent. `Prop. Ext' is the proportion of individuals nominated who are external to the community. \textit{(continued)}}\\
\toprule
Site & N Resp. & Prop. Zero & N Indiv. & N SU & Prop. Same SU & Prop. Ext.\\
\midrule
\endhead

\endfoot
\bottomrule
\endlastfoot
AH & 138 & 0.25 & 3.02 (2.95) & 2.51 (2.46) & 0.10 (0.25) & 0.04 (0.14)\\
AP & 49 & 0.18 & 1.18 (0.86) & 1.02 (0.88) & 0.00 (0.00) & 0.15 (0.34)\\
AR & 158 & 0.41 & 1.23 (1.44) & 0.90 (1.20) & 0.02 (0.09) & 0.20 (0.36)\\
AV & 119 & 0.00 & 8.08 (5.76) & 6.54 (4.85) & 0.22 (0.25) & 0.00 (0.00)\\
AZ & 441 & 0.03 & 3.46 (1.62) & 2.52 (1.52) & 0.23 (0.33) & 0.12 (0.25)\\
BD & 79 & 0.10 & 2.61 (1.63) & 1.47 (1.07) & 0.24 (0.36) & 0.29 (0.34)\\
BM & 59 & 0.08 & 2.17 (1.35) & 1.63 (1.51) & 0.05 (0.20) & 0.25 (0.39)\\
BY & 112 & 0.08 & 4.58 (2.92) & 2.93 (2.21) & 0.39 (0.35) & 0.06 (0.15)\\
CP & 100 & 0.21 & 0.99 (0.73) & 0.82 (0.52) & 0.87 (0.33) & 0.03 (0.17)\\
CR & 93 & 0.24 & 1.59 (1.45) & 0.95 (0.71) & 0.71 (0.43) & 0.02 (0.11)\\
DJ & 201 & 0.73 & 0.41 (0.92) & 0.38 (0.88) & 0.41 (0.47) & 0.00 (0.00)\\
EK & 49 & 0.16 & 2.06 (1.69) & 1.61 (1.41) & 0.02 (0.16) & 0.09 (0.27)\\
FF & 44 & 0.82 & 0.23 (0.52) & 0.14 (0.41) & 0.31 (0.46) & 0.38 (0.52)\\
FJ & 42 & 0.05 & 3.52 (2.10) & 3.19 (2.04) & 0.00 (0.03) & 0.08 (0.19)\\
HE & 63 & 0.13 & 2.83 (2.18) & 2.49 (1.92) & 0.00 (0.02) & 0.03 (0.11)\\
HI & 39 & 0.18 & 2.77 (2.06) & 1.44 (1.29) & 0.60 (0.42) & 0.07 (0.20)\\
IH & 90 & 0.42 & 4.73 (7.05) & 0.29 (0.64) & 0.05 (0.20) & 0.91 (0.21)\\
KA & 100 & 0.09 & 2.96 (1.37) & 1.91 (1.26) & 0.03 (0.10) & 0.34 (0.32)\\
KM & 72 & 0.44 & --- & 0.90 (1.15) & --- & 0.08 (0.24)\\
KO & 65 & 0.03 & 2.34 (1.15) & 1.97 (1.15) & 0.05 (0.19) & 0.14 (0.27)\\
KS & 53 & 0.00 & 4.40 (0.82) & 2.72 (1.21) & 0.35 (0.31) & 0.04 (0.09)\\
KT & 18 & 0.39 & 1.50 (1.86) & 1.22 (1.48) & 0.64 (0.43) & 0.00 (0.00)\\
LB & 30 & 0.33 & 1.30 (1.34) & 1.03 (1.00) & 0.00 (0.00) & 0.12 (0.31)\\
MA & 41 & 0.02 & 2.22 (1.39) & 1.76 (1.09) & 0.12 (0.25) & 0.12 (0.28)\\
MC & 98 & 0.54 & 0.90 (1.63) & 0.63 (1.38) & 0.06 (0.22) & 0.25 (0.38)\\
MG & 86 & 0.00 & 4.65 (1.22) & 4.41 (1.20) & 0.05 (0.12) & 0.00 (0.02)\\
MN & 38 & 0.24 & 1.92 (1.51) & 1.53 (1.52) & 0.00 (0.00) & 0.24 (0.37)\\
MY & 177 & 0.50 & 1.69 (2.06) & 1.24 (1.52) & 0.34 (0.37) & 0.04 (0.11)\\
NI & 98 & 0.00 & 3.86 (1.39) & 3.09 (1.48) & 0.16 (0.27) & 0.02 (0.08)\\
NP & 326 & 0.09 & 2.49 (1.81) & 1.58 (1.29) & 0.42 (0.42) & 0.19 (0.31)\\
PC & 66 & 0.39 & 1.94 (2.70) & 0.53 (1.00) & 0.17 (0.35) & 0.65 (0.41)\\
PE & 42 & 0.14 & 4.17 (3.44) & 3.14 (2.82) & 0.01 (0.03) & 0.19 (0.28)\\
PQ & 857 & 0.68 & 0.49 (0.89) & 0.34 (0.76) & 0.14 (0.34) & 0.33 (0.45)\\
PS & 91 & 0.52 & 1.34 (1.97) & 0.99 (1.46) & 0.06 (0.18) & 0.16 (0.28)\\
PT & 68 & 0.34 & 1.28 (1.30) & 1.09 (1.09) & 0.64 (0.40) & 0.00 (0.00)\\
RA & 54 & 0.22 & 3.00 (2.07) & 1.89 (1.73) & 0.01 (0.04) & 0.35 (0.34)\\
SH & 46 & 0.11 & 1.74 (1.16) & 1.28 (0.91) & 0.29 (0.40) & 0.12 (0.29)\\
SN & 63 & 0.06 & 3.68 (2.27) & 2.48 (1.74) & 0.04 (0.12) & 0.18 (0.26)\\
TE & 346 & 0.03 & 3.36 (1.42) & 2.14 (1.45) & 0.28 (0.37) & 0.18 (0.32)\\
TI & 74 & 0.46 & 1.03 (1.18) & 0.77 (0.96) & 0.15 (0.32) & 0.16 (0.35)\\
TM & 129 & 0.09 & 3.02 (1.76) & 1.83 (1.52) & 0.01 (0.08) & 0.31 (0.34)\\
TP & 94 & 0.11 & 2.48 (2.44) & 1.19 (0.85) & 0.80 (0.33) & 0.05 (0.20)\\
TS & 50 & 0.24 & 1.30 (1.02) & 1.06 (0.84) & 0.12 (0.28) & 0.10 (0.25)\\
TZ & 106 & 0.14 & 1.87 (1.85) & 1.10 (1.23) & 0.71 (0.43) & 0.11 (0.29)\\
UP & 470 & 0.06 & 3.88 (2.38) & 2.45 (1.85) & 0.37 (0.40) & 0.10 (0.21)\\
VT & 34 & 0.32 & 2.26 (2.21) & 1.85 (2.00) & 0.02 (0.10) & 0.19 (0.33)\\
WH & 36 & 0.78 & 0.58 (2.03) & 0.17 (0.45) & 0.25 (0.46) & 0.44 (0.50)\\
WL & 113 & 0.72 & 0.84 (1.82) & 0.32 (0.89) & 0.08 (0.26) & 0.55 (0.46)\\
YI & 69 & 0.20 & 2.10 (1.70) & 1.71 (1.36) & 0.02 (0.14) & 0.07 (0.23)\\
YN & 39 & 0.62 & 0.49 (0.68) & 0.28 (0.46) & 0.43 (0.50) & 0.30 (0.46)\\*
\end{longtable}
\endgroup{}

\begingroup\fontsize{8}{10}\selectfont

\begin{longtable}[t]{lrlllll}
\caption{\label{tab:nomination-summary-m_behav}Mean (and standard deviation) values for nominations made across all respondents, for the Behavioral support (men) name generator. `N Resp.' is the number of respondents. `Prop. Zero' is the proportion of respondents nominating no alters. `N Indiv' is the number of individuals nominated. `N SU' is the number of sharing units. `Prop. Same SU' is the proportion of individuals nominated who are members of the same sharing unit as the respondent. `Prop. Ext' is the proportion of individuals nominated who are external to the community.}\\
\toprule
Site & N Resp. & Prop. Zero & N Indiv. & N SU & Prop. Same SU & Prop. Ext.\\
\midrule
\endfirsthead
\caption[]{Mean (and standard deviation) values for nominations made across all respondents, for the Behavioral support (men) name generator. `N Resp.' is the number of respondents. `Prop. Zero' is the proportion of respondents nominating no alters. `N Indiv' is the number of individuals nominated. `N SU' is the number of sharing units. `Prop. Same SU' is the proportion of individuals nominated who are members of the same sharing unit as the respondent. `Prop. Ext' is the proportion of individuals nominated who are external to the community. \textit{(continued)}}\\
\toprule
Site & N Resp. & Prop. Zero & N Indiv. & N SU & Prop. Same SU & Prop. Ext.\\
\midrule
\endhead

\endfoot
\bottomrule
\endlastfoot
AH & 138 & 0.16 & 4.20 (3.17) & 3.72 (2.93) & 0.06 (0.18) & 0.04 (0.14)\\
AP & 40 & 0.07 & 2.12 (1.40) & 1.88 (1.45) & 0.00 (0.00) & 0.15 (0.30)\\
AR & 158 & 0.37 & 1.46 (1.64) & 1.15 (1.42) & 0.03 (0.15) & 0.16 (0.29)\\
AV & 84 & 0.00 & 9.48 (6.29) & 8.44 (5.84) & 0.13 (0.15) & 0.00 (0.00)\\
AZ & 441 & 0.10 & 3.08 (1.74) & 2.20 (1.59) & 0.22 (0.35) & 0.15 (0.29)\\
BD & 79 & 0.05 & 3.18 (1.77) & 0.92 (0.87) & 0.10 (0.24) & 0.63 (0.34)\\
BM & 59 & 0.00 & 5.00 (2.20) & 4.22 (2.13) & 0.00 (0.00) & 0.11 (0.26)\\
BY & 112 & 0.12 & 3.77 (2.74) & 2.56 (2.08) & 0.32 (0.35) & 0.08 (0.18)\\
CP & 100 & 0.22 & 1.19 (1.21) & 0.86 (0.75) & 0.71 (0.44) & 0.11 (0.28)\\
CR & 93 & 0.23 & 1.41 (1.24) & 1.05 (1.00) & 0.45 (0.47) & 0.05 (0.21)\\
DJ & 148 & 0.62 & 0.60 (0.99) & 0.59 (0.98) & 0.21 (0.40) & 0.00 (0.00)\\
EK & 49 & 0.22 & 1.82 (1.54) & 1.53 (1.16) & 0.01 (0.08) & 0.05 (0.12)\\
FF & 44 & 0.66 & 0.45 (0.73) & 0.23 (0.42) & 0.26 (0.41) & 0.44 (0.46)\\
FJ & 42 & 0.14 & 3.52 (2.54) & 3.05 (2.24) & 0.02 (0.07) & 0.07 (0.22)\\
HE & 63 & 0.17 & 2.48 (1.78) & 2.17 (1.71) & 0.00 (0.00) & 0.05 (0.19)\\
HI & 39 & 0.08 & 4.21 (2.52) & 2.03 (1.65) & 0.16 (0.29) & 0.32 (0.37)\\
IH & 63 & 0.22 & 3.87 (4.62) & 0.21 (0.51) & 0.04 (0.12) & 0.93 (0.14)\\
KA & 100 & 0.02 & 3.80 (0.97) & 2.30 (1.19) & 0.10 (0.22) & 0.34 (0.29)\\
KM & 72 & 0.51 & --- & 0.92 (1.18) & --- & 0.01 (0.08)\\
KO & 67 & 0.00 & 2.88 (1.33) & 2.06 (1.50) & 0.01 (0.07) & 0.29 (0.34)\\
KS & 53 & 0.00 & 4.75 (1.11) & 4.40 (1.84) & 0.03 (0.15) & 0.11 (0.17)\\
KT & 18 & 0.44 & 1.33 (1.85) & 0.83 (1.04) & 0.59 (0.46) & 0.03 (0.08)\\
LB & 30 & 0.60 & 0.87 (1.28) & 0.53 (1.11) & 0.02 (0.07) & 0.43 (0.45)\\
MA & 42 & 0.05 & 3.33 (1.22) & 2.60 (1.31) & 0.00 (0.00) & 0.23 (0.28)\\
MC & 91 & 0.31 & 1.47 (1.42) & 1.23 (1.19) & 0.00 (0.00) & 0.14 (0.30)\\
MG & 85 & 0.01 & 4.60 (1.46) & 4.24 (1.47) & 0.05 (0.12) & 0.03 (0.09)\\
MN & 38 & 0.26 & 2.42 (2.38) & 2.03 (1.98) & 0.00 (0.00) & 0.12 (0.23)\\
MY & 177 & 0.51 & 1.86 (2.37) & 1.36 (1.74) & 0.23 (0.31) & 0.03 (0.11)\\
NI & 98 & 0.02 & 3.89 (1.52) & 2.91 (1.54) & 0.23 (0.29) & 0.04 (0.10)\\
NP & 326 & 0.22 & 2.65 (2.16) & 1.81 (1.64) & 0.11 (0.27) & 0.25 (0.30)\\
PC & 59 & 0.19 & 3.07 (4.23) & 0.73 (1.10) & 0.05 (0.16) & 0.74 (0.33)\\
PE & 43 & 0.26 & 2.37 (1.81) & 1.77 (1.74) & 0.00 (0.00) & 0.25 (0.35)\\
PQ & 676 & 0.53 & 0.74 (1.03) & 0.52 (0.86) & 0.09 (0.30) & 0.29 (0.42)\\
PS & 91 & 0.42 & 1.69 (2.07) & 1.36 (1.75) & 0.20 (0.36) & 0.08 (0.23)\\
PT & 68 & 0.26 & 1.46 (1.26) & 1.25 (1.08) & 0.61 (0.39) & 0.01 (0.05)\\
RA & 54 & 0.20 & 2.78 (2.13) & 1.54 (1.48) & 0.05 (0.19) & 0.32 (0.37)\\
SH & 46 & 0.07 & 2.59 (1.61) & 1.83 (1.52) & 0.07 (0.17) & 0.29 (0.38)\\
SN & 63 & 0.08 & 3.86 (2.31) & 2.75 (1.97) & 0.04 (0.15) & 0.20 (0.25)\\
TE & 346 & 0.06 & 3.12 (1.61) & 2.11 (1.53) & 0.24 (0.36) & 0.19 (0.32)\\
TI & 74 & 0.38 & 1.68 (1.81) & 0.95 (1.27) & 0.07 (0.23) & 0.40 (0.41)\\
TM & 129 & 0.12 & 2.74 (1.72) & 1.52 (1.36) & 0.00 (0.00) & 0.39 (0.40)\\
TP & 94 & 0.12 & 1.81 (1.55) & 1.27 (1.05) & 0.69 (0.42) & 0.05 (0.18)\\
TS & 50 & 0.32 & 1.66 (1.36) & 1.36 (1.32) & 0.07 (0.21) & 0.21 (0.32)\\
TZ & 86 & 0.08 & 4.07 (2.72) & 1.36 (1.60) & 0.22 (0.36) & 0.50 (0.42)\\
UP & 470 & 0.07 & 4.27 (2.77) & 3.00 (2.29) & 0.24 (0.36) & 0.08 (0.20)\\
VT & 34 & 0.03 & 4.50 (2.68) & 2.76 (2.03) & 0.03 (0.10) & 0.33 (0.32)\\
WH & 30 & 0.27 & 2.73 (2.57) & 1.07 (1.44) & 0.09 (0.22) & 0.64 (0.34)\\
WL & 65 & 0.54 & 1.60 (2.23) & 0.37 (0.67) & 0.13 (0.35) & 0.65 (0.39)\\
YI & 69 & 0.22 & 2.19 (1.83) & 1.88 (1.74) & 0.00 (0.00) & 0.15 (0.33)\\
YN & 39 & 0.49 & 0.87 (1.06) & 0.41 (0.79) & 0.03 (0.11) & 0.59 (0.45)\\*
\end{longtable}
\endgroup{}

\begingroup\fontsize{8}{10}\selectfont

\begin{longtable}[t]{lrlllll}
\caption{\label{tab:nomination-summary-f_info}Mean (and standard deviation) values for nominations made across all respondents, for the Informational/affective support (women) name generator. `N Resp.' is the number of respondents. `Prop. Zero' is the proportion of respondents nominating no alters. `N Indiv' is the number of individuals nominated. `N SU' is the number of sharing units. `Prop. Same SU' is the proportion of individuals nominated who are members of the same sharing unit as the respondent. `Prop. Ext' is the proportion of individuals nominated who are external to the community.}\\
\toprule
Site & N Resp. & Prop. Zero & N Indiv. & N SU & Prop. Same SU & Prop. Ext.\\
\midrule
\endfirsthead
\caption[]{Mean (and standard deviation) values for nominations made across all respondents, for the Informational/affective support (women) name generator. `N Resp.' is the number of respondents. `Prop. Zero' is the proportion of respondents nominating no alters. `N Indiv' is the number of individuals nominated. `N SU' is the number of sharing units. `Prop. Same SU' is the proportion of individuals nominated who are members of the same sharing unit as the respondent. `Prop. Ext' is the proportion of individuals nominated who are external to the community. \textit{(continued)}}\\
\toprule
Site & N Resp. & Prop. Zero & N Indiv. & N SU & Prop. Same SU & Prop. Ext.\\
\midrule
\endhead

\endfoot
\bottomrule
\endlastfoot
AH & 138 & 0.13 & 4.13 (2.90) & 3.82 (2.75) & 0.01 (0.10) & 0.04 (0.09)\\
AP & 49 & 0.00 & 4.31 (1.14) & 3.71 (1.27) & 0.00 (0.00) & 0.12 (0.19)\\
AR & 158 & 0.20 & 2.97 (2.36) & 1.75 (1.67) & 0.00 (0.03) & 0.33 (0.36)\\
AV & 119 & 0.18 & 2.93 (2.99) & 2.62 (2.69) & 0.32 (0.38) & 0.00 (0.00)\\
AZ & 441 & 0.03 & 3.59 (1.71) & 2.72 (1.66) & 0.27 (0.36) & 0.08 (0.21)\\
BD & 79 & 0.00 & 8.19 (3.13) & 2.42 (1.77) & 0.05 (0.10) & 0.65 (0.25)\\
BM & 59 & 0.15 & 2.24 (1.47) & 1.63 (1.31) & 0.00 (0.00) & 0.07 (0.22)\\
BY & 112 & 0.04 & 5.59 (3.28) & 4.03 (2.53) & 0.08 (0.16) & 0.09 (0.20)\\
CP & 100 & 0.34 & 1.29 (1.41) & 0.54 (0.82) & 0.12 (0.28) & 0.52 (0.45)\\
CR & 93 & 0.08 & 3.19 (1.94) & 2.51 (1.56) & 0.15 (0.28) & 0.05 (0.17)\\
DJ & 201 & 0.06 & 2.07 (1.40) & 1.85 (1.16) & 0.21 (0.32) & 0.00 (0.00)\\
EK & 49 & 0.04 & 3.33 (1.98) & 2.98 (1.60) & 0.00 (0.00) & 0.00 (0.00)\\
FF & 44 & 0.43 & 1.20 (1.34) & 0.80 (1.19) & 0.00 (0.00) & 0.40 (0.44)\\
FJ & 42 & 0.05 & 3.45 (2.36) & 3.26 (2.44) & 0.00 (0.00) & 0.03 (0.10)\\
HE & 63 & 0.05 & 3.84 (3.01) & 3.14 (1.86) & 0.00 (0.00) & 0.01 (0.06)\\
HI & 39 & 0.03 & 3.69 (1.98) & 3.18 (1.94) & 0.01 (0.05) & 0.16 (0.29)\\
IH & 90 & 0.18 & 3.53 (3.20) & 0.46 (0.82) & 0.02 (0.12) & 0.83 (0.30)\\
KA & 100 & 0.00 & 3.98 (0.72) & 3.22 (1.10) & 0.02 (0.07) & 0.19 (0.24)\\
KM & 72 & 0.31 & --- & 1.60 (1.47) & --- & 0.04 (0.16)\\
KO & 65 & 0.05 & 2.58 (1.35) & 2.34 (1.35) & 0.05 (0.17) & 0.10 (0.21)\\
KS & 53 & 0.00 & 4.92 (1.36) & 4.81 (1.36) & 0.00 (0.03) & 0.02 (0.07)\\
KT & 18 & 0.06 & 4.44 (1.76) & 3.50 (1.89) & 0.06 (0.21) & 0.05 (0.09)\\
LB & 30 & 0.10 & 1.63 (0.93) & 0.87 (0.73) & 0.04 (0.19) & 0.42 (0.42)\\
MA & 41 & 0.05 & 4.07 (1.99) & 3.41 (1.72) & 0.06 (0.13) & 0.08 (0.18)\\
MC & 98 & 0.00 & 3.58 (1.89) & 2.97 (1.71) & 0.00 (0.03) & 0.17 (0.25)\\
MG & 86 & 0.00 & 4.51 (1.12) & 4.40 (1.19) & 0.00 (0.00) & 0.01 (0.06)\\
MN & 38 & 0.18 & 3.08 (2.17) & 2.58 (1.83) & 0.00 (0.00) & 0.11 (0.19)\\
MY & 177 & 0.50 & 2.30 (2.68) & 1.83 (2.19) & 0.03 (0.12) & 0.09 (0.21)\\
NI & 98 & 0.00 & 4.26 (1.43) & 3.69 (1.54) & 0.09 (0.21) & 0.01 (0.06)\\
NP & 326 & 0.06 & 3.63 (2.72) & 2.69 (2.08) & 0.07 (0.20) & 0.14 (0.25)\\
PC & 66 & 0.14 & 3.68 (2.67) & 1.38 (1.43) & 0.05 (0.15) & 0.49 (0.39)\\
PE & 42 & 0.02 & 8.95 (5.60) & 7.57 (4.79) & 0.00 (0.01) & 0.12 (0.18)\\
PQ & 857 & 0.00 & 5.02 (1.89) & 3.07 (1.65) & 0.36 (0.29) & 0.15 (0.20)\\
PS & 91 & 0.38 & 1.79 (1.82) & 1.64 (1.73) & 0.00 (0.00) & 0.06 (0.20)\\
PT & 68 & 0.32 & 1.62 (1.68) & 1.47 (1.52) & 0.03 (0.15) & 0.00 (0.00)\\
RA & 54 & 0.24 & 3.13 (2.35) & 1.80 (1.75) & 0.00 (0.00) & 0.42 (0.36)\\
SH & 46 & 0.07 & 3.04 (1.85) & 2.26 (1.77) & 0.06 (0.20) & 0.22 (0.32)\\
SN & 63 & 0.03 & 3.51 (1.84) & 2.60 (1.73) & 0.00 (0.03) & 0.18 (0.30)\\
TE & 346 & 0.03 & 3.40 (1.46) & 2.47 (1.51) & 0.32 (0.36) & 0.10 (0.23)\\
TI & 74 & 0.01 & 4.30 (2.35) & 3.30 (2.03) & 0.01 (0.03) & 0.12 (0.26)\\
TM & 129 & 0.12 & 2.83 (1.69) & 1.69 (1.41) & 0.01 (0.07) & 0.32 (0.31)\\
TP & 94 & 0.21 & 1.86 (2.00) & 1.30 (1.18) & 0.47 (0.47) & 0.01 (0.04)\\
TS & 50 & 0.16 & 2.30 (1.46) & 1.88 (1.35) & 0.03 (0.11) & 0.18 (0.27)\\
TZ & 106 & 0.06 & 3.52 (2.41) & 2.57 (2.03) & 0.12 (0.27) & 0.16 (0.30)\\
UP & 470 & 0.05 & 4.62 (2.52) & 2.78 (1.88) & 0.17 (0.27) & 0.21 (0.30)\\
VT & 34 & 0.12 & 2.94 (1.86) & 1.71 (1.62) & 0.00 (0.00) & 0.42 (0.41)\\
WH & 36 & 0.19 & 3.00 (2.93) & 1.19 (1.28) & 0.00 (0.03) & 0.42 (0.40)\\
WL & 113 & 0.02 & 4.94 (3.91) & 1.78 (1.68) & 0.05 (0.19) & 0.55 (0.35)\\
YI & 69 & 0.25 & 1.97 (1.53) & 1.64 (1.41) & 0.02 (0.11) & 0.12 (0.30)\\
YN & 39 & 0.90 & 0.23 (0.78) & 0.08 (0.35) & 0.00 (0.00) & 0.62 (0.48)\\*
\end{longtable}
\endgroup{}

\begingroup\fontsize{8}{10}\selectfont

\begin{longtable}[t]{lrlllll}
\caption{\label{tab:nomination-summary-m_info}Mean (and standard deviation) values for nominations made across all respondents, for the Informational/affective support (men) name generator. `N Resp.' is the number of respondents. `Prop. Zero' is the proportion of respondents nominating no alters. `N Indiv' is the number of individuals nominated. `N SU' is the number of sharing units. `Prop. Same SU' is the proportion of individuals nominated who are members of the same sharing unit as the respondent. `Prop. Ext' is the proportion of individuals nominated who are external to the community.}\\
\toprule
Site & N Resp. & Prop. Zero & N Indiv. & N SU & Prop. Same SU & Prop. Ext.\\
\midrule
\endfirsthead
\caption[]{Mean (and standard deviation) values for nominations made across all respondents, for the Informational/affective support (men) name generator. `N Resp.' is the number of respondents. `Prop. Zero' is the proportion of respondents nominating no alters. `N Indiv' is the number of individuals nominated. `N SU' is the number of sharing units. `Prop. Same SU' is the proportion of individuals nominated who are members of the same sharing unit as the respondent. `Prop. Ext' is the proportion of individuals nominated who are external to the community. \textit{(continued)}}\\
\toprule
Site & N Resp. & Prop. Zero & N Indiv. & N SU & Prop. Same SU & Prop. Ext.\\
\midrule
\endhead

\endfoot
\bottomrule
\endlastfoot
AH & 138 & 0.17 & 4.79 (3.34) & 4.54 (3.11) & 0.00 (0.01) & 0.02 (0.06)\\
AP & 40 & 0.12 & 3.35 (2.02) & 3.30 (1.98) & 0.00 (0.00) & 0.01 (0.03)\\
AR & 158 & 0.23 & 2.33 (2.03) & 1.61 (1.67) & 0.01 (0.07) & 0.25 (0.37)\\
AV & 84 & 0.10 & 4.15 (3.68) & 3.96 (3.58) & 0.17 (0.26) & 0.00 (0.00)\\
AZ & 441 & 0.08 & 3.23 (1.76) & 2.18 (1.44) & 0.30 (0.38) & 0.12 (0.24)\\
BD & 79 & 0.04 & 4.22 (2.53) & 1.23 (1.38) & 0.07 (0.15) & 0.68 (0.28)\\
BM & 59 & 0.03 & 2.49 (1.24) & 1.86 (1.20) & 0.00 (0.00) & 0.06 (0.21)\\
BY & 112 & 0.19 & 4.14 (3.29) & 3.21 (2.62) & 0.04 (0.10) & 0.09 (0.18)\\
CP & 100 & 0.50 & 0.98 (1.19) & 0.45 (0.76) & 0.06 (0.22) & 0.48 (0.45)\\
CR & 93 & 0.12 & 2.85 (2.09) & 2.27 (1.70) & 0.08 (0.21) & 0.05 (0.17)\\
DJ & 148 & 0.05 & 2.70 (2.26) & 2.47 (2.04) & 0.15 (0.26) & 0.00 (0.03)\\
EK & 49 & 0.29 & 1.80 (1.86) & 1.00 (0.98) & 0.48 (0.50) & 0.03 (0.12)\\
FF & 44 & 0.41 & 1.55 (1.82) & 0.98 (1.39) & 0.00 (0.00) & 0.32 (0.43)\\
FJ & 42 & 0.12 & 2.95 (1.74) & 2.62 (1.51) & 0.01 (0.06) & 0.04 (0.12)\\
HE & 63 & 0.11 & 3.98 (2.78) & 3.57 (2.58) & 0.01 (0.06) & 0.02 (0.09)\\
HI & 39 & 0.26 & 4.10 (3.52) & 3.87 (3.45) & 0.00 (0.00) & 0.05 (0.11)\\
IH & 63 & 0.08 & 4.75 (3.77) & 0.67 (1.03) & 0.07 (0.18) & 0.82 (0.25)\\
KA & 100 & 0.08 & 3.87 (1.36) & 2.77 (1.33) & 0.00 (0.03) & 0.29 (0.25)\\
KM & 72 & 0.35 & --- & 1.40 (1.52) & --- & 0.07 (0.23)\\
KO & 68 & 0.12 & 2.54 (1.50) & 1.94 (1.59) & 0.03 (0.15) & 0.28 (0.38)\\
KS & 53 & 0.00 & 5.11 (1.50) & 4.92 (1.95) & 0.02 (0.14) & 0.10 (0.22)\\
KT & 18 & 0.11 & 3.61 (1.91) & 2.78 (1.56) & 0.07 (0.22) & 0.05 (0.09)\\
LB & 30 & 0.47 & 0.83 (1.02) & 0.57 (0.73) & 0.06 (0.25) & 0.29 (0.42)\\
MA & 42 & 0.00 & 4.00 (1.40) & 2.43 (1.86) & 0.02 (0.10) & 0.38 (0.38)\\
MC & 91 & 0.09 & 1.75 (1.31) & 1.58 (1.16) & 0.01 (0.11) & 0.07 (0.21)\\
MG & 85 & 0.01 & 4.93 (1.21) & 4.73 (1.31) & 0.01 (0.03) & 0.02 (0.08)\\
MN & 38 & 0.13 & 3.76 (2.53) & 3.45 (2.25) & 0.00 (0.00) & 0.05 (0.09)\\
MY & 177 & 0.51 & 2.16 (2.88) & 1.73 (2.36) & 0.03 (0.13) & 0.13 (0.28)\\
NI & 98 & 0.01 & 4.40 (1.80) & 4.15 (1.84) & 0.05 (0.14) & 0.02 (0.07)\\
NP & 326 & 0.17 & 3.34 (2.57) & 2.39 (1.95) & 0.05 (0.18) & 0.23 (0.26)\\
PC & 59 & 0.08 & 4.80 (4.61) & 1.34 (1.55) & 0.01 (0.04) & 0.68 (0.31)\\
PE & 43 & 0.23 & 2.98 (2.25) & 2.49 (2.11) & 0.01 (0.04) & 0.17 (0.34)\\
PQ & 676 & 0.00 & 4.80 (1.89) & 3.17 (1.59) & 0.31 (0.29) & 0.14 (0.20)\\
PS & 91 & 0.32 & 2.63 (2.36) & 2.33 (2.13) & 0.01 (0.04) & 0.05 (0.12)\\
PT & 68 & 0.24 & 1.72 (1.49) & 1.60 (1.42) & 0.04 (0.15) & 0.00 (0.00)\\
RA & 54 & 0.20 & 3.19 (2.29) & 1.76 (1.70) & 0.03 (0.13) & 0.43 (0.37)\\
SH & 46 & 0.15 & 3.22 (2.22) & 1.28 (1.46) & 0.01 (0.08) & 0.58 (0.34)\\
SN & 63 & 0.08 & 3.83 (2.40) & 2.78 (2.26) & 0.01 (0.04) & 0.26 (0.31)\\
TE & 346 & 0.05 & 3.22 (1.54) & 2.12 (1.45) & 0.32 (0.38) & 0.17 (0.30)\\
TI & 74 & 0.07 & 4.43 (3.16) & 3.45 (2.88) & 0.02 (0.08) & 0.19 (0.34)\\
TM & 129 & 0.20 & 2.37 (1.81) & 1.80 (1.60) & 0.00 (0.05) & 0.21 (0.31)\\
TP & 94 & 0.36 & 1.19 (1.24) & 1.06 (1.17) & 0.15 (0.35) & 0.04 (0.19)\\
TS & 50 & 0.18 & 2.02 (1.50) & 1.80 (1.48) & 0.02 (0.16) & 0.11 (0.26)\\
TZ & 86 & 0.14 & 3.09 (2.72) & 2.08 (2.38) & 0.01 (0.10) & 0.30 (0.36)\\
UP & 470 & 0.06 & 5.09 (3.45) & 3.76 (3.03) & 0.08 (0.22) & 0.15 (0.26)\\
VT & 34 & 0.06 & 3.00 (1.76) & 1.62 (1.74) & 0.00 (0.00) & 0.46 (0.40)\\
WH & 30 & 0.23 & 4.27 (3.61) & 1.43 (1.50) & 0.00 (0.00) & 0.57 (0.30)\\
WL & 65 & 0.06 & 4.82 (3.40) & 2.02 (1.88) & 0.03 (0.18) & 0.51 (0.32)\\
YI & 69 & 0.17 & 2.17 (1.77) & 1.94 (1.60) & 0.02 (0.13) & 0.05 (0.17)\\
YN & 39 & 0.23 & 2.33 (1.74) & 1.82 (1.55) & 0.00 (0.00) & 0.26 (0.29)\\*
\end{longtable}
\endgroup{}

\begingroup\fontsize{8}{10}\selectfont

\begin{longtable}[t]{lrlllll}
\caption{\label{tab:nomination-summary-govt}Mean (and standard deviation) values for nominations made across all respondents, for the Government and NGO ties name generator. `N Resp.' is the number of respondents. `Prop. Zero' is the proportion of respondents nominating no alters. `N Indiv' is the number of individuals nominated. `N SU' is the number of sharing units. `Prop. Same SU' is the proportion of individuals nominated who are members of the same sharing unit as the respondent. `Prop. Ext' is the proportion of individuals nominated who are external to the community.}\\
\toprule
Site & N Resp. & Prop. Zero & N Indiv. & N SU & Prop. Same SU & Prop. Ext.\\
\midrule
\endfirsthead
\caption[]{Mean (and standard deviation) values for nominations made across all respondents, for the Government and NGO ties name generator. `N Resp.' is the number of respondents. `Prop. Zero' is the proportion of respondents nominating no alters. `N Indiv' is the number of individuals nominated. `N SU' is the number of sharing units. `Prop. Same SU' is the proportion of individuals nominated who are members of the same sharing unit as the respondent. `Prop. Ext' is the proportion of individuals nominated who are external to the community. \textit{(continued)}}\\
\toprule
Site & N Resp. & Prop. Zero & N Indiv. & N SU & Prop. Same SU & Prop. Ext.\\
\midrule
\endhead

\endfoot
\bottomrule
\endlastfoot
AH & 138 & 0.34 & 2.42 (2.75) & 0.72 (1.34) & 0.01 (0.05) & 0.75 (0.30)\\
AP & 89 & 0.03 & 3.01 (1.46) & 1.84 (0.64) & 0.00 (0.00) & 0.34 (0.23)\\
AR & 158 & 0.25 & 1.56 (1.68) & 0.96 (1.07) & 0.00 (0.00) & 0.33 (0.40)\\
AZ & 441 & 0.27 & 1.67 (1.46) & 0.86 (1.07) & 0.09 (0.26) & 0.45 (0.43)\\
BD & 79 & 0.20 & 1.32 (1.37) & 0.38 (0.74) & 0.01 (0.06) & 0.74 (0.40)\\
BM & 59 & 0.93 & 0.17 (0.65) & 0.00 (0.00) & 0.00 (0.00) & 1.00 (0.00)\\
CP & 100 & 0.76 & 0.38 (0.78) & 0.00 (0.00) & 0.00 (0.00) & 1.00 (0.00)\\
CR & 93 & 0.73 & 0.49 (1.02) & 0.12 (0.36) & 0.00 (0.00) & 0.60 (0.50)\\
EK & 49 & 0.78 & 0.29 (0.58) & 0.18 (0.49) & 0.00 (0.00) & 0.36 (0.50)\\
FF & 44 & 0.14 & 2.11 (1.48) & 0.07 (0.25) & 0.01 (0.03) & 0.98 (0.05)\\
FJ & 42 & 0.33 & 1.12 (1.02) & 0.14 (0.35) & 0.05 (0.20) & 0.83 (0.36)\\
HE & 63 & 0.65 & 0.89 (1.64) & 0.00 (0.00) & 0.00 (0.00) & 1.00 (0.00)\\
HI & 39 & 0.59 & 0.62 (0.94) & 0.03 (0.16) & 0.00 (0.00) & 0.94 (0.25)\\
KA & 100 & 0.00 & 3.80 (0.92) & 3.34 (1.23) & 0.00 (0.02) & 0.12 (0.23)\\
KM & 72 & 0.28 & --- & 0.94 (0.87) & --- & 0.12 (0.33)\\
KO & 66 & 0.23 & 1.64 (1.18) & 0.33 (0.66) & 0.00 (0.00) & 0.81 (0.30)\\
KS & 53 & 0.00 & 4.15 (0.53) & 0.11 (0.32) & 0.00 (0.00) & 0.97 (0.08)\\
KT & 18 & 1.00 & 0.00 (0.00) & 0.00 (0.00) & --- & ---\\
LB & 30 & 0.63 & 0.67 (1.12) & 0.50 (1.07) & 0.00 (0.00) & 0.32 (0.46)\\
MA & 42 & 0.10 & 2.40 (1.33) & 0.38 (0.70) & 0.00 (0.00) & 0.82 (0.31)\\
MC & 189 & 0.02 & 2.41 (2.45) & 1.48 (1.18) & 0.00 (0.04) & 0.31 (0.37)\\
MG & 101 & 0.15 & 5.53 (2.84) & 4.74 (2.53) & 0.01 (0.04) & 0.15 (0.09)\\
MN & 38 & 0.71 & 0.47 (0.89) & 0.18 (0.39) & 0.00 (0.00) & 0.57 (0.42)\\
MY & 177 & 0.71 & 0.78 (1.93) & 0.01 (0.11) & 0.00 (0.00) & 0.97 (0.15)\\
NI & 98 & 0.73 & 0.41 (0.74) & 0.00 (0.00) & 0.00 (0.00) & 1.00 (0.00)\\
NP & 326 & 0.50 & 1.15 (1.51) & 0.22 (0.59) & 0.02 (0.12) & 0.80 (0.35)\\
PE & 85 & 0.41 & 1.06 (1.14) & 0.00 (0.00) & 0.00 (0.00) & 1.00 (0.00)\\
PQ & 1533 & 0.18 & 1.69 (1.26) & 0.80 (0.88) & 0.00 (0.03) & 0.54 (0.38)\\
PS & 91 & 0.41 & 0.98 (1.15) & 0.21 (0.75) & 0.00 (0.03) & 0.88 (0.29)\\
PT & 68 & 0.26 & 1.63 (1.48) & 0.16 (0.41) & 0.00 (0.00) & 0.86 (0.32)\\
RA & 54 & 0.67 & 0.70 (1.24) & 0.37 (0.92) & 0.01 (0.05) & 0.56 (0.41)\\
SH & 46 & 0.35 & 1.41 (1.51) & 0.00 (0.00) & 0.00 (0.00) & 1.00 (0.00)\\
SN & 63 & 0.35 & 1.56 (1.59) & 0.59 (1.01) & 0.02 (0.11) & 0.56 (0.49)\\
TE & 346 & 0.30 & 1.51 (1.41) & 0.51 (0.81) & 0.06 (0.19) & 0.64 (0.41)\\
TI & 74 & 0.23 & 2.30 (2.00) & 0.15 (0.46) & 0.00 (0.00) & 0.96 (0.12)\\
TM & 129 & 0.75 & 0.50 (0.98) & 0.14 (0.45) & 0.00 (0.00) & 0.69 (0.42)\\
TP & 94 & 0.09 & 3.90 (2.42) & 3.13 (2.04) & 0.06 (0.20) & 0.16 (0.28)\\
TS & 50 & 0.92 & 0.12 (0.44) & 0.00 (0.00) & 0.00 (0.00) & 1.00 (0.00)\\
TZ & 192 & 0.82 & 0.30 (0.75) & 0.14 (0.41) & 0.00 (0.00) & 0.47 (0.47)\\
UP & 470 & 0.39 & 1.55 (1.69) & 0.27 (0.63) & 0.00 (0.04) & 0.82 (0.32)\\
VT & 34 & 0.50 & 0.85 (1.08) & 0.00 (0.00) & 0.00 (0.00) & 1.00 (0.00)\\
YI & 69 & 0.81 & 0.36 (0.92) & 0.04 (0.27) & 0.00 (0.00) & 0.94 (0.14)\\
YN & 39 & 0.18 & 2.31 (1.36) & 1.54 (1.19) & 0.04 (0.15) & 0.29 (0.34)\\*
\end{longtable}
\endgroup{}

\begingroup\fontsize{8}{10}\selectfont

\begin{longtable}[t]{lrlllll}
\caption{\label{tab:nomination-summary-external}Mean (and standard deviation) values for nominations made across all respondents, for the Support outside the community name generator. `N Resp.' is the number of respondents. `Prop. Zero' is the proportion of respondents nominating no alters. `N Indiv' is the number of individuals nominated. `N SU' is the number of sharing units. `Prop. Same SU' is the proportion of individuals nominated who are members of the same sharing unit as the respondent. `Prop. Ext' is the proportion of individuals nominated who are external to the community.}\\
\toprule
Site & N Resp. & Prop. Zero & N Indiv. & N SU & Prop. Same SU & Prop. Ext.\\
\midrule
\endfirsthead
\caption[]{Mean (and standard deviation) values for nominations made across all respondents, for the Support outside the community name generator. `N Resp.' is the number of respondents. `Prop. Zero' is the proportion of respondents nominating no alters. `N Indiv' is the number of individuals nominated. `N SU' is the number of sharing units. `Prop. Same SU' is the proportion of individuals nominated who are members of the same sharing unit as the respondent. `Prop. Ext' is the proportion of individuals nominated who are external to the community. \textit{(continued)}}\\
\toprule
Site & N Resp. & Prop. Zero & N Indiv. & N SU & Prop. Same SU & Prop. Ext.\\
\midrule
\endhead

\endfoot
\bottomrule
\endlastfoot
AH & 138 & 0.48 & 1.04 (1.51) & 0.09 (0.43) & 0.00 (0.00) & 0.94 (0.21)\\
AP & 89 & 0.03 & 2.52 (1.28) & 0.12 (0.45) & 0.00 (0.00) & 0.97 (0.09)\\
AR & 158 & 0.39 & 1.53 (2.21) & 1.02 (1.64) & 0.00 (0.00) & 0.30 (0.40)\\
AZ & 441 & 0.23 & 2.05 (1.75) & 0.51 (0.74) & 0.18 (0.34) & 0.62 (0.42)\\
BD & 79 & 0.00 & 7.16 (3.86) & 1.16 (1.30) & 0.06 (0.10) & 0.81 (0.21)\\
BM & 59 & 0.07 & 1.90 (0.96) & 0.02 (0.13) & 0.00 (0.00) & 0.99 (0.07)\\
BY & 112 & 0.46 & 1.16 (1.52) & 0.05 (0.26) & 0.01 (0.04) & 0.96 (0.16)\\
CP & 100 & 0.02 & 6.76 (3.32) & 2.38 (1.94) & 0.06 (0.14) & 0.58 (0.29)\\
CR & 93 & 0.27 & 0.94 (0.84) & 0.32 (0.49) & 0.00 (0.00) & 0.62 (0.47)\\
EK & 49 & 0.69 & 0.37 (0.60) & 0.06 (0.32) & 0.00 (0.00) & 0.90 (0.28)\\
FF & 44 & 0.59 & 0.59 (0.79) & 0.16 (0.43) & 0.00 (0.00) & 0.72 (0.43)\\
FJ & 42 & 0.38 & 0.86 (0.90) & 0.07 (0.26) & 0.00 (0.00) & 0.91 (0.27)\\
HE & 63 & 0.57 & 1.02 (1.44) & 0.95 (1.37) & 0.00 (0.00) & 0.06 (0.21)\\
HI & 39 & 0.46 & 1.08 (1.58) & 0.03 (0.16) & 0.00 (0.00) & 0.95 (0.22)\\
KA & 100 & 0.00 & 2.79 (0.87) & 0.78 (0.73) & 0.00 (0.03) & 0.68 (0.31)\\
KM & 72 & 0.31 & --- & 0.07 (0.26) & --- & 0.91 (0.28)\\
KO & 67 & 0.10 & 1.84 (1.20) & 0.13 (0.46) & 0.00 (0.00) & 0.95 (0.15)\\
KS & 53 & 0.00 & 4.38 (1.00) & 1.19 (1.43) & 0.00 (0.00) & 0.73 (0.32)\\
KT & 18 & 0.28 & 2.83 (2.41) & 0.67 (0.49) & 0.00 (0.00) & 0.63 (0.37)\\
LB & 30 & 0.20 & 3.27 (2.21) & 2.13 (1.72) & 0.02 (0.08) & 0.27 (0.31)\\
MA & 42 & 0.02 & 3.45 (1.15) & 1.71 (1.37) & 0.01 (0.05) & 0.53 (0.35)\\
MC & 189 & 0.01 & 2.20 (0.79) & 0.00 (0.00) & 0.00 (0.00) & 1.00 (0.00)\\
MG & 101 & 0.17 & 2.75 (1.47) & 2.72 (1.44) & 0.00 (0.02) & 0.00 (0.04)\\
MN & 38 & 0.21 & 1.58 (1.31) & 0.00 (0.00) & 0.00 (0.00) & 1.00 (0.00)\\
MY & 177 & 0.76 & 0.47 (1.08) & 0.01 (0.08) & 0.00 (0.00) & 0.98 (0.15)\\
NI & 98 & 0.29 & 1.74 (1.44) & 0.62 (0.81) & 0.00 (0.04) & 0.58 (0.41)\\
NP & 326 & 0.32 & 1.87 (1.89) & 0.16 (0.50) & 0.02 (0.10) & 0.89 (0.27)\\
PE & 85 & 0.49 & 0.67 (0.85) & 0.05 (0.26) & 0.00 (0.00) & 0.93 (0.26)\\
PQ & 1533 & 0.80 & 0.27 (0.62) & 0.00 (0.04) & 0.00 (0.00) & 0.99 (0.10)\\
PS & 91 & 0.09 & 2.34 (1.74) & 1.32 (1.42) & 0.11 (0.22) & 0.41 (0.40)\\
PT & 68 & 0.21 & 1.40 (1.02) & 0.60 (0.78) & 0.00 (0.00) & 0.54 (0.46)\\
RA & 54 & 0.06 & 2.96 (1.74) & 0.02 (0.14) & 0.00 (0.00) & 1.00 (0.02)\\
SH & 46 & 0.41 & 1.22 (1.50) & 0.00 (0.00) & 0.00 (0.00) & 1.00 (0.00)\\
SN & 63 & 0.57 & 1.00 (1.61) & 0.03 (0.18) & 0.05 (0.20) & 0.95 (0.20)\\
TE & 346 & 0.22 & 2.04 (1.62) & 0.44 (0.64) & 0.22 (0.37) & 0.65 (0.42)\\
TI & 74 & 0.09 & 3.07 (1.75) & 0.04 (0.26) & 0.00 (0.00) & 0.99 (0.08)\\
TM & 129 & 0.69 & 0.68 (1.22) & 0.26 (0.62) & 0.00 (0.00) & 0.57 (0.42)\\
TP & 94 & 0.39 & 1.47 (1.68) & 0.21 (0.41) & 0.05 (0.17) & 0.81 (0.31)\\
TS & 50 & 0.70 & 0.62 (1.09) & 0.00 (0.00) & 0.00 (0.00) & 1.00 (0.00)\\
TZ & 192 & 0.43 & 0.76 (1.03) & 0.55 (0.60) & 0.70 (0.45) & 0.16 (0.35)\\
UP & 470 & 0.13 & 3.25 (2.38) & 0.23 (0.65) & 0.01 (0.06) & 0.92 (0.19)\\
VT & 34 & 0.38 & 1.24 (1.18) & 0.00 (0.00) & 0.00 (0.00) & 1.00 (0.00)\\
YI & 69 & 0.30 & 1.45 (1.37) & 0.87 (1.01) & 0.13 (0.30) & 0.34 (0.44)\\
YN & 39 & 0.13 & 2.46 (1.57) & 1.26 (1.35) & 0.03 (0.17) & 0.51 (0.38)\\*
\end{longtable}
\endgroup{}

\clearpage
\subsection{Network prompts for each network layer}\label{si:name_gens}

Tables~\ref{tab:prompts-standardized-money-lending-question}-\ref{tab:prompts-other-important-external-ties} show the English-language version of the network prompt asked in each community. In some cases (especially for the gender-specific prompts), the intended respondent (shown above in Table~\ref{tab:su_summary}) influenced the phrasing of the social support network prompts. For example, in AH, where the intended respondent was the male head of household, the prompt on female-oriented labour/behavioural support was phrased around the \emph{wife's} relations, while the prompt on male-oriented labour/behavioural support was phrased around the respondent's own relations.

\begingroup\fontsize{8}{10}\selectfont

\begin{longtable}[t]{l>{\raggedright\arraybackslash}p{15cm}}
\caption{\label{tab:prompts-standardized-money-lending-question}Name generator prompts for borrowing money}\\
\toprule
Community & Prompt\\
\midrule
\endfirsthead
\caption[]{Name generator prompts for borrowing money \textit{(continued)}}\\
\toprule
Community & Prompt\\
\midrule
\endhead

\endfoot
\bottomrule
\endlastfoot
AH & In the event of an urgent need, from whom could you borrow \$175 kina?\\
AP & If someone in your house broke their leg and needed to go to Lethem right away to fix it and y'all did not have money, what are the names of the persons you would go to to borrow the money you need?\\
AR & If you needed money, who could you ask to lend you 20,000 pesos?\\
AV & Imagine you have to borrow 10.000 francs to pay for healthcare reasons, who in the community would lend you the money for this?\\
AZ & If you had an unexpected emergency expense, such as a hospital medical treatment, from whom could you get a loan of 2000 Rs [a weeks wages] or more?\\
BD & In the event of an urgent and unexpected need, such as a medical emergency, who could you or anyone in your khana borrow 2000 taka from?\\
BM & When you have an urgent and unexpected need, e.g. a medical emergency, from whom could you borrow the equivalent of \$1700 pesos?\\
BY & If you need 1000cfa, who could you ask for help with 1000cfa?\\
CP & In the case of an urgent or unexpected need, such as a medical emergency, from whom could you borrow 100 dollars?\\
CR & When you have an urgent and unexpected need, perhaps related to a medical emergency, from whom could you borrow the equivalent of one week's wages?\\
DJ & Imagine you have to borrow 10.000 francs to pay for healthcare reasons, who in the community would lend you the money for this?\\
EK & When you have an urgent and unexpected need (e.g. a medical emergency), from whom could you borrow the equivalent of Rs1,500?\\
FF & When you have an urgent and unexpected need, e.g. a medical emergency, from whom could you borrow 300 Namibian dollars?\\
FJ & If your household were experiencing trouble, such as a medical emergency, and you(plural) needed to borrow FJ\$300, which households might lend you(plural) that money?\\
HE & When you are in need of some help, from whom could you borrow the equivalent of K120?\\
HI & When you have an urgent and unexpected need, perhaps related to a medical emergency, from whom could you borrow the equivalent of one week's wages?\\
IH & Imagine that you needed to borrow some money in order to pay for an important and expensive medical treatment, who in this community do you think would give you 100,000 pesos to help you pay?\\
KA & When you have something urgent and unexpected, such as emergency medical expenses, from whom can you borrow a week's salary?\\
KM & In the event of an urgent need, from whom could you borrow the equivalent of a weeks wages?\\
KO & In the event of an urgent need, from whom could you BORROW the equivalent of a week's wages?\\
KS & When you have an urgent and unexpected need, perhaps related to a medical emergency, from whom could you borrow 250 soles?\\
KT & If you had a medical emergency, who would let you borrow 250 RM to seek help?\\
LB & From who would you borrow a week's worth of wages if you had an unexpected need?\\
MA & When you have an urgent and unexpected need, e.g. a medical emergency, from whom could you borrow 30,000 tshillings (cost of one adult goat)?\\
MC & If someone in your house broke their leg and needed to go to Lethem right away to fix it and y'all did not have money, what are the names of the persons you would go to to borrow the money you need?\\
MG & When you have an unexpected need (for example, a medical emergency), from whom can you ask to borrow 20,000 ariary or more?\\
MK & If you have a problem that would require you to borrow 20,000 MGA, who would you ask for the loan?\\
MN & If you or someone from your household had an emergency, such as a medical emergency, and you needed a loan of 600 Bolivianos, who, friends or family, could lend you this money?\\
MY & Who would you BORROW money from equivalent to a week's salary?\\
NI & In the case of an emergency, such as a medical situation involving you, who would let you borrow 1,500 cordobas?\\
NP & If you had an unexpected emergency expense, for example for a treatment at the hospital, who could lend you NPR 2,000 or more?\\
PC & Imagine that you needed to borrow some money in order to pay for an important and expensive medical treatment, who in this community do you think would give you 100,000 pesos to help you pay?\\
PE & To whom would you turn for a sum of money, say 20,000/-, if they had some sort of an emergency like an illness or accident?\\
PQ & In the event of an urgent need, who would lend you 400,000 pesos? For example, imagine you needed to borrow money to pay for an important and expensive medical treatment, who would lend you 400,000 pesos to help pay for the treatment?\\
PT & When you have an urgent and unexpected need, perhaps related to a medical emergency, from whom could you borrow the equivalent of one week's wages? [25 litres of gas]\\
RA & When you have an emergency, e.g. you or your child is sick, who can you borrow 25 Gourdes from?\\
SH & In the event of an urgent and unexpected need, such as a medical emergency, who could you borrow 2000 taka from?\\
SN & In the event of an urgent need, e.g. a medical emergency, from whom could you borrow 1000 MT?\\
TE & If you had an unexpected emergency expense, such as a hospital medical treatment, from whom could you get a loan of 2000 Rs [a weeks wages] or more?\\
TI & When you have sudden need, e.g. a medical emergency, from whom could you borrow 400 Bolivianos?\\
TM & When youre in urgent need, for instance, a medical emergency, to whom can you ask for a 1,000 peso loan?\\
TP & When you have an urgent and unexpected need, e.g. a medical emergency, from whom could you borrow \$175 SRD (Surinamese Dollars)?\\
TS & If you or someone from your household had an emergency, such as a medical emergency, and you needed a loan of 350 Bolivianos, whom of your friends and family could you ask for this loan?\\
TZ & In the event of an urgent need such as a medical or family emergency, from whom could you or other members of your household borrow 400 or more dirhams from?\\
UP & If you had an unexpected emergency expense, for example for a treatment at the hospital, who all could lend you INR 2,500 or more?\\
VT & If you needed about 8,000 vatu for an unexpected reason, who would you ask?\\
WH & Imagine that you needed to borrow some money in order to pay for an important and expensive medical treatment, who in this community do you think would give you 100,000 pesos to help you pay?\\
WL & Imagine that you needed to borrow some money in order to pay for an important and expensive medical treatment, who in this community do you think would give you 100,000 pesos to help you pay?\\
YI & In the event of an urgent need, from whom could you borrow 500 RMB?\\
YN & From who would you borrow a week's worth of wages if you had an unexpected need?\\*
\end{longtable}
\endgroup{}

\begingroup\fontsize{8}{10}\selectfont

\begin{longtable}[t]{l>{\raggedright\arraybackslash}p{15cm}}
\caption{\label{tab:prompts-double-sampled-money-lending-question}Name generator prompts for lending money}\\
\toprule
Community & Prompt\\
\midrule
\endfirsthead
\caption[]{Name generator prompts for lending money \textit{(continued)}}\\
\toprule
Community & Prompt\\
\midrule
\endhead

\endfoot
\bottomrule
\endlastfoot
AH & In the event of an urgent need, to whom would you lend \$175 kina?\\
AP & Lets say someone broke their leg and needed to go to Lethem right away to fix it and they did not have money, but someone in your house has the money they would need (maybe 15,000 dollars). What are the names of the persons who would come to your house to borrow that money from you?\\
AR & If someone needed money, to whom would you lend 20,000 pesos?\\
AV & To whom would you lend 10.000 francs for health expenditure in the community?\\
AZ & In the event of an unexpected emergency expense, such as a hospital medical treatment, who would come ask you for a loan of 2000 Rs or more?\\
BD & In the event of an urgent and unexpected need, such as a medical emergency, who would you or anyone in your khana lend 2000 taka to?\\
BM & In the event of an urgent need, to whom would you lend the equivalent of \$1700 pesos?\\
BY & If someone else needs 1000 cfa, who can ask you for help with 1000cfa?\\
CP & And to whom would you loan the same amount of money in the same situation?\\
CR & In the event of an urgent need, to whom would you lend the equivalent of a week's wages?\\
DJ & To whom would you lend 10.000 francs for health expenditure in the community?\\
EK & To whom would you lend the equivalent of Rs1,500 in the event of an urgent and unexpected need (e.g. a medical emergency)?\\
FF & In the event of an urgent need, to whom would you lend 300 Namibian dollars?\\
FJ & Which households might you(plural) lend FJ\$300 to if they were experiencing trouble?\\
HE & Suppose someone is in need of help, to whom would you lend about K120?\\
HI & In the event of an urgent need, to whom would you lend the equivalent of a weeks wages?\\
IH & (not asked)\\
KA & In an emergency, who would you lend your income for a week to?\\
KM & In the event of an urgent need, to whom would you lend the equivalent of a weeks wages?\\
KO & In the event of an urgent need, from whom would you LEND the equivalent of a week's wages?\\
KS & In an event of an urgent need, to whom would you lend 250 soles?\\
KT & Who would you lend 250 RM to in case of urgent need?\\
LB & To whom would you lend a week's worth of wages if they had an unexpected need?\\
MA & In the event of an urgent need, to whom would you lend 30,000 tshillings?\\
MC & Lets say someone broke their leg and needed to go to Lethem right away to fix it and they did not have money, but someone in your house has the money they would need (maybe 15,000 dollars). What are the names of the persons who would come to your house to borrow that money from you?\\
MG & To whom would you lend 20,000 ariary or more if they had an unexpected need?\\
MK & If there are others with problems that require them to borrow 20,000 MGA, who would you loan this sum to?\\
MN & Who (friends or family) could ask you for a loan of 600 bolivianos if they had an emergency, for example a medical emergency?\\
MY & Who would you LEND money to equivalent to a week's salary?\\
NI & In the case of an emergency, such as a medical situation involving you, to whom would you lend 1,500 cordobas?\\
NP & Who would come to ask you for a loan of NPR 2,000 or more, if they had an unexpected emergency expense, for example for a treatement at the hospital?\\
PC & (not asked)\\
PE & Who would lend you lend a sum of money, say 20,000/-, if you had some sort of an emergency like an illness or accident?\\
PQ & To which people in this community would you lend 400,000 pesos to help them if they had an urgent need? (e.g., to pay for an important and expensive medical treatment).\\
PT & In the event of an urgent need, to whom would you lend the equivalent of a week's wages? [25 litres of gas]\\
RA & When an emergency occurs, e.g. a child is sick, who would you lend 25 Gourdes to?\\
SH & In the event of an urgent and unexpected need, such as a medical emergency, who would you lend 2000 taka to?\\
SN & In the event of an urgent need to whom would you lend 1000 MT?\\
TE & In the event of an unexpected emergency expense, such as a hospital medical treatment, who would come ask you for a loan of 2000 Rs or more?\\
TI & If someone had a sudden need, for example, a medical emergency, to whom would you loan 400 Bolivianos?\\
TM & In case of urgent need, to whom would you give a loan of 1000 pesos?\\
TP & In the event of an urgent need, to whom would you lend the equivalent of \$175 SRD to?\\
TS & To whom would you and your spouse lend 350 bolivianos, if they needed it for an emergency such as a medical emergency?\\
TZ & In the event of an urgent need such as a medical or family emergency, to whom would you or other members of your household lend 400 or more dirhams?\\
UP & Who all would come to ask you for a loan of INR 2,500 or more, if they had an unexpected emergency expense, for example for a treatment at the hospital?\\
VT & If someone else in the community needed 8,000 vatu unexpectedly, who would you give it to?\\
WH & (not asked)\\
WL & (not asked)\\
YI & In the event of an urgent need, to whom would you lend 500 RMB?\\
YN & To whom would you lend a week's worth of wages if they had an unexpected need?\\*
\end{longtable}
\endgroup{}

\begingroup\fontsize{8}{10}\selectfont

\begin{longtable}[t]{l>{\raggedright\arraybackslash}p{15cm}}
\caption{\label{tab:prompts-household-level-exchange-question}Name generator prompts for borrowing goods}\\
\toprule
Community & Prompt\\
\midrule
\endfirsthead
\caption[]{Name generator prompts for borrowing goods \textit{(continued)}}\\
\toprule
Community & Prompt\\
\midrule
\endhead

\endfoot
\bottomrule
\endlastfoot
AH & If they have plenty of fish, which households often share fish with you?\\
AP & Who would give someone in your house little bit of beef, chicken, wild meat, eggs, farine?\\
AR & When you are lacking a basic necessity such as flour, rice, sugar, oil, or other food items, whom do you ask to borrow from?\\
AV & Think about the past month. We would like you to nominate the people who have given you food or money\\
AZ & For your basic essentials such as rice, sugar, oil and other groceries and household needs, who could you immediately ask and get?\\
BD & Who could you borrow any food items or small household items from if you and your khana needed?\\
BM & From what ranches do you request water?\\
BY & Who shared food with you?\\
CP & If it were necessary, whom could you ask to gift you a sack of potatoes?\\
CR & Which households often share food (bring to you) with you?  For example, in the last week, who brought food to you?\\
DJ & Think about the past month. We would like you to nominate the people who have given you food or money\\
EK & From which other HH would you borrow basic items (e.g. rice, agricultural tools and kitchen utensils)?\\
FF & If your household needs sugar, whom do you usually ask to give you some?\\
FJ & Which households have given food to you(plural) in the past month?\\
HE & Which households normally share food with you?\\
HI & Which compounds could you go to if you needed to borrow an axe?\\
IH & Thinking about the last month or two, can you gives the names of the people outside of this household helped you by loaning or giving you food, money, or something else that you needed, but didn't have at that time?\\
KA & Who are people outside of your household who often cooks food for you?\\
KM & Which other households in the village do you usually receive salmon/reindeer/potatoes from?\\
KO & What households do you get food from?\\
KS & Which households often share food with you, for example bananas or game meat?\\
KT & What families share food with your family?\\
LB & Who would you borrow farm equipment from?\\
MA & Which households often share food with you?\\
MC & Who would give someone in your house little bit of beef, chicken, wild meat, eggs, farine?\\
MG & Which households regularly share food/meals with you?\\
MK & Who shares with you when they have lots of food?\\
MN & In the past 30 days (from this day last month until today), who has invited you to eat or drink with them OR brought you food or produce? How many times?\\
MY & Who would you BORROW items from?\\
NI & Who gives meat to your family?\\
NP & If you need some basic essentials---for example oil, masala, uncooked rice or lentils---who could you ask and obtain from?\\
PC & Thinking about the last month or two, can you gives the names of the people outside of this household helped you by loaning or giving you food, money, or something else that you needed, but didn't have at that time?\\
PE & Which households commonly help you with food? For example, people who will bring food (prepared or unprepared) to your house?\\
PQ & Thinking back over the last month, who are the people outside this household who helped you by giving you food, lending you money, or anything else you needed but didn't have at the time?\\
PT & In the last month, from whom did you borrow day-by-day things, such as mate, hoe, pliers ?\\
RA & Who shares household items with you (e.g. chairs, machete, knife)?\\
SH & Which households would share food/meals with you and your household if you needed?\\
SN & Who from another household offered you food (cooked or uncooked) when you are at their household?\\
TE & For your basic essentials such as rice, sugar, oil and other groceries and household needs, who could you immediately ask and get?\\
TI & Who has shared food with you?\\
TM & What households (or people) normally share food (corn, beans, meat, fruits) with you?\\
TP & Which households often share meat with you?\\
TS & In the past 30 days (from this same day last month until today), who gave food to you or anyone else in your household?\\
TZ & Who are the people who have shared milk, butter, raib, or lben with you and/or other members of your household over the past year?\\
UP & If you need some basic essentials---for example oil, masala, uncooked rice or lentils---who all could you ask and obtain from?\\
VT & Who do you ask to use their things?\\
WH & Thinking about the last month or two, can you gives the names of the people outside of this household helped you by loaning or giving you food, money, or something else that you needed, but didn't have at that time?\\
WL & Thinking about the last month or two, can you gives the names of the people outside of this household helped you by loaning or giving you food, money, or something else that you needed, but didn't have at that time?\\
YI & Who could you ask to borrow small household items or farm equipment from? (e.g. rice cooker, tractor)?\\
YN & Who would you borrow farm equipment from?\\*
\end{longtable}
\endgroup{}

\begingroup\fontsize{8}{10}\selectfont

\begin{longtable}[t]{l>{\raggedright\arraybackslash}p{15cm}}
\caption{\label{tab:prompts-double-sampled-household-level-exchange-question}Name generator prompts for lending goods}\\
\toprule
Community & Prompt\\
\midrule
\endfirsthead
\caption[]{Name generator prompts for lending goods \textit{(continued)}}\\
\toprule
Community & Prompt\\
\midrule
\endhead

\endfoot
\bottomrule
\endlastfoot
AH & If you have plenty of fish, which households do you often share fish with?\\
AP & Who would someone in your house give little bit of beef, chicken, wild meat, eggs, or farine to?\\
AR & If someone is lacking a basic necessity such as flour, rice, sugar, oil, or other food items, to whom would you give it?\\
AV & Think about the past month. We would like you to nominate the people to whom you have given food or money\\
AZ & For their basic essentials such as rice, sugar, oil and other groceries and household needs, who could immediately ask and get from you?\\
BD & Who would you loan any food items or small household items to if they needed?\\
BM & To what ranches do you give water?\\
BY & Who do you share food with?\\
CP & Are there any families to whom you have gifted a sack of potatoes in this past year?\\
CR & Which households often share food with you?  For example, in the last week, who did you bring food to?\\
DJ & Think about the past month. We would like you to nominate the people to whom you have given food or money\\
EK & Which other households borrow basic items from your HH (e.g. rice, agricultural tools and kitchen utensils)?\\
FF & Which household usually ask you for sugar when they are in need of some?\\
FJ & Which households have you(plural) given food to in the past month?\\
HE & Which households do you normally share food with?\\
HI & To which compounds would you lend an axe if they needed to borrow one?\\
IH & Thinking about the last month or two, can you gives the names of the people outside of this household who you helped by loaning or giving food, money, or something else that they needed, but didn't have at that time?\\
KA & Who are people outside your household that you often make food for?\\
KM & Which other households in the village do you usually give salmon/reindeer/potatoes to?\\
KO & What households give food to you?\\
KS & Which households do you often share food with, for example bananas or game meat?\\
KT & To which families does your family give food?\\
LB & Who would you lend farm equipment to?\\
MA & Which households do you often share food with?\\
MC & Who would someone in your house give little bit of beef, chicken, wild meat, eggs, or farine to?\\
MG & With which households do you regularly share meals?\\
MK & When you gain a large quantity of food, who do you share it with?\\
MN & In the past 30 days (from this day last month until today), whom have you or your spouse invited to eat or drink with you OR brought food or produce to? How many times?\\
MY & Who would you LEND items to?\\
NI & To which families do you give meat?\\
NP & If they need some basic essentials---for example oil, masala, uncooked rice or lentils---who could ask you and obtain?\\
PC & Thinking about the last month or two, can you gives the names of the people outside of this household who you helped by loaning or giving food, money, or something else that they needed, but didn't have at that time?\\
PE & To which households do you often give food? For example, people to whom you will take food (prepared or unprepared) to their house?\\
PQ & Thinking back over the last month, who are the people outside this household who you helped by lending them food, money, or anything else they needed but didnt have at the time?\\
PT & In the last month, to whom did you lend day-by-day things, such as mate, hoe, pliers ?\\
RA & Who do you share househld items with (e.g. chairs, machete, knife)?\\
SH & Which households would your household share food/meals with?\\
SN & To whom from another household, when they are visiting your household, has your household frequently shared food (cooked/uncooked) recently? (If no response: Such as in the last month or this season?)\\
TE & For their basic essentials such as rice, sugar, oil and other groceries and household needs, who could immediately ask and get from you?\\
TI & Who have you given food to?\\
TM & With which households (or people) do you normally share food (corn, beans, meat, fruit)?\\
TP & Which households do you share meat with?\\
TS & In the past 30 days (from this same day last month until today), to whom have you or others in your household given food?\\
TZ & With whom have you and/ or other members of your household shared milk, butter, raib, or lben over the past year?\\
UP & If they need some basic essentials---for example oil, masala, uncooked rice or lentils---who all could ask you and obtain?\\
VT & Who asks you if they can use your things?\\
WH & Thinking about the last month or two, can you gives the names of the people outside of this household who you helped by loaning or giving food, money, or something else that they needed, but didn't have at that time?\\
WL & Thinking about the last month or two, can you gives the names of the people outside of this household who you helped by loaning or giving food, money, or something else that they needed, but didn't have at that time?\\
YI & Who would you lend small household items or farm equipment to if they needed?\\
YN & Who would you lend farm equipment to?\\*
\end{longtable}
\endgroup{}

\begingroup\fontsize{8}{10}\selectfont

\begin{longtable}[t]{l>{\raggedright\arraybackslash}p{15cm}}
\caption{\label{tab:prompts-female-oriented-Q-network}Name generator prompts for female-oriented labor/behavioral support}\\
\toprule
Community & Prompt\\
\midrule
\endfirsthead
\caption[]{Name generator prompts for female-oriented labor/behavioral support \textit{(continued)}}\\
\toprule
Community & Prompt\\
\midrule
\endhead

\endfoot
\bottomrule
\endlastfoot
AH & When your wife does women's work with others (e.g., gleaning, handicrafts), who does she do it with?\\
AP & Who would you weed farm with?\\
AR & (If the respondent is female) If you (and the other women in your household) need help with a task, who helps you? For example, caring for animals or crops.\\
AV & Think about the past month. We would like you to nominate the people that you have spent time working with on a daily or weekly basis\\
AZ & For you and/or the other women in your household, who happily helps you with tasks [question in Tamil implies physical assistance]\\
BD & If you or any other women in this khana needed help caring for your children or grand-children, who outside of this household could you ask for help?\\
BM & With whom would you get together with to sell items?\\
BY & Who do you (f) /your wife forage (literally, walk in the forest) with?\\
CP & If there were no women at home to prepare dinner, perhaps because she/they were sick or traveling, who would cook in her/their place?\\
CR & Who helps you in your work?\\
DJ & Think about the past month. We would like you to nominate the people that you have spent time working with on a daily or weekly basis\\
EK & Who could women in the HH call upon for help (e.g. with house chores so that they can work) at short notice?\\
FF & Who provides childcare for you so that you can attend to a task say in Fransfontein or Khorixas the town?\\
FJ & With whom do you and the other the women in this household work together, such as in gathering on the reef or collecting pandanus?\\
HE & Who goes most often to work in the garden with the women of the househould?\\
HI & Who do your wives work in their gardens with?\\
IH & Thinking about the last month, can you gives the names of the people you spent time working with (e.g., fishing with, working with in the field, working with in the same place of work)?\\
KA & Who watches your children (so you can work)?\\
KM & Which households in the village give help with childcare, processing food, or other household tasks?\\
KO & Who provides childcare (weaving other, ensete scraping) for you (so that you can work in fields, get water, etc)?\\
KS & Who provides an extra hand in the field when you (and other women in your household) need it (for sowing peanut, manioc or 'sachapapa')?\\
KT & Women: with whom do you cooperate?\\
LB & Who would help you prepare a meal for a community event like a wedding or funeral?\\
MA & Who provides childcare for your wives so they can go get water/firewood or go to church/market?\\
MC & Who would you weed farm with?\\
MG & With whom do you regularly fish or farm? [Asked only of women respondents; men respondents not asked about their wives.]\\
MK & (One joint question asked) If there was an agricultural wage labor opportunity that required a team of workers, who would you invite to join you?\\
MN & If [the female HH] needed help with work, such as childcare, cleaning clothes, or preparing food - for example, if [she] was ill, who could you call on for help (without paying them)?\\
MY & Who helps you? (women only, childcare \& cooking as examples)\\
NI & For women's work that requires a helping hand, who helps the women in your family?\\
NP & For [you and the other women / the women] in your household, who helps you with tasks? For example, cooking, cleaning, housework. Or working in khet-baari, for example by doing parma?\\
PC & Thinking about the last month, can you gives the names of the people you spent time working with (e.g., fishing with, working with in the field, working with in the same place of work)?\\
PE & With whom do you commonly work with, when you are farming, doing household tasks, or business, etc.)?\\
PQ & Thinking back over the past month, who are people you have spent time working with (i.e., fishing, working in the fields, working at the same workplace).\\
PT & Regarding the women here in household, when they go fishing, gather bait or hunting, with whom they go with?\\
RA & (Female HH) When you are sick, who comes to help you with cleaning and cooking?\\
SH & To a woman: If you needed help caring for your children, who outside of this household could you ask for help? To a man: If your wife needed help caring for your children, who outside of this household could she ask for help?\\
SN & If you (your wife) is sick, who from another household would help with her work?\\
TE & For you and/or the other women in your household, who happily helps you with tasks [question in Tamil implies physical assistance]\\
TI & Who provides childcare to your children, when the women in your house need help?\\
TM & Who usually helps the women in your house with work (for example, preparing food for a fiesta)?\\
TP & With whom do you cooperate on agricultural/field labor?\\
TS & When you need help with household work, for example to help look after your children, or to help prepare food or chicha, whom can you ask for help?\\
TZ & {}[If participant is a woman] Who helps you without pay in your daily work activities (i.e., agricultural work, attending to animals or housechores)\\
UP & For [you and the other women / the women] in your household, who all helps you with tasks? For example, cooking, cleaning, housework. Or working in kheti-baari.\\
VT & When you go to the market to sell, who do you go with?\\
WH & Thinking about the last month, can you gives the names of the people you spent time working with (e.g., fishing with, working with in the field, working with in the same place of work)?\\
WL & Thinking about the last month, can you gives the names of the people you spent time working with (e.g., fishing with, working with in the field, working with in the same place of work)?\\
YI & If your wife or any women in your household needed help with household work (e.g. feeding pigs, watching children, etc.), whom could they contact for help?\\
YN & Who would help you prepare a meal for a community event like a wedding or funeral?\\*
\end{longtable}
\endgroup{}

\begingroup\fontsize{8}{10}\selectfont

\begin{longtable}[t]{l>{\raggedright\arraybackslash}p{15cm}}
\caption{\label{tab:prompts-male-oriented-Q-network}Name generator prompts for male-oriented labor/behavioral support}\\
\toprule
Community & Prompt\\
\midrule
\endfirsthead
\caption[]{Name generator prompts for male-oriented labor/behavioral support \textit{(continued)}}\\
\toprule
Community & Prompt\\
\midrule
\endhead

\endfoot
\bottomrule
\endlastfoot
AH & With whom do you fish?\\
AP & Who would you go fishing with?\\
AR & (If the respondent is female) When the man/men in your household need help with a task, who helps them? For example, caring for animals or crops. (If the respondent is male) When you (and the other men in your household) need help with a task, who helps you? For example, caring for animals or crops.\\
AV & Think about the past month. We would like you to nominate the people that you have spent time working with on a daily or weekly basis\\
AZ & For you and/or the other men in your household, who happily helps you with tasks [question in Tamil implies physical assistance]\\
BD & Who outside this household can your husband or any other men in this khana ask for help if they have any work to do in the house or community?\\
BM & Who do you ask when livestock go missing?\\
BY & Who do you (m) /your husband (if divorced/widowed, brother living in the house, or teenage son) forage with?\\
CP & If there were no men at home to cut firewood, perhaps because he/they were sick or traveling, who would do it in his/their place?\\
CR & With whom does your husband work?\\
DJ & Think about the past month. We would like you to nominate the people that you have spent time working with on a daily or weekly basis\\
EK & Who could men in this HH call upon for help (e.g. to prepare the field, to do a manly job) at short notice?\\
FF & With whom do you search for your lost livestock?\\
FJ & With whom do the men in this household work together, such as in farming or diving?\\
HE & Who goes most often to work in the garden with the men of the househould?\\
HI & Who would you ask to help you dig a sandwell?\\
IH & Thinking about the last month, can you gives the names of the people you spent time working with (e.g., fishing with, working with in the field, working with in the same place of work)?\\
KA & With whom do you work in the fields?\\
KM & Which households in the village do you usually hunt or go ice fishing with?\\
KO & With whom do you work in the fields or take care of livestock?\\
KS & Who provides an extra hand for the day when the men in your household need it? With whom would they clear the forest before sowing?\\
KT & Men: with whom do you cooperate?\\
LB & Who would help you with home repairs if you needed?\\
MA & If you had to dig a waterhole for livestock who would help you?\\
MC & Who would you go fishing with?\\
MG & With whom do you (if man) / your husband (if married woman, otherwise not asked) regularly fish or farm?\\
MK & (One joint question asked) If there was an agricultural wage labor opportunity that required a team of workers, who would you invite to join you?\\
MN & Whom does [the male HH] engage in aine with for work purposes, such as to chop down [their] field, or harvest lumber, or for construction projects?\\
MY & Who helps you? (men only)\\
NI & For men's work that requires a helping hand, who helps the men in your family?\\
NP & For [you and the other men / the men] in your household, who helps you with tasks? For example, doing work around the house? Or working in khet-baari, for example by doing parma?\\
PC & Thinking about the last month, can you gives the names of the people you spent time working with (e.g., fishing with, working with in the field, working with in the same place of work)?\\
PE & With whom do you commonly work with, when you are farming, fishing or doing business?\\
PQ & Thinking back over the past month, who are people you have spent time working with (i.e., fishing, working in the fields, working at the same workplace).\\
PT & Regarding the men here in household, when they go fishing, gather bait or hunting, with whom they go with?\\
RA & (Male HH) Who helps you when you need assistance in your farm or around your house?\\
SH & To a man: Who outside this household can you ask if you need help repairing your boat or house? To a woman: Who outside this household can your husband ask if he needs help repairing the boat or house?\\
SN & If you (your husband) is sick, who from another household would help with his work?\\
TE & For you and/or the other men in your household, who happily helps you with tasks [question in Tamil implies physical assistance]\\
TI & With whom do the men in your house work?\\
TM & When you and the men in your house need help working in the cornfield and other jobs, who helps you?\\
TP & With whom do you clear fields with (if manioc beer not present)?\\
TS & When your husband needs help with work, for example to cut down his field, or to chop wood for lumber, or to build a house or other buildings, who can help him (paid labor excluded)?\\
TZ & {}[If participant is a man] Who helps you without pay in your daily work activities (i.e., wage labor, sale of livestock in the market or other market activities)\\
UP & For [you and the other men / the men] in your household, who all helps you with tasks? For example, doing work around the house? Or working in kheti-baari.\\
VT & Who would you help in the garden if they needed extra hands?\\
WH & Thinking about the last month, can you gives the names of the people you spent time working with (e.g., fishing with, working with in the field, working with in the same place of work)?\\
WL & Thinking about the last month, can you gives the names of the people you spent time working with (e.g., fishing with, working with in the field, working with in the same place of work)?\\
YI & If you or any other men in your household needed help with household work (e.g. feeding pigs, watching children, etc.), whom could they contact for help?\\
YN & Who would help you with home repairs if you needed?\\*
\end{longtable}
\endgroup{}

\begingroup\fontsize{8}{10}\selectfont

\begin{longtable}[t]{l>{\raggedright\arraybackslash}p{15cm}}
\caption{\label{tab:prompts-female-oriented-K-network}Name generator prompts for female-oriented informational/affective support}\\
\toprule
Community & Prompt\\
\midrule
\endfirsthead
\caption[]{Name generator prompts for female-oriented informational/affective support \textit{(continued)}}\\
\toprule
Community & Prompt\\
\midrule
\endhead

\endfoot
\bottomrule
\endlastfoot
AH & At a big community event or gathering (e.g., church, soccer game, community meeting), who would your wife usually sit with and talk with?\\
AP & Who would come to your house to gaffe or you would go to their house to gaffe?\\
AR & (If the respondent is female) With whom do you (and the other women in your household) talk during times of rest, when you are not very busy with household or work responsibilities? (If the respondent is male) With whom do the woman/women in the household talk during times of rest, when they are not very busy with household or work responsibilities?\\
AV & Think about the past month. We would like you to nominate the people that you have spent a lot of time socializing with\\
AZ & For you and/or the other women in your household, who do you happily and casually have conversations with?\\
BD & Who do you or any other women in this khana usually spend time talking with or visiting?\\
BM & When you have to leave for the city, who would you ask to watch your ranch?\\
BY & Who do you (f)/your wife like to gossip with in the evening?\\
CP & With whom do the women in your family visit on Sundays at the market?\\
CR & Which are the friends and relatives with whom you spend most of your free time?\\
DJ & Think about the past month. We would like you to nominate the people that you have spent a lot of time socializing with\\
EK & Who visits the women of this HH on Sundays?\\
FF & Who do women from this household sometimes visit on afternoons?\\
FJ & With whom do you and the other women in this household sometimes visit for conversation, for example after lunch?\\
HE & With whom do the women of the household share stories most often?\\
HI & Who do your wives visit regularly (at least once a month)?\\
IH & Thinking about the last month, can you gives the names of the people with whom you have hung out with regularly (e.g., chatting, drinking beers, talking about the day, playing games)?\\
KA & With whom do you hang out with on Sundays?\\
KM & Which other households in the village did you drink tea with in the past week? (Female HH Head)\\
KO & Who do you go to church/mosque with?\\
KS & With whom do you (and other women in your household) visit when you have spare time, when work in the fields is finished or the days you rest?\\
KT & Women: with whom do you visit in your free time?\\
LB & With whom do the women hang out after dinner?\\
MA & With whom do you go to church with?\\
MC & Who would come to your house to gaffe or you would go to their house to gaffe?\\
MG & When you have leisure time, who do you spend it with? [Asked only of women respondents; men respondents not asked about their wives.]\\
MK & (One joint question asked) If you need advice, who would you go to?\\
MN & Who are the people that [the female HH] likes to spend time with socially, to visit and chat with, or to go to parties and gatherings with?\\
MY & Who in the village to you visit weekly? (women only)\\
NI & With whom do the women in your family visit on Sundays?\\
NP & For [you and the other women / the women] in your household, who do you enjoy having casual conversations with?\\
PC & Thinking about the last month, can you gives the names of the people with whom you have hung out with regularly (e.g., chatting, drinking beers, talking about the day, playing games)?\\
PE & With whom do you hang out when you are not busy, or when you would like to chat and/or get advice?\\
PQ & Thinking back over the past month, who are the people you spend time with regularly? (This could be friends, family, or people you socialize most with).\\
PT & Regarding the women here in household, who do they visit during the weekend to polish their nails or to drink a mate?\\
RA & (Female HH) When you have freetime during the week, who do you go and visit?\\
SH & To a woman: Who are the people you usually spend time talking with?  To a man: Who are the people your wife usually spends time talking with?\\
SN & Who do you (the women in the household) like to exchange anecdotes and news with from other households?\\
TE & For you and/or the other women in your household, who do you happily and casually have conversations with?\\
TI & With whom do the women in your house visit?\\
TM & Who do you and the women in your house normally talk to (i.e. when you are not working)?\\
TP & With whom do you (women) sometimes visit on Sunday afternoons?\\
TS & In the past 30 days, who did you (or other adult women of the household) visit, to chat or drink chicha? Who came to visit you?\\
TZ & {}[Women only] Who are the people you like to sit outside with in the afternoon and chat?\\
UP & For [you and the other women / the women] in your household, who all do you enjoy having casual conversations with?\\
VT & Who do the women in your household consider to be their close friends?\\
WH & Thinking about the last month, can you gives the names of the people with whom you have hung out with regularly (e.g., chatting, drinking beers, talking about the day, playing games)?\\
WL & Thinking about the last month, can you gives the names of the people with whom you have hung out with regularly (e.g., chatting, drinking beers, talking about the day, playing games)?\\
YI & Who does your wife or other women in the household usually talk with or spend time together with?\\
YN & With whom do the women hang out after dinner?\\*
\end{longtable}
\endgroup{}

\begingroup\fontsize{8}{10}\selectfont

\begin{longtable}[t]{l>{\raggedright\arraybackslash}p{15cm}}
\caption{\label{tab:prompts-male-oriented-K-network}Name generator prompts for male-oriented informational/affective support}\\
\toprule
Community & Prompt\\
\midrule
\endfirsthead
\caption[]{Name generator prompts for male-oriented informational/affective support \textit{(continued)}}\\
\toprule
Community & Prompt\\
\midrule
\endhead

\endfoot
\bottomrule
\endlastfoot
AH & At a big community event or gathering (e.g., church, soccer game, community meeting), who would you usually sit with and talk to?\\
AP & Who would you talk to before selecting another toshao?\\
AR & (If the respondent is female) With whom do the man/men in your household talk during times of rest, when they are not very busy with household or work responsibilities? (If the respondent male) With whom do you (and the other men in your household) talk during times of rest, when you are not very busy with household or work responsibilities?\\
AV & Think about the past month. We would like you to nominate the people that you have spent a lot of time socializing with\\
AZ & For you and/or the other men in your household, who do you happily and casually have conversations with?\\
BD & Who do your husband or any other men in this khana usually spend time talking with or visiting?\\
BM & When you have to leave for the city, who would you ask to watch your ranch?\\
BY & Who do you (m)/your husband (rarely, teenage son/brother) like to gossip with in the evening?\\
CP & With whom do the men in your family visit on Sundays at the market?\\
CR & Which friends/ relatives does your husband mostly spend his free time?\\
DJ & Think about the past month. We would like you to nominate the people that you have spent a lot of time socializing with\\
EK & With whom do men of this HH go to religious services?\\
FF & Who do men from this household sometimes visit on afternoons?\\
FJ & With whom do the men in this household sometimes visit for conversation, for example in the evening?\\
HE & With whom do men of the household share stories most often?\\
HI & Who do you visit regularly (at least once a month)?\\
IH & Thinking about the last month, can you gives the names of the people with whom you have hung out with regularly (e.g., chatting, drinking beers, talking about the day, playing games)?\\
KA & With whom do you often chat with at the coffee shop?\\
KM & Which other households in the village did you drink tea with in the past week? (Male HH Head)\\
KO & Who do you go to church/mosque or tej bet/karibo bet?\\
KS & With whom do the men in your household get together to talk or share manioc beer when they have spare time?\\
KT & Men: with whom do you visit in your free time?\\
LB & With whom do the men hang out after dinner?\\
MA & Who do you discuss important matters with?\\
MC & Who would you talk to before selecting another toshao?\\
MG & When you (if man) / your husband (if married woman, otherwise not asked) has leisure time, who would he spend it with?\\
MK & (One joint question asked) If you need advice, who would you go to?\\
MN & Who are the people that [the male HH] likes to spend time with socially, to visit and chat with, or to go to parties and gatherings with?\\
MY & Who in the village to you visit weekly? (men only)\\
NI & With whom do the men in your family visit on Sundays?\\
NP & For [you and the other men / the men] in your household, who do you enjoy having casual conversations with?\\
PC & Thinking about the last month, can you gives the names of the people with whom you have hung out with regularly (e.g., chatting, drinking beers, talking about the day, playing games)?\\
PE & With whom do you typically walk to and from mosque, or generally hang out with as a friend?\\
PQ & Thinking back over the past month, who are the people you spend time with regularly? (This could be friends, family, or people you socialize most with).\\
PT & Regarding the men here in household, who do they visit during the weekend to play football, have mate or to talk?\\
RA & (Male HH) When you have freetime during the week, who do you go and visit?\\
SH & To a man: Who are the people you usually spend time talking with?  To a woman: Who are the people your husband usually spends time talking with?\\
SN & Who do you (the men in the household) like to exchange anecdotes and news with from other households?\\
TE & For you and/or the other men in your household, who do you happily and casually have conversations with?\\
TI & With whom do the men in your house visit?\\
TM & When you and the other men in your house need to learn or ask about something, who do you talk to?\\
TP & With whom do you (men) sometimes visit on Sunday afternoons?\\
TS & In the past 30 days, who did your husband (or other adult men of the household) visit, to chat or drink chicha? Who came to visit him?\\
TZ & {}[Men only] Who are the people you like to socialize with at the local coffeeshop?\\
UP & For [you and the other men / the men] in your household, who all do you enjoy having casual conversations with?\\
VT & Who do the men in your household consider to be their close friends?\\
WH & Thinking about the last month, can you gives the names of the people with whom you have hung out with regularly (e.g., chatting, drinking beers, talking about the day, playing games)?\\
WL & Thinking about the last month, can you gives the names of the people with whom you have hung out with regularly (e.g., chatting, drinking beers, talking about the day, playing games)?\\
YI & Who do you or other men in the household usually talk with or spend time together with?\\
YN & With whom do the men hang out after dinner?\\*
\end{longtable}
\endgroup{}

\begingroup\fontsize{8}{10}\selectfont

\begin{longtable}[t]{l>{\raggedright\arraybackslash}p{15cm}}
\caption{\label{tab:prompts-external-organizational-ties}Name generator prompts for support outside the community}\\
\toprule
Community & Prompt\\
\midrule
\endfirsthead
\caption[]{Name generator prompts for support outside the community \textit{(continued)}}\\
\toprule
Community & Prompt\\
\midrule
\endhead

\endfoot
\bottomrule
\endlastfoot
AH & Who are the government (including the local member) and NGO employees whom you or others in your household know well?\\
AP & What are the names of people who work for the government or an organization (like NRDDB, WWF or others) that someone in your house interacts with steady?\\
AR & Which people from the state or from organizations would help you when you need a favor? For example, in the case of the state: government officials, teachers, health workers, INTA staff; or, in the case of organizations: Warmi. (Clarify that these can be people from outside El Fuerte).\\
AV & (not asked)\\
AZ & Among those you know very well, who are those of high office, important people, government or NGO employees of high position? For example those in the police department, in politics, or working as a lawyer?\\
BD & Who are the government workers, ICDDRB employees, or other NGO employees you or anyone in your khana know and could ask for help?\\
BM & Who are the government and NGO employees who you know well?\\
BY & (not asked)\\
CP & Who are the politicians or employees of projects or agencies (e.g.,NGOs) whom you know well?\\
CR & Who are the government, NGO, and private industry employees whom you know well?\\
DJ & (not asked)\\
EK & Who are the government, church and NGO employees whom you know well? (i.e. whom you can call upon)\\
FF & Who are the government and NGO employees whom you know well?\\
FJ & Who do you(plural) know in government, the police, or the military that you(plural) could ask for financial help if needed?\\
HE & Who are the government and NGO employees whom you know well?\\
HI & Do you know someone in government that you could go to for help or advice if you needed it?\\
IH & (not asked)\\
KA & Who are government employees that you know well?\\
KM & Which people living in the village or elsewhere would you ask to help you navigate bureaucracy?\\
KO & Do you get help from people who work at government offices, local NGOs, churches?\\
KS & Who are the government, NGO employees and indigenous federations representatives whom you know well?\\
KT & Government agencies?\\
LB & Who are the government or NGO employees that you know well?\\
MA & Who are the government and NGO employees whom you know well?\\
MC & What are the names of people who work for the government or an organization (like NRDDB, WWF or others) that someone in your house interacts with steady?\\
MG & Who are the government and NGO employees you know personally and interact with frequently?\\
MK & (not asked)\\
MN & Are there representatives of the government, NGOs, or other institutions that can help you or others in your household, for example if you are in a conflict with someone or you need help with paperwork? Who? What organization are they from? How did they last help you?\\
MY & What government or NGO employees do you know well?\\
NI & Whom do you know who work for an organization and would provide assistance to you in a time of need?\\
NP & Who do you know well who currently works for an NGO, the government (sarkari karmachari), the police, or who has a high position in a political party---and who could help you?\\
PC & (not asked)\\
PE & Is there anyone in an institution (government or non-governmental), or anyone in a religious or cooperative institution, whom you know personally?\\
PQ & Do you have close ties to any people from the community council, the local government office, NGOs, or other important people in or outside of Palenque? What are their names?\\
PT & Who are the government and NGO employees whom you know well? You can call in case of any urgency\\
RA & Who are the government or NGO workers that you know well?\\
SH & Who are the government workers and ICDDRB employees you or your family know and could ask for help?\\
SN & Who are the government, NGO, and Park employees whom you know well?\\
TE & Among those you know very well, who are those of high office, important people, government or NGO employees of high position? For example those in the police department, in politics, or working as a lawyer?\\
TI & Who are visitors from the government or an organization whom you know?\\
TM & Who are the government workers, party members or association members that you know well and who support you when you need it?\\
TP & Who are the government and NGO employees whom you know well?\\
TS & Are there any representatives of the government, NGOs or other institutions like the Gran Consejo Tsiman that you can ask for help with problems (for example paperwork)?\\
TZ & Do you know any people in local associations, NGOs, the mouqataa, the jamaa or other administrative offices who could offer you and/or other household members support if needed?\\
UP & Among those you know very well, who are those of high office, important people, government or NGO employees of high position? For example those in the police department, in politics, or working as a lawyer?\\
VT & Who do you know that works for the government or an NGO?\\
WH & (not asked)\\
WL & (not asked)\\
YI & Who are the government or non-profit employees that you know well?\\
YN & Who are the government or NGO employees that you know well?\\*
\end{longtable}
\endgroup{}

\begingroup\fontsize{8}{10}\selectfont

\begin{longtable}[t]{l>{\raggedright\arraybackslash}p{15cm}}
\caption{\label{tab:prompts-other-important-external-ties}Name generator prompts for ties to government or NGO employees}\\
\toprule
Community & Prompt\\
\midrule
\endfirsthead
\caption[]{Name generator prompts for ties to government or NGO employees \textit{(continued)}}\\
\toprule
Community & Prompt\\
\midrule
\endhead

\endfoot
\bottomrule
\endlastfoot
AH & Who are the business or trade leaders whom you or other members of your family know well?\\
AP & Who are people from outside your village that someone in your house buys from or sells to?\\
AR & With which people do you maintain commercial relationships? For example, to whom do you sell your products, from whom do you buy products, and who helps you receive passengers/tourists? (Remember that these can be people who live in El Fuerte or outside the community.)\\
AV & (not asked)\\
AZ & Who are the people in Madurai, Chennai, or other outside or overseas places who could get things done for you?\\
BD & Who are the most important ties outside of this village for you or anyone in your khana?\\
BM & When you want to sell goods, who are the merchants that you work with?\\
BY & Since i was last here, who have you visited with in other villages?\\
CP & If there were a death in the family, on whom could you count to help you with the wake and the funeral?\\
CR & When you bring your fish products to sell/trade, who are the merchants/traders you work with?\\
DJ & (not asked)\\
EK & When you bring your produce to sell/trade, who are the merchants/traders you work with?\\
FF & When you want to bring your livestock for sale at the auction points, who do you ask for support?\\
FJ & Who are the merchants or business men, at local resorts or in town, that could help you(plural) find work?\\
HE & When you bring your produce to sell/trade who are the merchants/traders you work with?\\
HI & Who could you sell a goat to in town?\\
IH & (not asked)\\
KA & When you bring merchandise to be sold/exchanged, who are the buyers/merchants that cooperate with you?\\
KM & Which people living in the village or elsewhere would you ask to help you buy or sell something in the city?\\
KO & In case of need, do you have people outside Chano such as friends or family living in Kele town/Dila/Arba Minch, that could help you?\\
KS & Who are the people you count on when you sell your produce?, Who are the people you rely on when you or your close family travel to cities?\\
KT & NGOs? Merchants?\\
LB & If you needed a big favor, from whom would you ask for help outside Labai?\\
MA & Who are your livestock exchange partners? Livestock exchange partners are friends who live considerable distances away who share livestock in times of need, such as drought or disease?\\
MC & Who are people from outside your village that someone in your house buys from or sells to?\\
MG & When you bring your produce to sell/trade, who are the merchants/traders you work with and know personally?\\
MK & (not asked)\\
MN & Are there merchants that you or others in your household regularly engage in business with with? How many times in the past 12 months have you done business with this person?\\
MY & What business owners or commercial people do you know well?\\
NI & What merchants do you know who would do you a favor, if necessary?\\
NP & Who are the people in Chautara or Kathmandu, abroad, or elsewhere, who get things done for you?\\
PC & (not asked)\\
PE & When you have produce to to sell/trade, are there any specific individuals you work with who you know personally?\\
PQ & Who are economically important people outside of Palenque who you have close connections with? (e.g., tourism connections, employers, and traders in Cartagena or elsewhere)\\
PT & Who are the people you sell fish and bait?\\
RA & Who do you colaborate with in the market when you have items to sell or trade?\\
SH & Who are you or your family's most important ties outside of this bohor?\\
SN & If you were to need something from the city, who could you ask outside of Khodzue to buy it for you or bring it on their next visit?\\
TE & Who are the people in Madurai, Chennai, or other outside or overseas places who could get things done for you?\\
TI & Who are the merchants who buy your produce?\\
TM & Of the merchants who buy and sell anything, who are the ones you know well and feel you can ask for a favor?\\
TP & Who are friends and family whom you know well living in the city (e.g. Paramaribo)?\\
TS & Are there any merchants that you, your husband, or others in your household regularly do business with?\\
TZ & Who are the merchants and other middlemen you and/or other members of your household commonly do business with on market days or in town?\\
UP & Who are the people in Kanpur or Bithoor, abroad, or other outside and overseas places, who could get things done for you?\\
VT & Who do you know well outside of this community? Anyone who can do useful things for you?\\
WH & (not asked)\\
WL & (not asked)\\
YI & If you were outside of the Wujie area and needed help (i.e. were injured), who would you call?\\
YN & If you needed a big favor, from whom would you ask for help outside Yongning?\\*
\end{longtable}
\endgroup{}

\clearpage
\subsection{Network Overlap}\label{si:network_overlap}

\subsubsection{Alter overlap}\label{si:alter_overlap}

In the main text in Figure~\ref{fig:alter_overlap_pooled_pca}, we show the Jaccard similarities across each network layer, across all respondents in all communities. (Note that this means that the values presented there are across all individuals (or sharing units) named, giving more weight to communities with more respondents and where respondents made more nominations). As a point of comparison, the Jaccard similarity across our borrowing/lending money (on the one hand) and informational/affective ties (on the other) of between $0.11$ and $0.13$ is somewhat lower than the reported value of about $0.27$ between friendship and financial networks from undirected, individual-level networks from $176$ villages in Honduras \cite{papamichalis_deep_2026}.

\begin{figure}[h]
\begin{center}
\includegraphics[page=1, width = 1.1\linewidth]{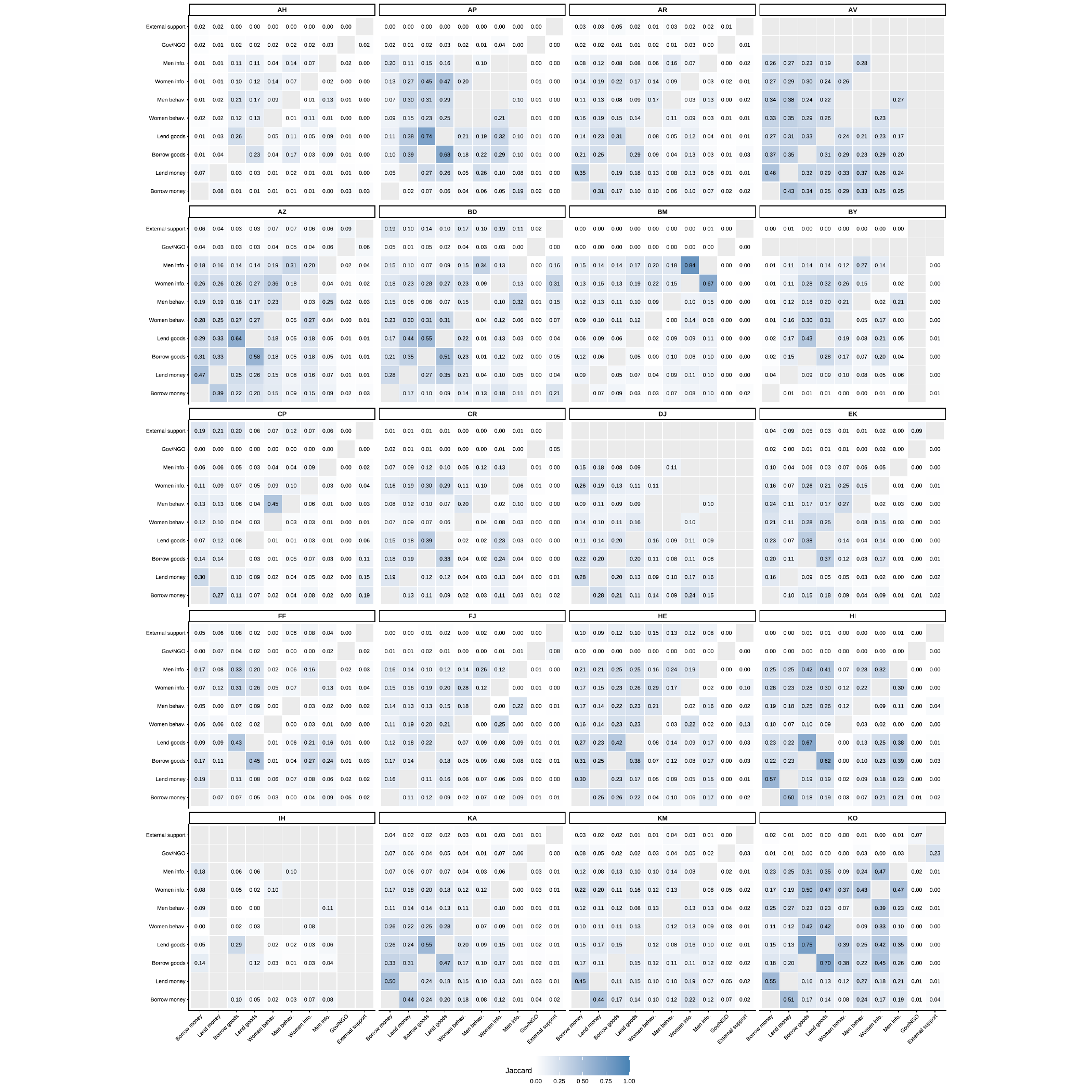}
\end{center}
\caption{Jaccard similarity of the set of alters respondents name across network prompts, considering all \emph{individuals} named (lower triangle) and all \emph{sharing units} named (upper triangle).}
\label{fig:alter_overlap_sitewise1}
\end{figure}

\begin{figure}[h]
\begin{center}
\includegraphics[page=2, width = 1.1\linewidth]{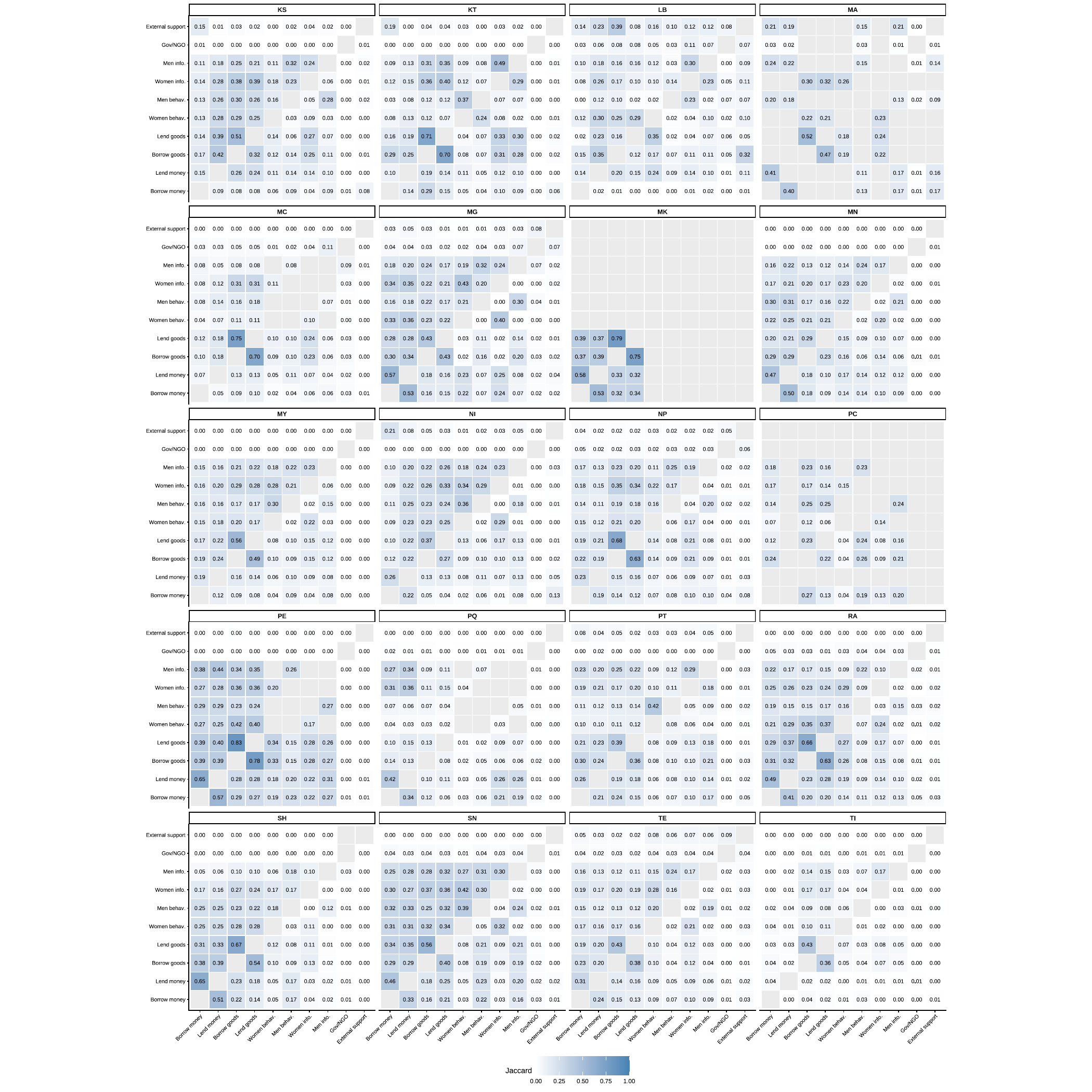}
\end{center}
\caption{Jaccard similarity of the set of alters respondents make across network prompts, considering all \emph{individuals} named (lower triangle) and all \emph{sharing units} named (upper triangle).}
\label{fig:alter_overlap_sitewise2}
\end{figure}

\begin{figure}[h]
\begin{center}
\includegraphics[page=3, width = 0.9\linewidth]{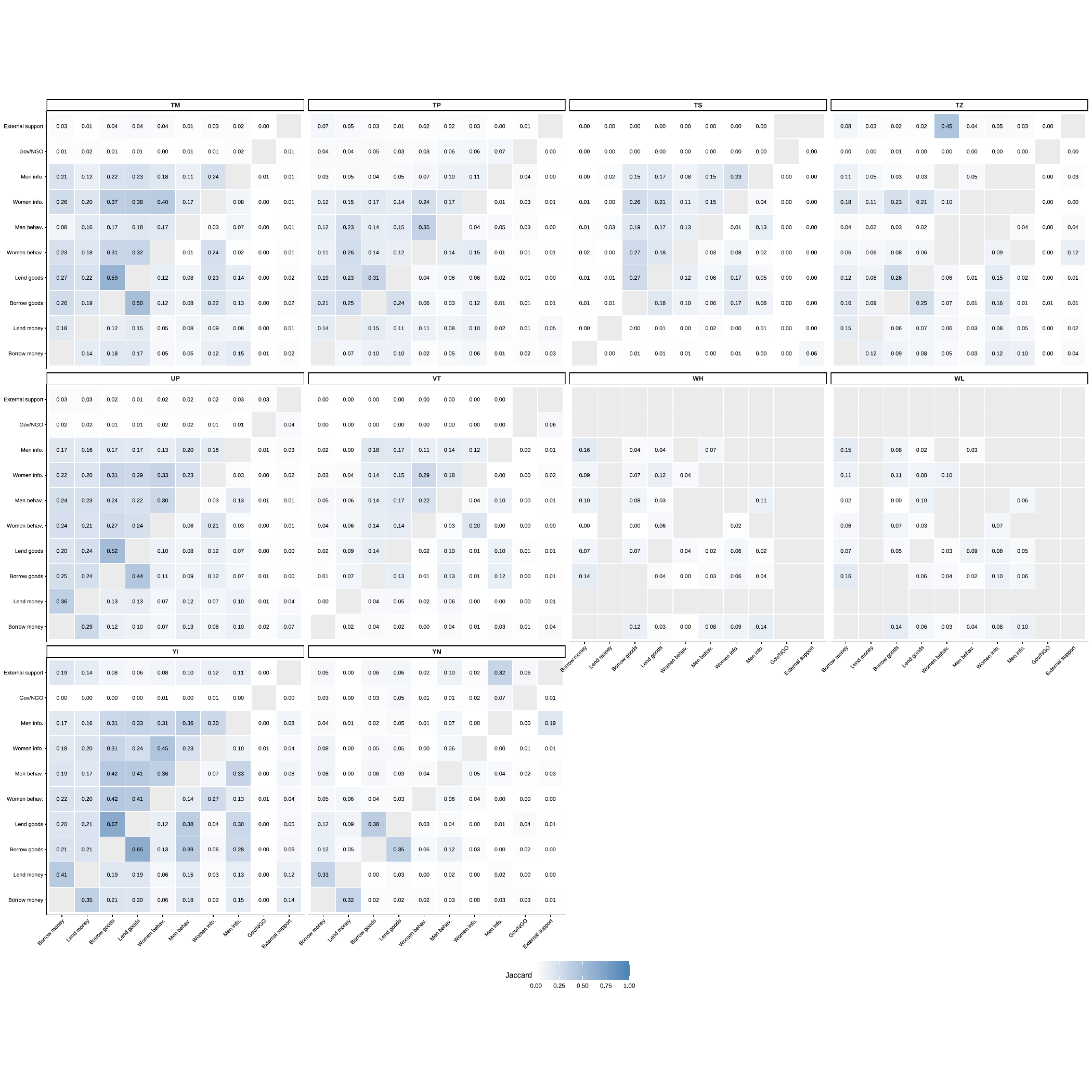}
\end{center}
\caption{Jaccard similarity of the set of alters respondents make across network prompts, considering all \emph{individuals} named (lower triangle) and all \emph{sharing units} named (upper triangle).}
\label{fig:alter_overlap_sitewise3}
\end{figure}

Figures~\ref{fig:alter_overlap_sitewise1}-\ref{fig:alter_overlap_sitewise3} show the Jaccard similarity for the set of alters named by respondents for each network prompt, separately for each community.

Note that in some communities, different questions were posed to different respondents. This holds most often for the gender-specific questions: as shown in Table~\ref{tab:respondents_by_site}, in some communities (namely, AP, AV, DJ, IH, MC, PC, PE, PQ, TZ, WH, and WL) the women-specific questions were posed to women and the men-specific questions were posed to men. In MA and KO, men were presented with the two monetary loan questions, the men-specific questions, and the two questions on external alters, while women were presented with the two borrowing/lending goods questions and the women-specific questions. In such cases, cells are left blank. (In KO, there are a few exceptions where people of different genders answered the questions, so we present similarities).

Similarly, in some communities, some questions were not posed as expected. The two questions on external alters were not posed in AV, DJ, IH, MK, PC, WH, and WL. The double-sampled question on lending money was not asked in IH, PC, WH, or WL. The question about government and NGO employees was not posed in BY. The domains that were generally asked with gender-specific questions were posed generally (without gender specification) in MK, meaning that the prompts used do not align with the others. In these cases, as well, cells are left blank.

\clearpage
\subsubsection{Dyad overlap}\label{si:edge_overlap}

\begin{figure}[h]
\begin{center}
\includegraphics[page=1, width = 0.6\linewidth]{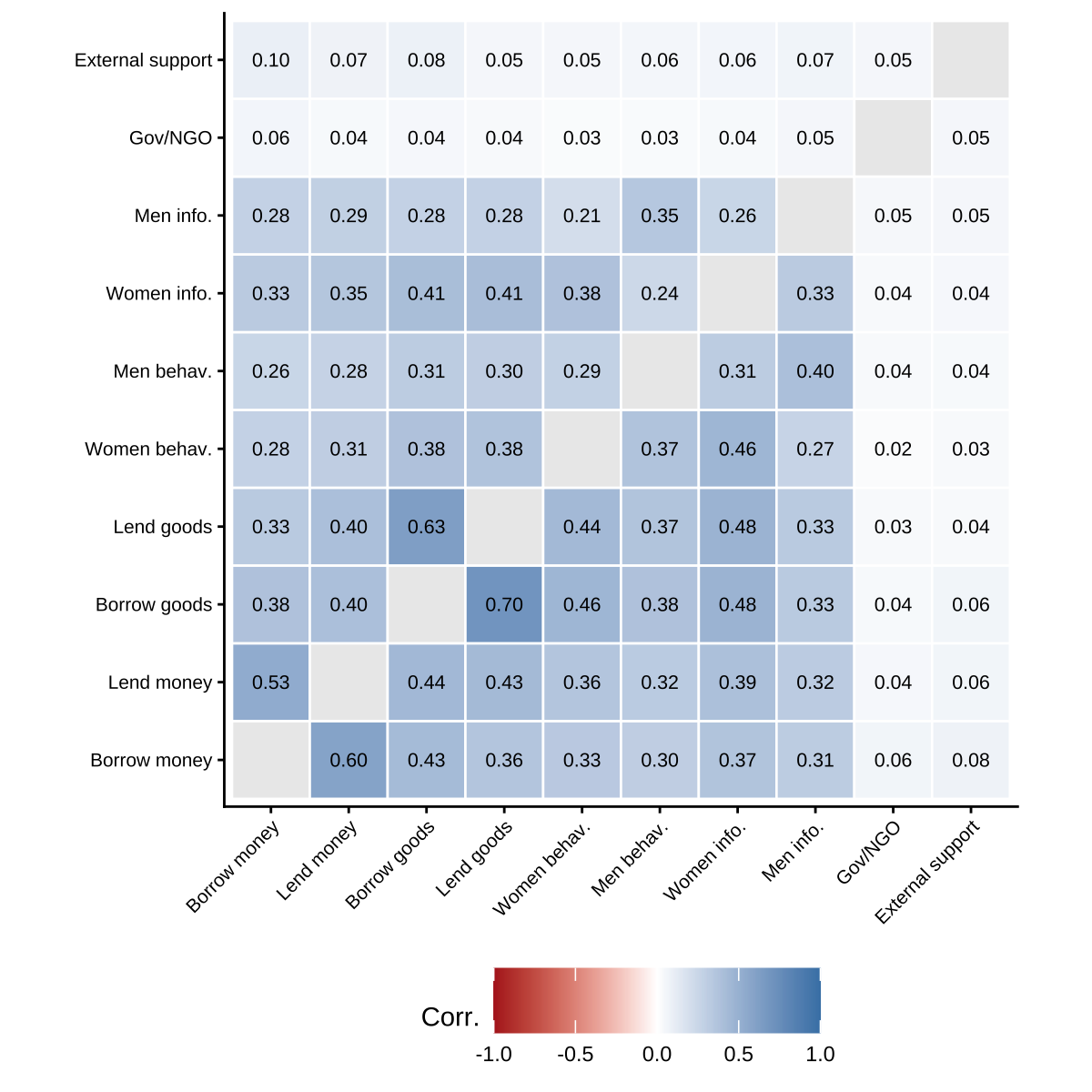}
\end{center}
\caption{Correlation of the dyads, pooled across all communities, considering \emph{weighted} (lower triangle) and \emph{binary} (upper triangle) edges.}
\label{fig:edge_overlap_pooled}
\end{figure}

\begin{figure}[h]
\begin{center}
\includegraphics[page=1, width = 1\linewidth]{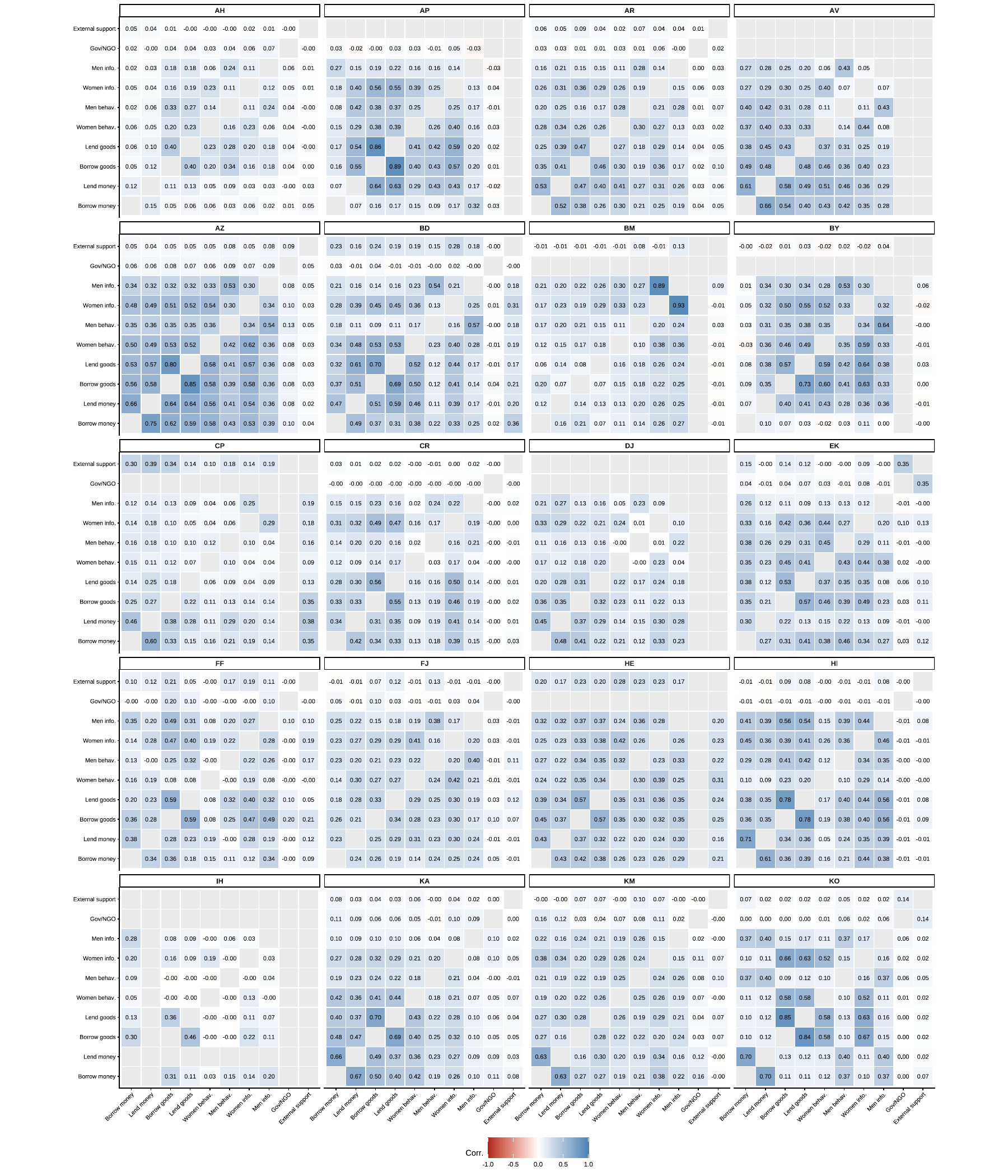}
\end{center}
\caption{Correlation of the dyads, considering \emph{weighted} (lower triangle) and \emph{binary} (upper triangle) edges.}
\label{fig:edge_overlap_sitewise1}
\end{figure}

\begin{figure}[h]
\begin{center}
\includegraphics[page=2, width = 1\linewidth]{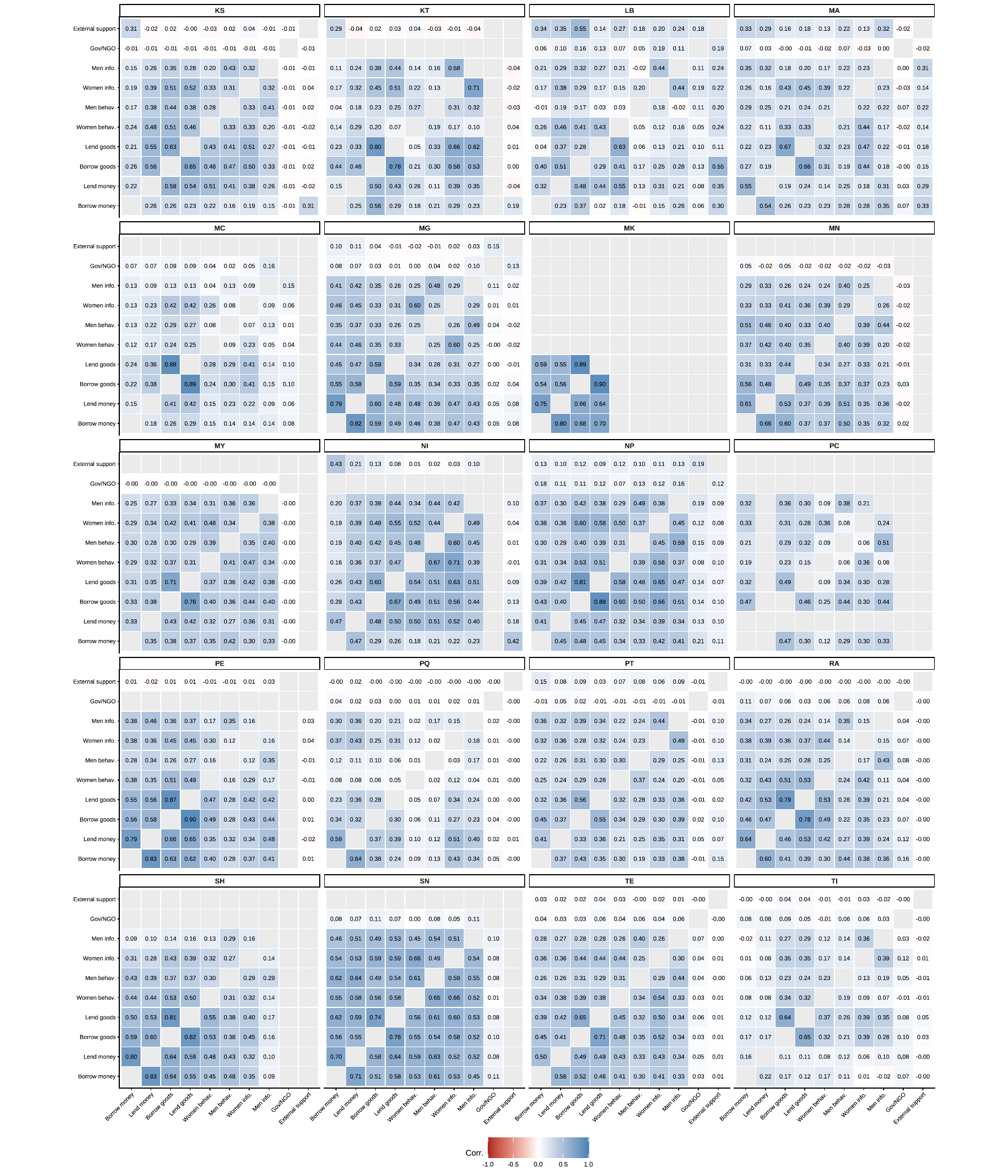}
\end{center}
\caption{Correlation of the dyads, considering \emph{weighted} (lower triangle) and \emph{binary} (upper triangle) edges.}
\label{fig:edge_overlap_sitewise2}
\end{figure}

\begin{figure}[h]
\begin{center}
\includegraphics[page=3, width = 0.9\linewidth]{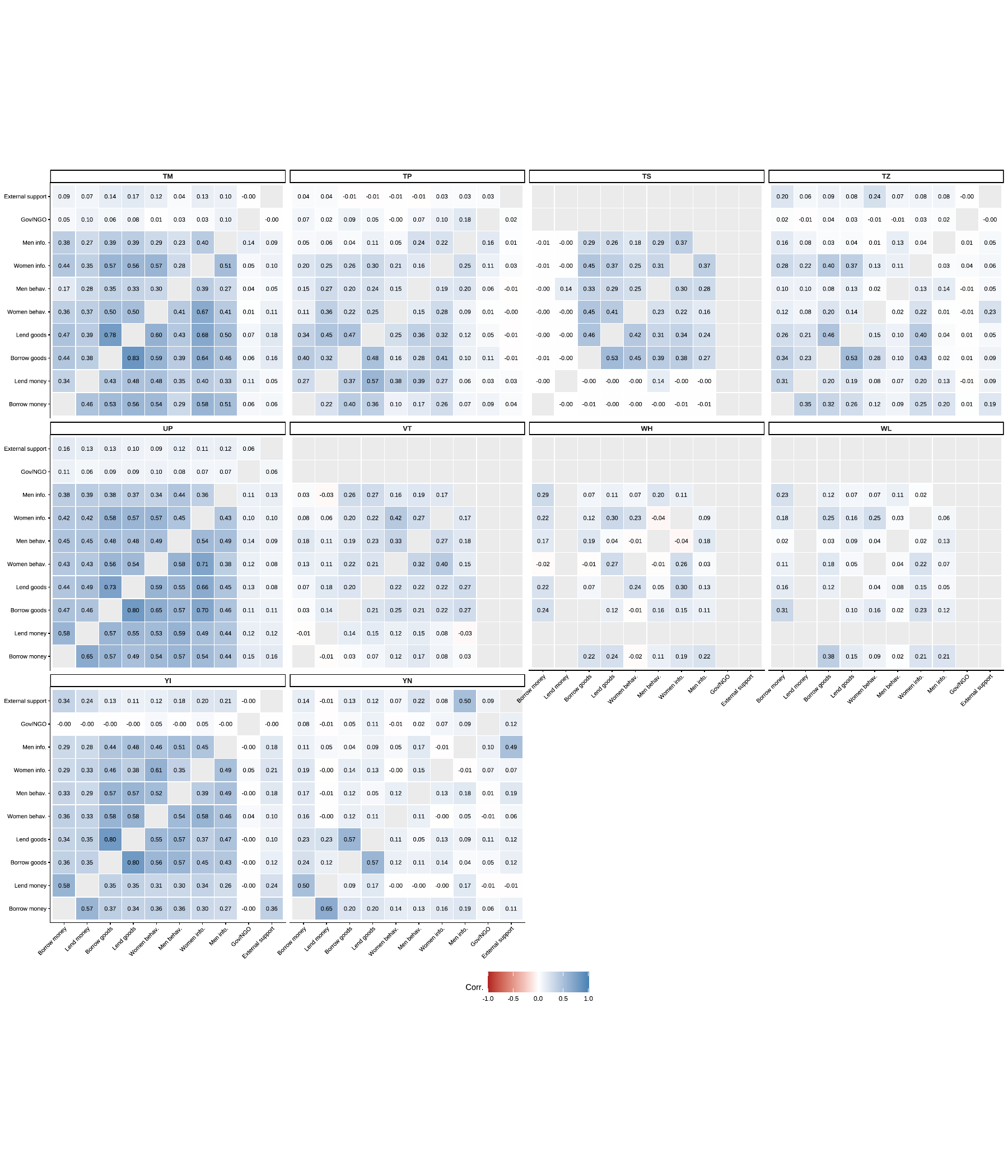}
\end{center}
\caption{Correlation of the dyads, considering \emph{weighted} (lower triangle) and \emph{binary} (upper triangle) edges.}
\label{fig:edge_overlap_sitewise3}
\end{figure}

Figure~\ref{fig:edge_overlap_pooled} shows the Pearson's correlation coefficient comparing across all dyads in each network layer, pooled across all communities. We construct a dataframe with a row for each dyad and a column for each network layer, across all communities. The entries in each column reflect the binary or weighted edge values for that particular network layer. We can then simply calculate the correlations between the entries in each pair of layers, i.e., the correlations of the $i \rightarrow j$ dyad in layer $k$ to the $i \rightarrow j$ dyad in layer $l$ for all pairs of layers. In Figures~\ref{fig:edge_overlap_sitewise1}-\ref{fig:edge_overlap_sitewise3}, we show the same, done individually for each community.

In some communities, some questions were not posed. The two questions on external alters were not posed in AV, DJ, IH, MK, PC, WH, and WL. The double-sampled question on lending money was not asked in IH, PC, WH, or WL. The question about government and NGO employees was not posed in BY. The domains that were generally asked with gender-specific questions were posed generally (without gender specification) in MK, meaning that the prompts used do not align with the others. In these cases, cells are left blank.
Similarly, in some communities, even when posed, the prompts focused on external ties resulted in no nominations to other community residents. (This holds for BM, CP, HE, KT, NI, PE, SH, TS, and VT for the question on government and NGO employees; and for AP, MC, MN, MY, SH, SN, TS, and VT for the question on other important support outside the community). In these cases, as well, cells are left blank.

\clearpage
\subsubsection{Principal Component Analysis}\label{si:pca}

\begin{figure}[h]
\begin{center}
\includegraphics[page=1, width = 1.1\linewidth]{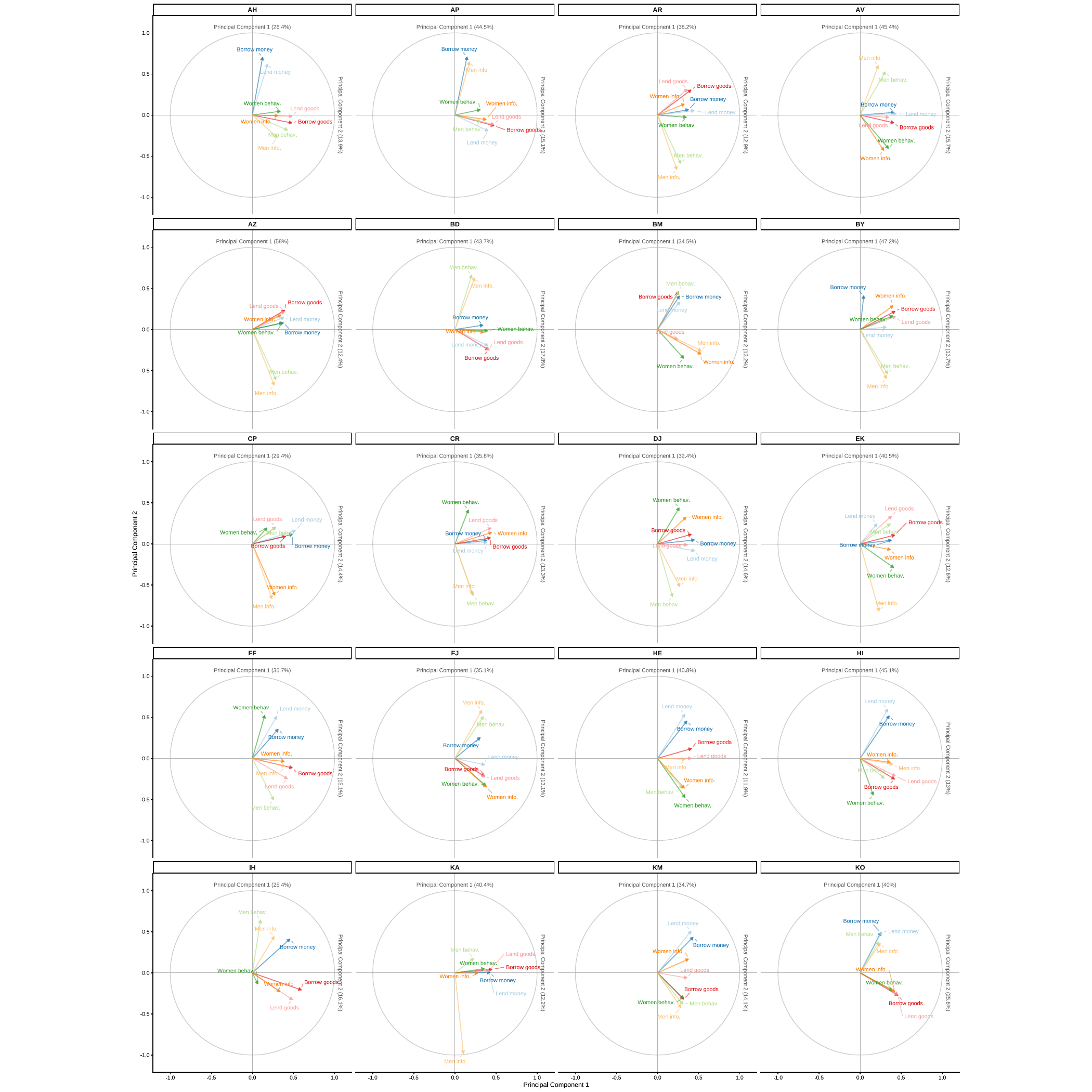}
\end{center}
\caption{Loadings on the first and second principal components of the sharing unit network layers, using weighted edges. }
\label{fig:pca_sitewise1}
\end{figure}

\begin{figure}[h]
\begin{center}
\includegraphics[page=2, width = 1.1\linewidth]{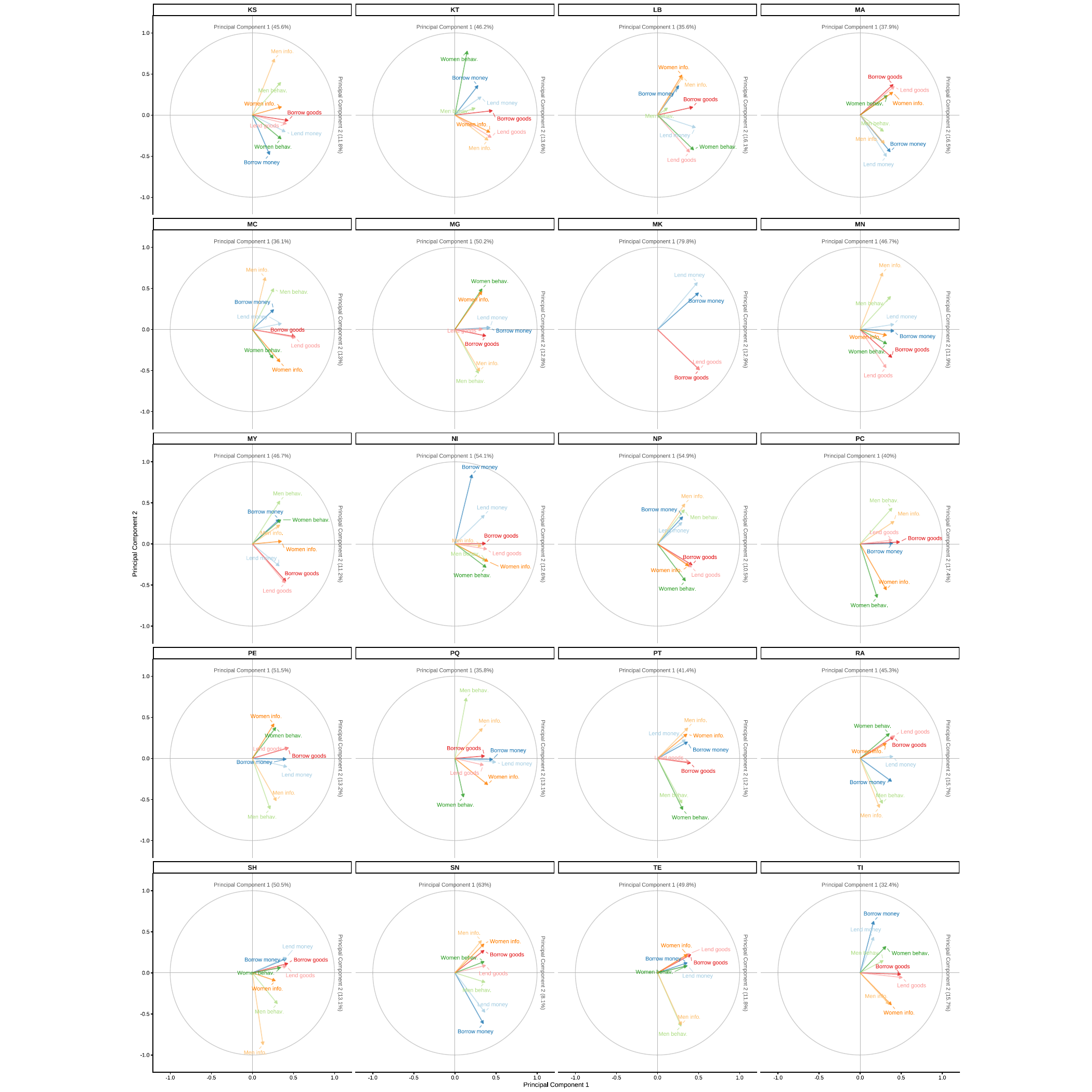}
\end{center}
\caption{Loadings on the first and second principal components of the sharing unit network layers, using weighted edges. }
\label{fig:pca_sitewise2}
\end{figure}

\begin{figure}[h]
\begin{center}
\includegraphics[page=3, width = 0.9\linewidth]{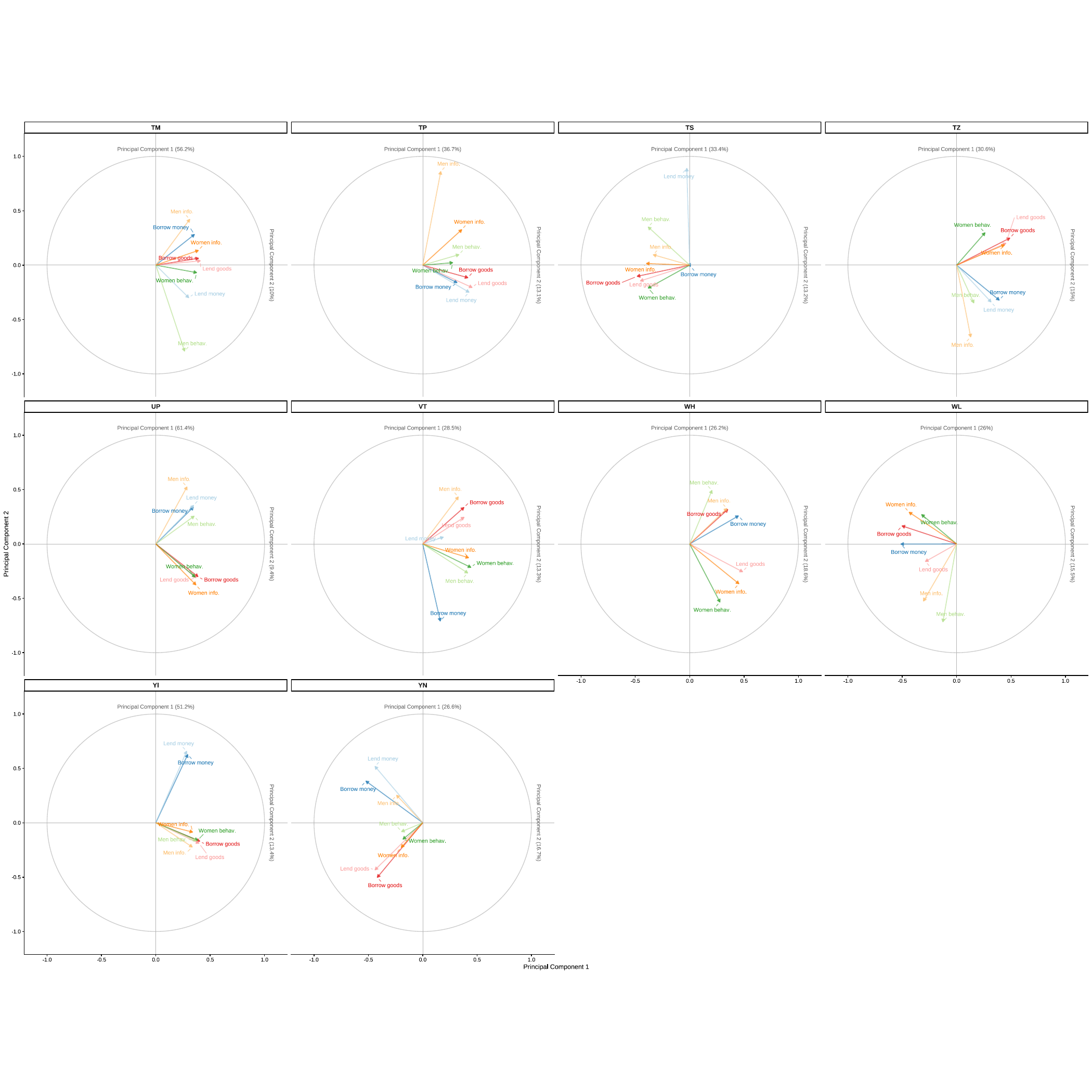}
\end{center}
\caption{Loadings on the first and second principal components of the sharing unit network layers, using weighted edges. }
\label{fig:pca_sitewise3}
\end{figure}

In Figures~\ref{fig:pca_sitewise1}-\ref{fig:pca_sitewise3}, we show the loadings on the first and second principal component for each community, constructed using weighted (as opposed to binary) edges. In Table~\ref{tab:pca_loadings} we show the variance explained by each principal component and the loading of each layer onto each principal component.

\begingroup\fontsize{8}{10}\selectfont

\begin{longtable}[t]{lrrrrrrrr}
\caption{PCA loadings by layer, weighted. Each entry is the correlation (loading) between a network layer and a principal component. Loadings near $\pm 1$ indicate the layer is strongly associated with that component; loadings near 0 indicate little association. Reading across a row shows how a layer's variance is distributed across components; reading down a column reveals the substantive meaning of each component. Components are ordered by descending variance explained.}\\
\toprule
Layer & PC1 & PC2 & PC3 & PC4 & PC5 & PC6 & PC7 & PC8\\
\midrule
\endfirsthead
\caption[]{PCA loadings by layer, weighted. Each entry is the correlation (loading) between a network layer and a principal component. Loadings near $\pm 1$ indicate the layer is strongly associated with that component; loadings near 0 indicate little association. Reading across a row shows how a layer's variance is distributed across components; reading down a column reveals the substantive meaning of each component. Components are ordered by descending variance explained. \textit{(continued)}}\\
\toprule
Layer & PC1 & PC2 & PC3 & PC4 & PC5 & PC6 & PC7 & PC8\\
\midrule
\endhead

\endfoot
\bottomrule
\endlastfoot
\addlinespace[0.5em]
\multicolumn{9}{l}{\textbf{AH}}\\
\hspace{1em}Borrow money & 0.12 & 0.70 & -0.03 & -0.43 & -0.55 & 0.07 & -0.02 & -0.01\\
\hspace{1em}Lend money & 0.18 & 0.62 & 0.32 & 0.22 & 0.63 & -0.15 & -0.02 & 0.06\\
\hspace{1em}Borrow goods & 0.48 & -0.10 & 0.15 & 0.27 & -0.20 & 0.31 & 0.05 & 0.72\\
\hspace{1em}Lend goods & 0.48 & -0.02 & 0.03 & 0.24 & -0.06 & 0.36 & 0.43 & -0.62\\
\hspace{1em}Women behav. & 0.34 & 0.05 & -0.56 & 0.20 & -0.09 & -0.68 & 0.22 & 0.08\\
\hspace{1em}Men behav. & 0.43 & -0.18 & 0.32 & 0.04 & -0.19 & -0.28 & -0.70 & -0.27\\
\hspace{1em}Women info. & 0.31 & -0.01 & -0.60 & -0.32 & 0.41 & 0.38 & -0.36 & 0.01\\
\end{longtable}
\endgroup{}
\label{tab:pca_loadings}

\clearpage
\subsection{Concordance}\label{si:concordance}

We examine the agreement (i.e., concordance) between survey respondents in two ways: first, considering the agreement between members of the same sharing unit (when there are multiple respondents per sharing unit), and second, considering the agreement between respondents for the two double-sampled questions.

\subsubsection{Concordance within Sharing Units}\label{si:concordance_su}

\begin{figure}[h]
\begin{center}
\includegraphics[width=\textwidth]{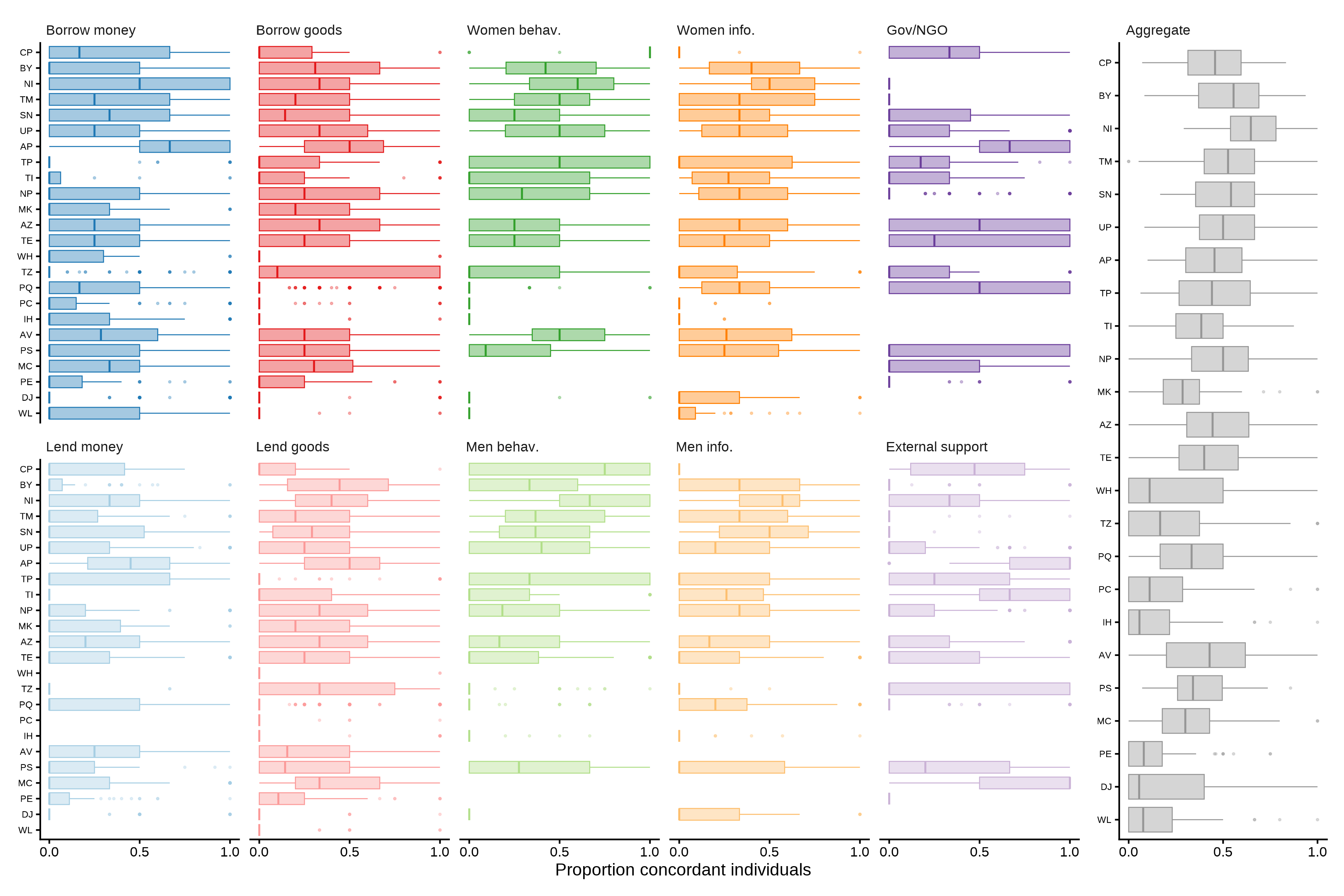}
\end{center}
\caption{Concordance in nominations \emph{of individuals} by members of the same sharing units, for each network prompt. Empty entries reflect communities where there were fewer than 10 cases of sharing units with multiple respondents, where questions were not asked, or where questions were not asked of multiple respondents (\textit{e.g.}, in communities where gender-aligned questions were asked only of one gender, even when both male and female household heads were sampled, there would be no opportunity for concordance in those questions). Communities are ordered by their concordance of nominations of sharing units in the prompts contributing to the aggregate network. }
\label{fig:concordance_su_layer_person}
\end{figure}

\begin{figure}[h]
\begin{center}
\includegraphics[width=\textwidth]{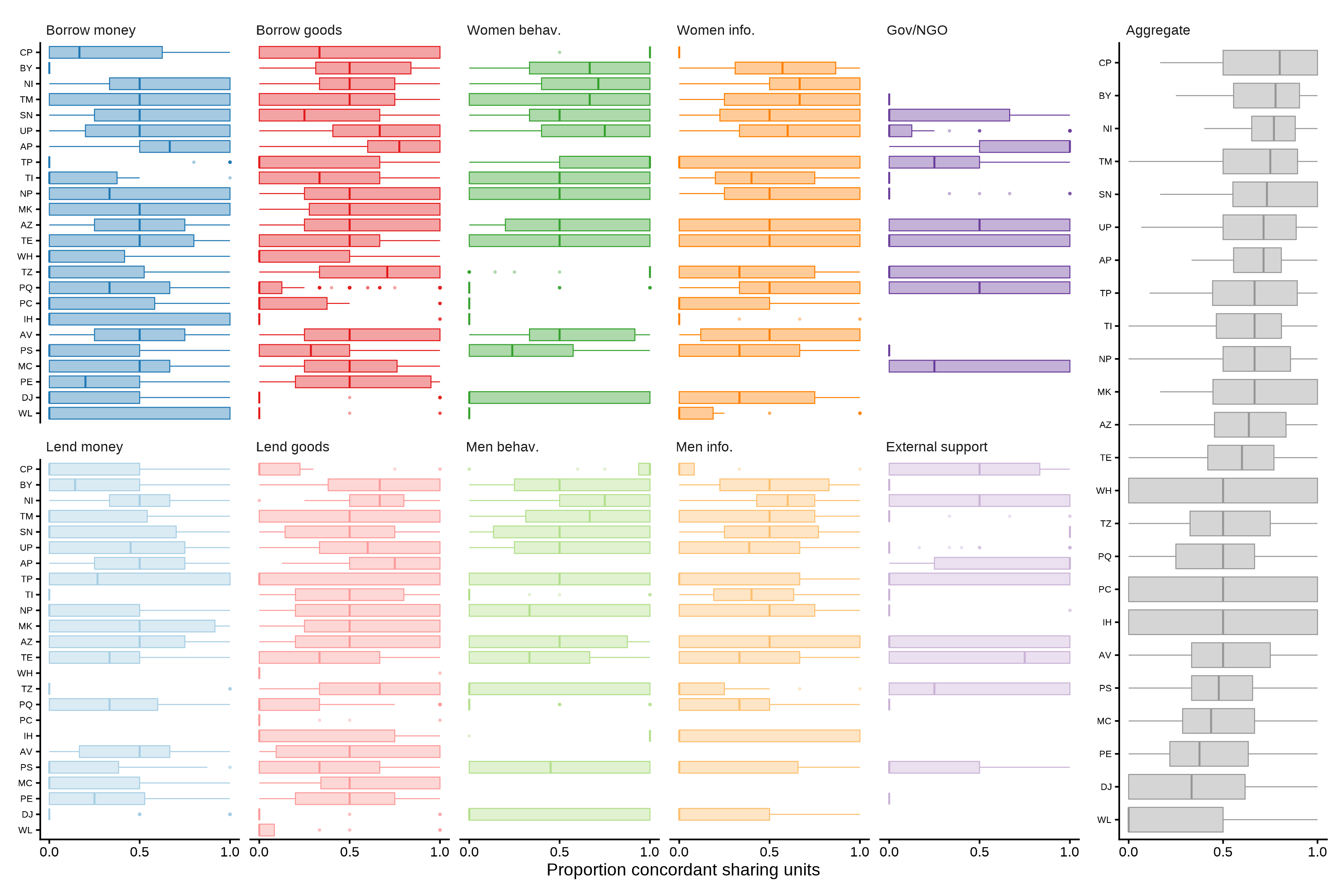}
\end{center}
\caption{Concordance in nominations \emph{of sharing units} by members of the same sharing units, for each layer. As compared to Figure~\ref{fig:concordance_su_layer_person}, here the standard for concordance is lower: if two sharing unit members name different members of another sharing unit, it would here be considered concordant. Empty entries reflect communities where there were fewer than 10 cases of sharing units with multiple respondents, where questions were not asked, or where questions were not asked of multiple respondents (\textit{e.g.}, in communities where gender-aligned questions were asked only of one gender, even when both male and female household heads were sampled, there would be no opportunity for concordance in those questions). Communities are ordered by their concordance of nominations of sharing units in the prompts contributing to the aggregate network. }
\label{fig:concordance_su_layer_su}
\end{figure}

We consider concordance of nominations by members of the same sharing unit for communities where there were at least 10 cases of sharing units with multiple respondents. We compute values from the set of nominations made by each respondent.
Figure~\ref{fig:concordance_su_layer_person} shows the concordance in nominations of \emph{individuals} for each network prompt, as well as for the aggregate network (so, where respondents need not be naming the same individual or sharing unit for the same network prompt, but for any of those contributing to the aggregate network). Similarly, Figure~\ref{fig:concordance_su_layer_su}  depicts concordance in nominations of \emph{sharing units} for each network prompt and the prompts contributing to the aggregate network.

\clearpage
\subsubsection{Concordance for Double-Sampled Networks}\label{si:concordance_double}

\begin{figure}[h]
\begin{center}
\includegraphics[width=0.5\textwidth]{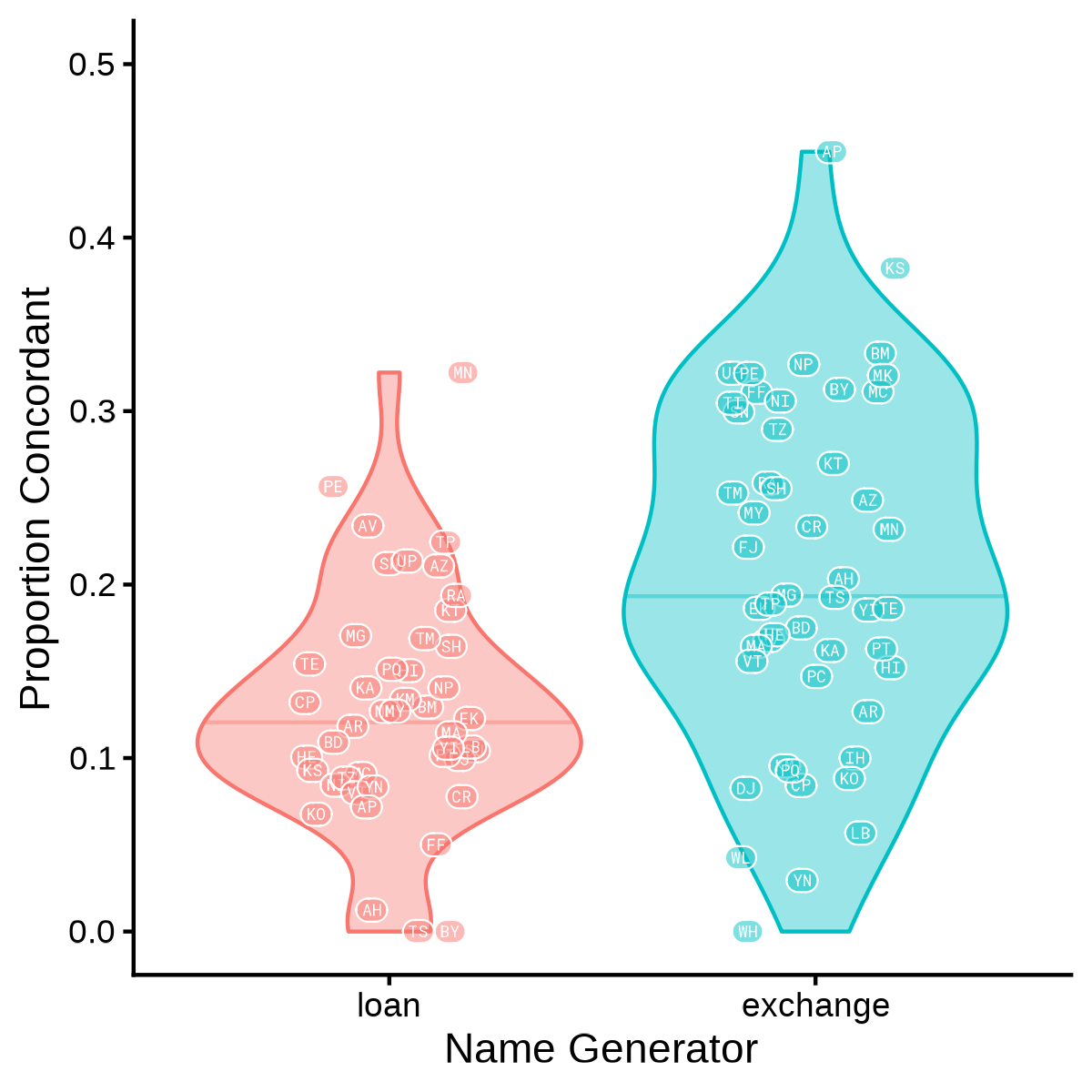}
\end{center}
\caption{Concordance in nominations of sharing units across the two double-sampled network layers (this corresponds to the ``Overall'' column in Table~\ref{tab:ds_concordance}.) }
\label{fig:concordance_ds}
\end{figure}

The double-sampled networks allow estimates of concordance from different sharing units. We compute this from the sharing unit-level networks.
If there is a tie from $i \rightarrow j$ in the borrow money layer (derived from sharing unit $i$'s responses and meaning that sharing unit $i$ says they would ask to borrow money from sharing unit $j$), then we should expect to have a tie from $j \rightarrow i$ in the lend money layer (derived sharing unit $j$'s responses and meaning that sharing unit $j$ says that they would lend money to sharing unit $i$).
Table~\ref{tab:ds_concordance} shows the proportion of concordant ties overall and for each contributing layer for the double-sampled monetary loan and household item exchange network layers.
The ``Request'' and ``Give'' columns show the proportion of ties in each individual layer that are contributing to concordant ties in the combined set of layers. For instance, if a Request layer had 20 ties, half of which were all present in a Give layer, and the Give had 100 ties, the values would be 9\% (10/110) concordance overall across all unique reported ties, 50\% (10/20) in the Request layer and 10\% (10/100) in the Give layer.
Figure~\ref{fig:concordance_ds} summarises what is presented in the ``Overall'' column in Table~\ref{tab:ds_concordance} for the two double-sampled networks. It is evident that there is greater concordance across the household item exchange network layers than across the monetary loan layers.

\begin{longtable}[t]{lrrrrrr}
\caption{\label{tab:ds_concordance}Proportion of concordant ties in the sharing-unit networks for each double-sampled relationship, in each community. ``Request'' refers to the network based on the name generator in which respondents reporting receiving a type of support, while ``Give'' refers to respondents reporting giving support to others. The column ``Overall'' reports the proportion of concordant ties in the union of the request and (transposed) give networks.}\\
\toprule
\multicolumn{1}{c}{ } & \multicolumn{3}{c}{Loan} & \multicolumn{3}{c}{Exchange} \\
\cmidrule(l{3pt}r{3pt}){2-4} \cmidrule(l{3pt}r{3pt}){5-7}
Community & Overall & Request & Give & Overall & Request & Give\\
\midrule
AH & 0.012 & 0.030 & 0.020 & 0.203 & 0.354 & 0.323\\
AP & 0.072 & 0.142 & 0.127 & 0.450 & 0.642 & 0.600\\
AR & 0.118 & 0.247 & 0.185 & 0.127 & 0.234 & 0.217\\
AV & 0.234 & 0.414 & 0.350 & 0.168 & 0.288 & 0.287\\
AZ & 0.211 & 0.353 & 0.344 & 0.249 & 0.401 & 0.395\\
\addlinespace
BD & 0.109 & 0.241 & 0.167 & 0.175 & 0.302 & 0.294\\
BM & 0.129 & 0.204 & 0.262 & 0.333 & 0.545 & 0.462\\
BY & 0.000 & 0.000 & 0.000 & 0.312 & 0.514 & 0.443\\
CP & 0.132 & 0.241 & 0.226 & 0.084 & 0.113 & 0.246\\
CR & 0.078 & 0.158 & 0.133 & 0.233 & 0.376 & 0.381\\
\addlinespace
DJ & 0.104 & 0.191 & 0.186 & 0.083 & 0.165 & 0.142\\
EK & 0.123 & 0.156 & 0.368 & 0.186 & 0.358 & 0.279\\
FF & 0.050 & 0.083 & 0.111 & 0.311 & 0.469 & 0.479\\
FJ & 0.099 & 0.191 & 0.171 & 0.222 & 0.440 & 0.308\\
HE & 0.101 & 0.170 & 0.197 & 0.171 & 0.302 & 0.283\\
\addlinespace
HI & 0.150 & 0.286 & 0.241 & 0.152 & 0.280 & 0.250\\
IH & --- & --- & --- & 0.100 & 0.167 & 0.200\\
KA & 0.140 & 0.253 & 0.240 & 0.162 & 0.271 & 0.287\\
KM & 0.134 & 0.211 & 0.268 & 0.096 & 0.193 & 0.159\\
KO & 0.068 & 0.126 & 0.128 & 0.088 & 0.157 & 0.168\\
\addlinespace
KS & 0.093 & 0.191 & 0.153 & 0.382 & 0.547 & 0.560\\
KT & 0.185 & 0.250 & 0.417 & 0.270 & 0.405 & 0.447\\
LB & 0.106 & 0.455 & 0.122 & 0.057 & 0.076 & 0.185\\
MA & 0.115 & 0.202 & 0.210 & 0.164 & 0.298 & 0.268\\
MC & 0.091 & 0.162 & 0.172 & 0.311 & 0.469 & 0.480\\
\addlinespace
MG & 0.171 & 0.285 & 0.298 & 0.194 & 0.330 & 0.320\\
MK & 0.127 & 0.272 & 0.192 & 0.321 & 0.500 & 0.472\\
MN & 0.322 & 0.630 & 0.397 & 0.232 & 0.364 & 0.390\\
MY & 0.127 & 0.281 & 0.188 & 0.241 & 0.405 & 0.373\\
NI & 0.084 & 0.247 & 0.114 & 0.306 & 0.556 & 0.405\\
\addlinespace
NP & 0.140 & 0.245 & 0.248 & 0.327 & 0.523 & 0.466\\
PC & --- & --- & --- & 0.147 & 0.254 & 0.259\\
PE & 0.256 & 0.408 & 0.408 & 0.322 & 0.491 & 0.483\\
PQ & 0.151 & 0.294 & 0.237 & 0.093 & 0.195 & 0.150\\
PT & 0.102 & 0.179 & 0.190 & 0.163 & 0.290 & 0.271\\
\addlinespace
RA & 0.194 & 0.322 & 0.327 & 0.259 & 0.390 & 0.434\\
SH & 0.164 & 0.300 & 0.267 & 0.255 & 0.414 & 0.400\\
SN & 0.212 & 0.340 & 0.361 & 0.300 & 0.464 & 0.458\\
TE & 0.154 & 0.272 & 0.263 & 0.186 & 0.327 & 0.302\\
TI & 0.000 & 0.000 & 0.000 & 0.305 & 0.513 & 0.429\\
\addlinespace
TM & 0.169 & 0.286 & 0.292 & 0.253 & 0.402 & 0.405\\
TP & 0.224 & 0.379 & 0.355 & 0.189 & 0.338 & 0.300\\
TS & 0.000 & 0.000 & 0.000 & 0.193 & 0.296 & 0.356\\
TZ & 0.088 & 0.114 & 0.286 & 0.289 & 0.527 & 0.391\\
UP & 0.214 & 0.346 & 0.358 & 0.322 & 0.506 & 0.469\\
\addlinespace
VT & 0.080 & 0.286 & 0.100 & 0.156 & 0.292 & 0.250\\
WH & --- & --- & --- & 0.000 & 0.000 & 0.000\\
WL & --- & --- & --- & 0.043 & 0.097 & 0.071\\
YI & 0.106 & 0.206 & 0.178 & 0.185 & 0.319 & 0.307\\
YN & 0.083 & 0.111 & 0.250 & 0.029 & 0.059 & 0.056\\
\bottomrule
\end{longtable}

\clearpage
\subsection{Summary Statistics for Social Support Networks}\label{si:summary_stats}

We present here key network summary statistics for each social support network layer, the aggregate network, and the close kin network. We do not present here statistics for networks derived from the two network prompts focused on external ties, as so many of the nominations are to external alters, making the structure of the remaining ties uninformative. Summary statistics were calculated in \textsf{R} using functions from the \textsf{igraph} package \cite{igraph_orig_2006, antonov2026igraph, igraph_package_2026}.

Table~\ref{tab:mean-median-net-summary-stats} shows the median, mean, and standard deviation of values across all ENDOW communities for the various measures, while Tables~\ref{tab:net-summary-stats-loan_req}-\ref{tab:net-summary-stats-m_info} show these same summary statistics for each community, one table for each network layer and Table~\ref{tab:net-summary-stats-sum} for the aggregate network (and Table~\ref{tab:net-summary-stats-primary_kin} for the close kin network).
Throughout, we reference other plausible benchmarks, which may either be reasonable points of comparison for the ENDOW data or be common referents that may inform scholars tacit expectations of what is ``normal'' for any particular measure.

The networks are directed and can be weighted (with edge weights reflecting the multiple nominations by and of individuals within a sharing unit), but as many of the summary statistics we present do not make use of edge weights, edges can generally be understood as binary, with an edge from sharing unit $i$ to sharing unit $j$ reflecting the fact that at least one member of $i$ nominated at least one member of $j$ for that network prompt (in the case of the individual network layers) or that at least one member of $i$ reported receiving support of some type from at least one member of $j$ (in the case of the aggregate network).
Sub-Section~\ref{si:sampling_impact} examines the consequences of the variable sampling of sharing units across communities and considers measures that draw on our edge weights in a bid to account for some of these differences.

\begin{landscape}
    \begingroup\fontsize{8}{10}\selectfont

\begin{longtable}[t]{lccccccccc}
\caption{\label{tab:mean-median-net-summary-stats}Median (above) and mean and standard deviation (below) of network summary statistics for each network layer. `Mean Deg.' shows the mean degree, while `Mean W. Deg.' uses edge weights. `(In-)Deg. Centr.' shows the (in-)degree centralization. `Deg. Assort.' shows the degree assortativity. `APL' shows the average path length. `Diam.' shows the diameter. `N Isol.' shows the number of isolates, and `Prop. Isol.' their proportion of all nodes. `N Comp.' shows the number of weakly-connected components. `Largest Comp.' shows the proportion of nodes in the largest component. `Global Clust.' shows the global clustering coefficient. `Av. Local Clust.' shows the average local clustering coefficient. `Recip.' shows reciprocity.}\\
\toprule
 & Borrow money & Lend money & Borrow goods & Lend goods & Women behav. & Men behav. & Women info. & Men info. & Aggregate\\
\midrule
\endfirsthead
\caption[]{Median (above) and mean and standard deviation (below) of network summary statistics for each network layer. `Mean Deg.' shows the mean degree, while `Mean W. Deg.' uses edge weights. `(In-)Deg. Centr.' shows the (in-)degree centralization. `Deg. Assort.' shows the degree assortativity. `APL' shows the average path length. `Diam.' shows the diameter. `N Isol.' shows the number of isolates, and `Prop. Isol.' their proportion of all nodes. `N Comp.' shows the number of weakly-connected components. `Largest Comp.' shows the proportion of nodes in the largest component. `Global Clust.' shows the global clustering coefficient. `Av. Local Clust.' shows the average local clustering coefficient. `Recip.' shows reciprocity. \textit{(continued)}}\\
\toprule
 & Borrow money & Lend money & Borrow goods & Lend goods & Women behav. & Men behav. & Women info. & Men info. & Aggregate\\
\midrule
\endhead

\endfoot
\bottomrule
\endlastfoot
Density & 0.02 & 0.03 & 0.03 & 0.03 & 0.02 & 0.03 & 0.03 & 0.03 & 0.1\\
 & 0.03 (0.03) & 0.04 (0.03) & 0.05 (0.04) & 0.05 (0.04) & 0.03 (0.03) & 0.03 (0.03) & 0.05 (0.05) & 0.05 (0.04) & 0.12 (0.08)\\
Mean Deg. & 1.58 & 1.95 & 2.35 & 2.33 & 1.45 & 1.64 & 2.49 & 2.23 & 6.61\\
 & 1.88 (1.41) & 2.15 (1.5) & 2.53 (1.48) & 2.68 (1.68) & 1.62 (1.33) & 1.89 (1.42) & 2.7 (1.56) & 2.54 (1.66) & 6.92 (3.34)\\
Mean W. Deg. & 1.31 & 1.54 & 2.02 & 2.14 & 1.2 & 1.61 & 2.34 & 1.93 & 10.75\\
 & 1.5 (1.05) & 1.71 (1.11) & 2.1 (1.1) & 2.2 (1.24) & 1.53 (1.16) & 1.79 (1.52) & 2.5 (1.35) & 2.23 (1.16) & 11.57 (5.96)\\
Deg. Centr. & 0.07 & 0.06 & 0.07 & 0.07 & 0.05 & 0.06 & 0.07 & 0.09 & 0.16\\
 & 0.1 (0.09) & 0.08 (0.06) & 0.08 (0.05) & 0.08 (0.05) & 0.07 (0.05) & 0.08 (0.05) & 0.09 (0.05) & 0.11 (0.07) & 0.17 (0.08)\\
In-Deg. Centr. & 0.11 & 0.08 & 0.1 & 0.09 & 0.07 & 0.08 & 0.11 & 0.14 & 0.23\\
 & 0.18 (0.19) & 0.11 (0.09) & 0.12 (0.07) & 0.11 (0.07) & 0.08 (0.06) & 0.11 (0.08) & 0.13 (0.08) & 0.17 (0.1) & 0.25 (0.15)\\
Deg. Assort. & -0.05 & -0.01 & 0.02 & 0.02 & 0.05 & 0.06 & 0.03 & 0.04 & 0\\
 & -0.09 (0.16) & -0.02 (0.14) & 0 (0.13) & 0.02 (0.17) & 0.02 (0.14) & 0.06 (0.15) & 0.03 (0.09) & 0.02 (0.13) & 0.01 (0.07)\\
APL & 2.68 & 3.06 & 3.69 & 3.6 & 2.26 & 2.95 & 3.34 & 3.06 & 2.61\\
 & 2.76 (1.3) & 3.18 (1.31) & 3.48 (1.35) & 3.5 (1.45) & 2.74 (1.39) & 2.98 (1.44) & 3.58 (1.4) & 3.38 (1.45) & 2.82 (0.82)\\
Diam. & 7.5 & 8.5 & 10 & 8.5 & 6 & 7 & 9 & 8 & 6\\
 & 7.26 (4.01) & 8.33 (3.94) & 9.18 (4.54) & 9.28 (4.66) & 7.18 (4.18) & 7.69 (4.48) & 9.16 (4.22) & 8.94 (4.19) & 6.78 (2.46)\\
N Isol. & 5 & 4.5 & 2 & 2 & 10 & 10 & 5 & 8 & 0\\
 & 13.28 (17.7) & 11.3 (16.27) & 11.08 (31.66) & 10.22 (25.79) & 29.43 (73.09) & 26.94 (66.22) & 10.51 (14.9) & 14 (19.81) & 0.8 (1.95)\\
Prop. Isol. & 0.11 & 0.08 & 0.03 & 0.03 & 0.2 & 0.13 & 0.08 & 0.12 & 0\\
 & 0.2 (0.23) & 0.18 (0.23) & 0.1 (0.15) & 0.1 (0.15) & 0.27 (0.27) & 0.24 (0.23) & 0.13 (0.16) & 0.16 (0.13) & 0.01 (0.03)\\
N Comp. & 8.5 & 7 & 3.5 & 3.5 & 13 & 13 & 7 & 9 & 1\\
 & 16.26 (19.48) & 14.11 (18.39) & 13.96 (35.13) & 12.92 (29.37) & 34.18 (79.6) & 31.06 (70.61) & 12.55 (16.63) & 16 (20.52) & 2.02 (2.64)\\
Largest Comp. & 0.83 & 0.91 & 0.96 & 0.97 & 0.72 & 0.8 & 0.9 & 0.88 & 1\\
 & 0.7 (0.32) & 0.75 (0.31) & 0.82 (0.27) & 0.83 (0.25) & 0.59 (0.35) & 0.65 (0.33) & 0.83 (0.24) & 0.79 (0.21) & 0.97 (0.11)\\
Global Clust. & 0.18 & 0.21 & 0.26 & 0.24 & 0.25 & 0.25 & 0.26 & 0.23 & 0.34\\
 & 0.18 (0.11) & 0.22 (0.13) & 0.26 (0.12) & 0.26 (0.11) & 0.24 (0.14) & 0.25 (0.14) & 0.28 (0.12) & 0.25 (0.13) & 0.36 (0.13)\\
Av. Local Clust. & 0.28 & 0.25 & 0.36 & 0.31 & 0.31 & 0.3 & 0.31 & 0.28 & 0.41\\
 & 0.27 (0.18) & 0.27 (0.15) & 0.33 (0.15) & 0.32 (0.14) & 0.3 (0.17) & 0.3 (0.17) & 0.32 (0.14) & 0.3 (0.16) & 0.42 (0.14)\\
Recip. & 0.19 & 0.23 & 0.31 & 0.31 & 0.27 & 0.26 & 0.32 & 0.22 & 0.46\\
 & 0.18 (0.12) & 0.25 (0.12) & 0.32 (0.14) & 0.31 (0.16) & 0.26 (0.16) & 0.28 (0.15) & 0.32 (0.12) & 0.24 (0.11) & 0.45 (0.1)\\*
\end{longtable}
\endgroup{}

\end{landscape}

\clearpage
\subsubsection{Mean Degree}\label{si:mean_degree}
\begin{figure}[h]
\begin{center}
\includegraphics[width = \textwidth]{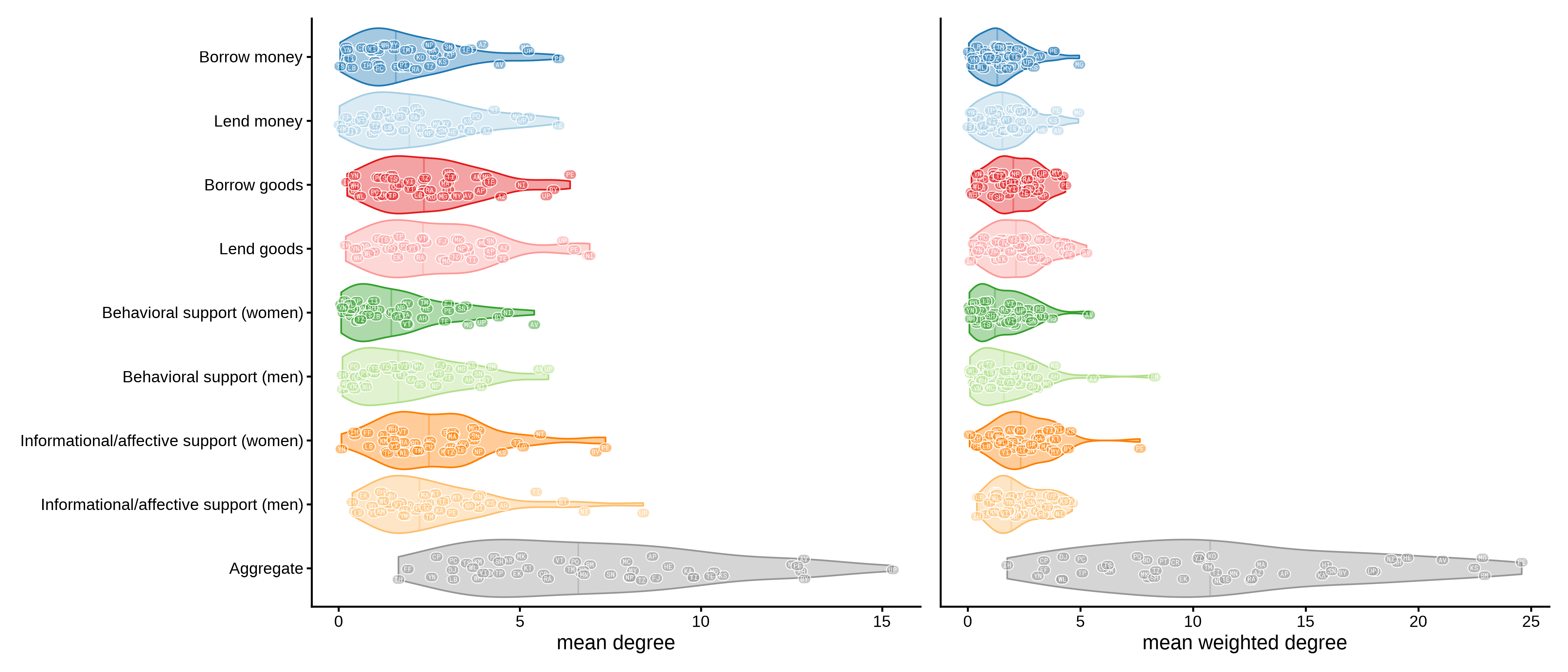}
\end{center}
\caption{Distribution of (left) mean degree and (right) mean weighted degree, for each network layer. Weighted degree is done on the basis of edge weights, which consider both the total number of nominations to individuals in sharing unit $j$ by sharing unit $i$, and the number of respondents able to make nominations from sharing unit $i$.}
\label{fig:mean_degree}
\end{figure}

Figure~\ref{fig:mean_degree} shows the distribution of mean degree for each network layer. This shows both mean degree (\textit{i.e.}, the number of sharing units named for the network prompt) and the mean weighted degree, drawing on an edge weight which divides the number of individuals in sharing unit $j$ named as alters by sharing unit $i$ by the number of individuals in sharing unit $i$ who were respondents for that network prompt. Specifically, if two individuals in sharing unit $i$ are interviewed, and one names $3$ people in sharing unit $j$ for the borrowing money question, and another names $2$ people, then the borrow money layer gets a weight of $(3+2)/2=2.5$; for the aggregate network, we sum the edge weights of each constituent layer.

Recall that degree here reflects the number of connections \emph{to other sharing units in the community}, and can be contrasted with what is shown in Figure~\ref{fig:nomination_counts_violin}, which shows the number of connections to all individuals and to sharing units reported by each respondent.
In each network layer, the number of sharing unit alters is relatively low, typically in the single digits (medians range between roughly $1.5$ and $2.4$).
It is evident that the mean degree in the aggregate network across the communities varies widely, from only $1.7$ connections on average in the Colombian community of IH to the maximum of $15.3$ in the Indian community of UP (with a median across communities of $6.6$).
This reflects variation in nominations, as well as the relative reliance on co-residents for support. In IH, for example, $86\%$ of nominations were to alters external to the community.
Weighted degree shows similar patterns across layers to that of mean degree, though is higher. This reflects that respondents (whether one or multiple) from a sharing unit may name multiple people as alters from another sharing unit. The mean weighted degree distributions follow very similar patterns.

In Sub-Section~\ref{si:sampling_impact} we compare the number of respondents in sharing units with their out-degree (Figure~\ref{fig:out_deg_vs_num_surv}) and weighted out-degree, i.e., out-strength (Figures~\ref{fig:weighted_out_deg_vs_num_surv}) within each community, as well as community-level average measures (in Figure~\ref{fig:degree_vs_sampling}), which show that the sharing units and communities with more sampled respondents often have higher values in the unweighted measures. In contrast, the weighted measures do not vary with the number of sampled respondents, suggesting that they better account for sampling variation.

Recall the statement by Koskinen and Snijders \cite{koskinen_multilevel_2023} that an average degree of approximately $4$ is ``quite usual'' for friendship networks: our most comparable network to this is the informational/affective support networks, which shows a median mean degree of $2.5$ for the women's network and $2.2$ for the men's, substantially lower (see Table~\ref{tab:mean-median-net-summary-stats}). Of course, this is to sharing units; if we consider the median of mean nominations of individuals by respondents, we get $3.4$ and $3.2$ respectively across the communities (see Figure~\ref{fig:nomination_counts_violin}), somewhat closer.

For comparison to our mean degree measures, in 2011, the mean number of US-based Facebook friends for US-based users was $214$ \cite{ugander_anatomy_2011}.
In the canonical Add Health network dataset (from over $100$ American schools), students named on average $3.8$ friends (SD $3.7$) within their school \cite{fowler_model_2009}. 
The dataset of $82$ Dutch classrooms reports a mean out-degree of about $4$ \cite{koskinen_multilevel_2023}.
The reported median mean degree for the undirected, \emph{individual-level} friendship and financial networks across the $176$ Honduran villages are $5.02/4.53$ and $2.41/2.26$ for wave 1/wave 3, respectively \cite{papamichalis_deep_2026}.
Aggregate individual-level networks from $20$ communities in Uganda and $12$ in Kenya had an average degree of $5.8$ and $1.6$, respectively, across communities \cite{chen_social_2021}.
All of these cases, though, are of individual respondents, whereas our networks are of sharing units.
The aggregated Karnataka household networks had an average degree of almost $15$ \cite{banerjee_diffusion_2013}, broadly comparable to our own values for the aggregate network.
Household-level networks of help with the rice harvest from 51 villages in Thailand \cite{entwisle_networks_2007} are comparable to our behavioural support networks; they report an average of $10.8$ immediate connections (with a standard deviation of $16.9$).

\clearpage
\subsubsection{Degree Distribution}\label{si:deg_distrib}
\begin{figure}[h]
\begin{center}
\includegraphics[width = 0.9\textwidth]{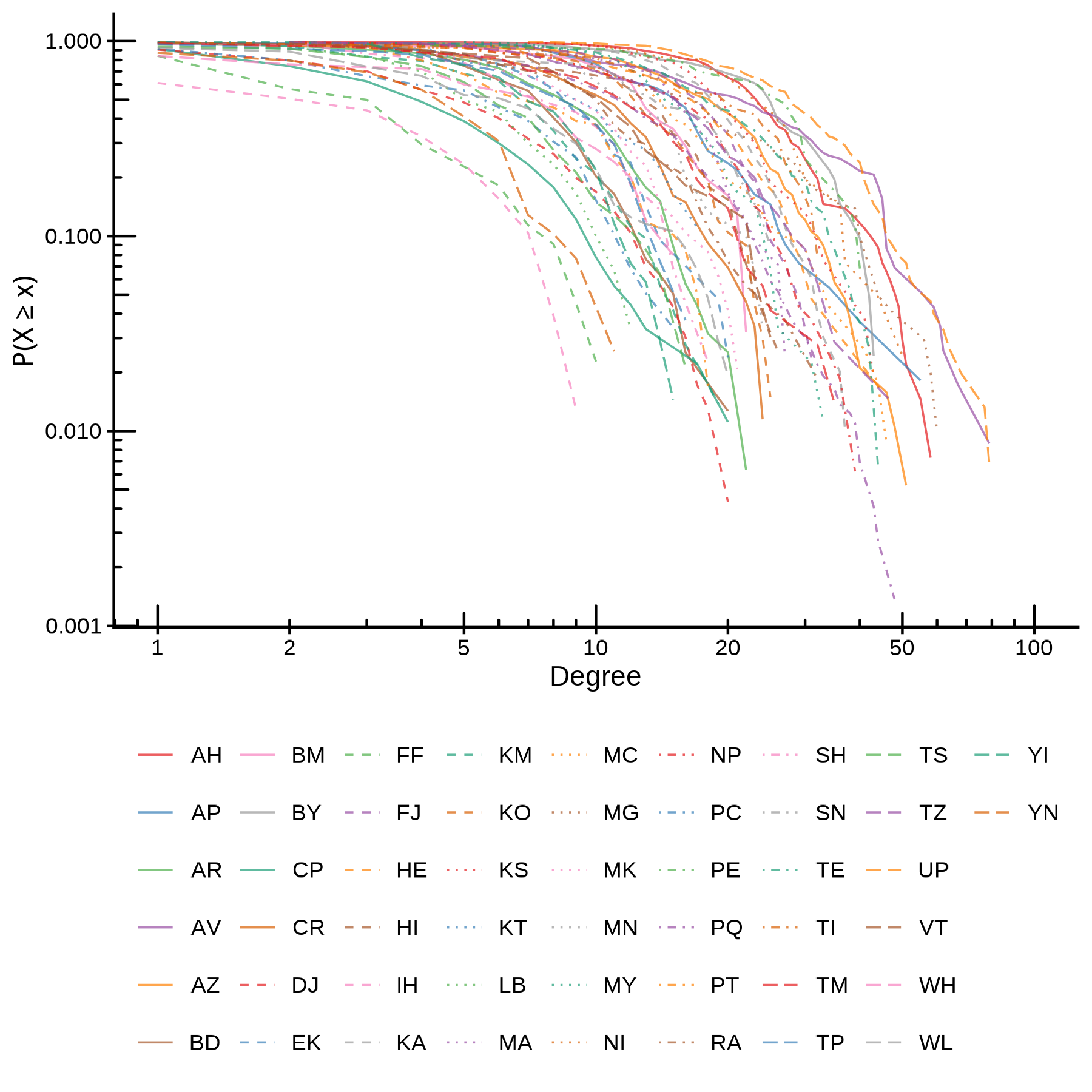}
\end{center}
\caption{Degree distributions for the aggregate network.}
\label{fig:degree_distrib}
\end{figure}

\begin{figure}[h]
\begin{center}
\includegraphics[width = 0.9\textwidth]{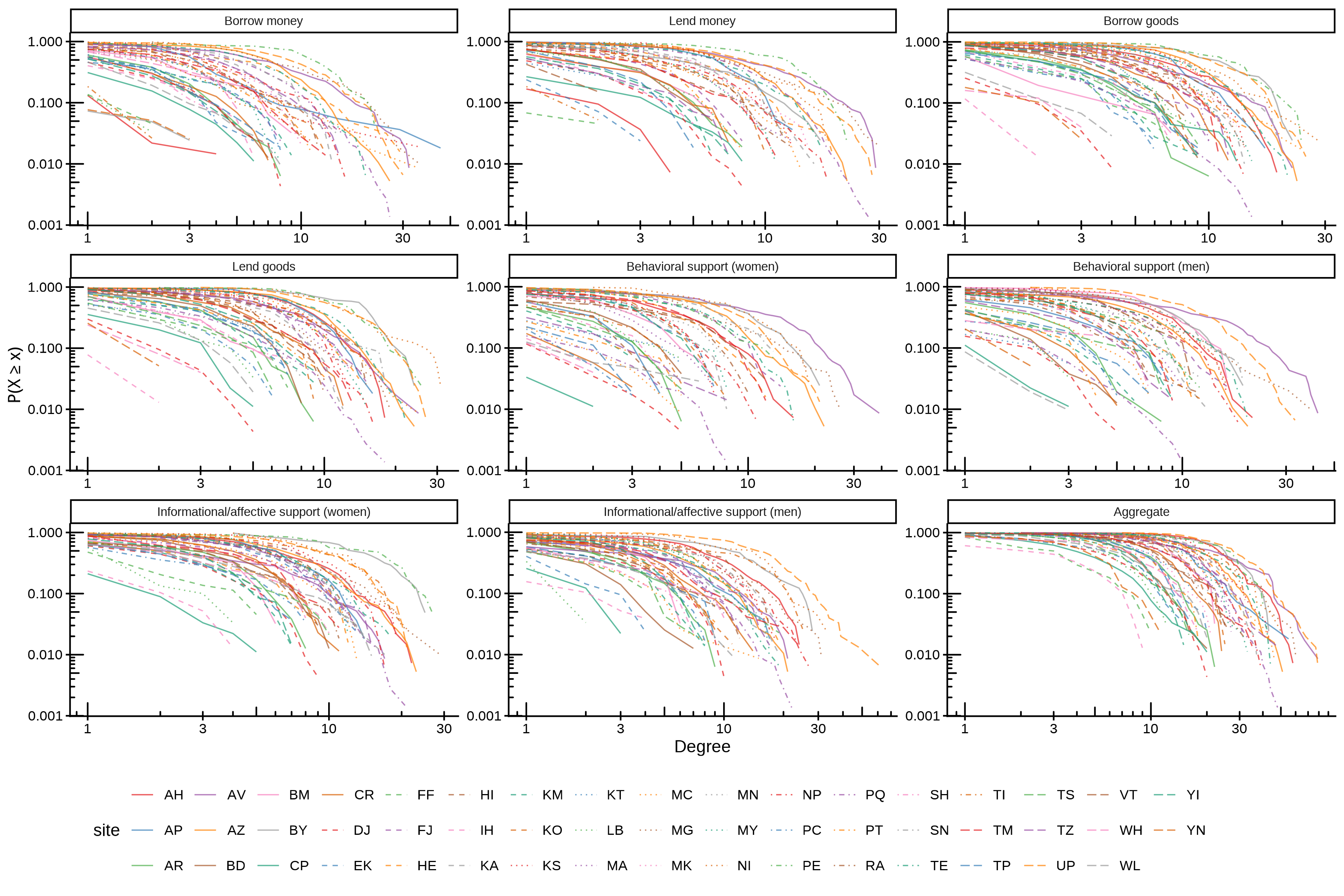}
\end{center}
\caption{Degree distributions for the aggregate network.}
\label{fig:degree_distrib_layers}
\end{figure}

Figure~\ref{fig:degree_distrib} shows the degree distributions for the aggregate network, while Figure~\ref{fig:degree_distrib_layers} shows the degree distributions for each network layer (excluding the two external-focused networks). While the distributions clearly differ in scale by community, they do largely show a right-skewed degree distribution with a heavy tail \cite{newman2003structure}. Few of these, however, are likely to be strictly defined as being scale-free \cite{broido2019scale}.

\clearpage
\subsubsection{In-Degree Centralization}\label{si:deg_centralization}

\begin{figure}[h]
\begin{center}
\includegraphics[width = \textwidth]{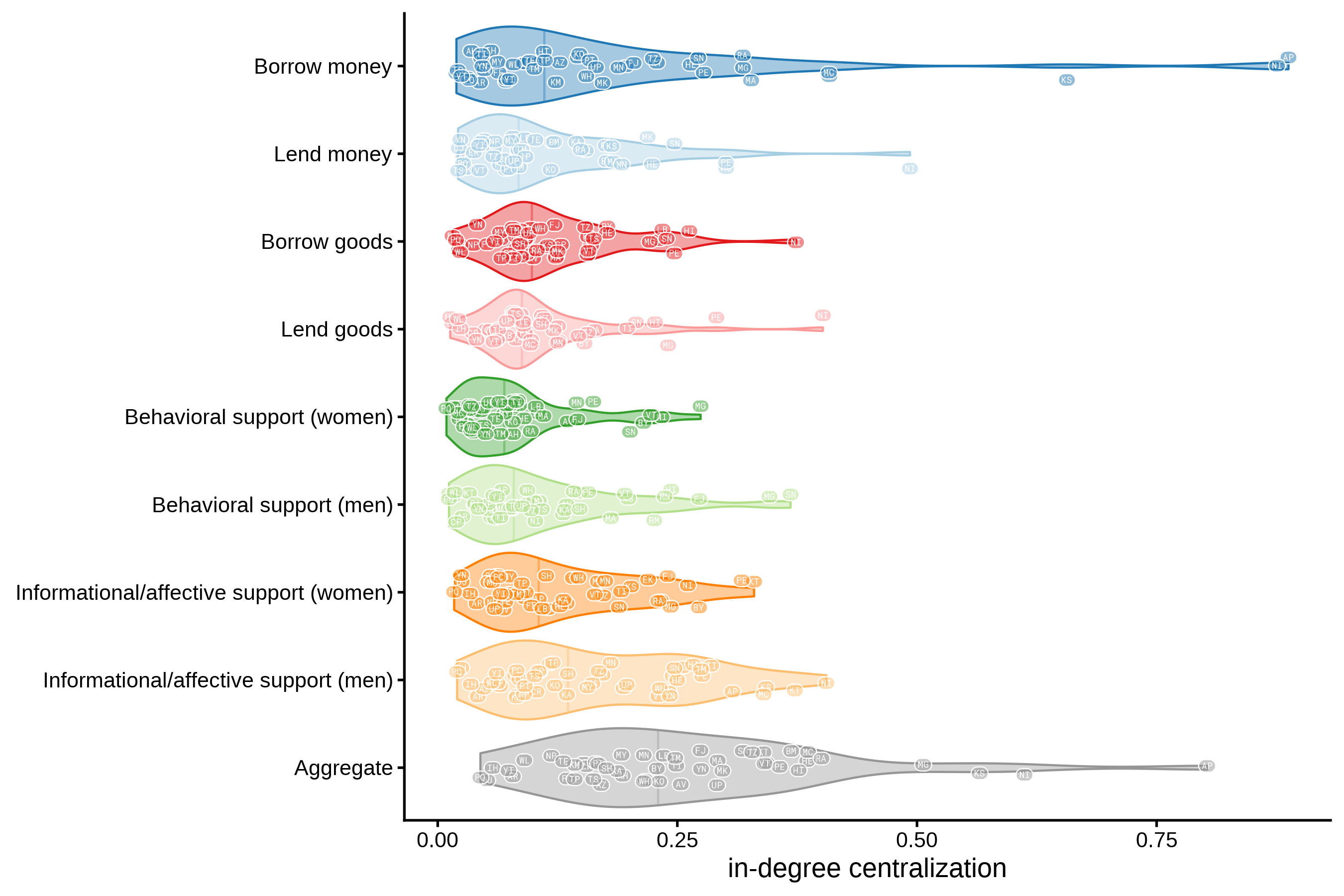}
\end{center}
\caption{Distribution of in-degree centralization, for each network layer.}
\label{fig:in_degree_centralization}
\end{figure}

Overall, in-degree centralization is seemingly low and not too variable across communities.
Figure~\ref{fig:in_degree_centralization} shows the distribution of in-degree centralization (not considering any edge weights) for each network layer. Two notable outliers are the communities of NI and AP: these have high in-degree centralization in the borrow money networks and are often high in other layers, all of which contributes to them having the highest values in the aggregate network. In both, much of this is driven by one or a few sharing units that almost all others report turning to for monetary loans, and influential sharing units are evident in other layers, too.
The network with the highest median degree and in-degree centralization is the men's informational/affective support network, which could suggest a propensity for men's socialization to be more concentrated on prominent figures in particular sharing units.

\clearpage
\subsubsection{Degree Assortativity}\label{si:degree_assort}

\begin{figure}[h]
\begin{center}
\includegraphics[width = \textwidth]{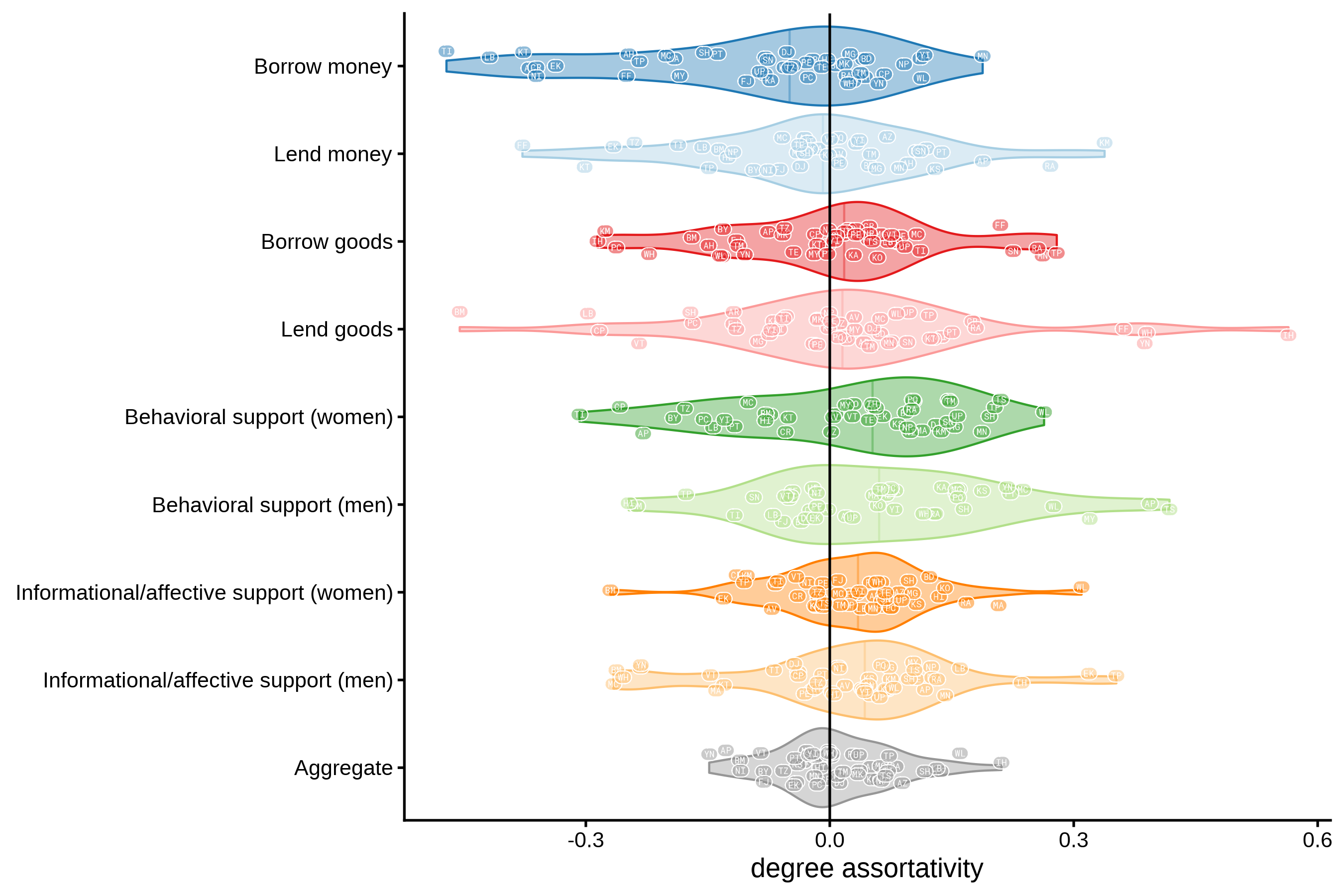}
\end{center}
\caption{Distribution of degree assortativity, for each network layer.}
\label{fig:deg_assort}
\end{figure}

Figure~\ref{fig:deg_assort} shows assortativity by degree \cite{newman_mixing_2003}. This outcome can range from -1 to 1, with 0 reflecting no assortment by degree. It is notable, then, that we observe negative values for the two monetary loan layers (as shown in Table~\ref{tab:mean-median-net-summary-stats}, the median across communities of -0.05 and -0.01 for borrow money and lend money, respectively), though as always there are many exceptions.
Generally, there is slight but not substantial positive assortment; indeed, in the aggregate network, the median value across communities is almost exactly 0. As social networks are expected to display positive degree assortativity \cite{newman_assortative_2002, newman_why_2003, jackson_meeting_2007}, this means that we find more evidence of degree \emph{dis}assortativity than might have been expected.

\clearpage
\subsubsection{Density}\label{si:density}
\begin{figure}[h]
\begin{center}
\includegraphics[width = \textwidth]{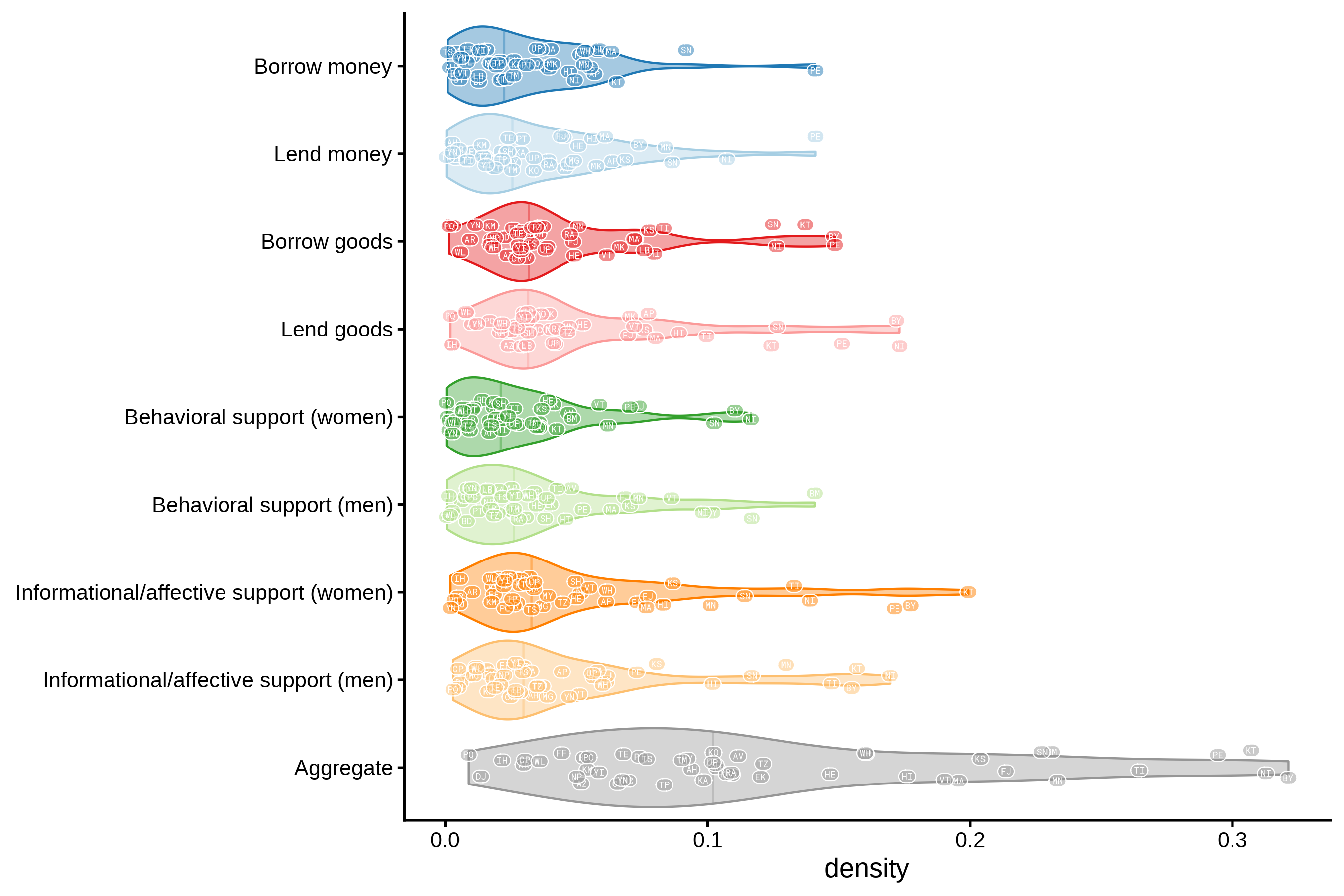}
\end{center}
\caption{Distribution of edge density, for each network layer.}
\label{fig:density}
\end{figure}

\begin{figure}[h]
\begin{center}
\includegraphics[width = 0.5\textwidth]{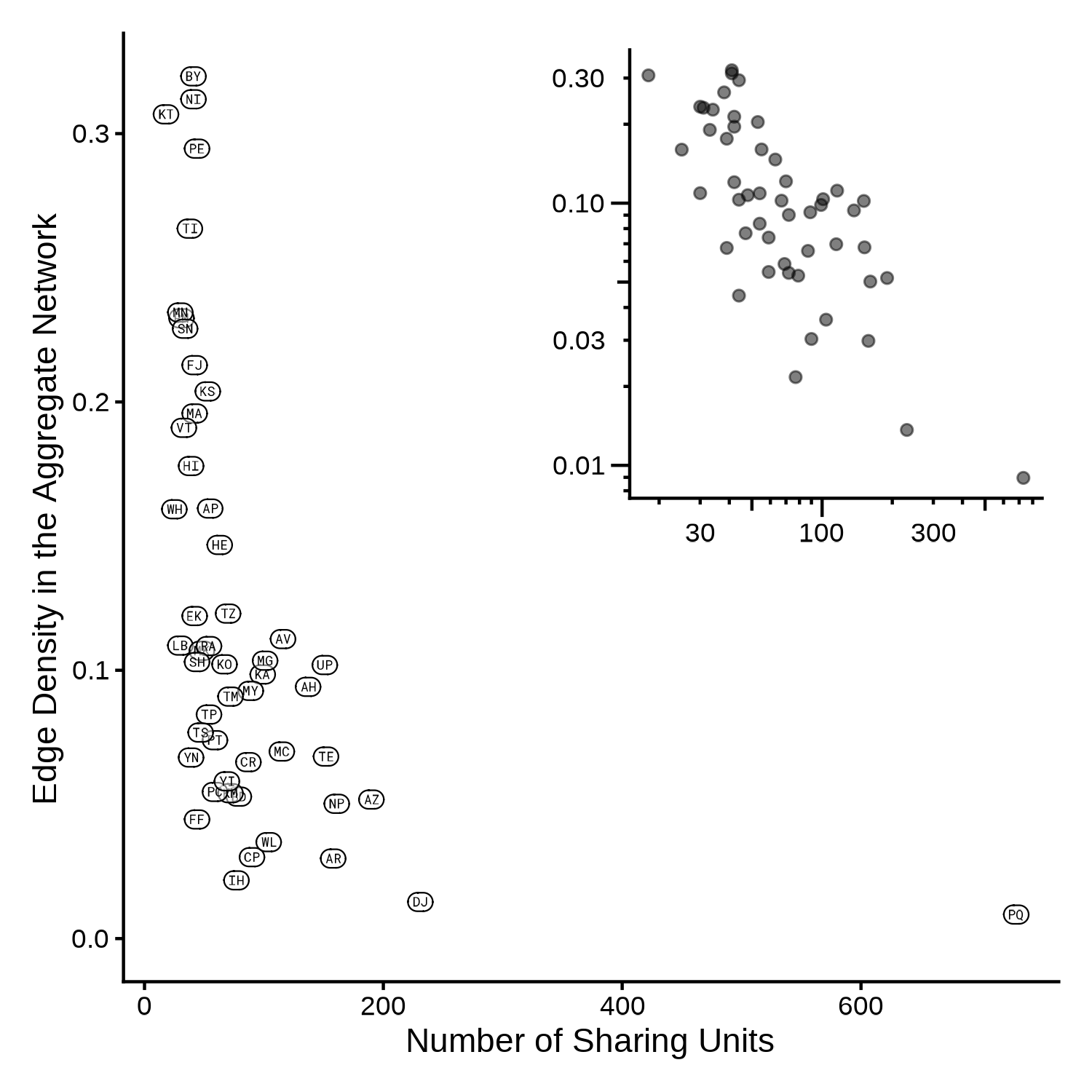}
\end{center}
\caption{Association between community size (number of sharing units) and edge density in the aggregate network. The inset figure shows the same bivariate association, with axes on the base 10 log scale.}
\label{fig:density_vs_size}
\end{figure}

Figure~\ref{fig:density} shows the distribution of edge density (not considering any edge weights) for each network layer. This will, of course, be influenced by the number of nodes (i.e., sharing units) in the network (see Figure~\ref{fig:density_vs_size} for their bivariate association). Indeed, in the aggregate network we see one of the highest density measures in the smallest community (KT, with only 18 sharing units and a density of 0.31) and the lowest density in the largest community (PQ, with 730 sharing units and a density of 0.01).
As observed in the 51 Thai villages \cite{entwisle_networks_2007}, however, even villages of comparable size can vary substantially in their edge density.
We find relatively little difference in density by network layer, with the obvious exception that the aggregate network is much denser. With the aggregation of edges in this network, it is not surprising that this shows a much higher average density across communities.
As one partial point of comparison, the reported median densities for the undirected, \emph{individual-level} friendship and financial networks across the 176 Honduran villages are 0.062/0.058 and 0.030/0.029 for wave 1/wave 3, respectively (with a median of 80 nodes, i.e., people) \cite{papamichalis_deep_2026}.

\clearpage
\subsubsection{Average Path Length}\label{si:av_path_length}
\begin{figure}[h]
\begin{center}
\includegraphics[width = \textwidth]{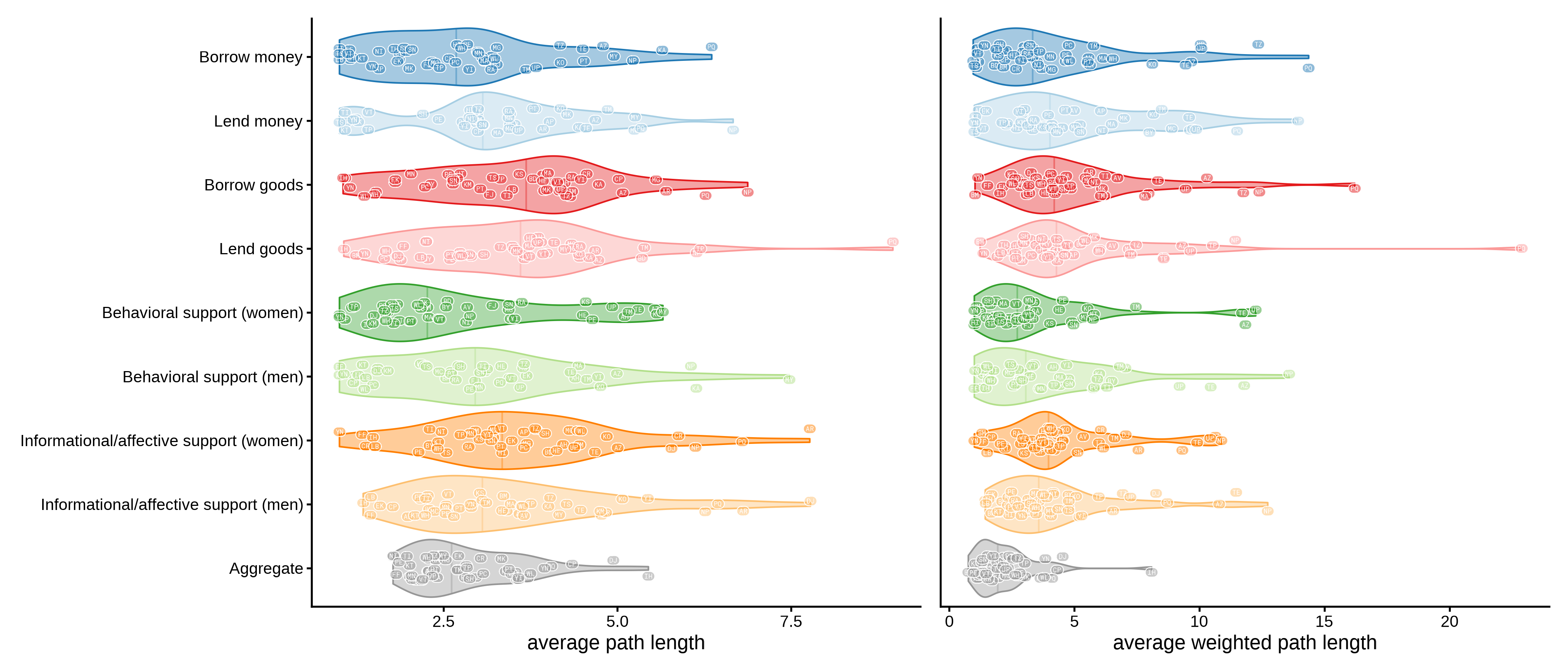}
\end{center}
\caption{Distribution of average path length, for each network layer, either without (left) or with (right) edge weights.}
\label{fig:avg_path_length}
\end{figure}

Figure~\ref{fig:avg_path_length} shows the distribution of average path length (and considering only existing paths) for each network layer. Shorter average path length would imply greater reachability and connectivity across the network. We present both the standard unweighted version (left) and a version using edge weights (right), where, to reflect that shorter values should imply greater proximity, we use the reciprocal of our standard weights (meaning $1/\text{weight}$).
Overall, (as shown in Table~\ref{tab:mean-median-net-summary-stats}) in the aggregate network we observe a median average path length of $2.61$, implying that it would take roughly 3 steps to connect most sharing units. The median diameter across the aggregate networks is $6$, suggesting that some sharing units may be still relatively distant from one another, even in these relatively small communities. It is also worth noting that there are sharing units in a few communities that are isolates, connected to no other sharing units (though they may have support from others outside the community). For particular network layers, there are more often distinct components in the network, making the interpretation and comparability of these measures more difficult. Still, the vast majority of sharing units are (at least weakly) connected into the largest connected component in the aggregate network and at only a few steps remove from one another.

\begin{figure}[h]
\begin{center}
\includegraphics[width = 0.5\textwidth]{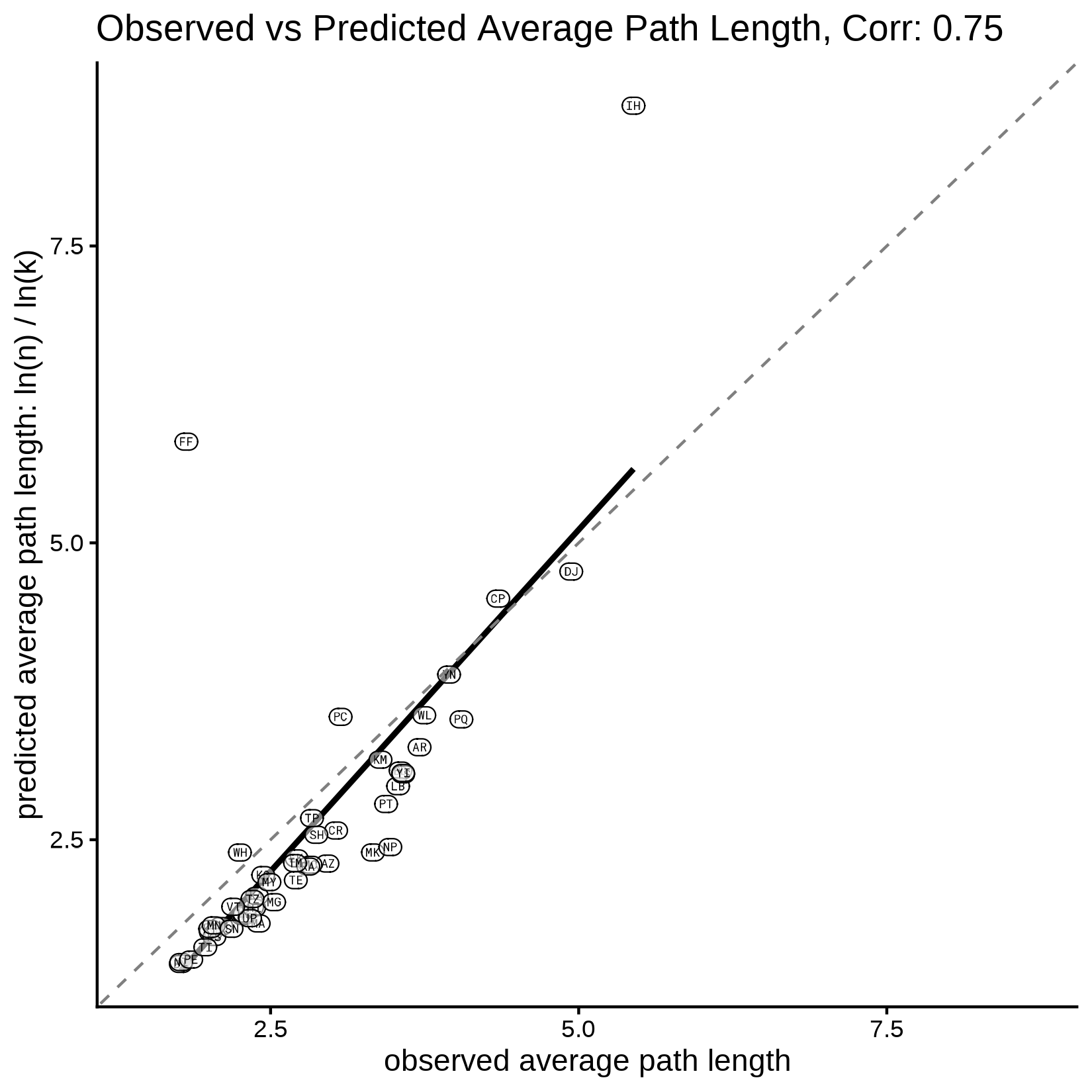}
\end{center}
\caption{Association between the observed and expected average path length in the aggregate network, if we follow the expectation for a random graph \cite{watts_collective_1998}. The dotted line would show perfect equivalence, and the solid line shows the linear regression line, with its associated correlation coefficient shown in the title. }
\label{fig:obs_exp_apl}
\end{figure}

In random graphs, the average path length should roughly be $\ln(n) / \ln(k)$, where $n$ is the number of nodes and $k$ is the mean degree \cite{watts_collective_1998}. As shown in Figure~\ref{fig:obs_exp_apl}, the average path length in the aggregate network is consistently well above this expected level. Given the skewed degree distributions and clustering in our networks (see Sub-Section~\ref{si:transitivity} below), this discrepancy is to be expected. The outliers FF and IH are likely due to their many disconnected components (see Figures~\ref{fig:networks_agg1} and \ref{fig:networks_agg2}).

The canonical empirical reference here is the classic ``small world'' experiment of Travers and Milgram \cite{travers_experimental_1969}, who found that it took $6.2$ steps, on average, for letters to successfully reach the intended recipient (going from person to person, starting in Omaha and Wichita and ending in Boston), leading to the common association of this paper with the idea of ``six degrees of separation".
A larger and global version of this experiment using email again found that successful chains took between five and seven steps \cite{dodds_experimental_2003}.
The ``small worlds'' concept was taken up in a classic model by Watts and Strogatz \cite{watts_collective_1998}, which again provides an empirical benchmark: the film actors collaboration network they reference had a characteristic path length of $3.65$.
An oft-cited reference is that the global Facebook social network showed (in 2011, with over 700 million users) an average path length (for users in the giant component) of $4.7$ \cite{ugander_anatomy_2011}.
Of course, the communities under study here are not at the national or global scale of these analyses, so the smaller average path length should be expected.
The closer comparison is again the aggregated Karnataka household networks, which had an average path length of $2.8$ \cite{banerjee_diffusion_2013}, very comparable to what we observe.
The Ugandan and Kenyan communities have a reported average path length (as with ours, based only on existing paths) of $7$, decidedly higher, but it is worth noting that these are substantially larger individual-level networks, with an average of over 5000 nodes (individuals) \cite{chen_social_2021}.

Isolates and multiple components are more frequent in the individual network layers, making interpretation of the average path length (and yet moreso the diameter) more difficult. The average path length as calculated here shows the average length of \emph{existing} paths. So, while we show a short average path length in the behavioural support networks, this does not imply easy reachability across these networks: as shown in Table~\ref{tab:mean-median-net-summary-stats}, there is often a relatively high number of isolates in these networks (a median of $13-20\%$ across communities).

Average \emph{weighted} path length distributions clearly have a different range, but follow broadly similar patterns (with, \textit{e.g.}, the behavioural support networks again showing as lower than the other layers, and the aggregate network being lowest overall). In Figure~\ref{fig:edge_weight_comparison} below, the weighted and unweighted measures are reasonably correlated, with the weighted values typically being lower than the unweighted versions.

\clearpage
\subsubsection{Reciprocity}\label{si:reciprocity}

\begin{figure}[h]
\begin{center}
\includegraphics[width=0.8\textwidth]{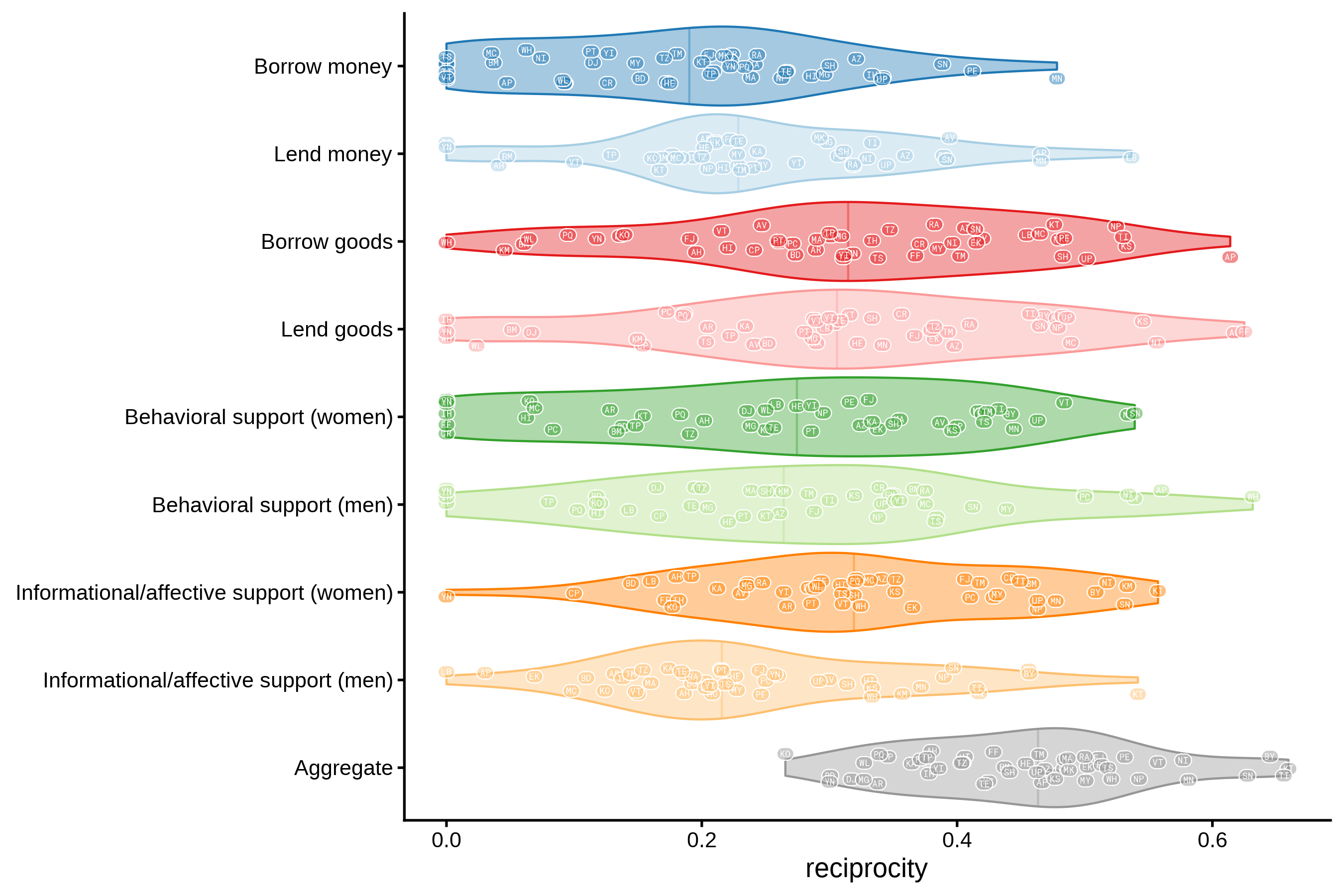}
\end{center}
\caption{Distribution of reciprocity scores, for each network layer.}
\label{fig:reciprocity_violin}
\end{figure}

Figure~\ref{fig:reciprocity_violin} shows the reciprocity across communities for each network layer. Substantial variation is evident across layers and communities.
Across the network layers, median reciprocity ranges between $0.2$ and $0.3$.
In some cases, reciprocity is much lower: in each of the network layers, at least one community has no reciprocal relationships, though these are typically very sparse networks with few within--community nominations.
Appropriately, reciprocity is generally higher in the layers meant to elicit exchanges of common household items between sharing units (median of $0.3$ across communities).
It is lowest in the monetary loan networks (median of $0.2$ across communities), which are also those with relatively higher proportions of nominations outside the community, meaning that these are often rather sparse.

Koskinen and Snijders \cite{koskinen_multilevel_2023} suggest that reciprocity of $0.60$ is ``quite usual'' for friendship networks.
For our informational/affective support networks, median reciprocity values where $0.32$ for the women's network and $0.22$ for the men's, substantially lower.
Our values are closer to the levels of reciprocity in the $75$ Karnataka networks \cite{ready_measuring_2021}, though these are at the individual-level:
on average reciprocity was $18\%$ in the money layer, $19\%$ in the household items layer, $18\%$ in the visit layer, and $35\%$ in the advice layer.
Notably, our aggregate network shows much higher levels of reciprocity across communities, with a median value of $0.46$.

\begin{figure}[h]
\begin{center}
\includegraphics[width = \textwidth]{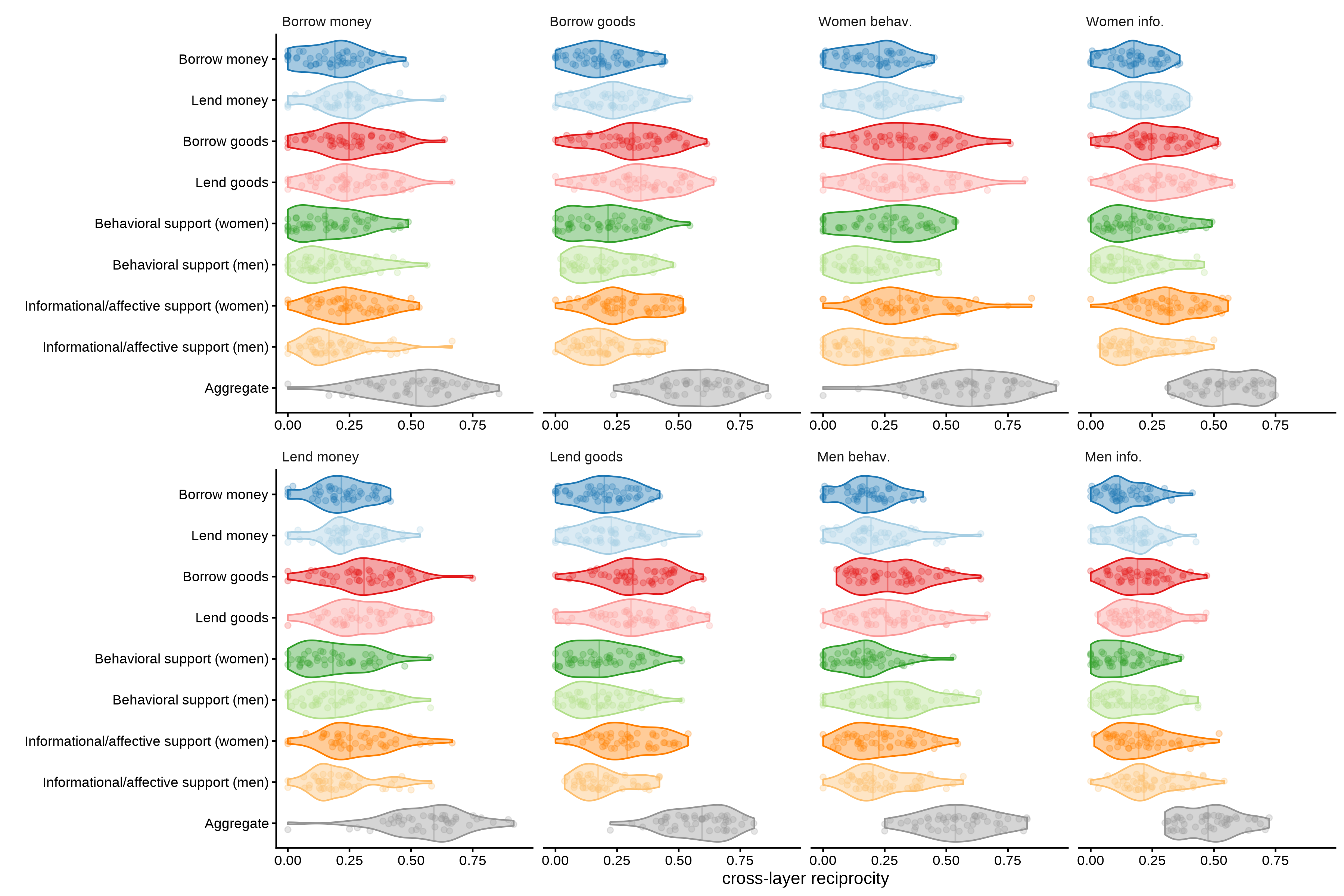}
\end{center}
\caption{Proportion of edges in each network layer that are reciprocated in each other network layer, and in any of the main network layer on requesting support (``Aggregate'').}
\label{fig:recip_across_layers}
\end{figure}

Reciprocity can additionally be considered across network layers.
Figure~\ref{fig:recip_across_layers} shows the distribution across communities of the proportion of edges from sharing unit $i$ to sharing unit $j$ that are reciprocated by $j$ in each other layer, or in any of the layers that are in the aggregate network, i.e., the main layers for requesting support that are focused within the community. So, for example, the top left plot for ``Borrow money'' shows first the proportion of nominations for the borrow money layer from sharing unit $i$ to sharing unit $j$ that are reciprocated with a nomination by $j$ of $i$ directly within the borrow money layer (which is equivalent to what is shown for that network layer in Figure~\ref{fig:reciprocity_violin}), then the proportion that are reciprocated within the lend money layer, etc.

The final ``aggregate'' violins are informative: for effectively all network layers, reciprocity is above $50\%$ in the majority of communities. So, while there may not be reciprocity within a particular relationship type, connections between sharing units are generally reciprocated in some way.
The men-focused information/affective support layer is often one of the least reciprocated across network layers.
We can also see some patterns that align with our concordance analysis in Section~\ref{si:concordance_double}: the highest cross-site median for the borrow goods layer is the lend goods layer (and equivalently for lend goods), but this is not the case for the borrow/lend money layers.

\clearpage
\subsubsection{Transitivity}\label{si:transitivity}
\begin{figure}[h]
\begin{center}
\includegraphics[width=\textwidth]{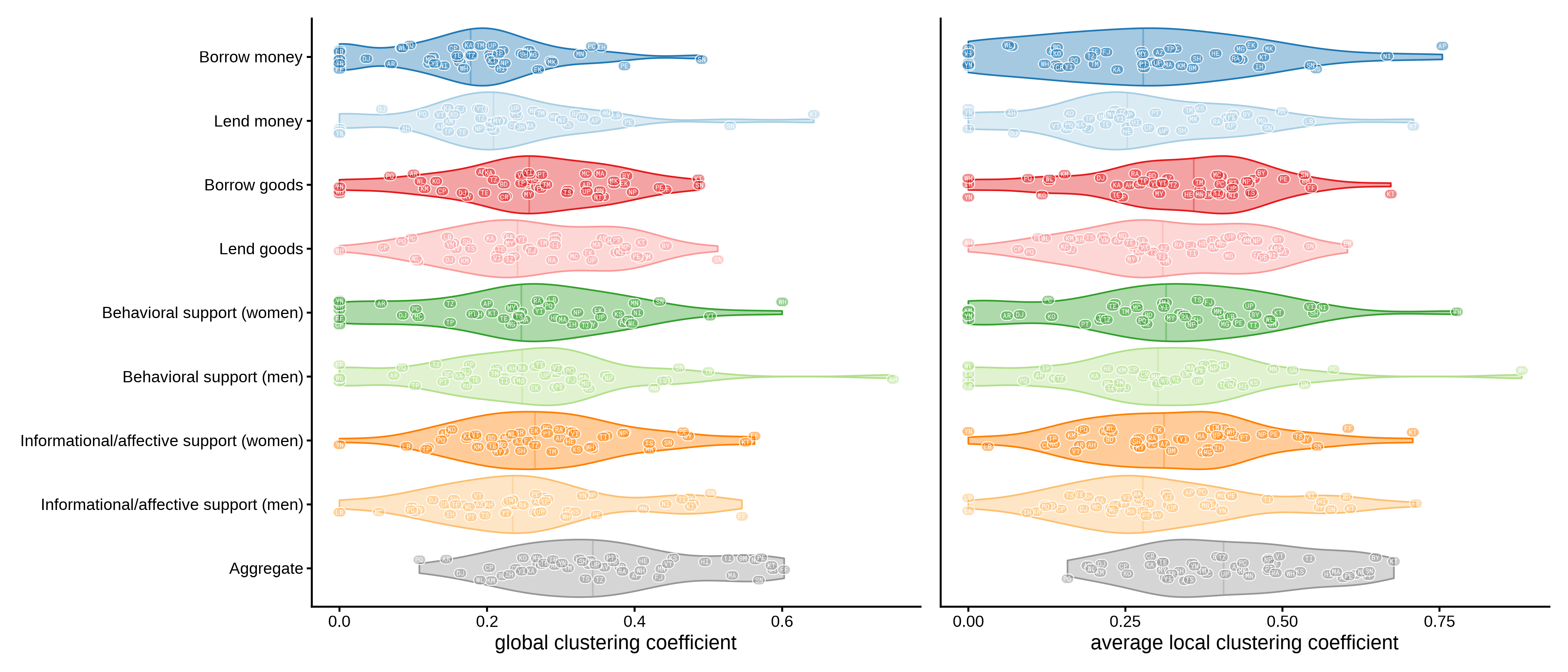}
\end{center}
\caption{Distribution of the (left) global and (right) average local clustering coefficient, for each network layer.}
\label{fig:clustering}
\end{figure}

Figure~\ref{fig:clustering} shows measures of transitivity, specifically the global (left) and average local (right) clustering coefficient (not considering any edge weights) for each network layer.
Across network types, transitivity is lowest for the borrow money network (median global clustering coefficient of $0.18$ and average local clustering coefficient of $0.28$ across communities) and highest for the borrow goods network (median global clustering coefficient of $0.26$ and average local clustering coefficient of $0.36$ across communities).

The seminal ``small worlds'' paper \cite{watts_collective_1998} provides a first possible benchmark for these measures: Watts and Strogatz use as their one indicative social network IMDB film actors collaboration network, for which the global clustering coefficient is $0.20$ and an average local clustering coefficient as $0.79$, with the latter clearly being much higher than what we observe.
A study of the Facebook network (in 2011) shows that individuals' individual clustering coefficient decreases monotonically with number of friends, and that those with one hundred friends (so, broadly comparable to the size of many of the ENDOW communities) had a clustering coefficient of 0.14, on average \cite{ugander_anatomy_2011}.
A much closer analogue to the networks under study here are the Karnataka village networks \cite{banerjee_diffusion_2013}; they construct an undirected, unweighted network of sampled households (where two households are linked if anyone from either household reported any one of the twelve relationships they elicited), and report global clustering coefficient of $0.26$ (which they report as the ``clustering rate'' or ``transitivity ratio'' of $26\%$). This would be most comparable to our aggregate network, where we find a median global clustering coefficient across sites of $0.34$, somewhat higher.
Drawing on undirected, individual-level networks from 176 Honduran villages, \cite{papamichalis_deep_2026} report the median average local clustering coefficient across the villages as being $0.29$/$0.32$ (for wave 1 versus wave 3) for a friendship network (joining together responses to three prompts), and $0.23$/$0.24$ for a double-sampled financial network (joining together responses to ``Who would you feel comfortable asking to borrow 200 lempiras from if you needed them for the day?'' and ``Who do you think would be comfortable asking you to borrow 200 lempiras for the day?'').
Aggregate individual-level networks from 32 communities in Kenya and Uganda had transitivity of $0.1$ and $0.2$ \cite{chen_social_2021}.

In Sub-Section~\ref{si:sampling_impact} below, we show that while within communities, sharing units with more respondents sometimes appear to have lower local clustering (see Figure~\ref{fig:clust_vs_num_surv}), across communities there is little suggestion that places with more respondents had lower average local clustering (see Figure~\ref{fig:clust_vs_sampling}). Compared to our mean degree measures, then, we can be less concerned that cross-site differences in average local clustering is driven by sampling variation. If we nonetheless construct a weighted local clustering coefficient (following \cite{barrat_architecture_2004}) using our edge weights that account for number of respondents, we see that the resulting values are very closely correlated with the unweighted versions (see Figure~\ref{fig:edge_weight_comparison}).

\clearpage
\subsubsection{Consequences of Sharing Unit Sampling}\label{si:sampling_impact}

As noted in the main text and shown in Table~\ref{tab:respondents_by_site}, different sharing unit members responded to the network prompt prompts on behalf of their sharing unit. With our simple rules for aggregation of responses, a potential concern is that some of the observed differences within and between communities is due to this variation in sampling.
The construction of our edge weights is meant to help address this issue, by including how many respondents there were for each sharing unit.

\begin{figure}[h]
\begin{center}
\includegraphics[width = \textwidth]{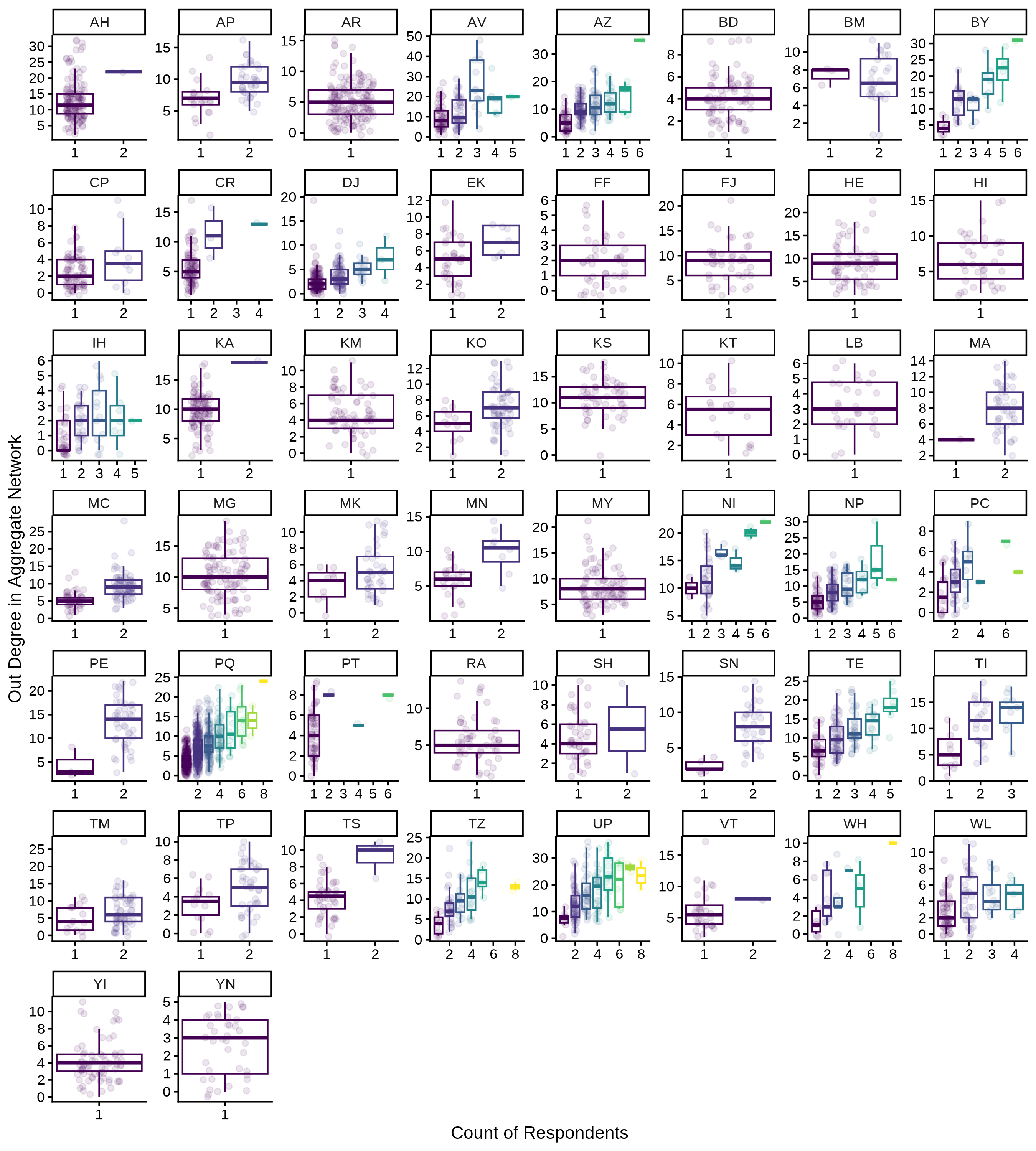}
\end{center}
\caption{Sharing unit's out-degree in the aggregate network (unweighted) versus the number of respondents from each sharing unit. Colours correspond to the count of respondents, with lighter colours reflecting greater number of respondents. }
\label{fig:out_deg_vs_num_surv}
\end{figure}

\begin{figure}[h]
\begin{center}
\includegraphics[width = \textwidth]{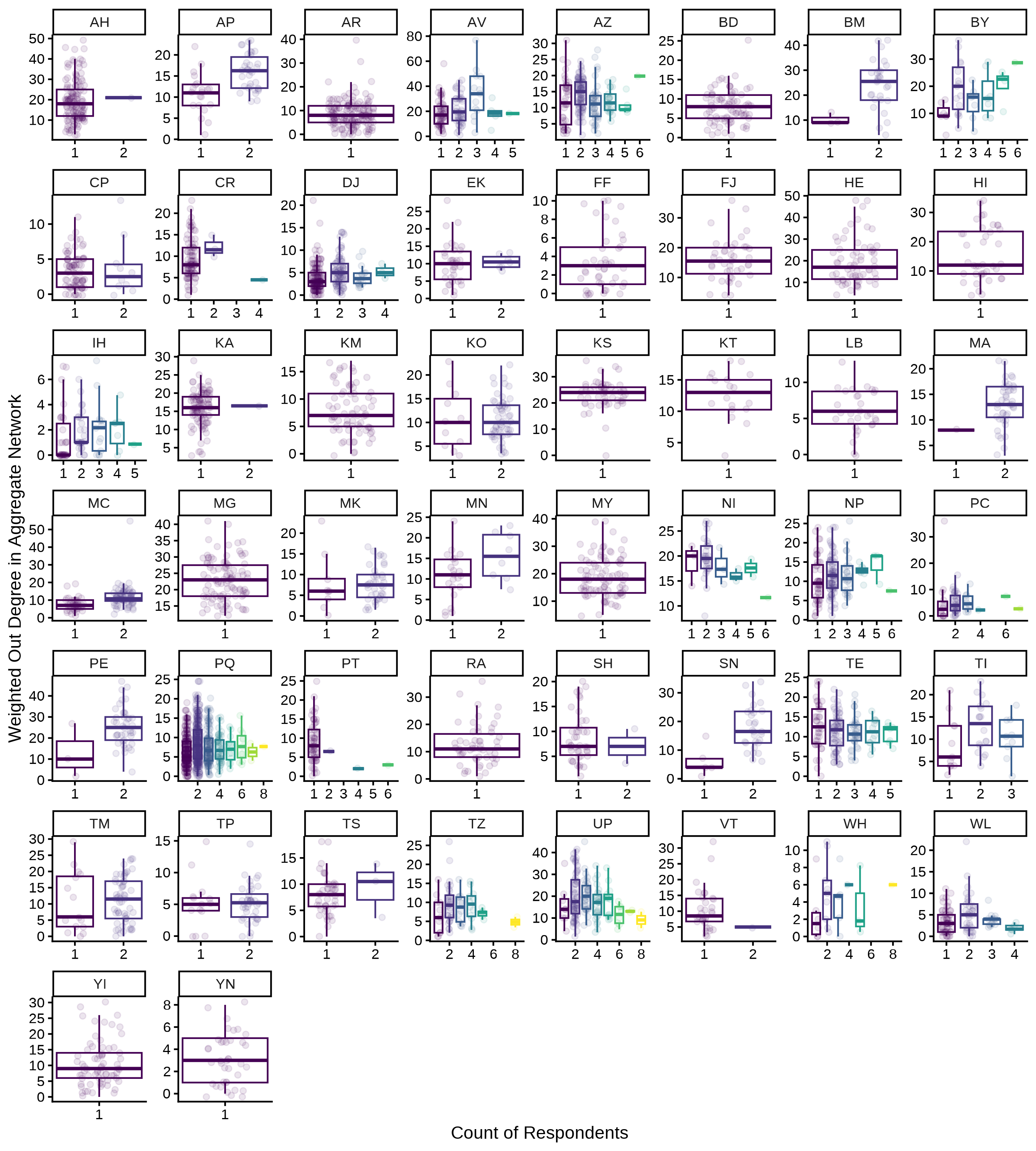}
\end{center}
\caption{Sharing unit's weighted out-degree in the aggregate network versus the number of respondents from each sharing unit. Colours correspond to the count of respondents, with lighter colours reflecting greater number of respondents. }
\label{fig:weighted_out_deg_vs_num_surv}
\end{figure}

\begin{figure}[h]
\begin{center}
\includegraphics[width = \textwidth]{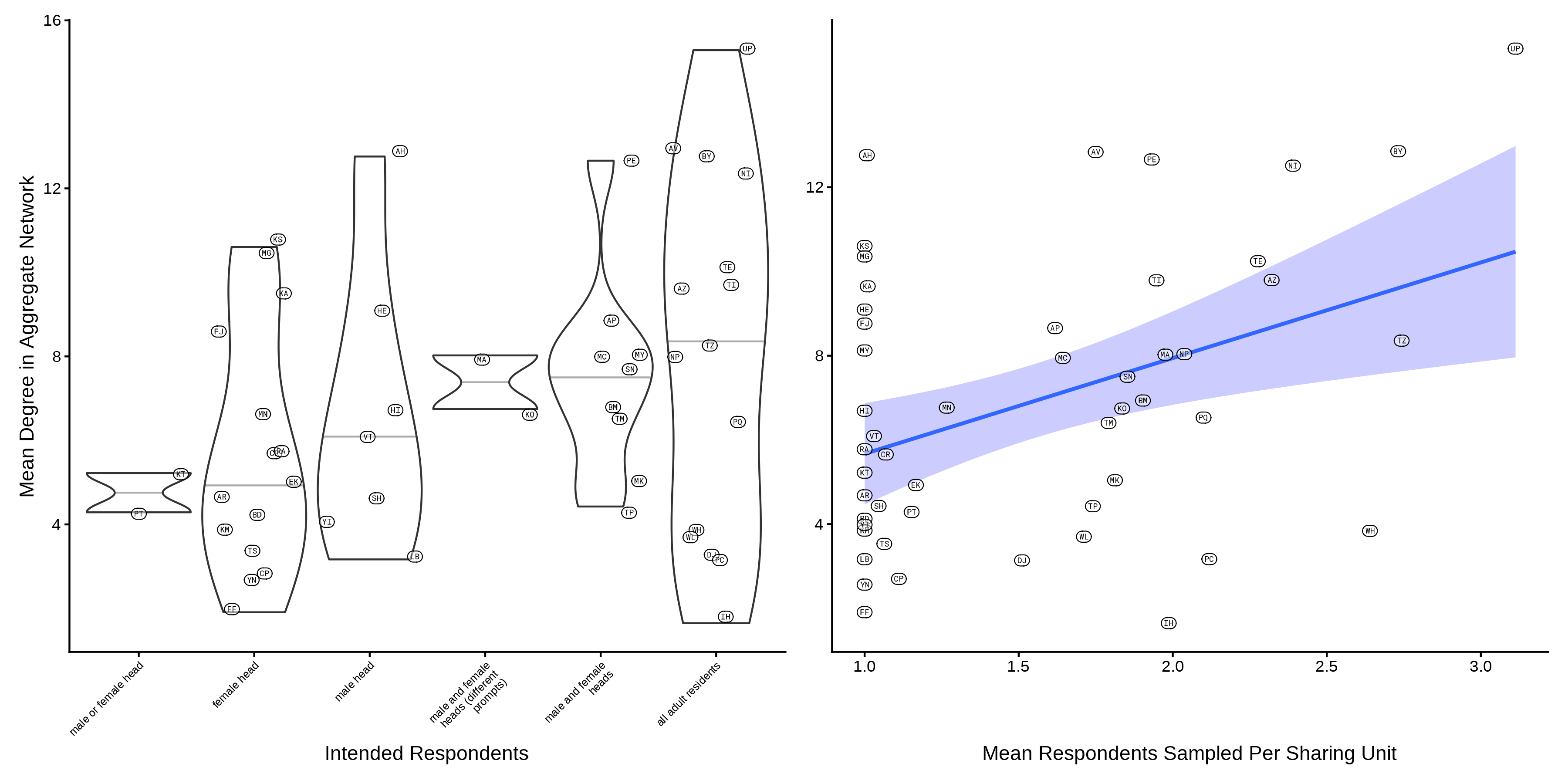}
\includegraphics[width = \textwidth]{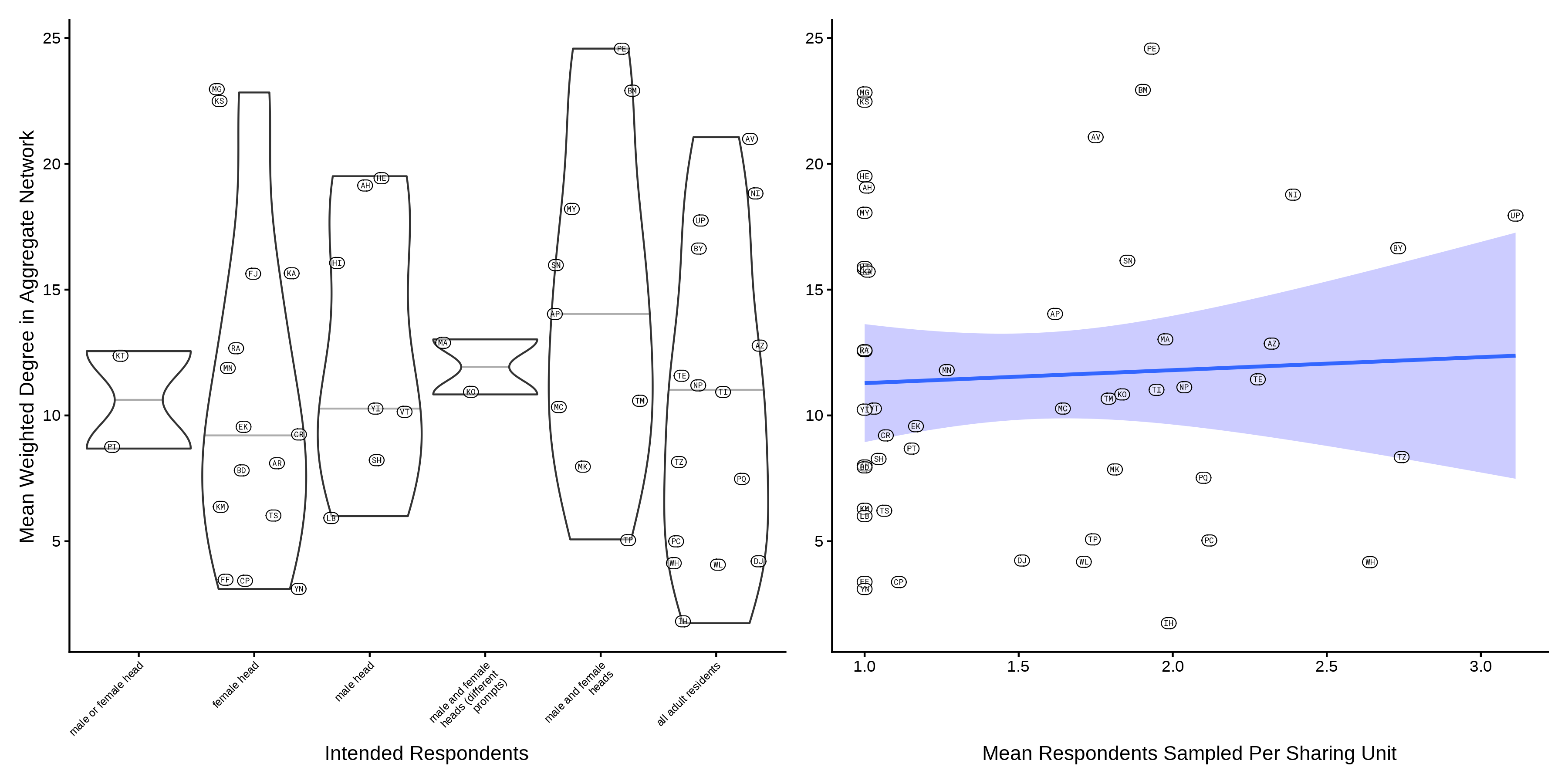}
\end{center}
\caption{Mean degree (top) and mean weighted degree (bottom) by the intended respondents (left) and the mean number of respondents per sharing unit (right) in each community.}
\label{fig:degree_vs_sampling}
\end{figure}

As can be seen in Figure~\ref{fig:out_deg_vs_num_surv}, sharing units from which more respondents answered the social support survey questions typically end up with higher out degree, as they have more opportunities to name alters in other sharing units. By contrast, consider \ref{fig:weighted_out_deg_vs_num_surv}, where edge weights account not only for the total number of nominations to other sharing units, but also the number of respondents per sharing unit. Using the edge weights helps account for variation in sampling within sharing units both within and between communities.
A similar pattern is apparent when considering mean degree across communities, as it varies by the intended respondent(s) and the mean number of respondents in that community, as shown in Figure~\ref{fig:degree_vs_sampling}. There again, while mean degree increases in communities with a greater number of respondents in each sharing unit, this is not the case when considering mean weighted degree, drawing on edge weights as we have constructed them here (to account for the number of respondents and so the varying number of opportunities for reporting a tie).

\begin{figure}[h]
\begin{center}
\includegraphics[width = \textwidth]{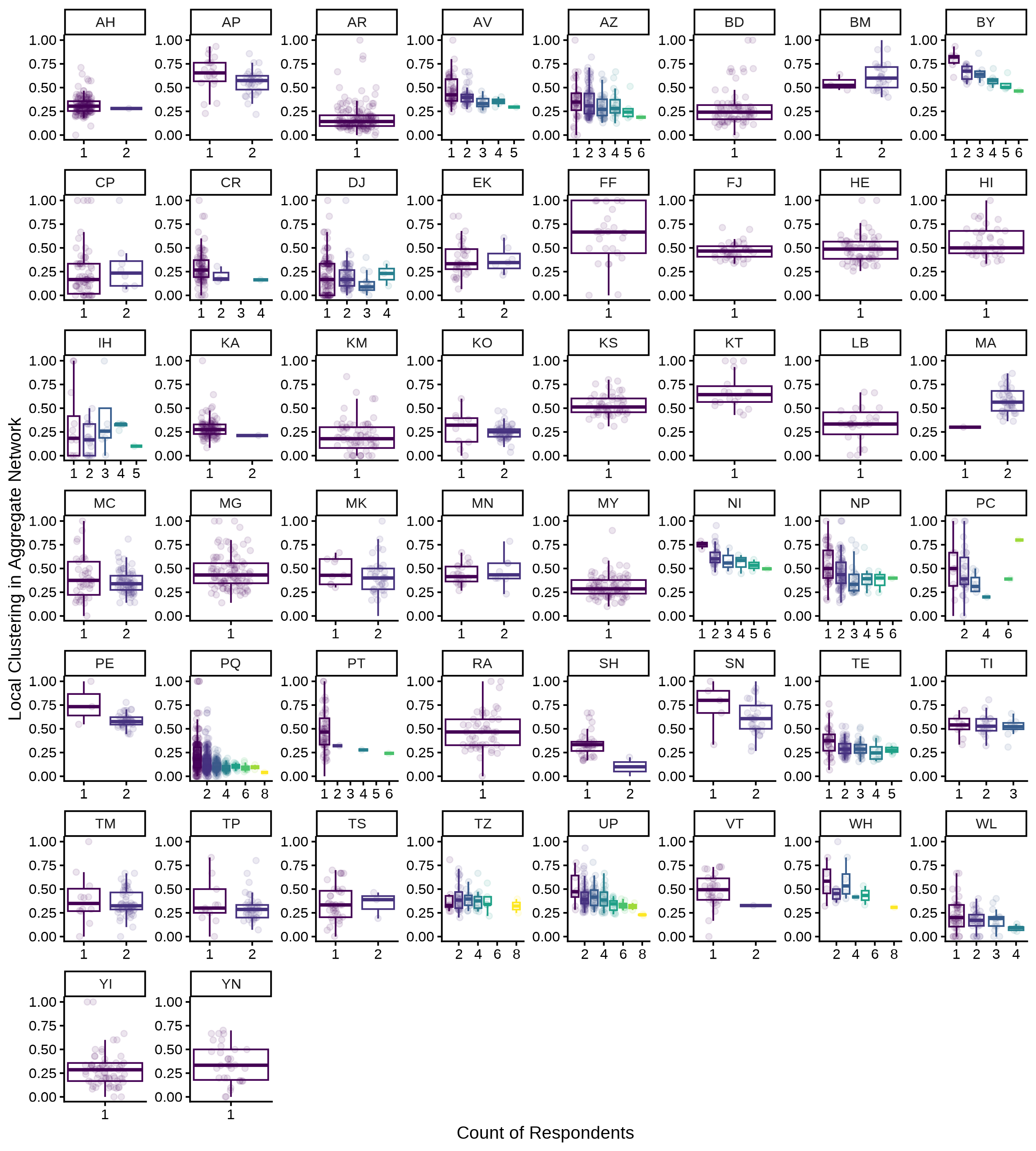}
\end{center}
\caption{Sharing unit's local clustering coefficient in the aggregate network (unweighted) versus the number of respondents from each sharing unit. Colours correspond to the count of respondents, with lighter colours reflecting greater number of respondents. }
\label{fig:clust_vs_num_surv}
\end{figure}

\begin{figure}[h]
\begin{center}
\includegraphics[width = \textwidth]{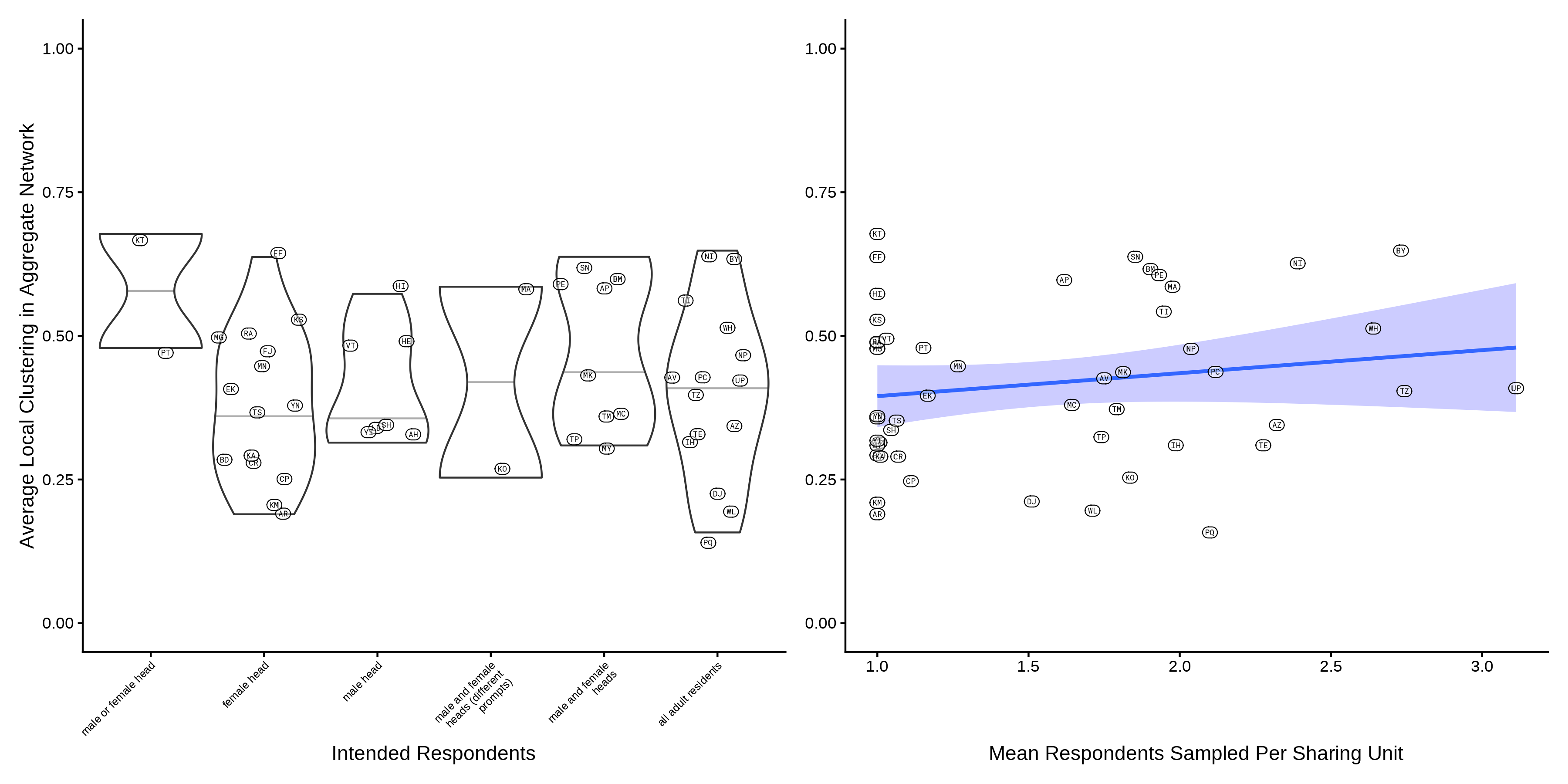}
\end{center}
\caption{Average local clustering coefficient by the intended respondents (left) and the mean number of respondents per sharing unit (right) in each community.}
\label{fig:clust_vs_sampling}
\end{figure}

Similarly, for the local clustering coefficient, sharing units with more respondents often have lower local clustering, as shown in Figure~\ref{fig:clust_vs_num_surv}. This is closely tied to the out-degree pattern just shown: with a greater number of reported connections, there are more potential triads to consider, fewer of which will be closed, thereby lowering the local clustering coefficient. However, as shown in Figure~\ref{fig:clust_vs_sampling}, when we look across communities there is no consistent association between a community's respondent sampling and its average local clustering coefficient.

\begin{figure}[h]
\begin{center}
\includegraphics{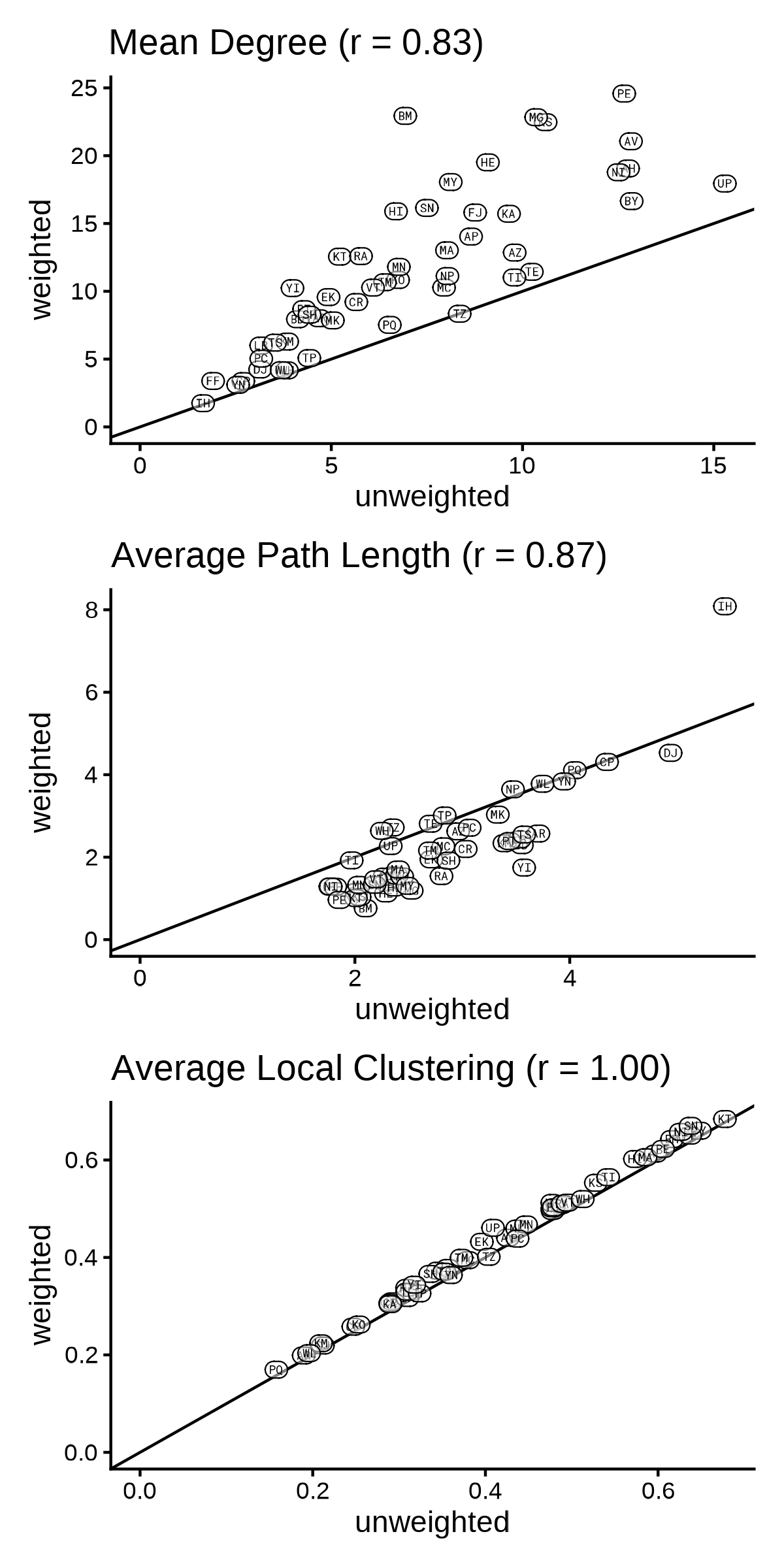}
\end{center}
\caption{Bivariate correlations between measures using weighted and unweighted edges on the aggregate network. The diagonal is the identity line.}
\label{fig:edge_weight_comparison}
\end{figure}

Where it is possible to contrast measures that consider a binary versus weighted edges, generally the two are closely correlated. Figure~\ref{fig:edge_weight_comparison} shows the bivariate associations between weighted and unweighted mean degree, average path length, and average local clustering coefficient in the aggregate network. Note that we shift how the weights are treated across the various measures. For weighted degree (i.e., `strength'), greater edge weights imply greater closeness or strength. The same holds for the weighted local clustering coefficient, as defined by \citep{barrat_architecture_2004}. In contrast, for weighted average path length, greater edge weights imply greater distance, so we take the reciprocal of our edge weights.

Overall, while there certainly are some differences between the unweighted and weighted versions, they track one another quite closely.
As most measures and reported values with comparable data sets use a binary (as opposed to weighted) edge, we largely proceed with those simpler measures. However, it is important to recognise the consequences of sampling coverage for comparisons across sharing units and across communities.

\clearpage
\subsubsection{Variance Decomposition}\label{si:var_decomp}

To better understand the substantial variance that we observe across network layers and communities, we fit models with various network-level summary statistics as the outcome variables, including the network layer as a fixed effect and the community as a random effect. Specifically, we estimate:
\[
y_{s\ell} = \alpha + \gamma_\ell + u_s + \varepsilon_{s\ell},
\qquad
u_s \sim \mathcal{N}(0,\tau^2),
\qquad
\varepsilon_{s\ell} \sim \mathcal{N}(0,\sigma^2),
\]
where $s$ indexes sites and $\ell$ indexes network layers.

In a second set of models, we also include a series of community-level covariates that plausibly influence the summary statistic, as they are likely to influence network density and, by extension, other measures (e.g., reciprocity and transitivity, as well as density and mean degree):
\begin{enumerate}
    \item the number of (sampled) sharing units: larger communities have lower density (see Figure~\ref{fig:density_vs_size}).
    \item the mean number of respondents for that layer: more respondents could allow for a greater number of nominations and so higher density (see Section~\ref{si:sampling_impact}).
    \item the proportion of nominations in that layer that are to alters external to the community: if there is greater reliance on those outside the community, the network within the community may be consequently less dense.
\end{enumerate}

So the model becomes:
\[
y_{s\ell}
= \alpha + \gamma_\ell
+ \beta_1\widetilde{\log N_s}
+ \beta_2\widetilde{\log R_{s\ell}}
+ \beta_3\widetilde{E_{s\ell}}
+ u_s + \varepsilon_{s\ell},
\qquad
u_s \sim \mathcal{N}(0,\tau^2),
\qquad
\varepsilon_{s\ell} \sim \mathcal{N}(0,\sigma^2).
\]
where $N_s$ is the number of sampled sharing units, $R_{s\ell}$ is the mean number of respondents per sharing unit for the relevant
prompt, and $E_{s\ell}$ is the proportion of nominations directed
outside the community. Tildes denote that the variables are centred at the overall sample mean.

The dependent variables we consider are: mean degree (more specifically, the \emph{log} of mean degree), reciprocity, the global clustering coefficient, the average local clustering coefficient, in-degree centralization, and degree assortativity. We do not consider density, for the reasons outlined above. We also do not consider average path length for individual layers, as these networks are often disconnected and the measure averages only over reachable pairs, complicating comparisons across communities.
Along these, we additionally consider two measures drawing on the edge weights: mean weighted degree (specifically, its \emph{natural logarithm}) and average weighted local clustering.
Finally, in addition to the straight outcome variables (and the log transformations for the mean degree measures), we additionally perform an empirical logit transformation for the outcome variables that range between 0 and 1 (reciprocity, global clustering, average local clustering, and in-degree centralization).

In Table~\ref{tab:variance_decomp} we show the results for these models, specifically we highlight the large amount of between community variance. That can be seen most clearly in the intraclass correlation coefficients (ICCs), which here capture the between-community share of remaining variance. (The ICC here is equal to $\frac{\tau^2}{\tau^2 + \sigma^2}$.)

\begin{figure}[h]
\begin{center}
\includegraphics[width = 0.7\textwidth]{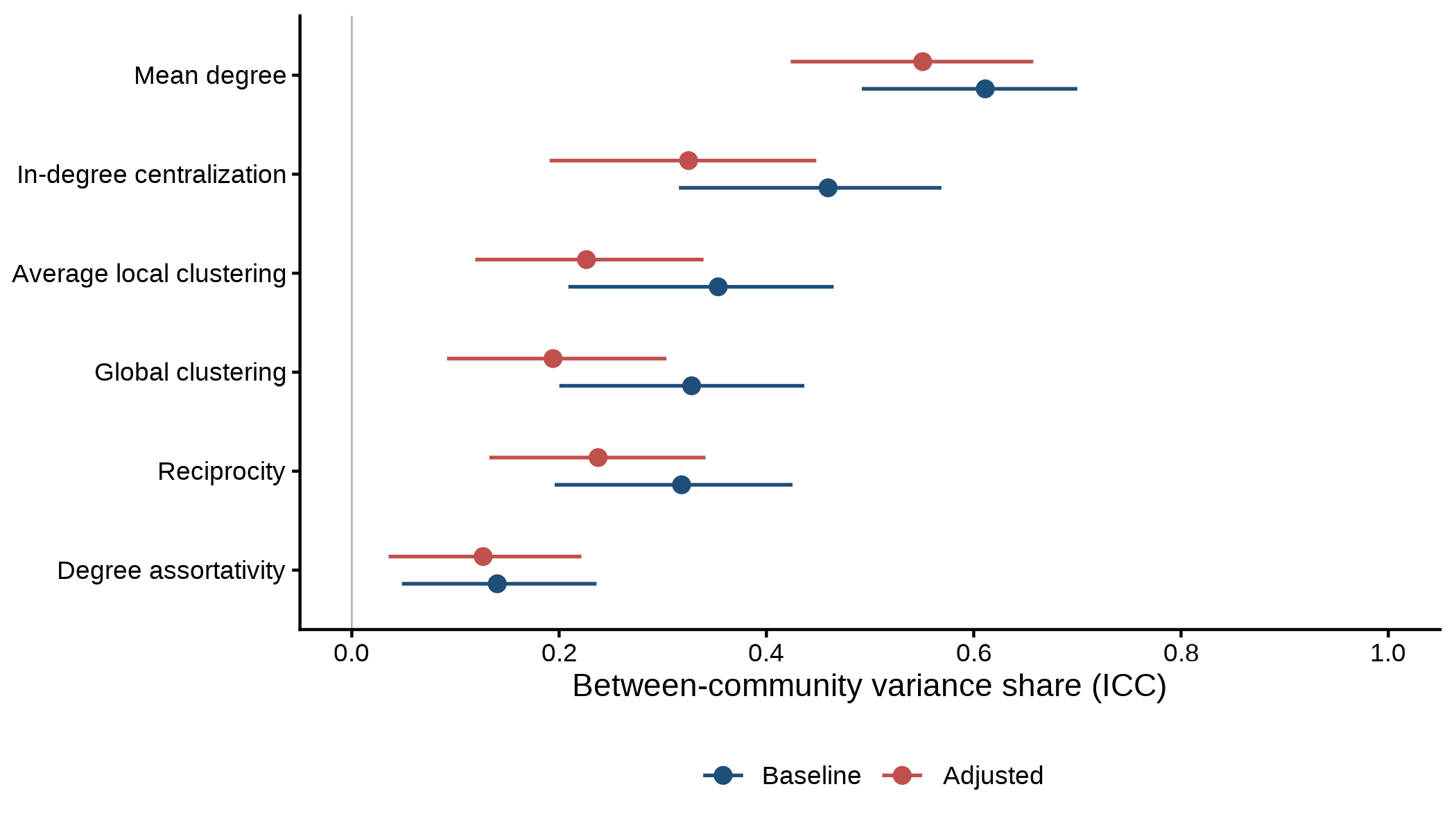}
\end{center}
\caption{Share of the variance attributable to between-community differences, based on regressions predicting each network summary statistic with layer fixed effects and community random effects, plus (for the ``adjusted'' models) terms that may more mechanically influence the summary statistic.}
\label{fig:var_decomp}
\end{figure}

\begin{figure}
\begin{center}
\includegraphics{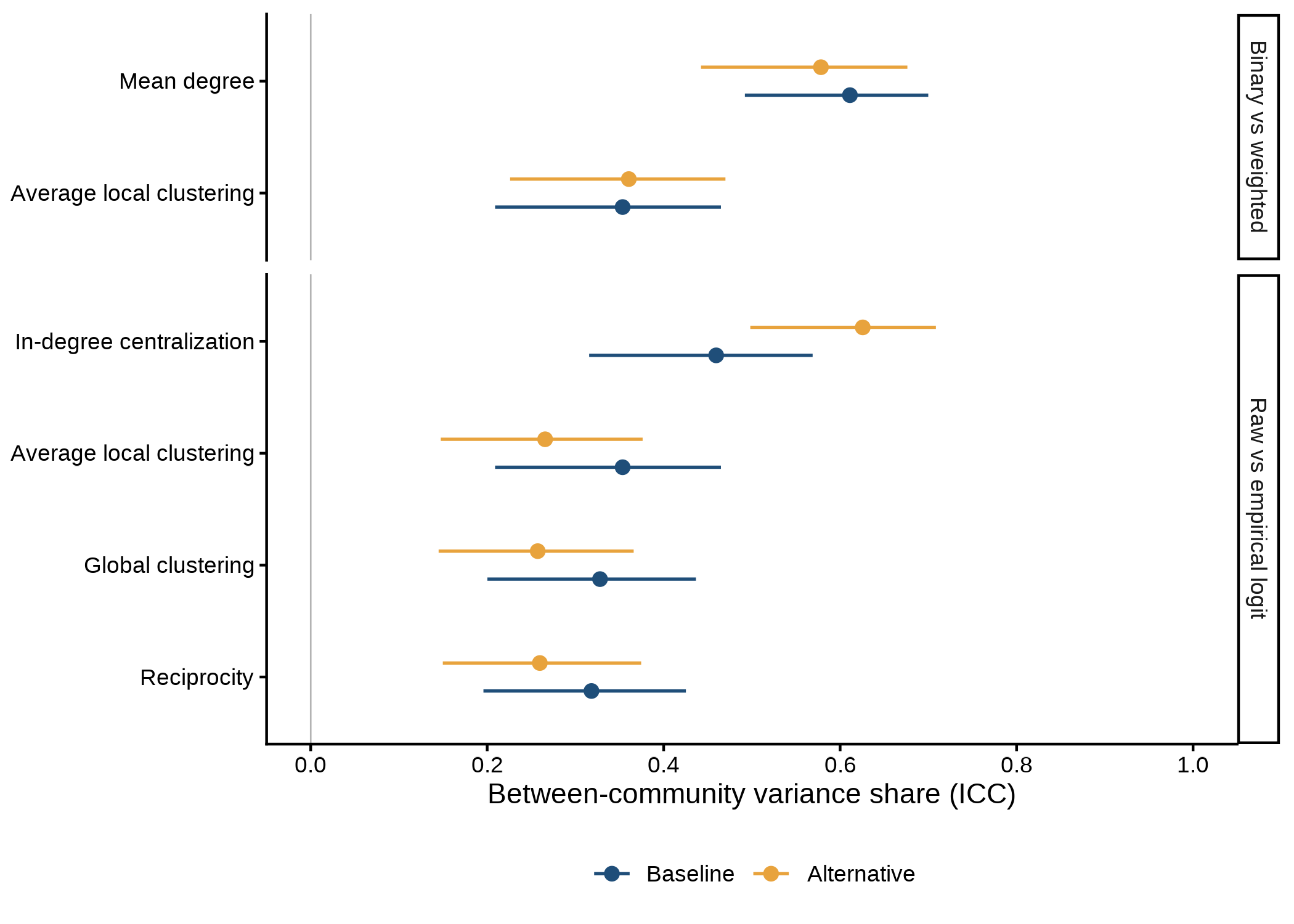}
\end{center}
\caption{Variance decomposition using alternative measures or models. }
\label{fig:variance_decomp_alt}
\end{figure}

Figure~\ref{fig:var_decomp} shows our primary models with both the baseline models and the adjusted ones that additionally control for the three variables outlined above. Figure~\ref{fig:variance_decomp_alt} shows the same primary models and the alternative versions (either using edge weights, or using the empirical logit transformation of the outcome variable). In Table~\ref{tab:agg_variance_decomp} we show more descriptively the between-community variance of these same measures, plus average path length, for the aggregate network. Again, we show this both as a baseline and having additionally controlled for the same covariates.

In all cases, we can see that there is a substantial amount of between-community variation. This variation is attenuated but still remains after the control variables are included.

\begin{table}[!h]
\centering
\caption{\label{tab:variance_decomp}Variance decomposition of network statistics across the ENDOW communities and the eight main network layers. Specifically, eight primary (binary) statistics and two robustness sets covering the same statistics where applicable -- on the bounded-scale empirical logit, and using their weighted-edge analogues. For each statistic, the `Base' row reports a model with layer fixed effects and a community random intercept; the `Adj.' row reports the same model additionally controlling for (centred) log community size, log mean respondents per prompt, and the share of nominations directed outside the community. `Between-community SD' is the standard deviation of the community random intercept, net of layer differences; `Residual SD' is the standard deviation of the community-by-layer residual; `ICC' is the intraclass correlation coefficient, capturing the between-community share of their combined variance. Bracketed values are 95\% parametric-bootstrap intervals.}
\centering
\fontsize{8}{10}\selectfont
\begin{tabular}[t]{llccc}
\toprule
 &  & Between-community SD [95\% CI] & Residual SD & ICC [95\% CI]\\
\midrule
\addlinespace[0.5em]
\multicolumn{5}{l}{\textbf{Primary (binary networks)}}\\
\hspace{1em}Mean degree & Base & 0.70 [0.55, 0.85] & 0.56 & 0.61 [0.49, 0.70]\\
\hspace{1em} & Adj. & 0.54 [0.42, 0.65] & 0.49 & 0.55 [0.42, 0.66]\\
\hspace{1em}Reciprocity & Base & 0.08 [0.06, 0.10] & 0.11 & 0.32 [0.20, 0.43]\\
\hspace{1em} & Adj. & 0.06 [0.04, 0.08] & 0.11 & 0.24 [0.13, 0.34]\\
\hspace{1em}Global clustering & Base & 0.07 [0.05, 0.09] & 0.10 & 0.33 [0.20, 0.44]\\
\hspace{1em} & Adj. & 0.05 [0.03, 0.06] & 0.10 & 0.19 [0.09, 0.30]\\
\hspace{1em}Average local clustering & Base & 0.09 [0.07, 0.12] & 0.13 & 0.35 [0.21, 0.46]\\
\hspace{1em} & Adj. & 0.07 [0.05, 0.09] & 0.12 & 0.23 [0.12, 0.34]\\
\hspace{1em}In-degree centralization & Base & 0.07 [0.05, 0.08] & 0.07 & 0.46 [0.32, 0.57]\\
\hspace{1em} & Adj. & 0.05 [0.04, 0.06] & 0.07 & 0.32 [0.19, 0.45]\\
\hspace{1em}Degree assortativity & Base & 0.05 [0.03, 0.07] & 0.13 & 0.14 [0.05, 0.24]\\
\hspace{1em} & Adj. & 0.05 [0.03, 0.07] & 0.13 & 0.13 [0.04, 0.22]\\
\addlinespace[0.5em]
\multicolumn{5}{l}{\textbf{Robustness: empirical logit}}\\
\hspace{1em}Reciprocity (emp. logit) & Base & 0.42 [0.30, 0.53] & 0.71 & 0.26 [0.15, 0.37]\\
\hspace{1em} & Adj. & 0.34 [0.22, 0.45] & 0.69 & 0.19 [0.09, 0.29]\\
\hspace{1em}Global clustering (emp. logit) & Base & 0.40 [0.28, 0.50] & 0.67 & 0.26 [0.14, 0.37]\\
\hspace{1em} & Adj. & 0.28 [0.17, 0.37] & 0.66 & 0.15 [0.06, 0.25]\\
\hspace{1em}Average local clustering (emp. logit) & Base & 0.46 [0.33, 0.57] & 0.77 & 0.27 [0.15, 0.38]\\
\hspace{1em} & Adj. & 0.34 [0.23, 0.46] & 0.75 & 0.17 [0.07, 0.28]\\
\hspace{1em}In-degree centralization (emp. logit) & Base & 0.67 [0.53, 0.82] & 0.52 & 0.63 [0.50, 0.71]\\
\hspace{1em} & Adj. & 0.41 [0.31, 0.51] & 0.47 & 0.43 [0.29, 0.54]\\
\addlinespace[0.5em]
\multicolumn{5}{l}{\textbf{Robustness: weighted edges}}\\
\hspace{1em}Mean degree (weighted) & Base & 0.67 [0.52, 0.81] & 0.57 & 0.58 [0.44, 0.68]\\
\hspace{1em} & Adj. & 0.55 [0.43, 0.68] & 0.50 & 0.55 [0.42, 0.65]\\
\hspace{1em}Average local clustering (weighted) & Base & 0.10 [0.07, 0.12] & 0.13 & 0.36 [0.23, 0.47]\\
\hspace{1em} & Adj. & 0.07 [0.05, 0.09] & 0.12 & 0.23 [0.13, 0.33]\\
\bottomrule
\end{tabular}
\end{table}

\begin{table}[!h]
\centering
\caption{\label{tab:agg_variance_decomp}Between-community dispersion of the aggregate network, one observation per community. `Mean [P10, P90]' summarises the empirical distribution of the statistic across communities (showing the 10th to 90th percentile in brackets). `Between-community SD' is its standard deviation. `SD After Controls' is the residual standard deviation from an OLS regression of the statistic on (log) community size, (log) mean respondents per prompt, and the share of nominations directed outside the community.}
\centering
\fontsize{8}{10}\selectfont
\begin{tabular}[t]{lccc}
\toprule
 & Mean [P10, P90] & Between-community SD & SD after controls\\
\midrule
Mean degree & 1.81 [1.15, 2.53] & 0.53 & 0.42\\
Reciprocity & 0.45 [0.33, 0.58] & 0.10 & 0.08\\
Global clustering & 0.36 [0.21, 0.56] & 0.13 & 0.08\\
Average local clustering & 0.42 [0.24, 0.62] & 0.14 & 0.09\\
In-degree centralization & 0.25 [0.09, 0.39] & 0.15 & 0.13\\
Degree assortativity & 0.01 [-0.08, 0.09] & 0.07 & 0.07\\
Average path length & 2.82 [2.01, 3.77] & 0.82 & 0.64\\
\bottomrule
\end{tabular}
\end{table}

\clearpage
\subsubsection{Network-wise summary statistics}

Tables~\ref{tab:net-summary-stats-loan_req}-\ref{tab:net-summary-stats-m_info} show summary statistics for each community, one table for each network layer. Table~\ref{tab:net-summary-stats-sum} shows the same for the aggregate network and Table~\ref{tab:net-summary-stats-primary_kin} for the close kin network. We do not include tables for the two networks focused on external ties, as so few nominations were to other sharing unit members.

\begin{landscape}
\begingroup\fontsize{8}{10}\selectfont

\begin{longtable}[t]{lrrrrrllrrlrrrrrrllr}
\caption{\label{tab:net-summary-stats-loan_req}Network summary statistics by community for the Borrow money layer. `Mean Deg.' shows the mean degree, while `Mean W. Deg.' uses edge weights. `In-Deg.' and `Out-Deg.' show the range. `(In-)Deg. Centr.' shows the (in-)degree centralization. `Deg. Assort.' shows the degree assortativity. `APL' shows the average path length. `Diam.' shows the diameter. `N Isol.' shows the number of isolates, and `Prop. Isol.' their proportion of all nodes. `N Comp.' shows the number of weakly-connected components. `Largest Comp.' shows the proportion of nodes in the largest component. `Global Clust.' shows the global clustering coefficient. `Av. Local Clust.' shows the average local clustering coefficient. `Recip.' shows reciprocity.}\\
\toprule
Site & Nodes & Edges & Density & \shortstack{Mean\\ Deg.} & \shortstack{Mean\\ W. Deg.} & \shortstack{In-\\ Deg.} & \shortstack{Out-\\ Deg.} & \shortstack{Deg.\\ Centr.} & \shortstack{In-Deg.\\ Centr.} & \shortstack{Deg.\\ Assort.} & APL & Diam. & N Isol. & \shortstack{Prop.\\ Isol.} & N Comp. & \shortstack{Largest\\ Comp.} & \shortstack{Global\\ Clust.} & \shortstack{Av. Loc.\\ Clust.} & Recip.\\
\midrule
\endfirsthead
\caption[]{Network summary statistics by community for the Borrow money layer. `Mean Deg.' shows the mean degree, while `Mean W. Deg.' uses edge weights. `In-Deg.' and `Out-Deg.' show the range. `(In-)Deg. Centr.' shows the (in-)degree centralization. `Deg. Assort.' shows the degree assortativity. `APL' shows the average path length. `Diam.' shows the diameter. `N Isol.' shows the number of isolates, and `Prop. Isol.' their proportion of all nodes. `N Comp.' shows the number of weakly-connected components. `Largest Comp.' shows the proportion of nodes in the largest component. `Global Clust.' shows the global clustering coefficient. `Av. Local Clust.' shows the average local clustering coefficient. `Recip.' shows reciprocity. \textit{(continued)}}\\
\toprule
Site & Nodes & Edges & Density & \shortstack{Mean\\ Deg.} & \shortstack{Mean\\ W. Deg.} & \shortstack{In-\\ Deg.} & \shortstack{Out-\\ Deg.} & \shortstack{Deg.\\ Centr.} & \shortstack{In-Deg.\\ Centr.} & \shortstack{Deg.\\ Assort.} & APL & Diam. & N Isol. & \shortstack{Prop.\\ Isol.} & N Comp. & \shortstack{Largest\\ Comp.} & \shortstack{Global\\ Clust.} & \shortstack{Av. Loc.\\ Clust.} & Recip.\\
\midrule
\endhead

\endfoot
\bottomrule
\endlastfoot
AH & 137 & 33 & 0.00 & 0.24 & 0.26 & 0 - 5 & 0 - 2 & 0.02 & 0.03 & -0.25 & 1.15 & 2 & 97 & 0.71 & 104 & 0.12 & 0 & 0 & 0.00\\
AP & 55 & 169 & 0.06 & 3.07 & 2.62 & 0 - 51 & 0 - 7 & 0.43 & 0.89 & -0.37 & 1.57 & 5 & 0 & 0.00 & 1 & 1.00 & 0.16 & 0.75 & 0.05\\
AR & 158 & 170 & 0.01 & 1.08 & 1.15 & 0 - 8 & 0 - 7 & 0.02 & 0.04 & -0.02 & 3.09 & 10 & 34 & 0.22 & 39 & 0.73 & 0.07 & 0.13 & 0.22\\
AV & 116 & 515 & 0.04 & 4.44 & 3.17 & 0 - 27 & 0 - 17 & 0.14 & 0.20 & -0.08 & 3.19 & 8 & 3 & 0.03 & 5 & 0.96 & 0.2 & 0.31 & 0.34\\
AZ & 190 & 754 & 0.02 & 3.97 & 2.37 & 0 - 28 & 0 - 15 & 0.07 & 0.13 & 0.04 & 4.79 & 12 & 3 & 0.02 & 4 & 0.98 & 0.21 & 0.3 & 0.32\\
BD & 79 & 79 & 0.01 & 1.00 & 1.08 & 0 - 8 & 0 - 3 & 0.04 & 0.09 & 0.05 & 1.86 & 4 & 14 & 0.18 & 16 & 0.77 & 0.12 & 0.14 & 0.15\\
BM & 31 & 54 & 0.06 & 1.74 & 1.97 & 0 - 14 & 0 - 5 & 0.20 & 0.41 & 0.01 & 2.33 & 5 & 1 & 0.03 & 2 & 0.97 & 0.21 & 0.36 & 0.04\\
BY & 41 & 9 & 0.01 & 0.22 & 0.11 & 0 - 1 & 0 - 4 & 0.05 & 0.02 & - & 1.18 & 2 & 29 & 0.71 & 32 & 0.12 & 0 & 0 & 0.00\\
\end{longtable}
\endgroup{}

\begingroup\fontsize{8}{10}\selectfont

\begin{longtable}[t]{lrrrrrllrrlrrrrrrllr}
\caption{\label{tab:net-summary-stats-loan_giv}Network summary statistics by community for the Lend money layer. `Mean Deg.' shows the mean degree, while `Mean W. Deg.' uses edge weights. `In-Deg.' and `Out-Deg.' show the range. `(In-)Deg. Centr.' shows the (in-)degree centralization. `Deg. Assort.' shows the degree assortativity. `APL' shows the average path length. `Diam.' shows the diameter. `N Isol.' shows the number of isolates, and `Prop. Isol.' their proportion of all nodes. `N Comp.' shows the number of weakly-connected components. `Largest Comp.' shows the proportion of nodes in the largest component. `Global Clust.' shows the global clustering coefficient. `Av. Local Clust.' shows the average local clustering coefficient. `Recip.' shows reciprocity.}\\
\toprule
Site & Nodes & Edges & Density & \shortstack{Mean\\ Deg.} & \shortstack{Mean\\ W. Deg.} & \shortstack{In-\\ Deg.} & \shortstack{Out-\\ Deg.} & \shortstack{Deg.\\ Centr.} & \shortstack{In-Deg.\\ Centr.} & \shortstack{Deg.\\ Assort.} & APL & Diam. & N Isol. & \shortstack{Prop.\\ Isol.} & N Comp. & \shortstack{Largest\\ Comp.} & \shortstack{Global\\ Clust.} & \shortstack{Av. Loc.\\ Clust.} & Recip.\\
\midrule
\endfirsthead
\caption[]{Network summary statistics by community for the Lend money layer. `Mean Deg.' shows the mean degree, while `Mean W. Deg.' uses edge weights. `In-Deg.' and `Out-Deg.' show the range. `(In-)Deg. Centr.' shows the (in-)degree centralization. `Deg. Assort.' shows the degree assortativity. `APL' shows the average path length. `Diam.' shows the diameter. `N Isol.' shows the number of isolates, and `Prop. Isol.' their proportion of all nodes. `N Comp.' shows the number of weakly-connected components. `Largest Comp.' shows the proportion of nodes in the largest component. `Global Clust.' shows the global clustering coefficient. `Av. Local Clust.' shows the average local clustering coefficient. `Recip.' shows reciprocity. \textit{(continued)}}\\
\toprule
Site & Nodes & Edges & Density & \shortstack{Mean\\ Deg.} & \shortstack{Mean\\ W. Deg.} & \shortstack{In-\\ Deg.} & \shortstack{Out-\\ Deg.} & \shortstack{Deg.\\ Centr.} & \shortstack{In-Deg.\\ Centr.} & \shortstack{Deg.\\ Assort.} & APL & Diam. & N Isol. & \shortstack{Prop.\\ Isol.} & N Comp. & \shortstack{Largest\\ Comp.} & \shortstack{Global\\ Clust.} & \shortstack{Av. Loc.\\ Clust.} & Recip.\\
\midrule
\endhead

\endfoot
\bottomrule
\endlastfoot
AH & 137 & 49 & 0.00 & 0.36 & 0.41 & 0 - 4 & 0 - 4 & 0.02 & 0.03 & 0.1 & 1.23 & 2 & 82 & 0.60 & 91 & 0.23 & 0.09 & 0.07 & 0.04\\
AP & 55 & 189 & 0.06 & 3.44 & 2.59 & 0 - 8 & 0 - 10 & 0.07 & 0.08 & 0.19 & 4.02 & 11 & 0 & 0.00 & 1 & 1.00 & 0.35 & 0.42 & 0.47\\
AR & 158 & 227 & 0.01 & 1.44 & 1.63 & 0 - 6 & 0 - 8 & 0.02 & 0.03 & -0.03 & 3.93 & 12 & 15 & 0.09 & 21 & 0.83 & 0.14 & 0.19 & 0.20\\
AV & 116 & 609 & 0.05 & 5.25 & 3.99 & 0 - 19 & 0 - 21 & 0.10 & 0.12 & 0.01 & 2.98 & 7 & 1 & 0.01 & 3 & 0.97 & 0.22 & 0.31 & 0.39\\
AZ & 190 & 774 & 0.02 & 4.07 & 2.41 & 0 - 20 & 0 - 11 & 0.06 & 0.08 & 0.07 & 4.68 & 13 & 3 & 0.02 & 4 & 0.98 & 0.21 & 0.25 & 0.36\\
BD & 79 & 114 & 0.02 & 1.44 & 1.48 & 0 - 7 & 0 - 4 & 0.03 & 0.07 & 0.05 & 2.98 & 9 & 2 & 0.03 & 4 & 0.95 & 0.19 & 0.21 & 0.21\\
BM & 31 & 42 & 0.05 & 1.35 & 1.39 & 0 - 5 & 0 - 6 & 0.09 & 0.12 & -0.14 & 3.00 & 8 & 2 & 0.06 & 3 & 0.94 & 0.24 & 0.42 & 0.05\\
BY & 41 & 121 & 0.07 & 2.95 & 1.35 & 0 - 10 & 0 - 16 & 0.16 & 0.18 & -0.09 & 2.92 & 8 & 2 & 0.05 & 3 & 0.95 & 0.32 & 0.44 & 0.25\\
\end{longtable}
\endgroup{}

\begingroup\fontsize{8}{10}\selectfont

\begin{longtable}[t]{lrrrrrllrrrrrrrrrrrr}
\caption{\label{tab:net-summary-stats-exchange_req}Network summary statistics by community for the Borrow goods layer. `Mean Deg.' shows the mean degree, while `Mean W. Deg.' uses edge weights. `In-Deg.' and `Out-Deg.' show the range. `(In-)Deg. Centr.' shows the (in-)degree centralization. `Deg. Assort.' shows the degree assortativity. `APL' shows the average path length. `Diam.' shows the diameter. `N Isol.' shows the number of isolates, and `Prop. Isol.' their proportion of all nodes. `N Comp.' shows the number of weakly-connected components. `Largest Comp.' shows the proportion of nodes in the largest component. `Global Clust.' shows the global clustering coefficient. `Av. Local Clust.' shows the average local clustering coefficient. `Recip.' shows reciprocity.}\\
\toprule
Site & Nodes & Edges & Density & \shortstack{Mean\\ Deg.} & \shortstack{Mean\\ W. Deg.} & \shortstack{In-\\ Deg.} & \shortstack{Out-\\ Deg.} & \shortstack{Deg.\\ Centr.} & \shortstack{In-Deg.\\ Centr.} & \shortstack{Deg.\\ Assort.} & APL & Diam. & N Isol. & \shortstack{Prop.\\ Isol.} & N Comp. & \shortstack{Largest\\ Comp.} & \shortstack{Global\\ Clust.} & \shortstack{Av. Loc.\\ Clust.} & Recip.\\
\midrule
\endfirsthead
\caption[]{Network summary statistics by community for the Borrow goods layer. `Mean Deg.' shows the mean degree, while `Mean W. Deg.' uses edge weights. `In-Deg.' and `Out-Deg.' show the range. `(In-)Deg. Centr.' shows the (in-)degree centralization. `Deg. Assort.' shows the degree assortativity. `APL' shows the average path length. `Diam.' shows the diameter. `N Isol.' shows the number of isolates, and `Prop. Isol.' their proportion of all nodes. `N Comp.' shows the number of weakly-connected components. `Largest Comp.' shows the proportion of nodes in the largest component. `Global Clust.' shows the global clustering coefficient. `Av. Local Clust.' shows the average local clustering coefficient. `Recip.' shows reciprocity. \textit{(continued)}}\\
\toprule
Site & Nodes & Edges & Density & \shortstack{Mean\\ Deg.} & \shortstack{Mean\\ W. Deg.} & \shortstack{In-\\ Deg.} & \shortstack{Out-\\ Deg.} & \shortstack{Deg.\\ Centr.} & \shortstack{In-Deg.\\ Centr.} & \shortstack{Deg.\\ Assort.} & APL & Diam. & N Isol. & \shortstack{Prop.\\ Isol.} & N Comp. & \shortstack{Largest\\ Comp.} & \shortstack{Global\\ Clust.} & \shortstack{Av. Loc.\\ Clust.} & Recip.\\
\midrule
\endhead

\endfoot
\bottomrule
\endlastfoot
AH & 137 & 522 & 0.03 & 3.81 & 3.93 & 0 - 17 & 0 - 10 & 0.05 & 0.10 & -0.15 & 4.24 & 12 & 3 & 0.02 & 4 & 0.98 & 0.19 & 0.26 & 0.20\\
AP & 55 & 215 & 0.07 & 3.91 & 3.35 & 0 - 10 & 0 - 9 & 0.10 & 0.11 & -0.08 & 3.93 & 9 & 1 & 0.02 & 2 & 0.98 & 0.33 & 0.45 & 0.61\\
AR & 158 & 235 & 0.01 & 1.49 & 1.57 & 0 - 15 & 0 - 6 & 0.04 & 0.09 & 0.08 & 5.70 & 19 & 16 & 0.10 & 17 & 0.90 & 0.10 & 0.13 & 0.29\\
AV & 116 & 413 & 0.03 & 3.56 & 2.73 & 0 - 15 & 0 - 14 & 0.07 & 0.10 & 0.04 & 3.94 & 11 & 0 & 0.00 & 2 & 0.98 & 0.17 & 0.29 & 0.25\\
AZ & 190 & 852 & 0.02 & 4.48 & 2.83 & 0 - 22 & 0 - 20 & 0.09 & 0.09 & 0.02 & 5.07 & 14 & 0 & 0.00 & 1 & 1.00 & 0.25 & 0.32 & 0.41\\
BD & 79 & 139 & 0.02 & 1.76 & 1.82 & 0 - 9 & 0 - 4 & 0.05 & 0.09 & 0.02 & 3.78 & 11 & 2 & 0.03 & 3 & 0.97 & 0.22 & 0.29 & 0.27\\
BM & 31 & 33 & 0.04 & 1.06 & 1.16 & 0 - 8 & 0 - 2 & 0.10 & 0.23 & -0.17 & 1.08 & 2 & 4 & 0.13 & 7 & 0.48 & 0.23 & 0.54 & 0.06\\
BY & 41 & 243 & 0.15 & 5.93 & 3.96 & 1 - 13 & 0 - 14 & 0.18 & 0.18 & -0.13 & 2.31 & 5 & 0 & 0.00 & 1 & 1.00 & 0.39 & 0.47 & 0.42\\
\end{longtable}
\endgroup{}

\begingroup\fontsize{8}{10}\selectfont

\begin{longtable}[t]{lrrrrrllrrrrrrrrrrrr}
\caption{\label{tab:net-summary-stats-exchange_giv}Network summary statistics by community for the Lend goods layer. `Mean Deg.' shows the mean degree, while `Mean W. Deg.' uses edge weights. `In-Deg.' and `Out-Deg.' show the range. `(In-)Deg. Centr.' shows the (in-)degree centralization. `Deg. Assort.' shows the degree assortativity. `APL' shows the average path length. `Diam.' shows the diameter. `N Isol.' shows the number of isolates, and `Prop. Isol.' their proportion of all nodes. `N Comp.' shows the number of weakly-connected components. `Largest Comp.' shows the proportion of nodes in the largest component. `Global Clust.' shows the global clustering coefficient. `Av. Local Clust.' shows the average local clustering coefficient. `Recip.' shows reciprocity.}\\
\toprule
Site & Nodes & Edges & Density & \shortstack{Mean\\ Deg.} & \shortstack{Mean\\ W. Deg.} & \shortstack{In-\\ Deg.} & \shortstack{Out-\\ Deg.} & \shortstack{Deg.\\ Centr.} & \shortstack{In-Deg.\\ Centr.} & \shortstack{Deg.\\ Assort.} & APL & Diam. & N Isol. & \shortstack{Prop.\\ Isol.} & N Comp. & \shortstack{Largest\\ Comp.} & \shortstack{Global\\ Clust.} & \shortstack{Av. Loc.\\ Clust.} & Recip.\\
\midrule
\endfirsthead
\caption[]{Network summary statistics by community for the Lend goods layer. `Mean Deg.' shows the mean degree, while `Mean W. Deg.' uses edge weights. `In-Deg.' and `Out-Deg.' show the range. `(In-)Deg. Centr.' shows the (in-)degree centralization. `Deg. Assort.' shows the degree assortativity. `APL' shows the average path length. `Diam.' shows the diameter. `N Isol.' shows the number of isolates, and `Prop. Isol.' their proportion of all nodes. `N Comp.' shows the number of weakly-connected components. `Largest Comp.' shows the proportion of nodes in the largest component. `Global Clust.' shows the global clustering coefficient. `Av. Local Clust.' shows the average local clustering coefficient. `Recip.' shows reciprocity. \textit{(continued)}}\\
\toprule
Site & Nodes & Edges & Density & \shortstack{Mean\\ Deg.} & \shortstack{Mean\\ W. Deg.} & \shortstack{In-\\ Deg.} & \shortstack{Out-\\ Deg.} & \shortstack{Deg.\\ Centr.} & \shortstack{In-Deg.\\ Centr.} & \shortstack{Deg.\\ Assort.} & APL & Diam. & N Isol. & \shortstack{Prop.\\ Isol.} & N Comp. & \shortstack{Largest\\ Comp.} & \shortstack{Global\\ Clust.} & \shortstack{Av. Loc.\\ Clust.} & Recip.\\
\midrule
\endhead

\endfoot
\bottomrule
\endlastfoot
AH & 137 & 573 & 0.03 & 4.18 & 4.30 & 0 - 11 & 0 - 10 & 0.04 & 0.05 & 0.02 & 3.86 & 8 & 2 & 0.01 & 3 & 0.99 & 0.22 & 0.27 & 0.30\\
AP & 55 & 230 & 0.08 & 4.18 & 3.46 & 0 - 11 & 0 - 9 & 0.09 & 0.13 & 0.04 & 3.66 & 9 & 1 & 0.02 & 2 & 0.98 & 0.36 & 0.47 & 0.62\\
AR & 158 & 254 & 0.01 & 1.61 & 1.71 & 0 - 14 & 0 - 6 & 0.04 & 0.08 & -0.12 & 4.68 & 16 & 10 & 0.06 & 12 & 0.92 & 0.11 & 0.18 & 0.20\\
AV & 116 & 415 & 0.03 & 3.58 & 2.67 & 0 - 14 & 0 - 20 & 0.10 & 0.09 & 0.03 & 3.66 & 8 & 3 & 0.03 & 4 & 0.97 & 0.16 & 0.24 & 0.24\\
AZ & 190 & 865 & 0.02 & 4.55 & 2.80 & 0 - 20 & 0 - 15 & 0.07 & 0.08 & 0.01 & 4.72 & 13 & 2 & 0.01 & 3 & 0.99 & 0.26 & 0.31 & 0.40\\
BD & 79 & 143 & 0.02 & 1.81 & 1.85 & 0 - 8 & 0 - 6 & 0.03 & 0.08 & 0.06 & 5.35 & 17 & 4 & 0.05 & 6 & 0.92 & 0.23 & 0.25 & 0.25\\
BM & 31 & 39 & 0.04 & 1.26 & 1.29 & 0 - 4 & 0 - 7 & 0.08 & 0.09 & -0.46 & 1.24 & 3 & 5 & 0.16 & 8 & 0.35 & 0.42 & 0.60 & 0.05\\
BY & 41 & 282 & 0.17 & 6.88 & 5.27 & 2 - 13 & 0 - 17 & 0.14 & 0.15 & -0.01 & 2.27 & 5 & 0 & 0.00 & 1 & 1.00 & 0.44 & 0.49 & 0.47\\
\end{longtable}
\endgroup{}

\begingroup\fontsize{8}{10}\selectfont

\begin{longtable}[t]{lrrrrrllrrlrrrrrrrrr}
\caption{\label{tab:net-summary-stats-f_behav}Network summary statistics by community for the Behavioral support (women) layer. `Mean Deg.' shows the mean degree, while `Mean W. Deg.' uses edge weights. `In-Deg.' and `Out-Deg.' show the range. `(In-)Deg. Centr.' shows the (in-)degree centralization. `Deg. Assort.' shows the degree assortativity. `APL' shows the average path length. `Diam.' shows the diameter. `N Isol.' shows the number of isolates, and `Prop. Isol.' their proportion of all nodes. `N Comp.' shows the number of weakly-connected components. `Largest Comp.' shows the proportion of nodes in the largest component. `Global Clust.' shows the global clustering coefficient. `Av. Local Clust.' shows the average local clustering coefficient. `Recip.' shows reciprocity.}\\
\toprule
Site & Nodes & Edges & Density & \shortstack{Mean\\ Deg.} & \shortstack{Mean\\ W. Deg.} & \shortstack{In-\\ Deg.} & \shortstack{Out-\\ Deg.} & \shortstack{Deg.\\ Centr.} & \shortstack{In-Deg.\\ Centr.} & \shortstack{Deg.\\ Assort.} & APL & Diam. & N Isol. & \shortstack{Prop.\\ Isol.} & N Comp. & \shortstack{Largest\\ Comp.} & \shortstack{Global\\ Clust.} & \shortstack{Av. Loc.\\ Clust.} & Recip.\\
\midrule
\endfirsthead
\caption[]{Network summary statistics by community for the Behavioral support (women) layer. `Mean Deg.' shows the mean degree, while `Mean W. Deg.' uses edge weights. `In-Deg.' and `Out-Deg.' show the range. `(In-)Deg. Centr.' shows the (in-)degree centralization. `Deg. Assort.' shows the degree assortativity. `APL' shows the average path length. `Diam.' shows the diameter. `N Isol.' shows the number of isolates, and `Prop. Isol.' their proportion of all nodes. `N Comp.' shows the number of weakly-connected components. `Largest Comp.' shows the proportion of nodes in the largest component. `Global Clust.' shows the global clustering coefficient. `Av. Local Clust.' shows the average local clustering coefficient. `Recip.' shows reciprocity. \textit{(continued)}}\\
\toprule
Site & Nodes & Edges & Density & \shortstack{Mean\\ Deg.} & \shortstack{Mean\\ W. Deg.} & \shortstack{In-\\ Deg.} & \shortstack{Out-\\ Deg.} & \shortstack{Deg.\\ Centr.} & \shortstack{In-Deg.\\ Centr.} & \shortstack{Deg.\\ Assort.} & APL & Diam. & N Isol. & \shortstack{Prop.\\ Isol.} & N Comp. & \shortstack{Largest\\ Comp.} & \shortstack{Global\\ Clust.} & \shortstack{Av. Loc.\\ Clust.} & Recip.\\
\midrule
\endhead

\endfoot
\bottomrule
\endlastfoot
AH & 137 & 317 & 0.02 & 2.31 & 2.57 & 0 - 13 & 0 - 9 & 0.05 & 0.08 & 0.01 & 5.10 & 13 & 14 & 0.10 & 15 & 0.90 & 0.19 & 0.23 & 0.20\\
AP & 55 & 50 & 0.02 & 0.91 & 0.91 & 0 - 3 & 0 - 3 & 0.03 & 0.04 & -0.23 & 1.69 & 6 & 11 & 0.20 & 19 & 0.33 & 0.20 & 0.21 & 0.40\\
AR & 158 & 125 & 0.01 & 0.79 & 0.92 & 0 - 5 & 0 - 7 & 0.02 & 0.03 & -0.07 & 2.01 & 7 & 53 & 0.34 & 56 & 0.64 & 0.06 & 0.06 & 0.13\\
AV & 116 & 626 & 0.05 & 5.40 & 5.38 & 0 - 21 & 0 - 39 & 0.22 & 0.14 & 0 & 2.84 & 7 & 4 & 0.03 & 7 & 0.93 & 0.25 & 0.34 & 0.39\\
AZ & 190 & 666 & 0.02 & 3.51 & 2.09 & 0 - 18 & 0 - 13 & 0.06 & 0.08 & 0 & 5.54 & 17 & 7 & 0.04 & 8 & 0.96 & 0.24 & 0.28 & 0.32\\
BD & 79 & 87 & 0.01 & 1.10 & 1.16 & 0 - 4 & 0 - 4 & 0.02 & 0.04 & 0.09 & 2.55 & 7 & 11 & 0.14 & 15 & 0.72 & 0.27 & 0.29 & 0.14\\
BM & 31 & 45 & 0.05 & 1.45 & 3.03 & 0 - 4 & 0 - 7 & 0.09 & 0.08 & -0.08 & 2.26 & 5 & 6 & 0.19 & 7 & 0.81 & 0.24 & 0.27 & 0.13\\
BY & 41 & 181 & 0.11 & 4.41 & 2.66 & 0 - 13 & 0 - 13 & 0.18 & 0.21 & -0.19 & 2.53 & 5 & 0 & 0.00 & 1 & 1.00 & 0.34 & 0.46 & 0.44\\
\end{longtable}
\endgroup{}

\begingroup\fontsize{8}{10}\selectfont

\begin{longtable}[t]{lrrrrrllrrlrrrrrrrrr}
\caption{\label{tab:net-summary-stats-m_behav}Network summary statistics by community for the Behavioral support (men) layer. `Mean Deg.' shows the mean degree, while `Mean W. Deg.' uses edge weights. `In-Deg.' and `Out-Deg.' show the range. `(In-)Deg. Centr.' shows the (in-)degree centralization. `Deg. Assort.' shows the degree assortativity. `APL' shows the average path length. `Diam.' shows the diameter. `N Isol.' shows the number of isolates, and `Prop. Isol.' their proportion of all nodes. `N Comp.' shows the number of weakly-connected components. `Largest Comp.' shows the proportion of nodes in the largest component. `Global Clust.' shows the global clustering coefficient. `Av. Local Clust.' shows the average local clustering coefficient. `Recip.' shows reciprocity.}\\
\toprule
Site & Nodes & Edges & Density & \shortstack{Mean\\ Deg.} & \shortstack{Mean\\ W. Deg.} & \shortstack{In-\\ Deg.} & \shortstack{Out-\\ Deg.} & \shortstack{Deg.\\ Centr.} & \shortstack{In-Deg.\\ Centr.} & \shortstack{Deg.\\ Assort.} & APL & Diam. & N Isol. & \shortstack{Prop.\\ Isol.} & N Comp. & \shortstack{Largest\\ Comp.} & \shortstack{Global\\ Clust.} & \shortstack{Av. Loc.\\ Clust.} & Recip.\\
\midrule
\endfirsthead
\caption[]{Network summary statistics by community for the Behavioral support (men) layer. `Mean Deg.' shows the mean degree, while `Mean W. Deg.' uses edge weights. `In-Deg.' and `Out-Deg.' show the range. `(In-)Deg. Centr.' shows the (in-)degree centralization. `Deg. Assort.' shows the degree assortativity. `APL' shows the average path length. `Diam.' shows the diameter. `N Isol.' shows the number of isolates, and `Prop. Isol.' their proportion of all nodes. `N Comp.' shows the number of weakly-connected components. `Largest Comp.' shows the proportion of nodes in the largest component. `Global Clust.' shows the global clustering coefficient. `Av. Local Clust.' shows the average local clustering coefficient. `Recip.' shows reciprocity. \textit{(continued)}}\\
\toprule
Site & Nodes & Edges & Density & \shortstack{Mean\\ Deg.} & \shortstack{Mean\\ W. Deg.} & \shortstack{In-\\ Deg.} & \shortstack{Out-\\ Deg.} & \shortstack{Deg.\\ Centr.} & \shortstack{In-Deg.\\ Centr.} & \shortstack{Deg.\\ Assort.} & APL & Diam. & N Isol. & \shortstack{Prop.\\ Isol.} & N Comp. & \shortstack{Largest\\ Comp.} & \shortstack{Global\\ Clust.} & \shortstack{Av. Loc.\\ Clust.} & Recip.\\
\midrule
\endhead

\endfoot
\bottomrule
\endlastfoot
AH & 137 & 491 & 0.03 & 3.58 & 3.83 & 0 - 14 & 0 - 10 & 0.06 & 0.08 & 0 & 4.39 & 12 & 6 & 0.04 & 7 & 0.96 & 0.23 & 0.28 & 0.26\\
AP & 55 & 75 & 0.03 & 1.36 & 1.38 & 0 - 5 & 0 - 6 & 0.06 & 0.07 & 0.39 & 3.06 & 7 & 14 & 0.25 & 18 & 0.55 & 0.36 & 0.39 & 0.56\\
AR & 158 & 154 & 0.01 & 0.97 & 1.03 & 0 - 5 & 0 - 9 & 0.02 & 0.03 & 0.08 & 2.61 & 8 & 47 & 0.30 & 50 & 0.68 & 0.07 & 0.11 & 0.19\\
AV & 116 & 641 & 0.05 & 5.53 & 5.56 & 0 - 21 & 0 - 33 & 0.16 & 0.13 & 0.02 & 2.71 & 6 & 8 & 0.07 & 9 & 0.93 & 0.29 & 0.37 & 0.35\\
AZ & 190 & 567 & 0.02 & 2.98 & 1.63 & 0 - 14 & 0 - 12 & 0.05 & 0.06 & 0.07 & 4.99 & 14 & 17 & 0.09 & 18 & 0.91 & 0.18 & 0.23 & 0.26\\
BD & 79 & 51 & 0.01 & 0.65 & 0.66 & 0 - 5 & 0 - 3 & 0.03 & 0.06 & 0.16 & 1.22 & 3 & 26 & 0.33 & 36 & 0.35 & 0.15 & 0.22 & 0.12\\
BM & 31 & 131 & 0.14 & 4.23 & 8.29 & 0 - 11 & 0 - 8 & 0.16 & 0.23 & -0.24 & 2.54 & 5 & 0 & 0.00 & 1 & 1.00 & 0.44 & 0.54 & 0.37\\
BY & 41 & 167 & 0.10 & 4.07 & 2.22 & 0 - 10 & 0 - 18 & 0.16 & 0.15 & -0.04 & 3.09 & 8 & 2 & 0.05 & 3 & 0.95 & 0.33 & 0.42 & 0.54\\
\end{longtable}
\endgroup{}

\begingroup\fontsize{8}{10}\selectfont

\begin{longtable}[t]{lrrrrrllrrlrrrrrrrrr}
\caption{\label{tab:net-summary-stats-f_info}Network summary statistics by community for the Informational/affective support (women) layer. `Mean Deg.' shows the mean degree, while `Mean W. Deg.' uses edge weights. `In-Deg.' and `Out-Deg.' show the range. `(In-)Deg. Centr.' shows the (in-)degree centralization. `Deg. Assort.' shows the degree assortativity. `APL' shows the average path length. `Diam.' shows the diameter. `N Isol.' shows the number of isolates, and `Prop. Isol.' their proportion of all nodes. `N Comp.' shows the number of weakly-connected components. `Largest Comp.' shows the proportion of nodes in the largest component. `Global Clust.' shows the global clustering coefficient. `Av. Local Clust.' shows the average local clustering coefficient. `Recip.' shows reciprocity.}\\
\toprule
Site & Nodes & Edges & Density & \shortstack{Mean\\ Deg.} & \shortstack{Mean\\ W. Deg.} & \shortstack{In-\\ Deg.} & \shortstack{Out-\\ Deg.} & \shortstack{Deg.\\ Centr.} & \shortstack{In-Deg.\\ Centr.} & \shortstack{Deg.\\ Assort.} & APL & Diam. & N Isol. & \shortstack{Prop.\\ Isol.} & N Comp. & \shortstack{Largest\\ Comp.} & \shortstack{Global\\ Clust.} & \shortstack{Av. Loc.\\ Clust.} & Recip.\\
\midrule
\endfirsthead
\caption[]{Network summary statistics by community for the Informational/affective support (women) layer. `Mean Deg.' shows the mean degree, while `Mean W. Deg.' uses edge weights. `In-Deg.' and `Out-Deg.' show the range. `(In-)Deg. Centr.' shows the (in-)degree centralization. `Deg. Assort.' shows the degree assortativity. `APL' shows the average path length. `Diam.' shows the diameter. `N Isol.' shows the number of isolates, and `Prop. Isol.' their proportion of all nodes. `N Comp.' shows the number of weakly-connected components. `Largest Comp.' shows the proportion of nodes in the largest component. `Global Clust.' shows the global clustering coefficient. `Av. Local Clust.' shows the average local clustering coefficient. `Recip.' shows reciprocity. \textit{(continued)}}\\
\toprule
Site & Nodes & Edges & Density & \shortstack{Mean\\ Deg.} & \shortstack{Mean\\ W. Deg.} & \shortstack{In-\\ Deg.} & \shortstack{Out-\\ Deg.} & \shortstack{Deg.\\ Centr.} & \shortstack{In-Deg.\\ Centr.} & \shortstack{Deg.\\ Assort.} & APL & Diam. & N Isol. & \shortstack{Prop.\\ Isol.} & N Comp. & \shortstack{Largest\\ Comp.} & \shortstack{Global\\ Clust.} & \shortstack{Av. Loc.\\ Clust.} & Recip.\\
\midrule
\endhead

\endfoot
\bottomrule
\endlastfoot
AH & 137 & 511 & 0.03 & 3.73 & 3.84 & 0 - 22 & 0 - 11 & 0.09 & 0.13 & 0.05 & 4.22 & 11 & 8 & 0.06 & 9 & 0.94 & 0.18 & 0.20 & 0.18\\
AP & 55 & 182 & 0.06 & 3.31 & 3.36 & 0 - 9 & 0 - 7 & 0.09 & 0.11 & 0.03 & 3.65 & 9 & 4 & 0.07 & 5 & 0.93 & 0.30 & 0.39 & 0.43\\
AR & 158 & 255 & 0.01 & 1.61 & 1.73 & 0 - 8 & 0 - 7 & 0.02 & 0.04 & 0.07 & 7.77 & 21 & 26 & 0.16 & 27 & 0.84 & 0.14 & 0.18 & 0.27\\
AV & 116 & 243 & 0.02 & 2.09 & 1.91 & 0 - 10 & 0 - 19 & 0.09 & 0.07 & -0.07 & 3.83 & 10 & 28 & 0.24 & 31 & 0.72 & 0.22 & 0.27 & 0.23\\
AZ & 190 & 734 & 0.02 & 3.86 & 2.32 & 0 - 21 & 0 - 12 & 0.07 & 0.09 & 0.09 & 5.00 & 13 & 7 & 0.04 & 8 & 0.96 & 0.24 & 0.31 & 0.34\\
BD & 79 & 166 & 0.03 & 2.10 & 2.24 & 0 - 8 & 0 - 7 & 0.05 & 0.08 & 0.12 & 4.01 & 9 & 4 & 0.05 & 5 & 0.95 & 0.21 & 0.23 & 0.14\\
BM & 31 & 48 & 0.05 & 1.55 & 4.00 & 0 - 5 & 0 - 5 & 0.10 & 0.12 & -0.27 & 2.95 & 8 & 4 & 0.13 & 5 & 0.87 & 0.31 & 0.32 & 0.46\\
BY & 41 & 291 & 0.18 & 7.10 & 4.45 & 1 - 18 & 1 - 19 & 0.16 & 0.27 & -0.01 & 2.30 & 5 & 0 & 0.00 & 1 & 1.00 & 0.47 & 0.54 & 0.51\\
\end{longtable}
\endgroup{}

\begingroup\fontsize{8}{10}\selectfont

\begin{longtable}[t]{lrrrrrllrrrrrrrrrrrr}
\caption{\label{tab:net-summary-stats-m_info}Network summary statistics by community for the Informational/affective support (men) layer. `Mean Deg.' shows the mean degree, while `Mean W. Deg.' uses edge weights. `In-Deg.' and `Out-Deg.' show the range. `(In-)Deg. Centr.' shows the (in-)degree centralization. `Deg. Assort.' shows the degree assortativity. `APL' shows the average path length. `Diam.' shows the diameter. `N Isol.' shows the number of isolates, and `Prop. Isol.' their proportion of all nodes. `N Comp.' shows the number of weakly-connected components. `Largest Comp.' shows the proportion of nodes in the largest component. `Global Clust.' shows the global clustering coefficient. `Av. Local Clust.' shows the average local clustering coefficient. `Recip.' shows reciprocity.}\\
\toprule
Site & Nodes & Edges & Density & \shortstack{Mean\\ Deg.} & \shortstack{Mean\\ W. Deg.} & \shortstack{In-\\ Deg.} & \shortstack{Out-\\ Deg.} & \shortstack{Deg.\\ Centr.} & \shortstack{In-Deg.\\ Centr.} & \shortstack{Deg.\\ Assort.} & APL & Diam. & N Isol. & \shortstack{Prop.\\ Isol.} & N Comp. & \shortstack{Largest\\ Comp.} & \shortstack{Global\\ Clust.} & \shortstack{Av. Loc.\\ Clust.} & Recip.\\
\midrule
\endfirsthead
\caption[]{Network summary statistics by community for the Informational/affective support (men) layer. `Mean Deg.' shows the mean degree, while `Mean W. Deg.' uses edge weights. `In-Deg.' and `Out-Deg.' show the range. `(In-)Deg. Centr.' shows the (in-)degree centralization. `Deg. Assort.' shows the degree assortativity. `APL' shows the average path length. `Diam.' shows the diameter. `N Isol.' shows the number of isolates, and `Prop. Isol.' their proportion of all nodes. `N Comp.' shows the number of weakly-connected components. `Largest Comp.' shows the proportion of nodes in the largest component. `Global Clust.' shows the global clustering coefficient. `Av. Local Clust.' shows the average local clustering coefficient. `Recip.' shows reciprocity. \textit{(continued)}}\\
\toprule
Site & Nodes & Edges & Density & \shortstack{Mean\\ Deg.} & \shortstack{Mean\\ W. Deg.} & \shortstack{In-\\ Deg.} & \shortstack{Out-\\ Deg.} & \shortstack{Deg.\\ Centr.} & \shortstack{In-Deg.\\ Centr.} & \shortstack{Deg.\\ Assort.} & APL & Diam. & N Isol. & \shortstack{Prop.\\ Isol.} & N Comp. & \shortstack{Largest\\ Comp.} & \shortstack{Global\\ Clust.} & \shortstack{Av. Loc.\\ Clust.} & Recip.\\
\midrule
\endhead

\endfoot
\bottomrule
\endlastfoot
AH & 137 & 623 & 0.03 & 4.55 & 4.63 & 0 - 19 & 0 - 10 & 0.06 & 0.11 & 0.07 & 3.59 & 13 & 9 & 0.07 & 10 & 0.93 & 0.20 & 0.23 & 0.19\\
AP & 55 & 132 & 0.04 & 2.40 & 2.42 & 0 - 19 & 0 - 7 & 0.16 & 0.31 & 0.12 & 1.98 & 5 & 14 & 0.25 & 15 & 0.75 & 0.27 & 0.35 & 0.03\\
AR & 158 & 228 & 0.01 & 1.44 & 1.61 & 0 - 8 & 0 - 8 & 0.02 & 0.04 & -0.04 & 6.81 & 17 & 21 & 0.13 & 23 & 0.85 & 0.10 & 0.11 & 0.13\\
AV & 116 & 287 & 0.02 & 2.47 & 2.32 & 0 - 12 & 0 - 20 & 0.08 & 0.08 & 0.02 & 3.65 & 11 & 36 & 0.31 & 38 & 0.67 & 0.27 & 0.30 & 0.30\\
AZ & 190 & 561 & 0.02 & 2.95 & 1.62 & 0 - 12 & 0 - 12 & 0.04 & 0.05 & 0.07 & 4.77 & 12 & 14 & 0.07 & 15 & 0.93 & 0.19 & 0.27 & 0.22\\
BD & 79 & 73 & 0.01 & 0.92 & 0.97 & 0 - 6 & 0 - 8 & 0.06 & 0.07 & -0.23 & 2.17 & 5 & 19 & 0.24 & 22 & 0.71 & 0.11 & 0.12 & 0.11\\
BM & 31 & 57 & 0.06 & 1.84 & 4.48 & 0 - 7 & 0 - 5 & 0.13 & 0.17 & -0.26 & 3.36 & 8 & 1 & 0.03 & 3 & 0.90 & 0.31 & 0.39 & 0.46\\
BY & 41 & 254 & 0.15 & 6.20 & 3.25 & 0 - 16 & 0 - 19 & 0.26 & 0.25 & 0.01 & 2.29 & 6 & 1 & 0.02 & 2 & 0.98 & 0.47 & 0.56 & 0.46\\
\end{longtable}
\endgroup{}

\begingroup\fontsize{8}{10}\selectfont

\begin{longtable}[t]{lrrrrrllrrrrrrrrrrrr}
\caption{\label{tab:net-summary-stats-sum}Network summary statistics by community for the Aggregate layer. `Mean Deg.' shows the mean degree, while `Mean W. Deg.' uses edge weights. `In-Deg.' and `Out-Deg.' show the range. `(In-)Deg. Centr.' shows the (in-)degree centralization. `Deg. Assort.' shows the degree assortativity. `APL' shows the average path length. `Diam.' shows the diameter. `N Isol.' shows the number of isolates, and `Prop. Isol.' their proportion of all nodes. `N Comp.' shows the number of weakly-connected components. `Largest Comp.' shows the proportion of nodes in the largest component. `Global Clust.' shows the global clustering coefficient. `Av. Local Clust.' shows the average local clustering coefficient. `Recip.' shows reciprocity.}\\
\toprule
Site & Nodes & Edges & Density & \shortstack{Mean\\ Deg.} & \shortstack{Mean\\ W. Deg.} & \shortstack{In-\\ Deg.} & \shortstack{Out-\\ Deg.} & \shortstack{Deg.\\ Centr.} & \shortstack{In-Deg.\\ Centr.} & \shortstack{Deg.\\ Assort.} & APL & Diam. & N Isol. & \shortstack{Prop.\\ Isol.} & N Comp. & \shortstack{Largest\\ Comp.} & \shortstack{Global\\ Clust.} & \shortstack{Av. Loc.\\ Clust.} & Recip.\\
\midrule
\endfirsthead
\caption[]{Network summary statistics by community for the Aggregate layer. `Mean Deg.' shows the mean degree, while `Mean W. Deg.' uses edge weights. `In-Deg.' and `Out-Deg.' show the range. `(In-)Deg. Centr.' shows the (in-)degree centralization. `Deg. Assort.' shows the degree assortativity. `APL' shows the average path length. `Diam.' shows the diameter. `N Isol.' shows the number of isolates, and `Prop. Isol.' their proportion of all nodes. `N Comp.' shows the number of weakly-connected components. `Largest Comp.' shows the proportion of nodes in the largest component. `Global Clust.' shows the global clustering coefficient. `Av. Local Clust.' shows the average local clustering coefficient. `Recip.' shows reciprocity. \textit{(continued)}}\\
\toprule
Site & Nodes & Edges & Density & \shortstack{Mean\\ Deg.} & \shortstack{Mean\\ W. Deg.} & \shortstack{In-\\ Deg.} & \shortstack{Out-\\ Deg.} & \shortstack{Deg.\\ Centr.} & \shortstack{In-Deg.\\ Centr.} & \shortstack{Deg.\\ Assort.} & APL & Diam. & N Isol. & \shortstack{Prop.\\ Isol.} & N Comp. & \shortstack{Largest\\ Comp.} & \shortstack{Global\\ Clust.} & \shortstack{Av. Loc.\\ Clust.} & Recip.\\
\midrule
\endhead

\endfoot
\bottomrule
\endlastfoot
AH & 137 & 1748 & 0.09 & 12.76 & 19.06 & 0 - 39 & 2 - 32 & 0.16 & 0.19 & -0.01 & 2.35 & 5 & 0 & 0.00 & 1 & 1.00 & 0.29 & 0.31 & 0.38\\
AP & 55 & 476 & 0.16 & 8.65 & 14.04 & 0 - 52 & 1 - 16 & 0.39 & 0.80 & -0.13 & 2.28 & 5 & 0 & 0.00 & 1 & 1.00 & 0.40 & 0.60 & 0.47\\
AR & 158 & 740 & 0.03 & 4.68 & 8.01 & 0 - 17 & 0 - 14 & 0.05 & 0.08 & 0.05 & 3.71 & 8 & 4 & 0.03 & 5 & 0.97 & 0.14 & 0.19 & 0.34\\
AV & 116 & 1489 & 0.11 & 12.84 & 21.06 & 0 - 42 & 1 - 48 & 0.26 & 0.25 & 0.00 & 2.29 & 6 & 0 & 0.00 & 1 & 1.00 & 0.36 & 0.43 & 0.48\\
AZ & 190 & 1861 & 0.05 & 9.79 & 12.85 & 0 - 42 & 1 - 35 & 0.13 & 0.17 & 0.09 & 2.96 & 6 & 0 & 0.00 & 1 & 1.00 & 0.27 & 0.34 & 0.47\\
BD & 79 & 326 & 0.05 & 4.13 & 7.94 & 0 - 16 & 1 - 9 & 0.08 & 0.15 & -0.04 & 3.55 & 8 & 0 & 0.00 & 1 & 1.00 & 0.22 & 0.29 & 0.30\\
BM & 31 & 215 & 0.23 & 6.94 & 22.94 & 0 - 18 & 1 - 11 & 0.21 & 0.37 & -0.11 & 2.10 & 5 & 0 & 0.00 & 1 & 1.00 & 0.55 & 0.62 & 0.44\\
BY & 41 & 527 & 0.32 & 12.85 & 16.65 & 4 - 22 & 2 - 29 & 0.30 & 0.23 & -0.08 & 1.77 & 4 & 0 & 0.00 & 1 & 1.00 & 0.59 & 0.65 & 0.65\\
\end{longtable}
\endgroup{}

\end{landscape}

\clearpage
\subsection{Network Plots}\label{si:network_plots}

\subsubsection{Social Support Network Layers}\label{si:network_plots_each}

\begin{figure}[h]
\begin{center}
\includegraphics[width = \textwidth]{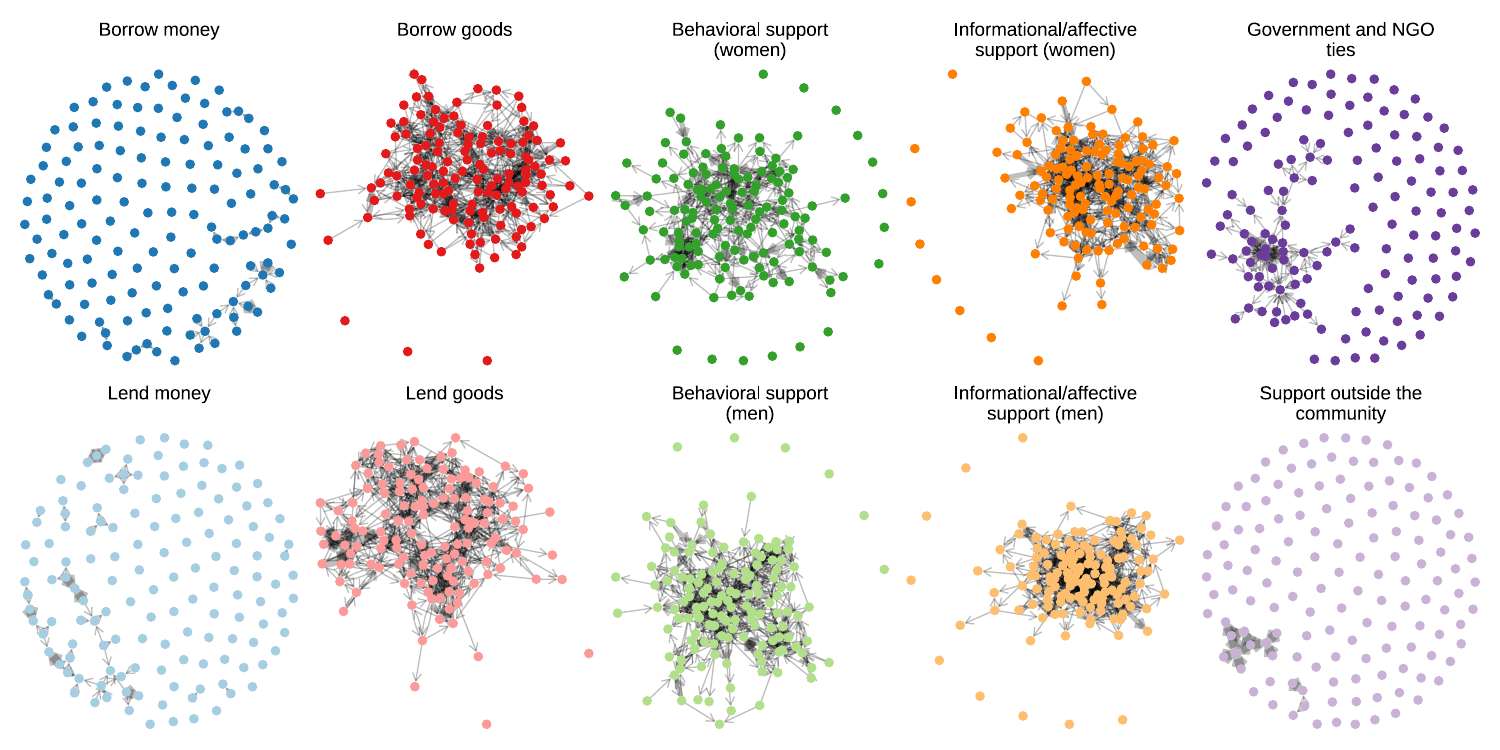}
\end{center}
\caption{Sharing unit-level networks for community AH}
\label{fig:networks_ah}
\end{figure}

\begin{figure}[h]
\begin{center}
\includegraphics[width = \textwidth]{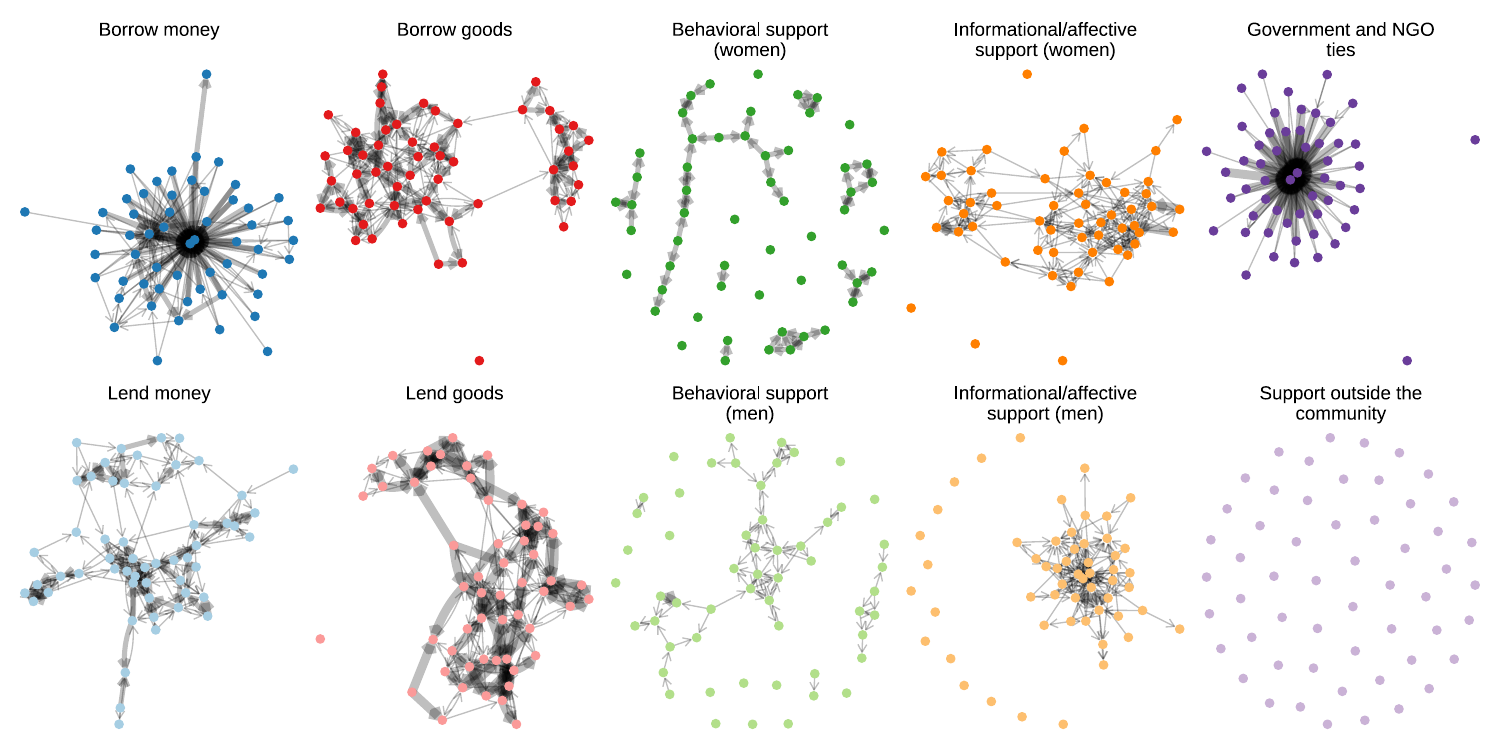}
\end{center}
\caption{Sharing unit-level networks for community AP}
\label{fig:networks_ap}
\end{figure}

\begin{figure}
\begin{center}
\includegraphics[width = \textwidth]{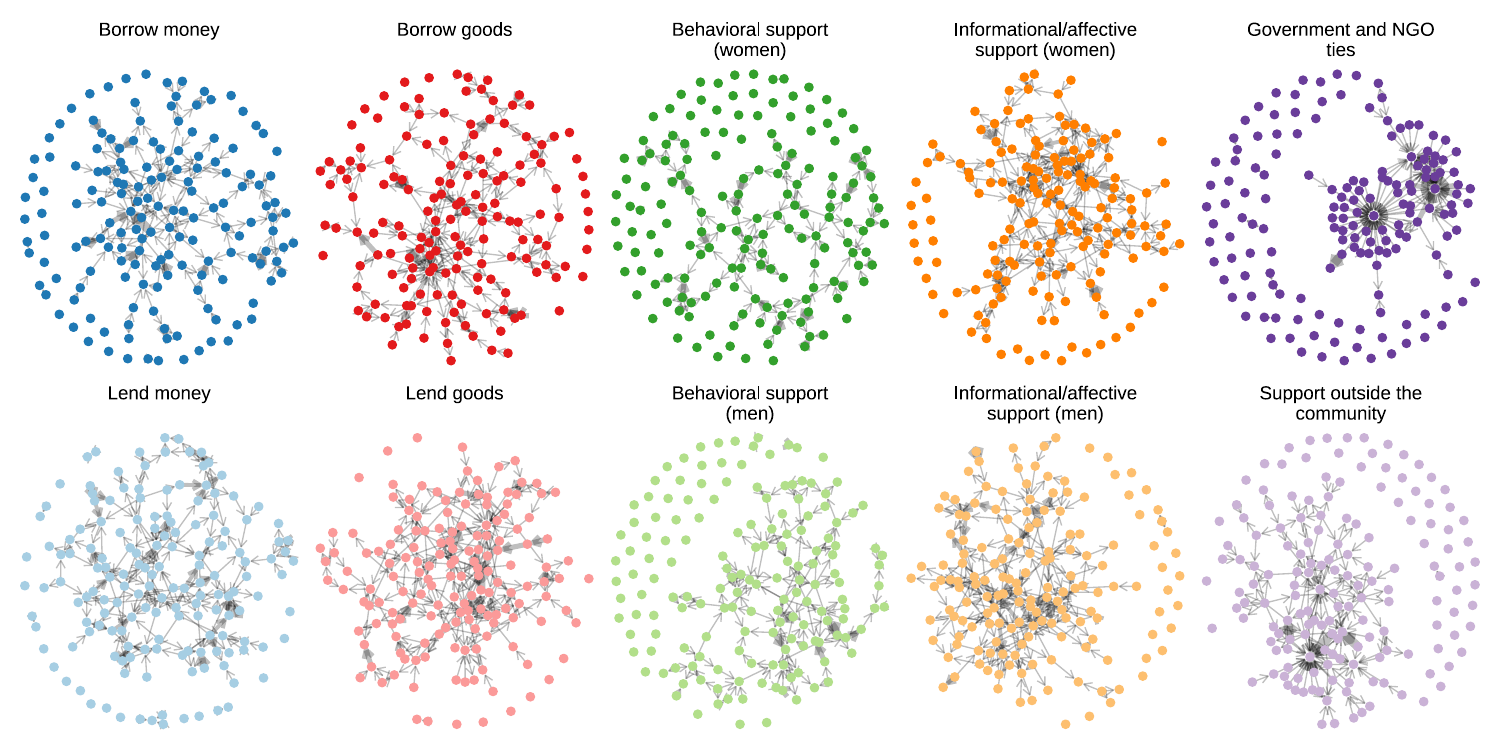}
\end{center}
\caption{Sharing unit-level networks for community AR}
\label{fig:networks_ar}
\end{figure}

\begin{figure}
\begin{center}
\includegraphics[width = \textwidth]{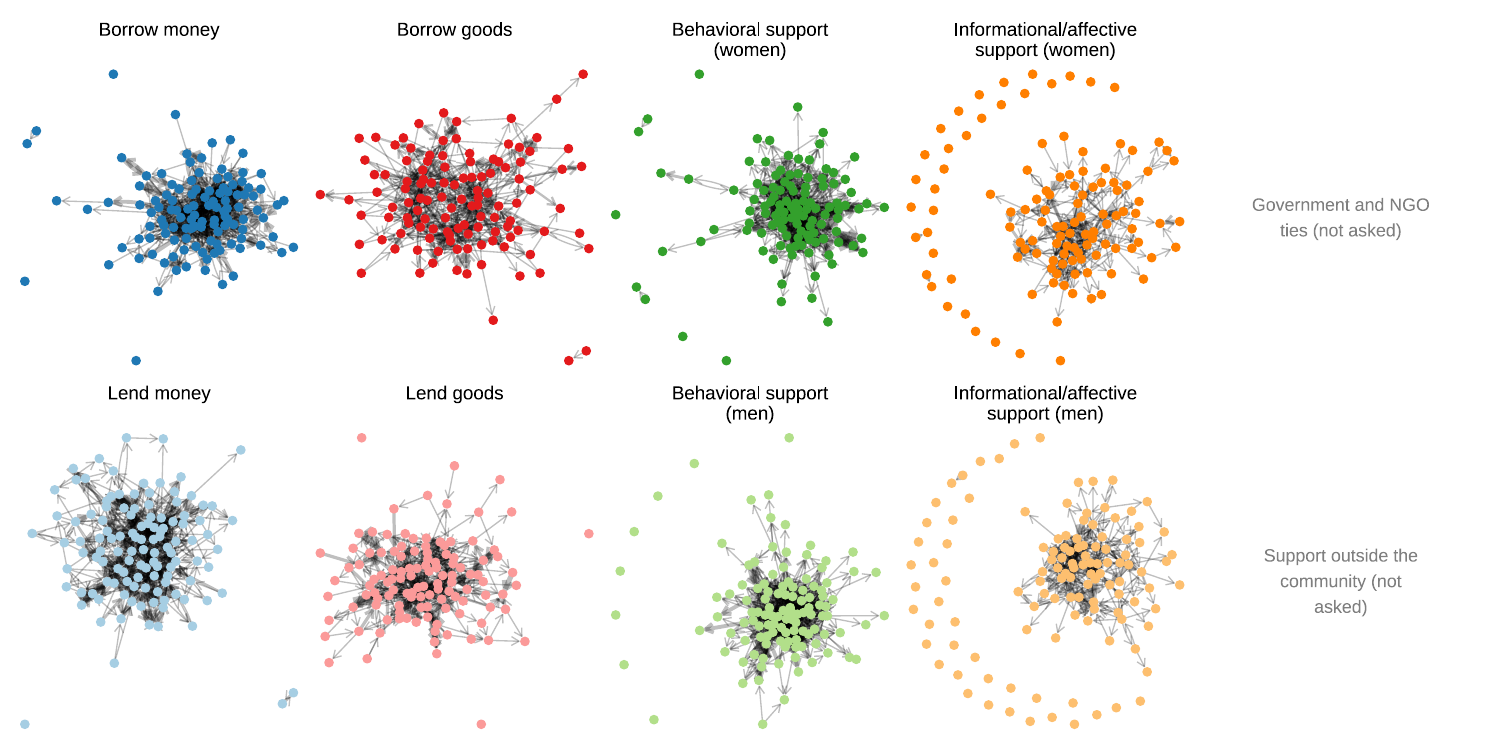}
\end{center}
\caption{Sharing unit-level networks for community AV}
\label{fig:networks_av}
\end{figure}

\begin{figure}
\begin{center}
\includegraphics[width = \textwidth]{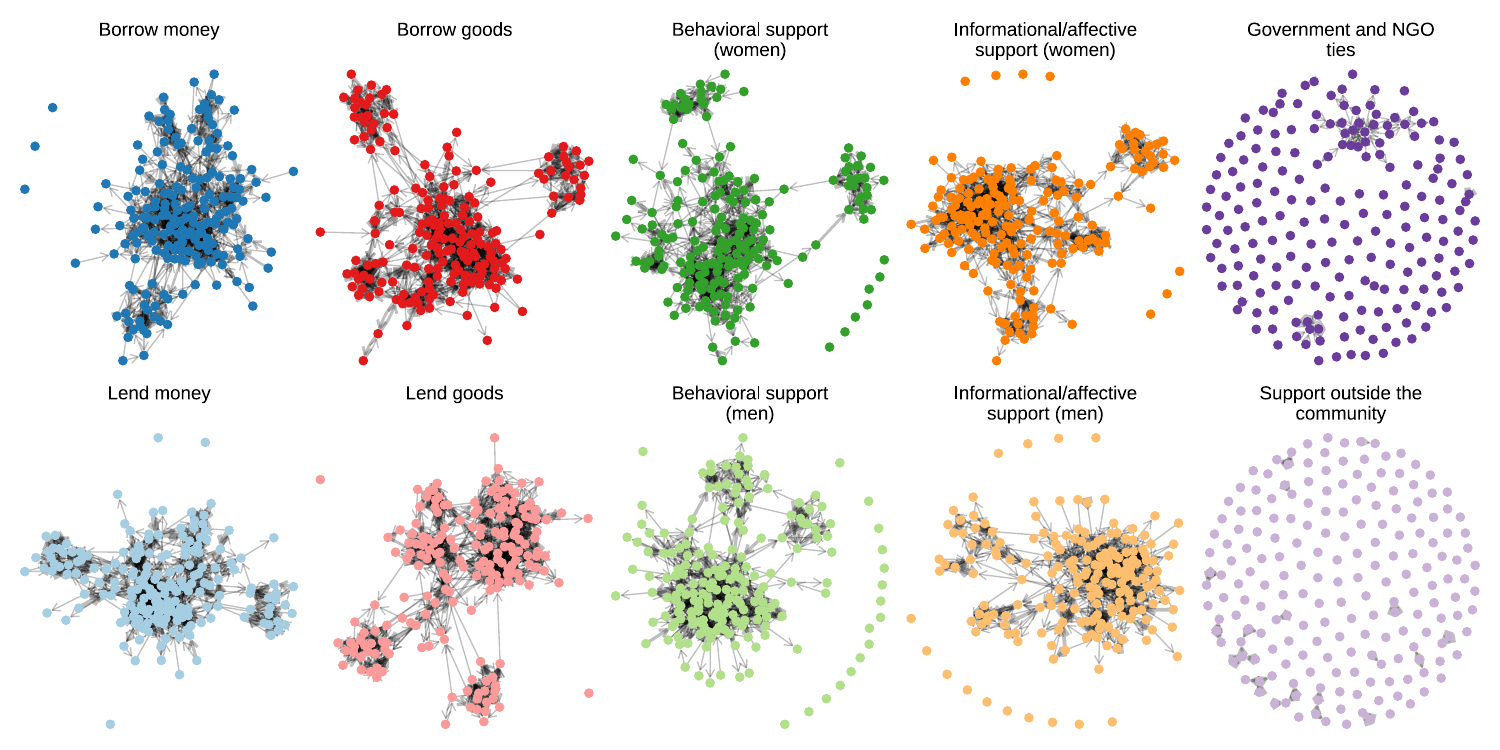}
\end{center}
\caption{Sharing unit-level networks for community AZ}
\label{fig:networks_az}
\end{figure}

\begin{figure}
\begin{center}
\includegraphics[width = \textwidth]{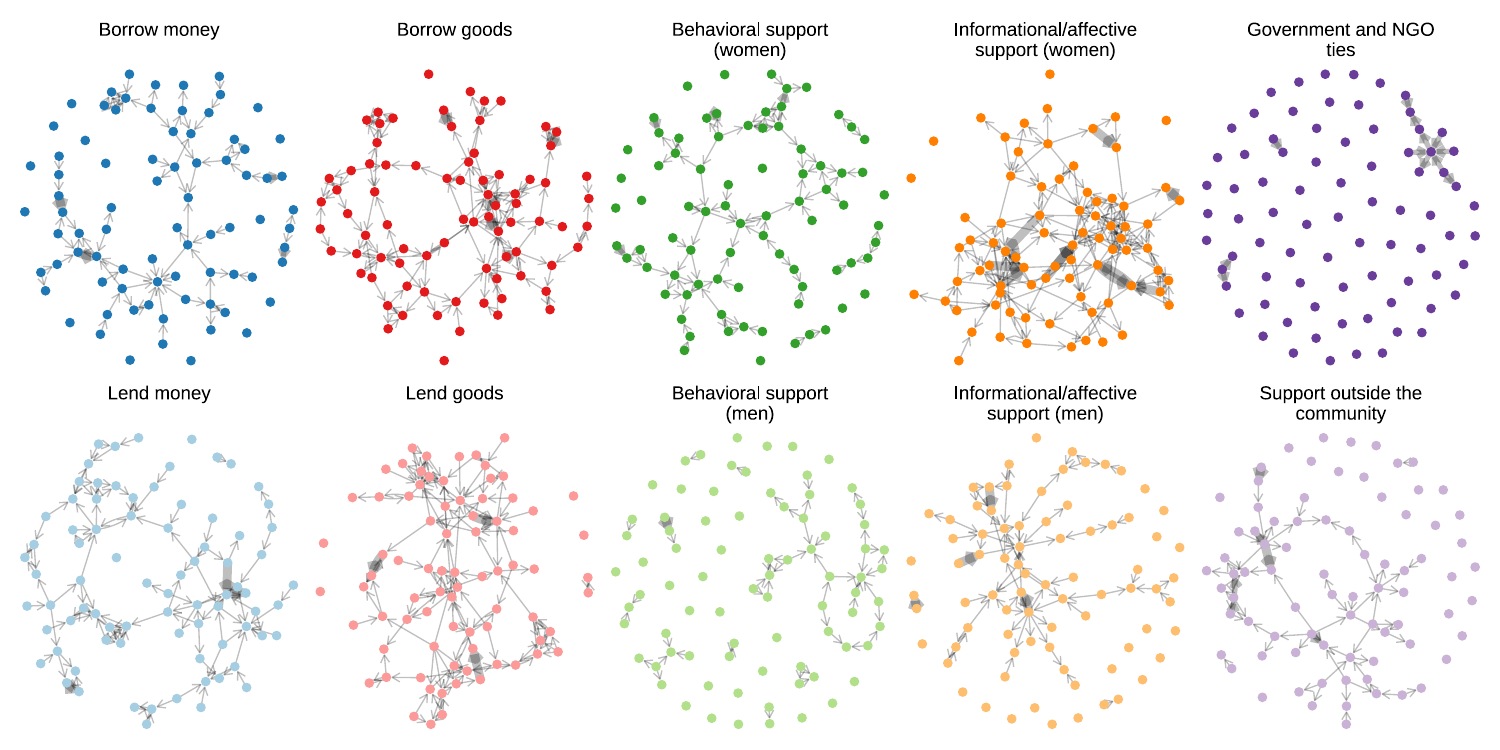}
\end{center}
\caption{Sharing unit-level networks for community BD}
\label{fig:networks_bd}
\end{figure}

\begin{figure}
\begin{center}
\includegraphics[width = \textwidth]{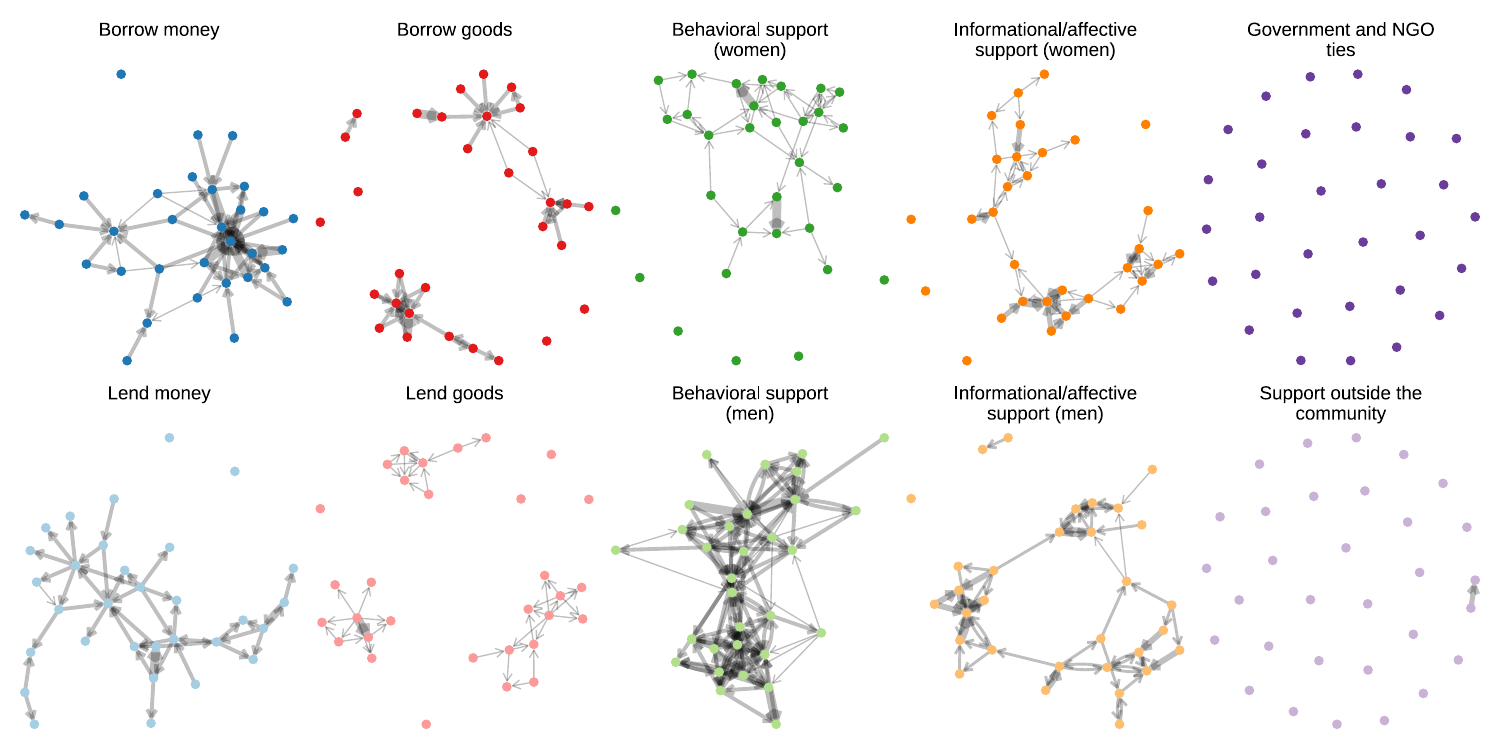}
\end{center}
\caption{Sharing unit-level networks for community BM}
\label{fig:networks_bm}
\end{figure}

\begin{figure}
\begin{center}
\includegraphics[width = \textwidth]{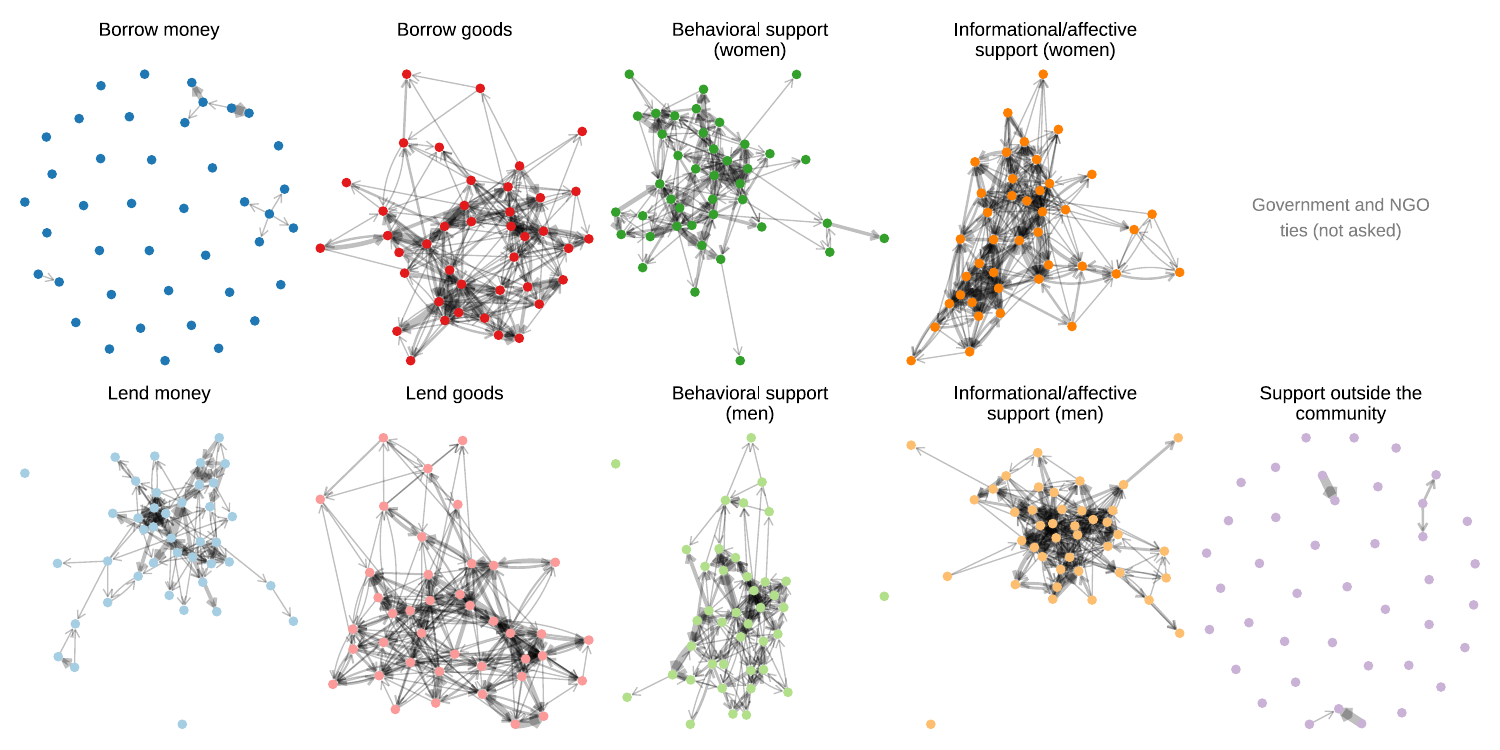}
\end{center}
\caption{Sharing unit-level networks for community BY}
\label{fig:networks_by}
\end{figure}

\begin{figure}
\begin{center}
\includegraphics[width = \textwidth]{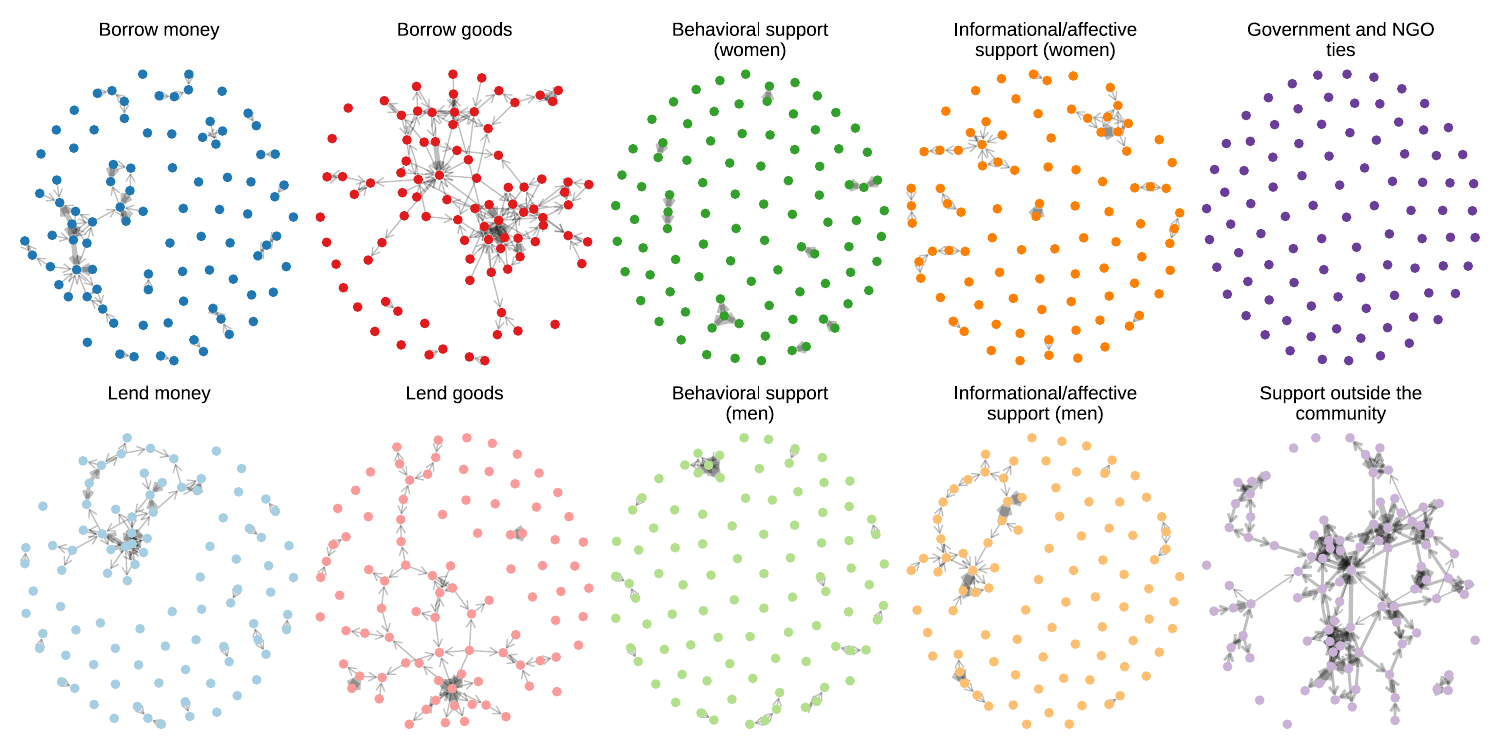}
\end{center}
\caption{Sharing unit-level networks for community CP}
\label{fig:networks_cp}
\end{figure}

\begin{figure}
\begin{center}
\includegraphics[width = \textwidth]{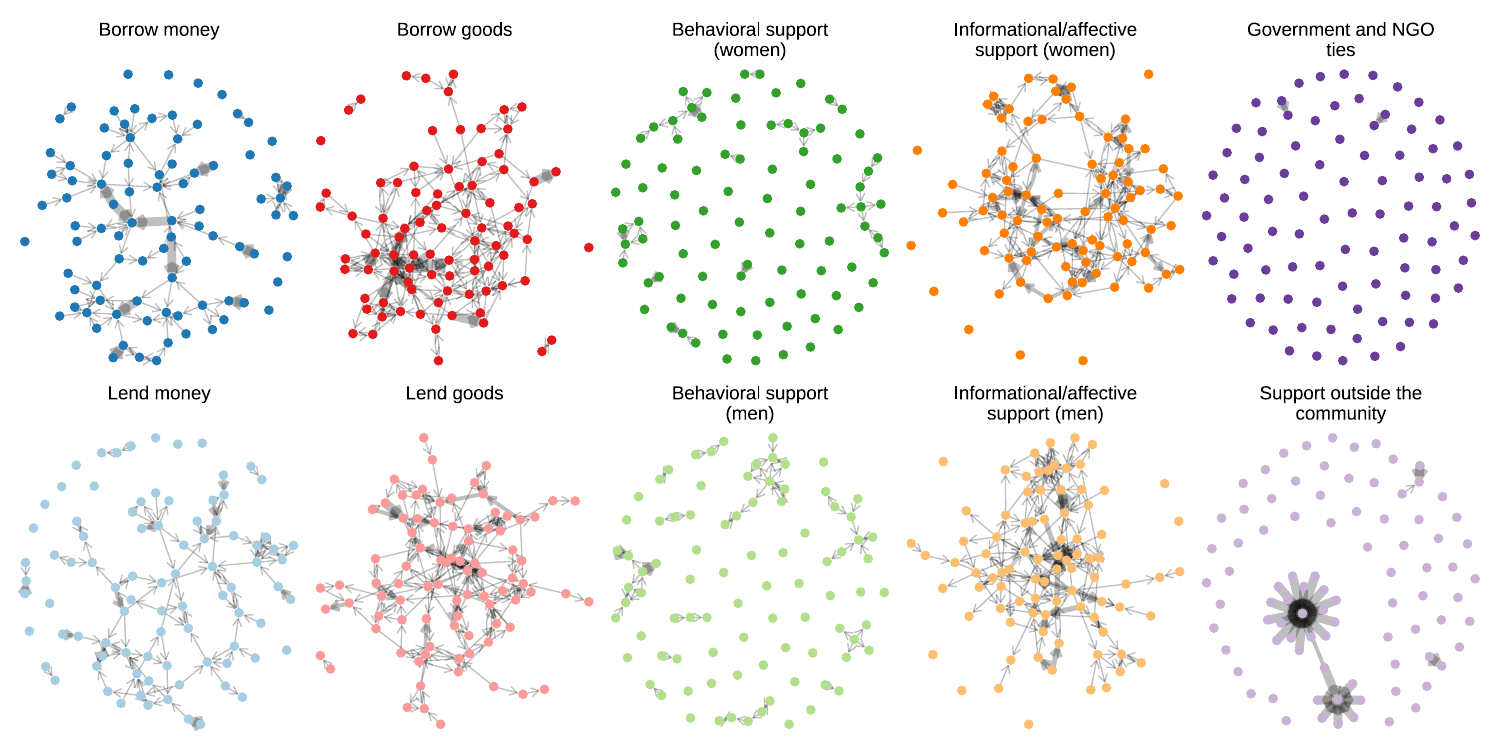}
\end{center}
\caption{Sharing unit-level networks for community CR}
\label{fig:networks_cr}
\end{figure}

\begin{figure}
\begin{center}
\includegraphics[width = \textwidth]{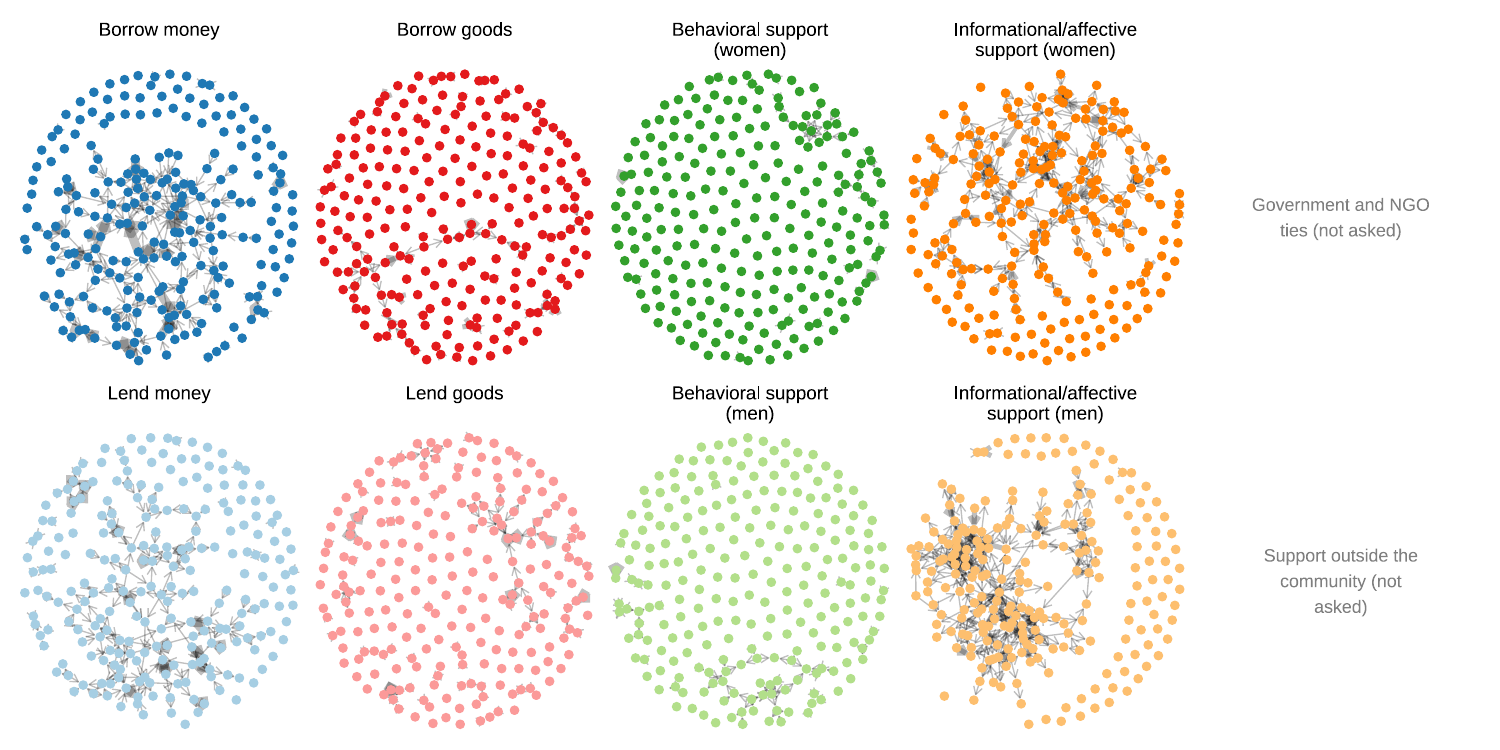}
\end{center}
\caption{Sharing unit-level networks for community DJ}
\label{fig:networks_dj}
\end{figure}

\begin{figure}
\begin{center}
\includegraphics[width = \textwidth]{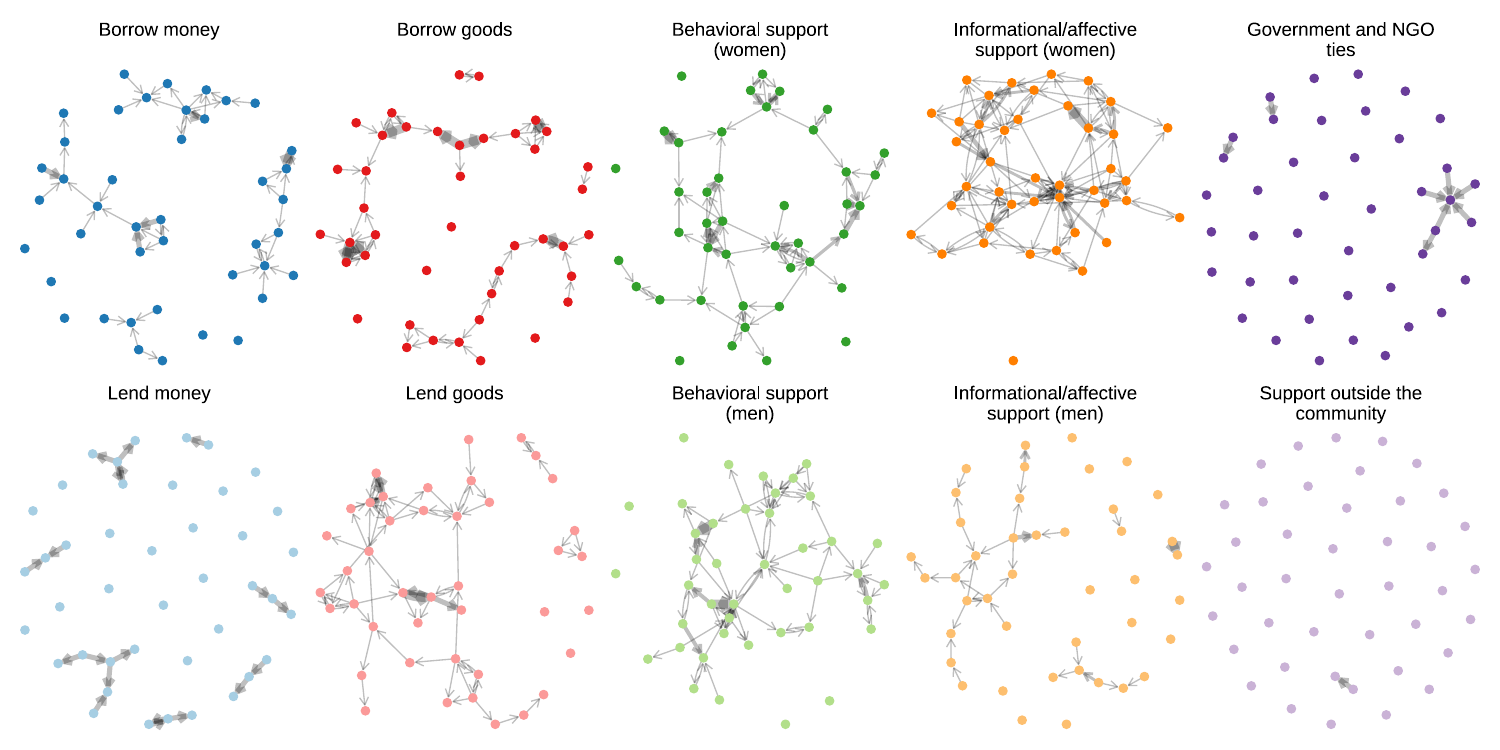}
\end{center}
\caption{Sharing unit-level networks for community EK}
\label{fig:networks_ek}
\end{figure}

\begin{figure}
\begin{center}
\includegraphics[width = \textwidth]{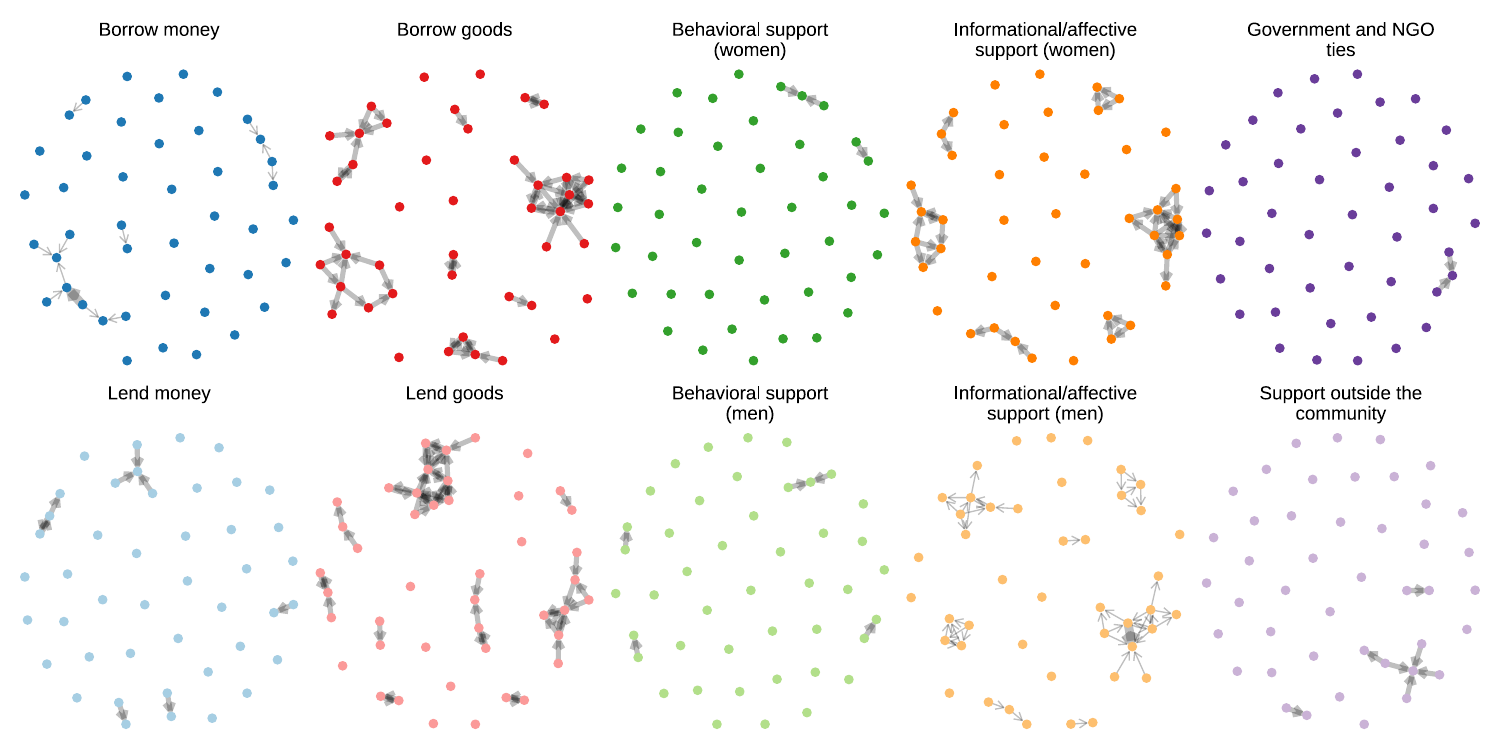}
\end{center}
\caption{Sharing unit-level networks for community FF}
\label{fig:networks_ff}
\end{figure}

\begin{figure}
\begin{center}
\includegraphics[width = \textwidth]{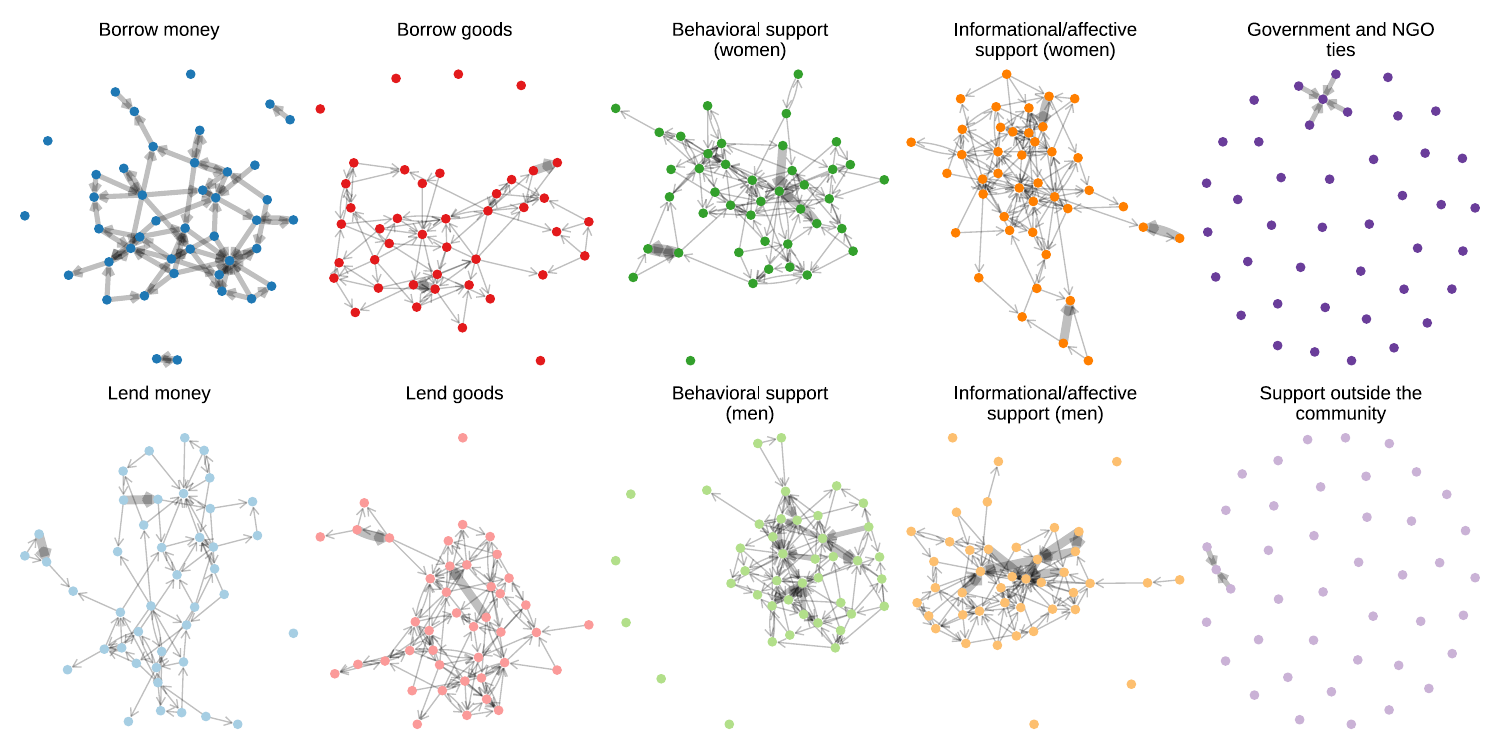}
\end{center}
\caption{Sharing unit-level networks for community FJ}
\label{fig:networks_fj}
\end{figure}

\begin{figure}
\begin{center}
\includegraphics[width = \textwidth]{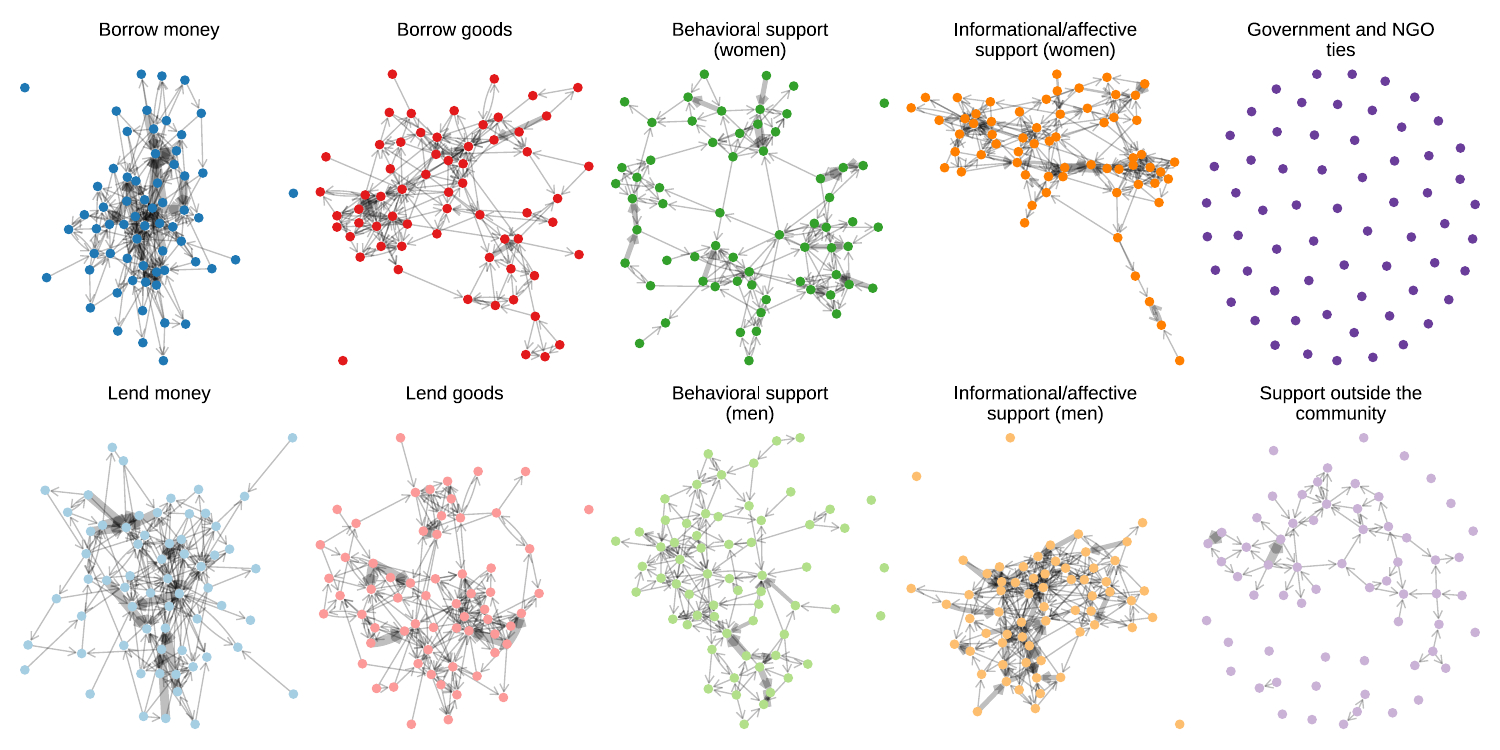}
\end{center}
\caption{Sharing unit-level networks for community HE}
\label{fig:networks_he}
\end{figure}

\begin{figure}
\begin{center}
\includegraphics[width = \textwidth]{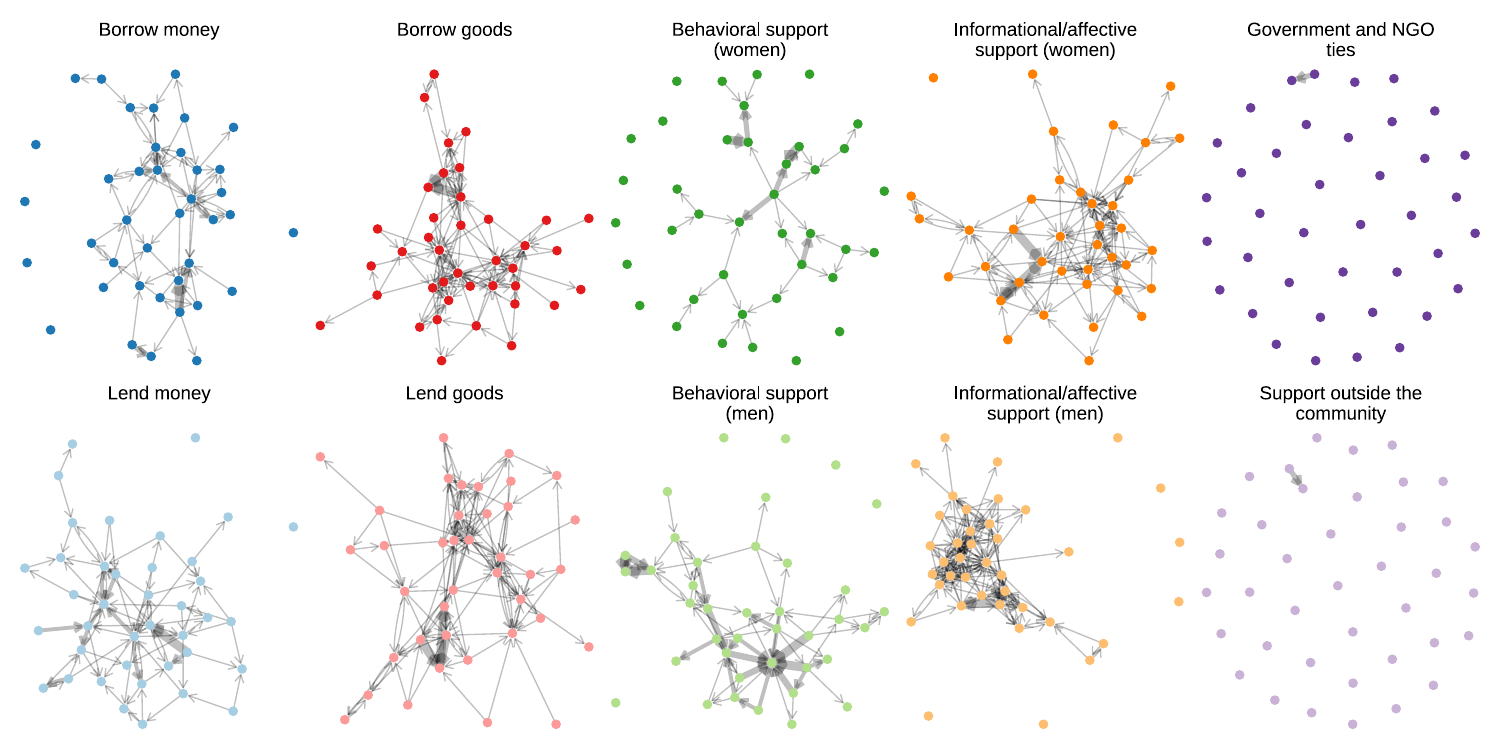}
\end{center}
\caption{Sharing unit-level networks for community HI}
\label{fig:networks_hi}
\end{figure}

\begin{figure}
\begin{center}
\includegraphics[width = \textwidth]{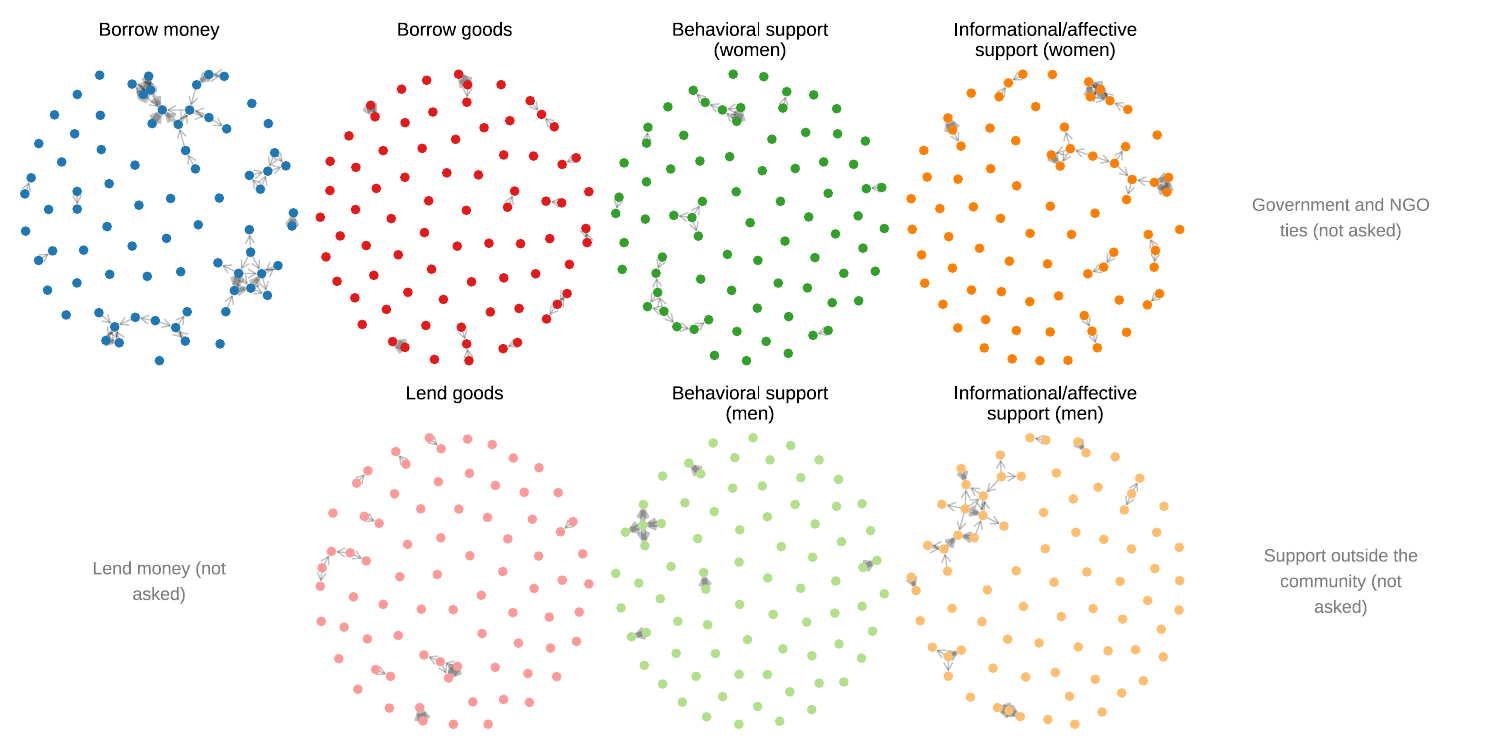}
\end{center}
\caption{Sharing unit-level networks for community IH}
\label{fig:networks_ih}
\end{figure}

\begin{figure}
\begin{center}
\includegraphics[width = \textwidth]{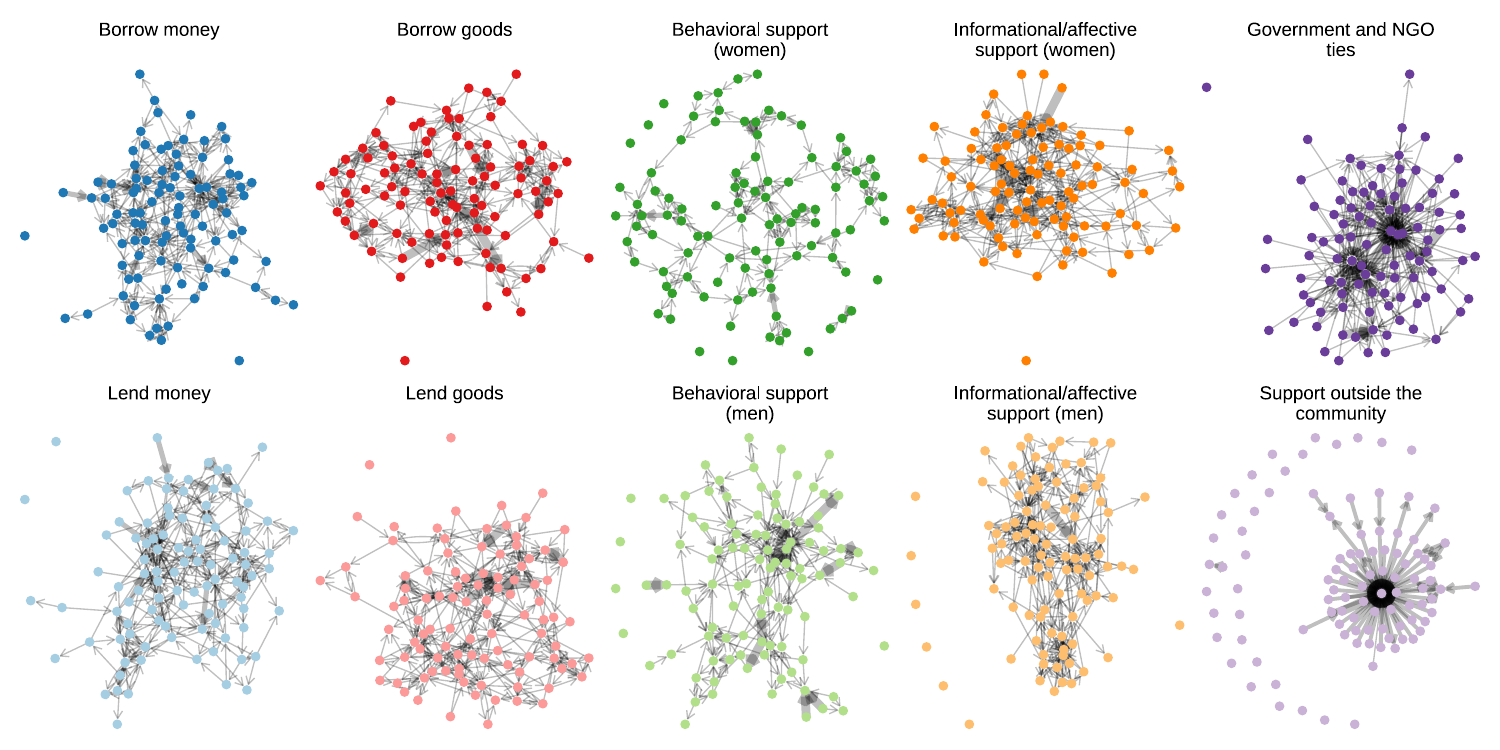}
\end{center}
\caption{Sharing unit-level networks for community KA}
\label{fig:networks_ka}
\end{figure}

\begin{figure}
\begin{center}
\includegraphics[width = \textwidth]{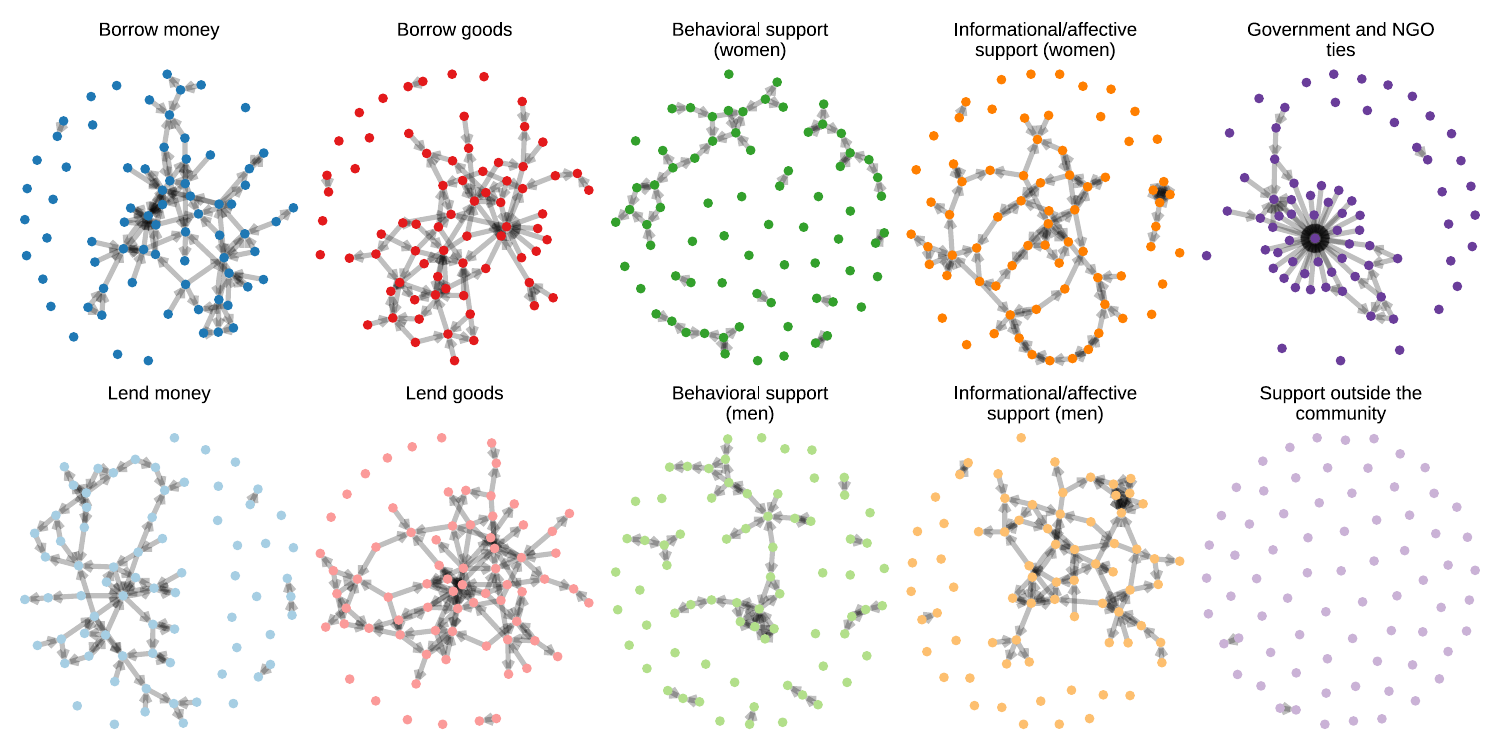}
\end{center}
\caption{Sharing unit-level networks for community KM}
\label{fig:networks_km}
\end{figure}

\begin{figure}
\begin{center}
\includegraphics[width = \textwidth]{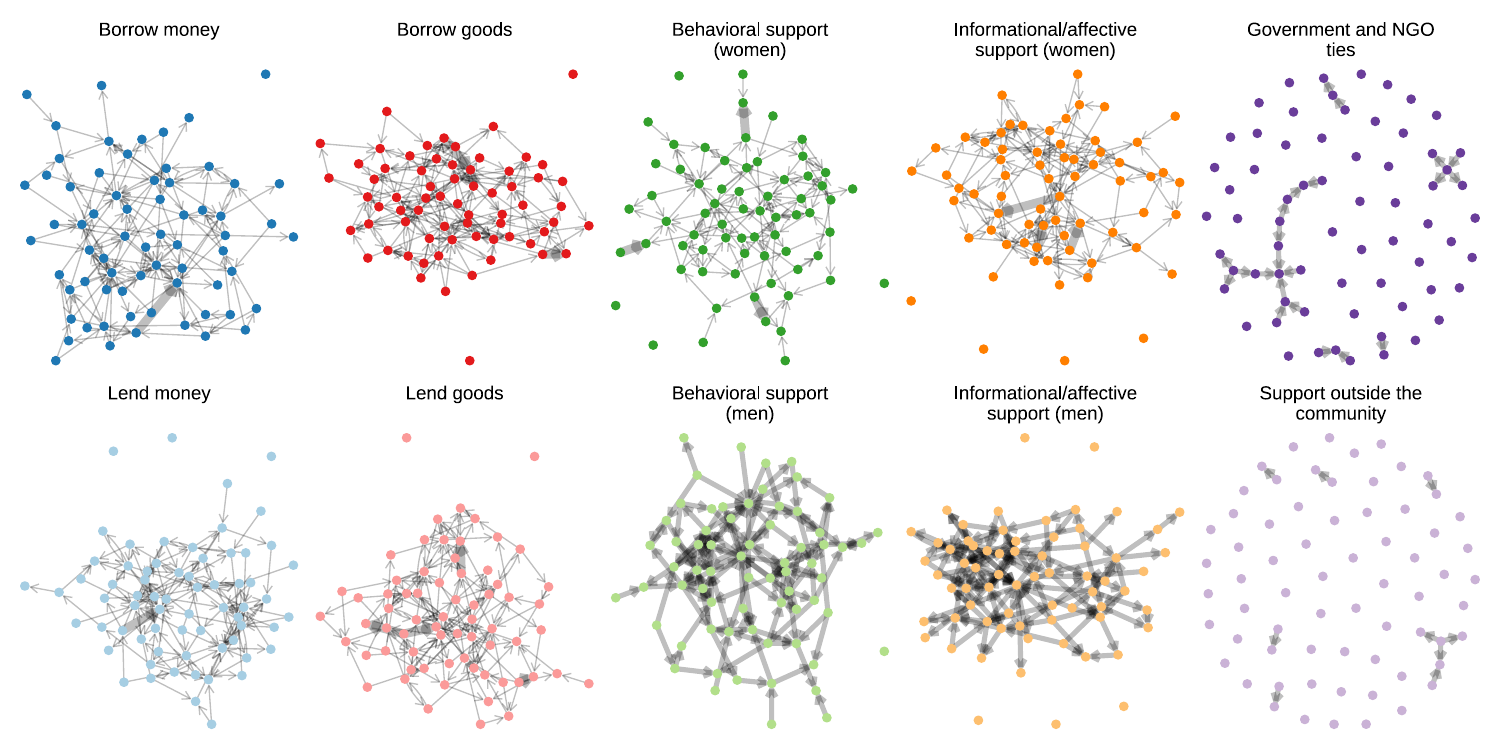}
\end{center}
\caption{Sharing unit-level networks for community KO}
\label{fig:networks_ko}
\end{figure}

\begin{figure}
\begin{center}
\includegraphics[width = \textwidth]{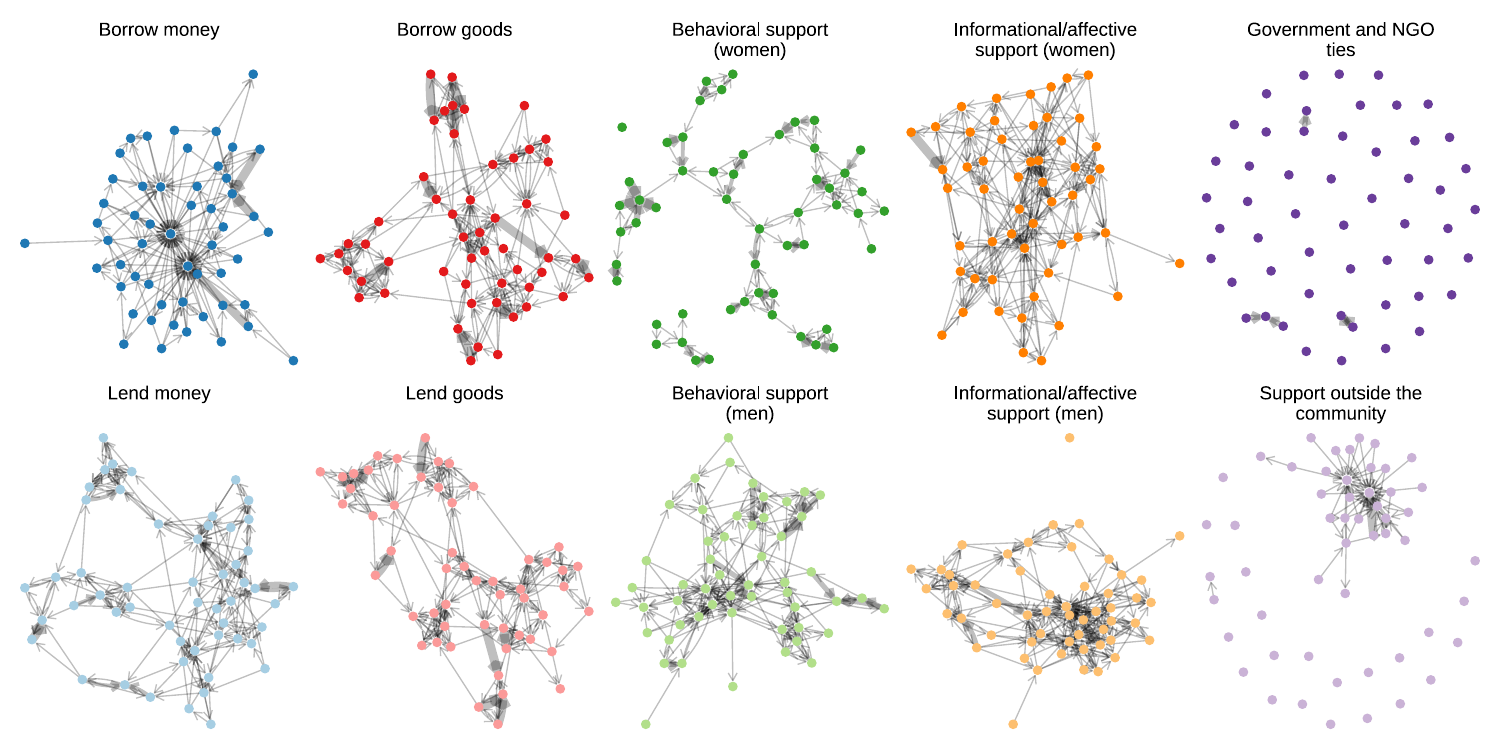}
\end{center}
\caption{Sharing unit-level networks for community KS}
\label{fig:networks_ks}
\end{figure}

\begin{figure}
\begin{center}
\includegraphics[width = \textwidth]{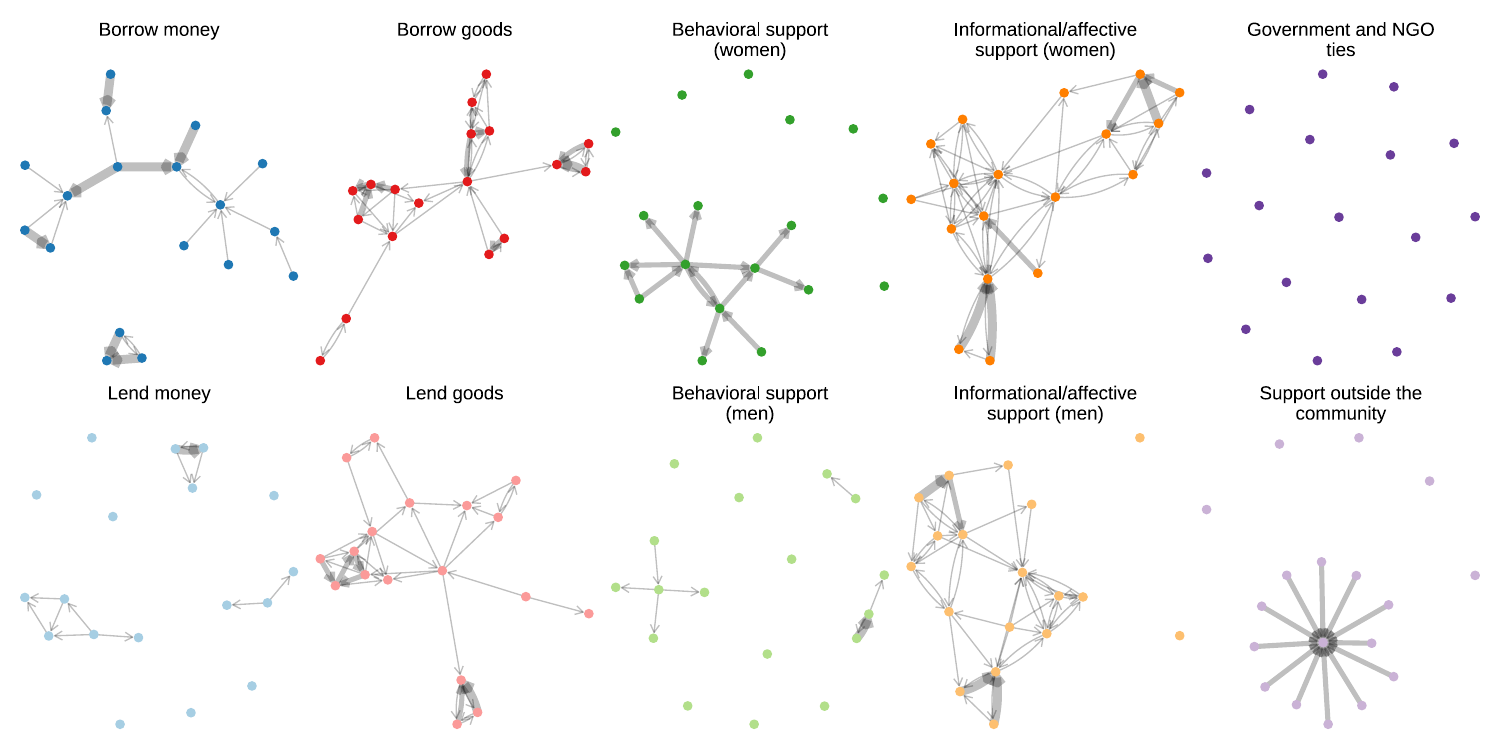}
\end{center}
\caption{Sharing unit-level networks for community KT}
\label{fig:networks_kt}
\end{figure}

\begin{figure}
\begin{center}
\includegraphics[width = \textwidth]{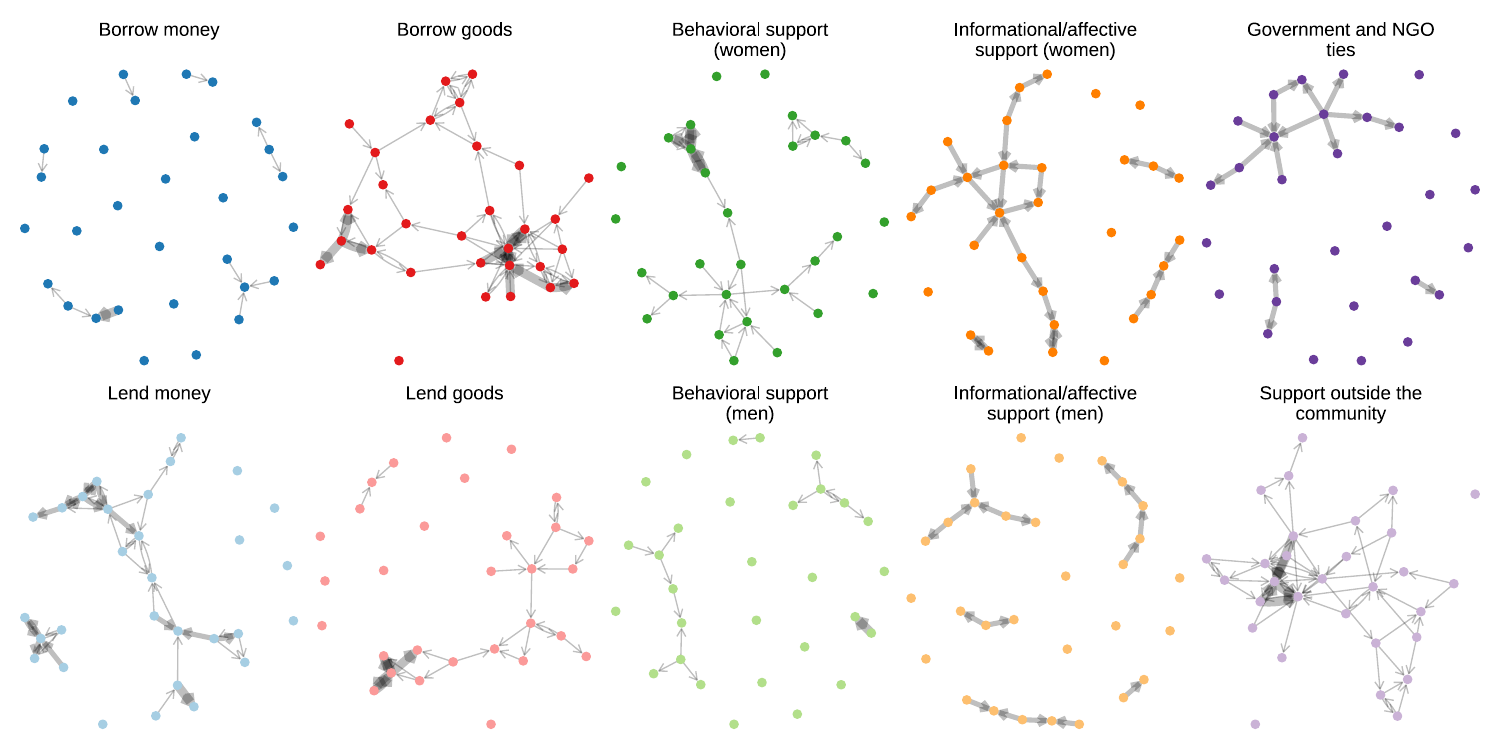}
\end{center}
\caption{Sharing unit-level networks for community LB}
\label{fig:networks_lb}
\end{figure}

\begin{figure}
\begin{center}
\includegraphics[width = \textwidth]{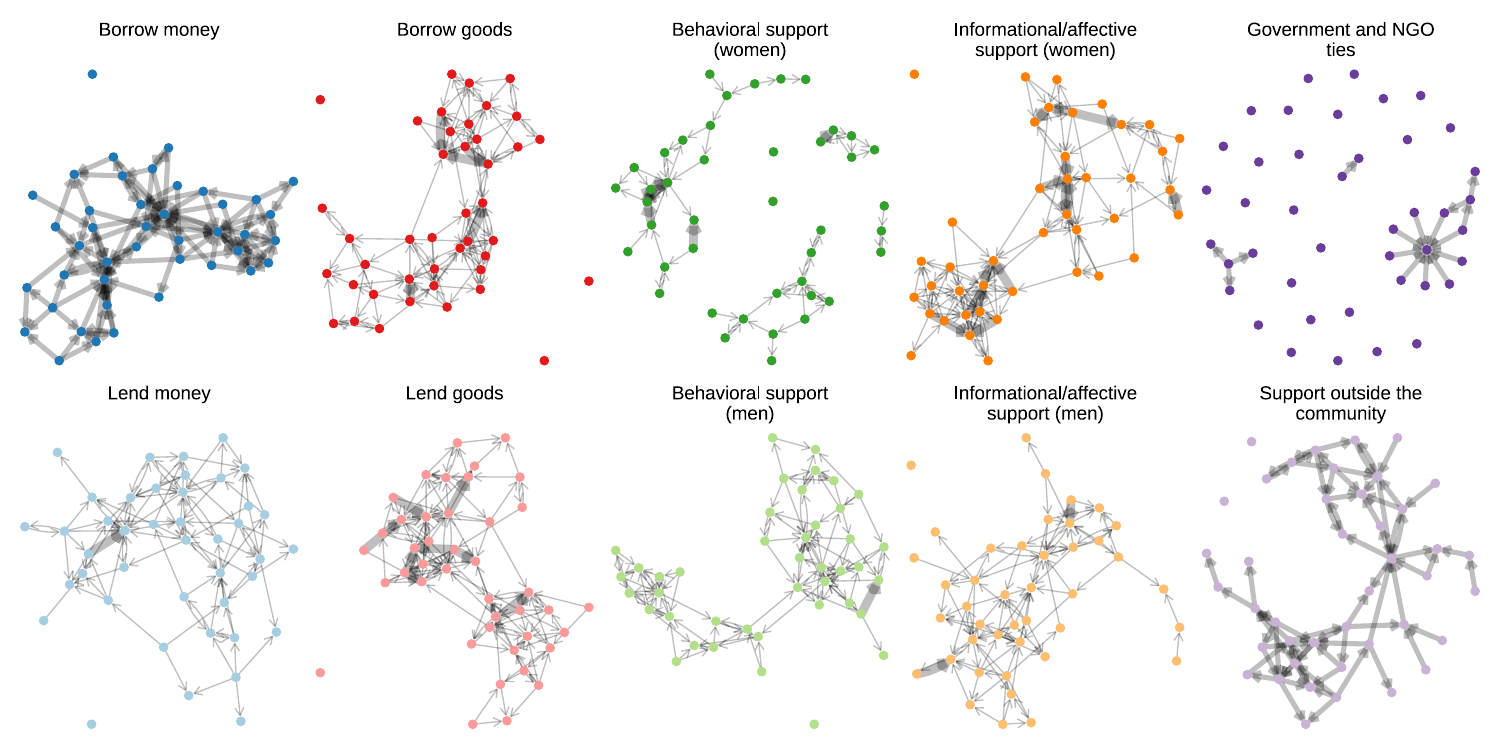}
\end{center}
\caption{Sharing unit-level networks for community MA}
\label{fig:networks_ma}
\end{figure}

\begin{figure}
\begin{center}
\includegraphics[width = \textwidth]{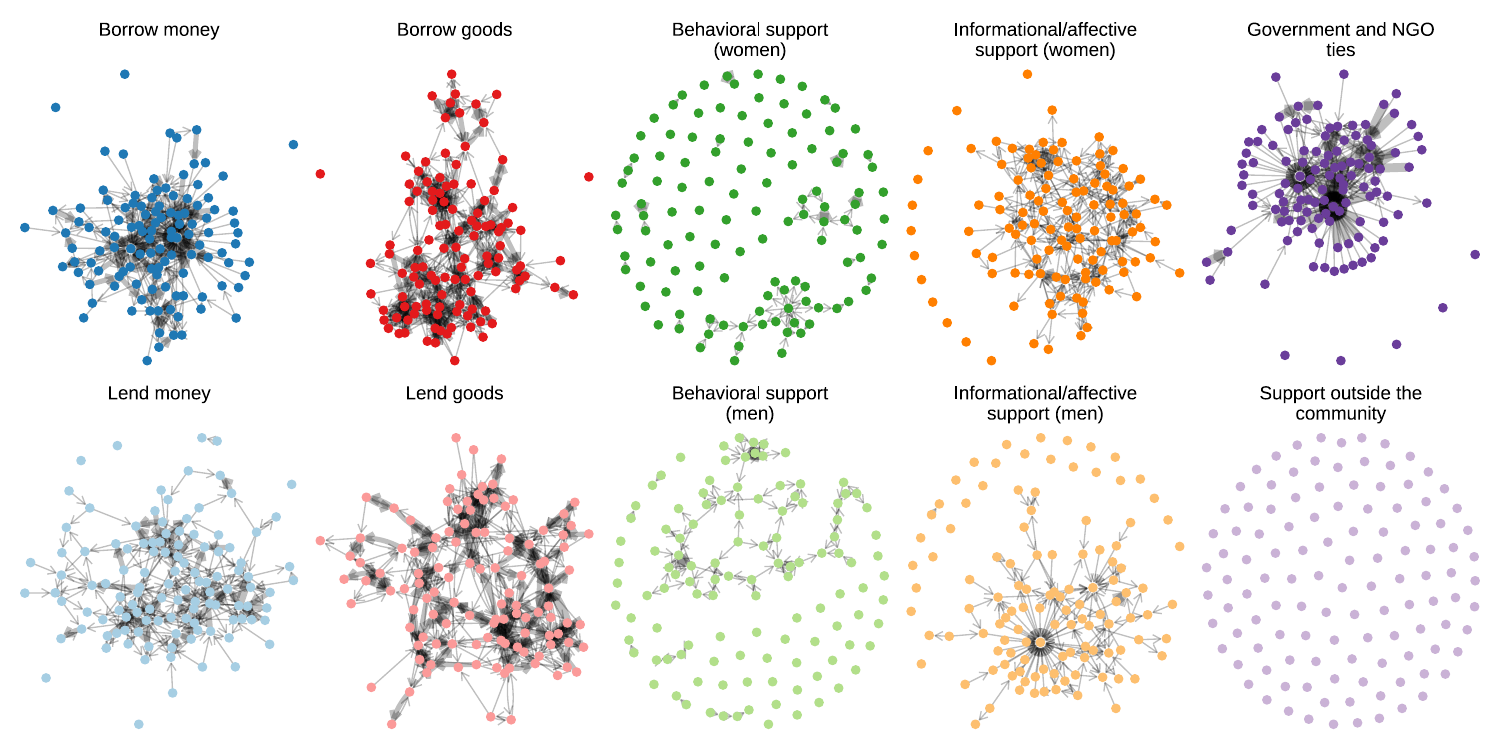}
\end{center}
\caption{Sharing unit-level networks for community MC}
\label{fig:networks_mc}
\end{figure}

\begin{figure}
\begin{center}
\includegraphics[width = \textwidth]{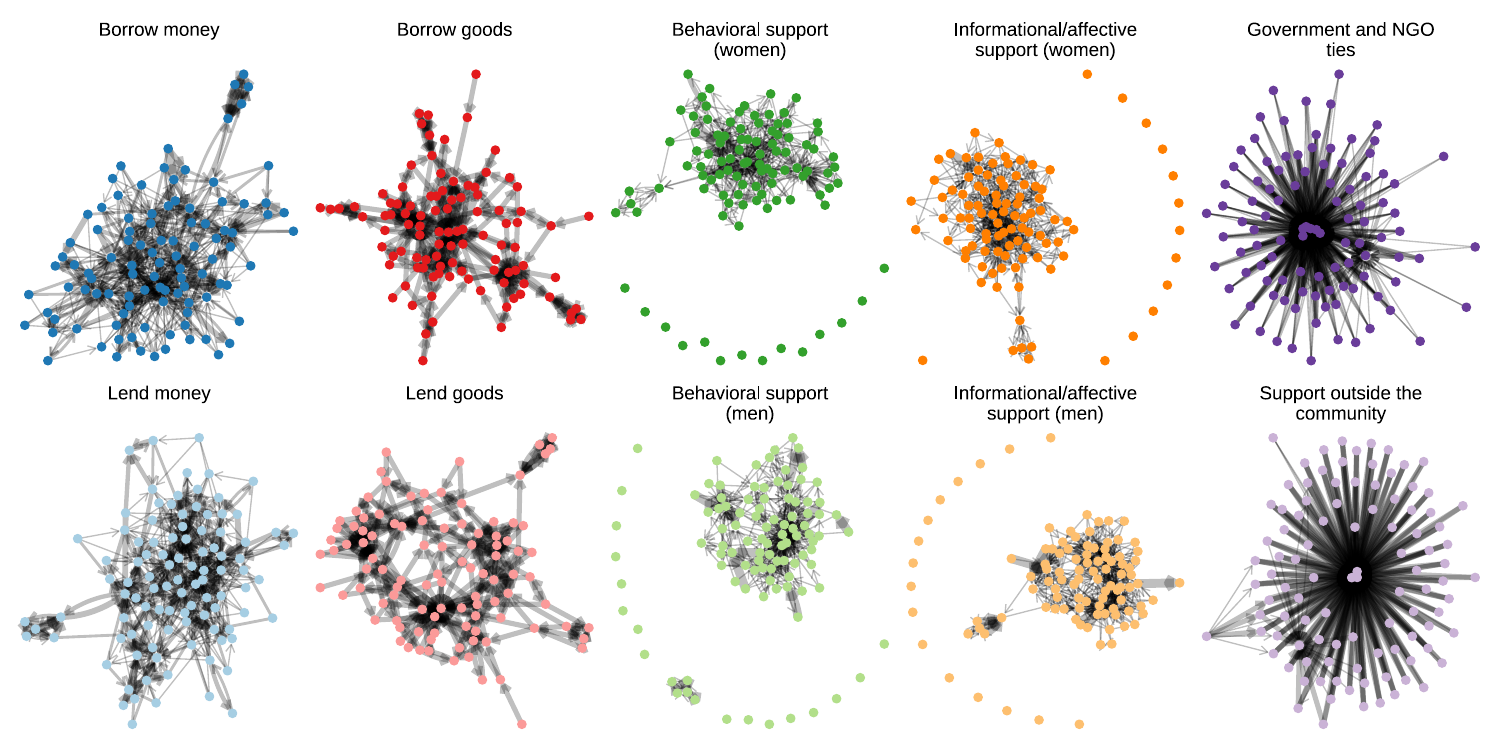}
\end{center}
\caption{Sharing unit-level networks for community MG}
\label{fig:networks_mg}
\end{figure}

\begin{figure}
\begin{center}
\includegraphics[width = \textwidth]{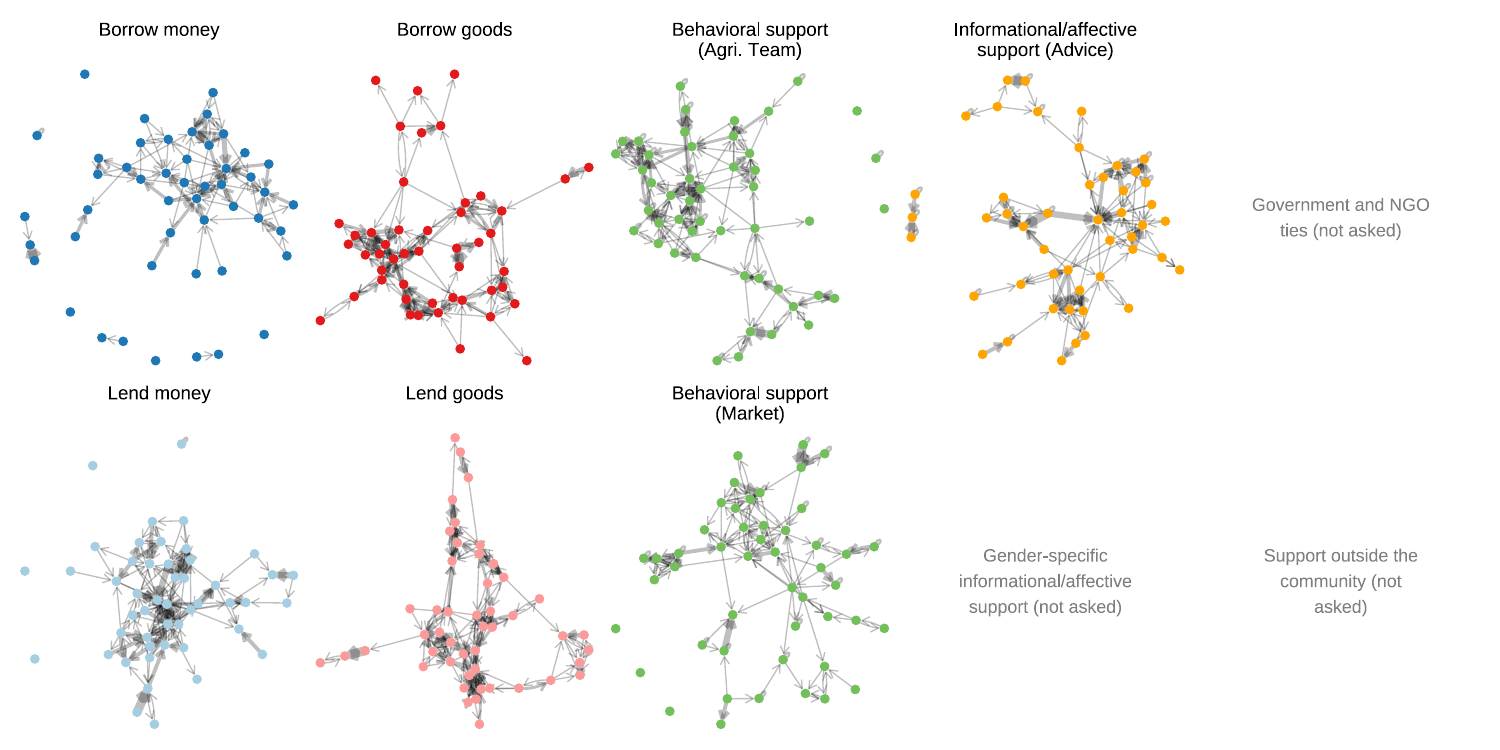}
\end{center}
\caption{Sharing unit-level networks for community MK. (Note that the behavioural and informational/effective prompts were not gender specific. When constructing the aggregate network, only the ``Agri. Team'' prompt is used, not the ``Market'' prompt, for behavioural support.}
\label{fig:networks_mk}
\end{figure}

\begin{figure}
\begin{center}
\includegraphics[width = \textwidth]{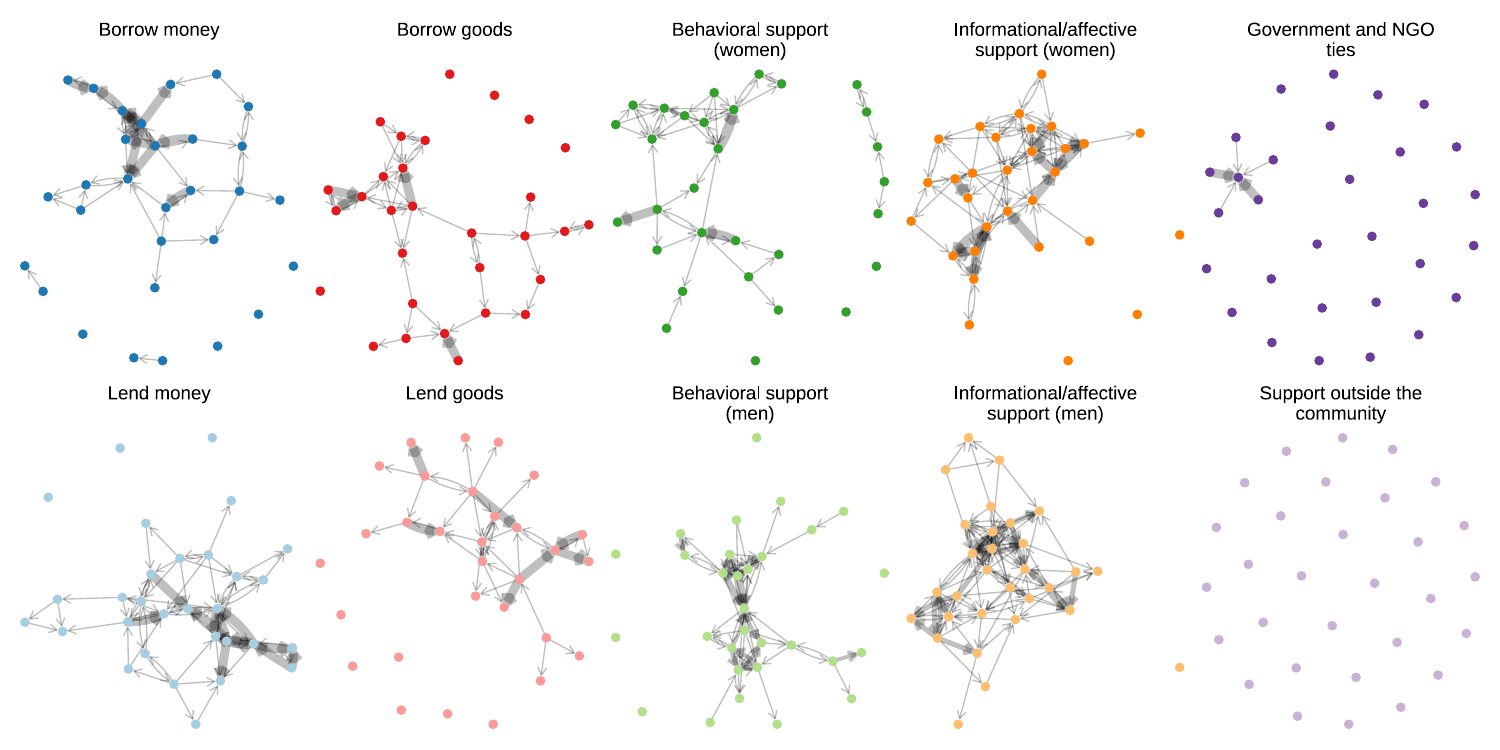}
\end{center}
\caption{Sharing unit-level networks for community MN}
\label{fig:networks_mn}
\end{figure}

\begin{figure}
\begin{center}
\includegraphics[width = \textwidth]{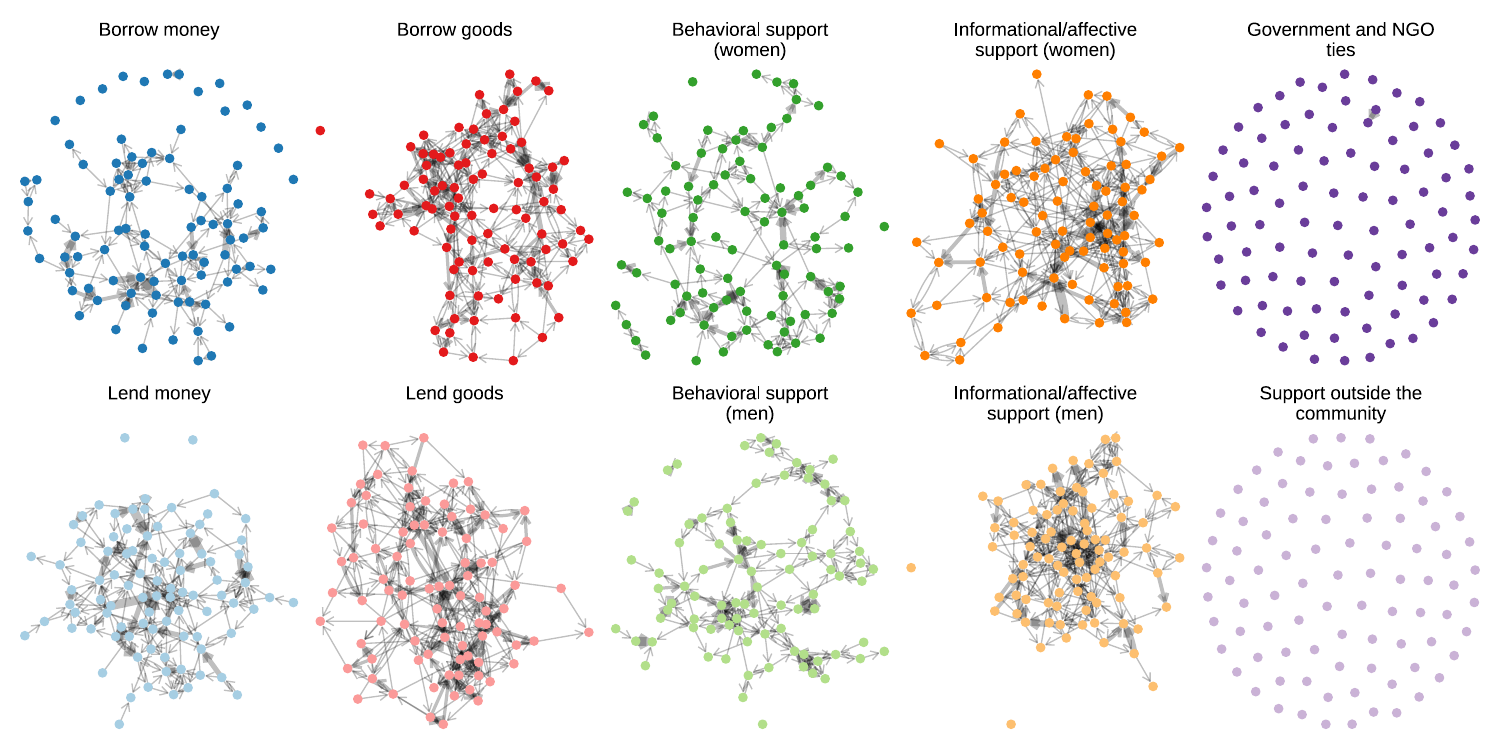}
\end{center}
\caption{Sharing unit-level networks for community MY}
\label{fig:networks_my}
\end{figure}

\begin{figure}
\begin{center}
\includegraphics[width = \textwidth]{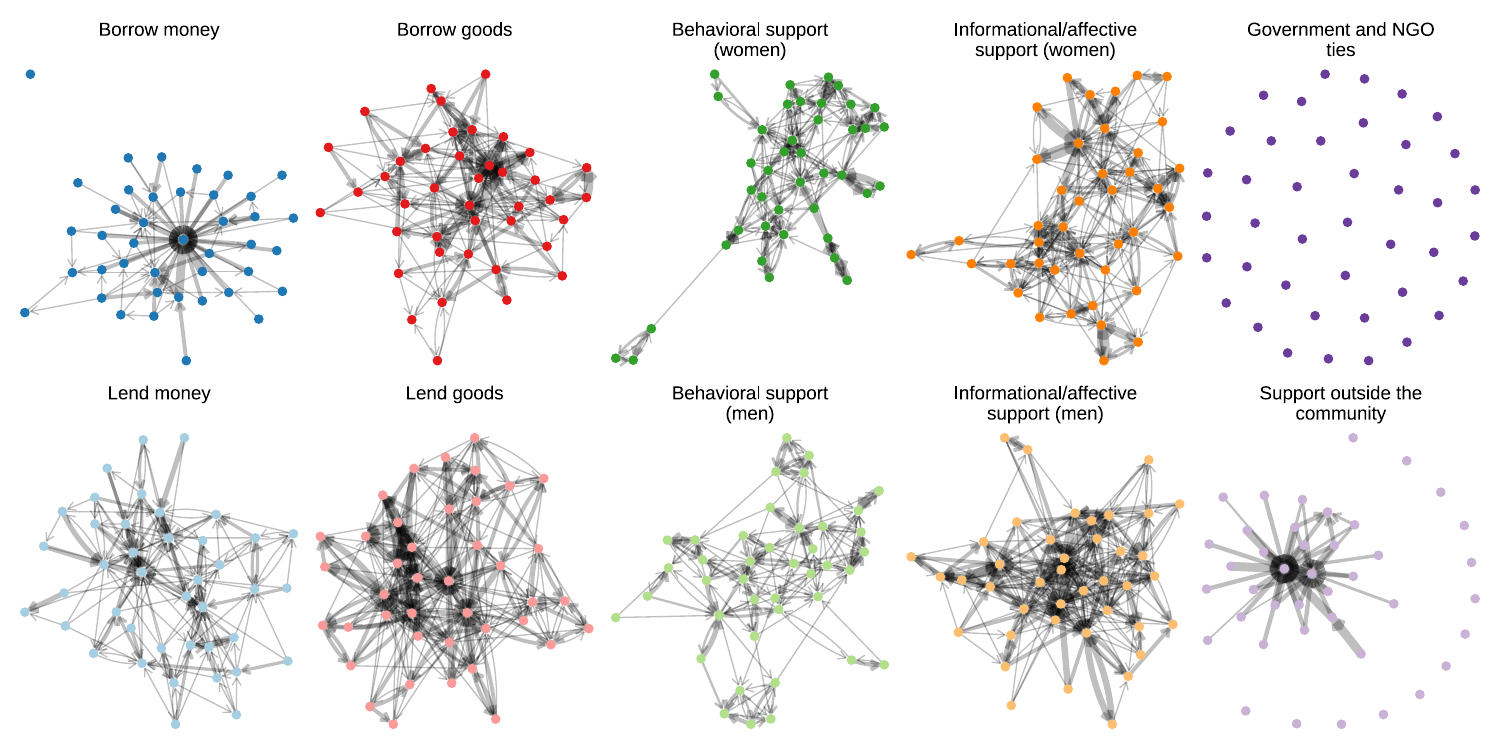}
\end{center}
\caption{Sharing unit-level networks for community NI}
\label{fig:networks_ni}
\end{figure}

\begin{figure}
\begin{center}
\includegraphics[width = \textwidth]{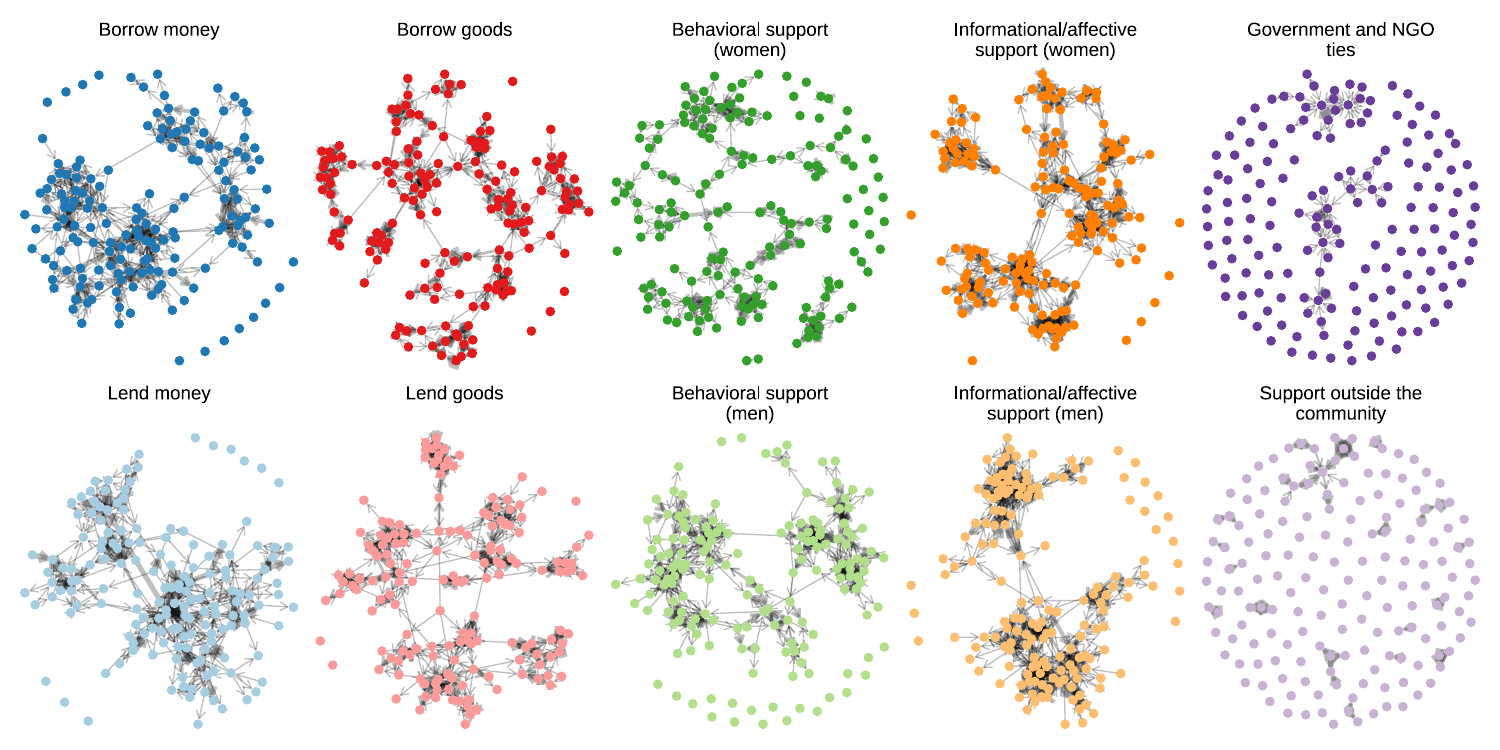}
\end{center}
\caption{Sharing unit-level networks for community NP}
\label{fig:networks_np}
\end{figure}

\begin{figure}
\begin{center}
\includegraphics[width = \textwidth]{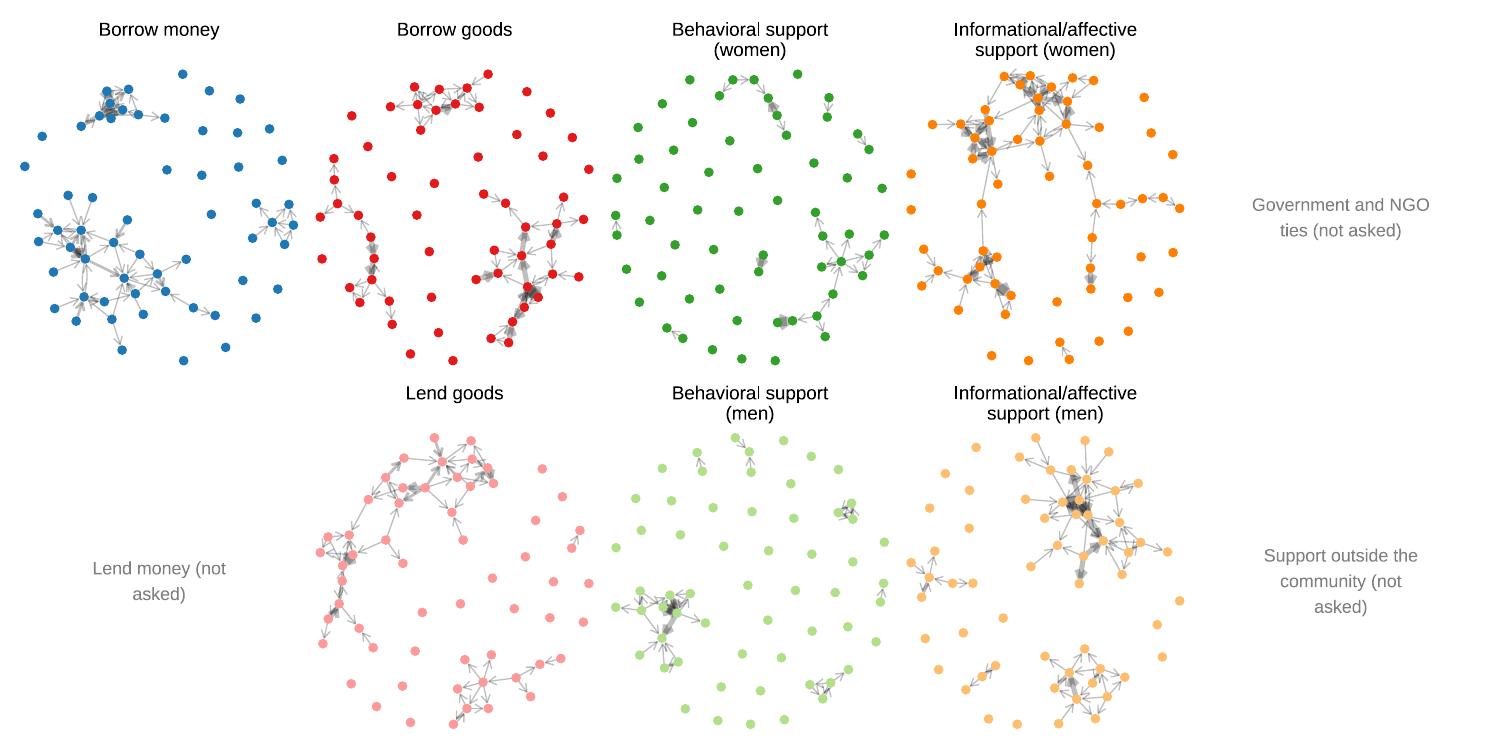}
\end{center}
\caption{Sharing unit-level networks for community PC}
\label{fig:networks_pc}
\end{figure}

\begin{figure}
\begin{center}
\includegraphics[width = \textwidth]{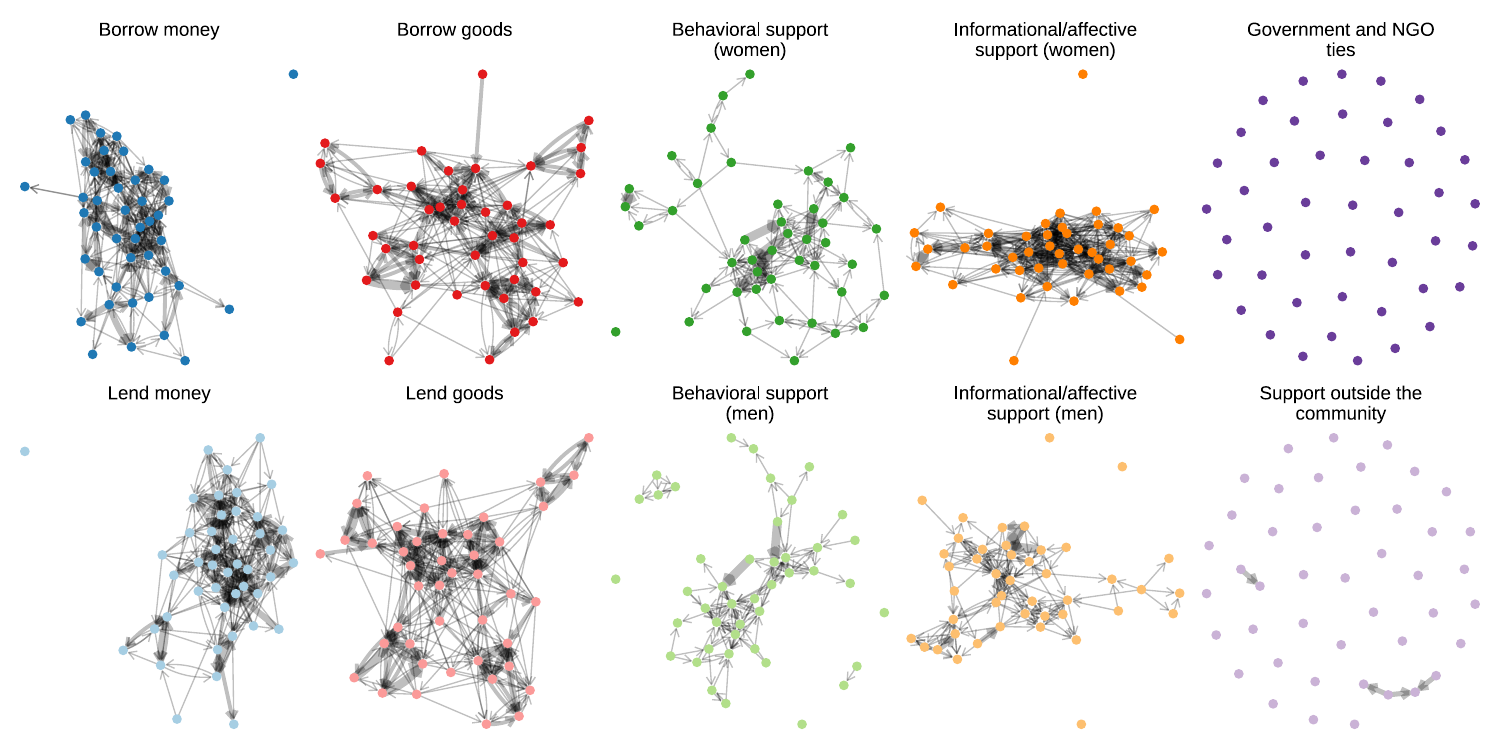}
\end{center}
\caption{Sharing unit-level networks for community PE}
\label{fig:networks_pe}
\end{figure}

\begin{figure}
\begin{center}
\includegraphics[width = \textwidth]{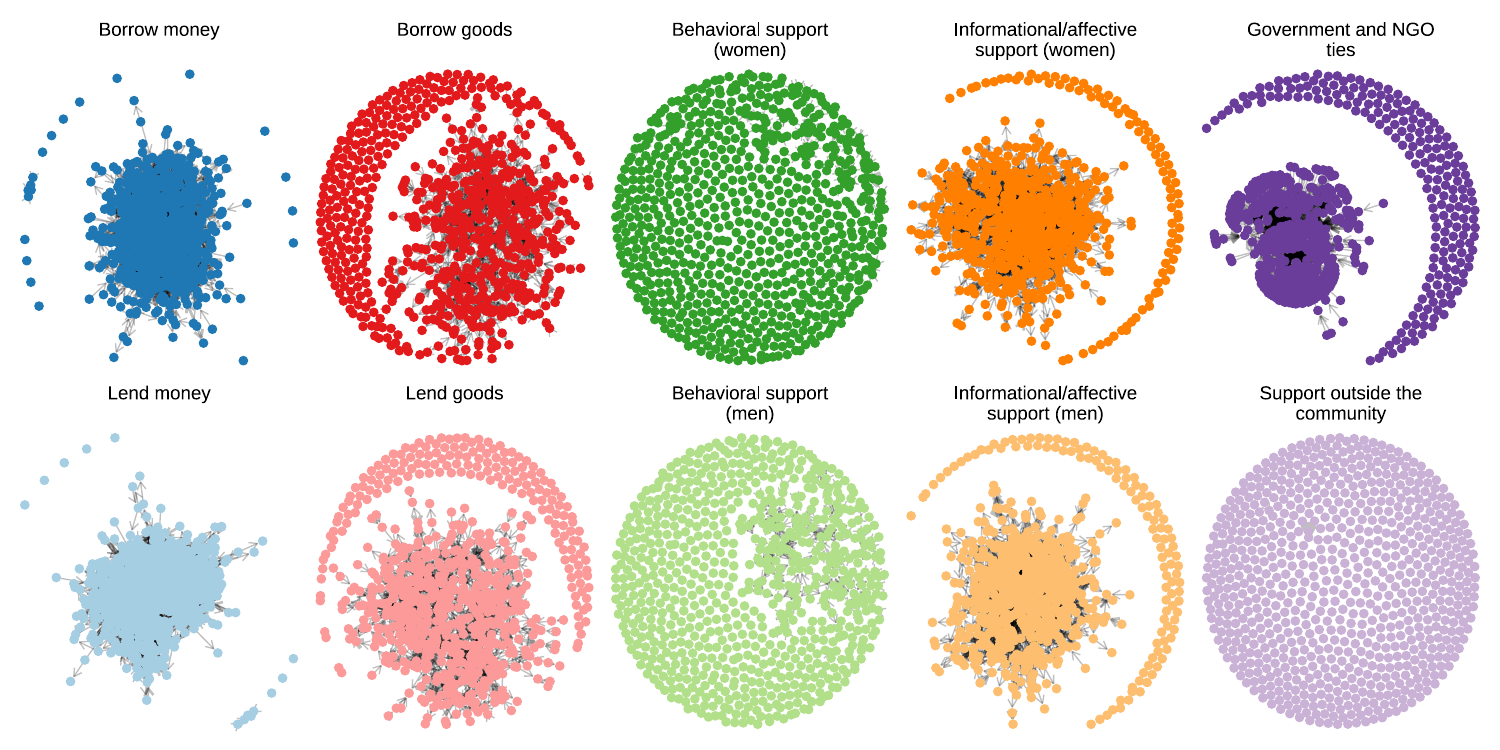}
\end{center}
\caption{Sharing unit-level networks for community PQ}
\label{fig:networks_pq}
\end{figure}

\begin{figure}
\begin{center}
\includegraphics[width = \textwidth]{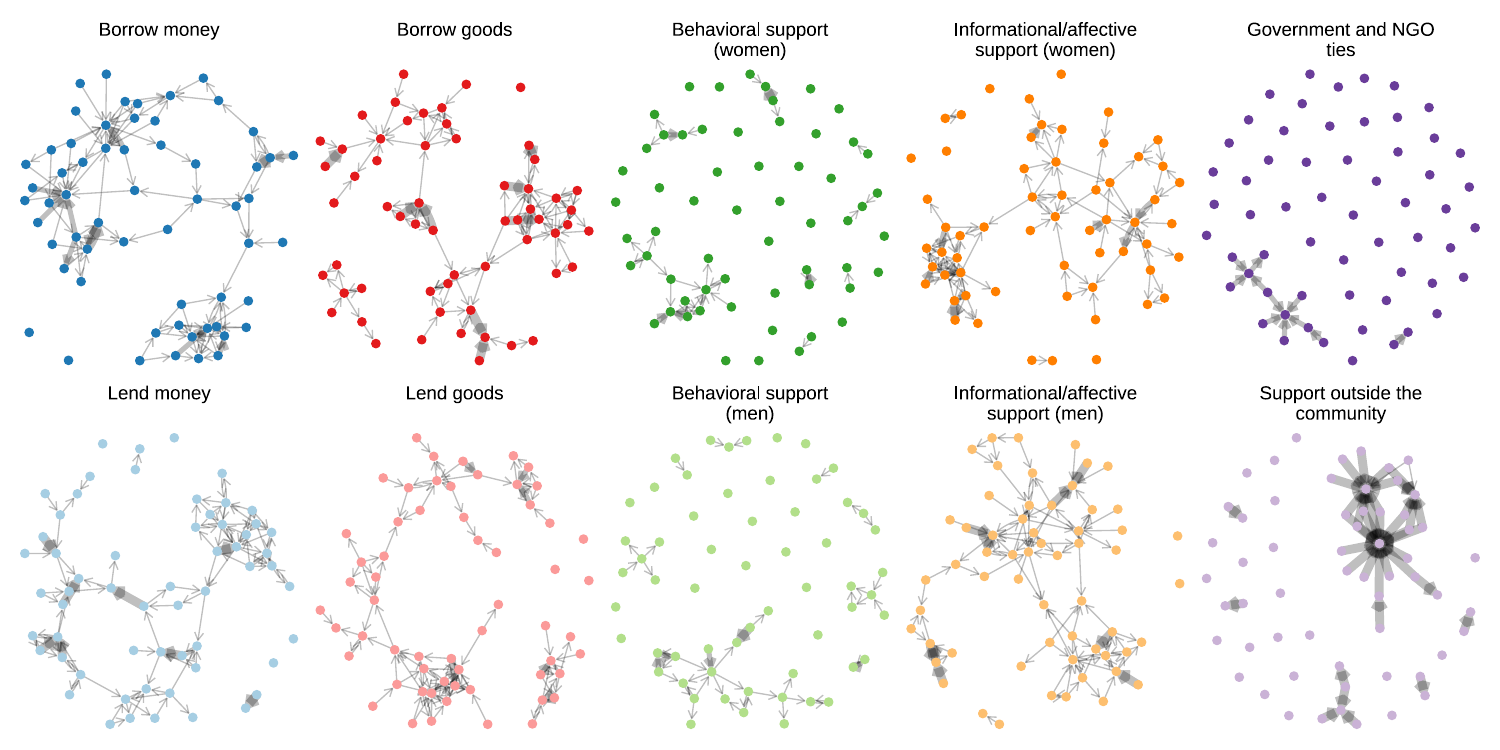}
\end{center}
\caption{Sharing unit-level networks for community PT}
\label{fig:networks_pt}
\end{figure}

\begin{figure}
\begin{center}
\includegraphics[width = \textwidth]{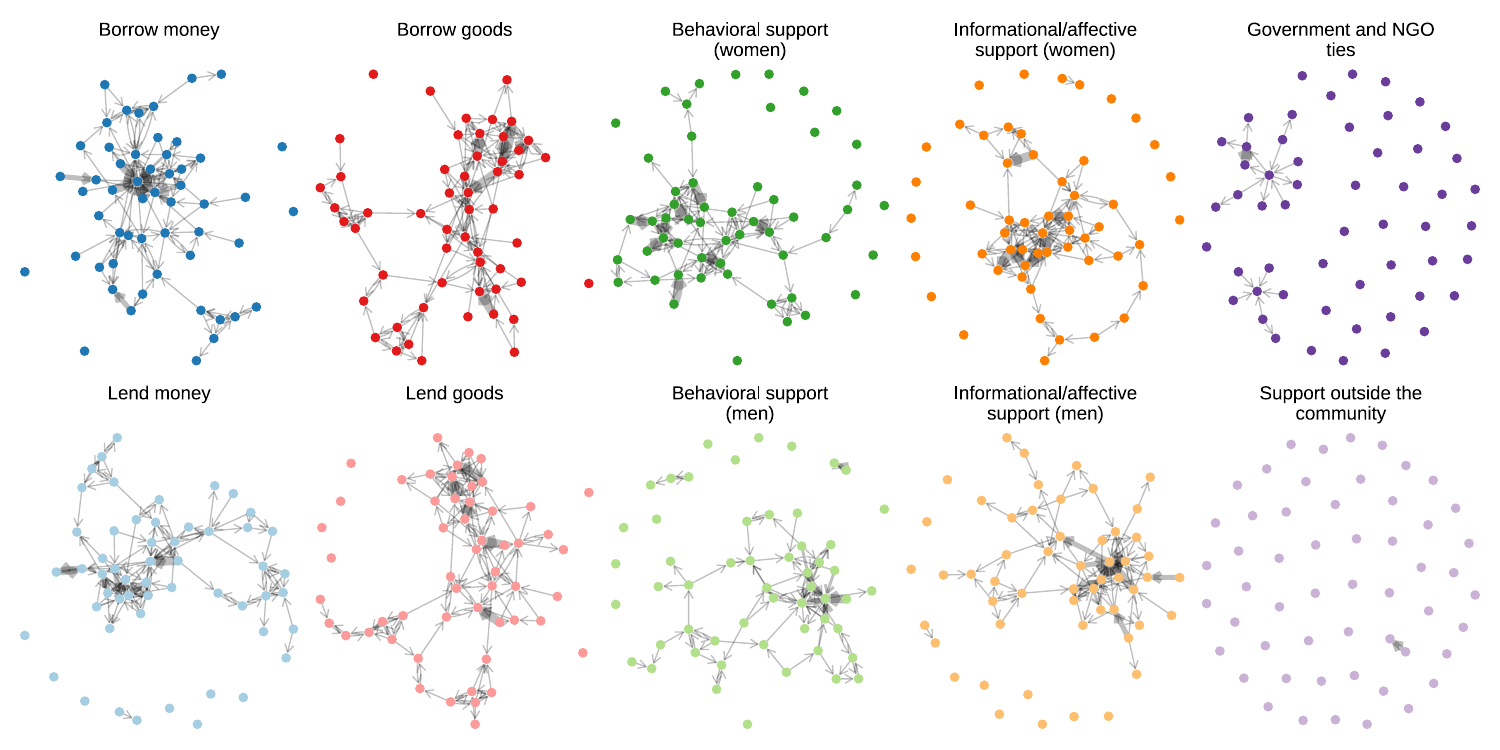}
\end{center}
\caption{Sharing unit-level networks for community RA}
\label{fig:networks_ra}
\end{figure}

\begin{figure}
\begin{center}
\includegraphics[width = \textwidth]{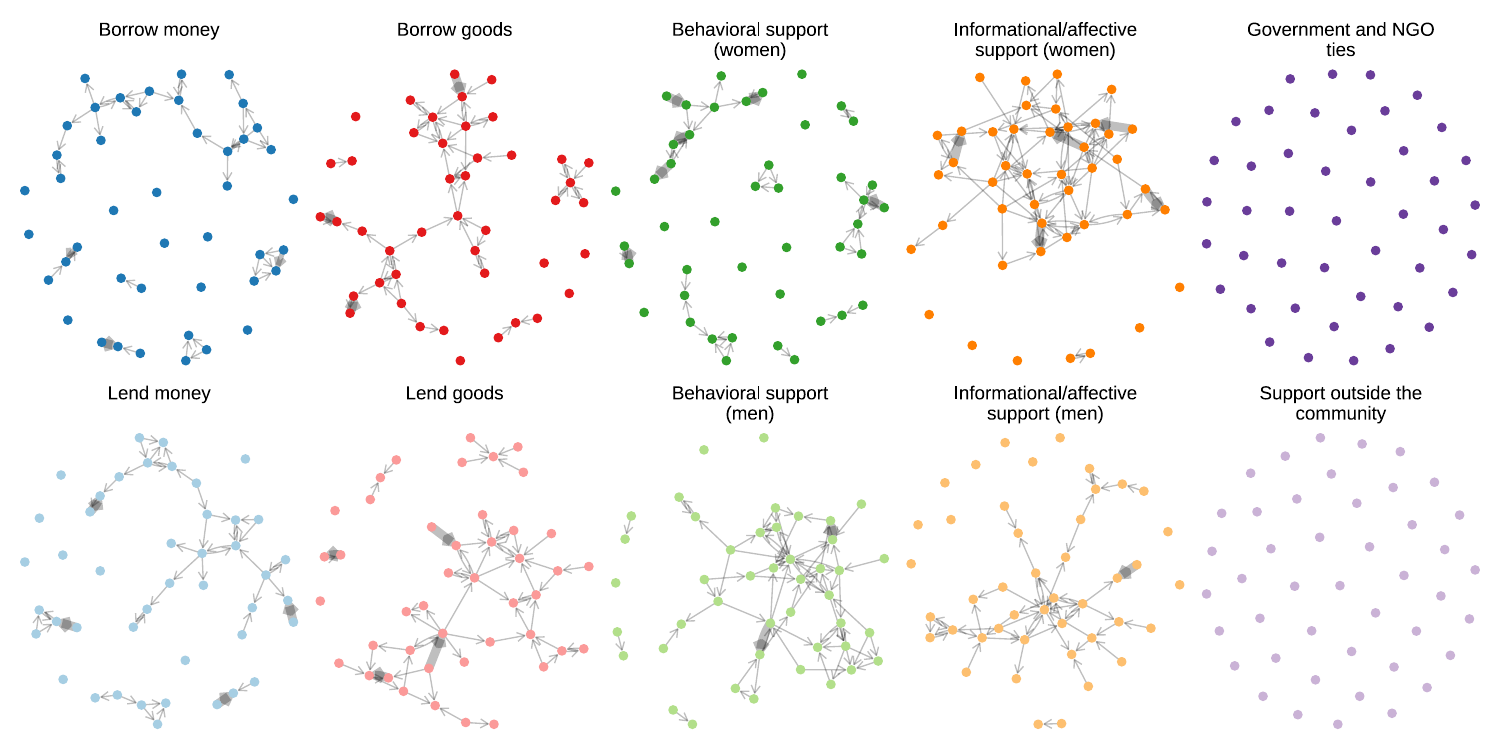}
\end{center}
\caption{Sharing unit-level networks for community SH}
\label{fig:networks_sh}
\end{figure}

\begin{figure}
\begin{center}
\includegraphics[width = \textwidth]{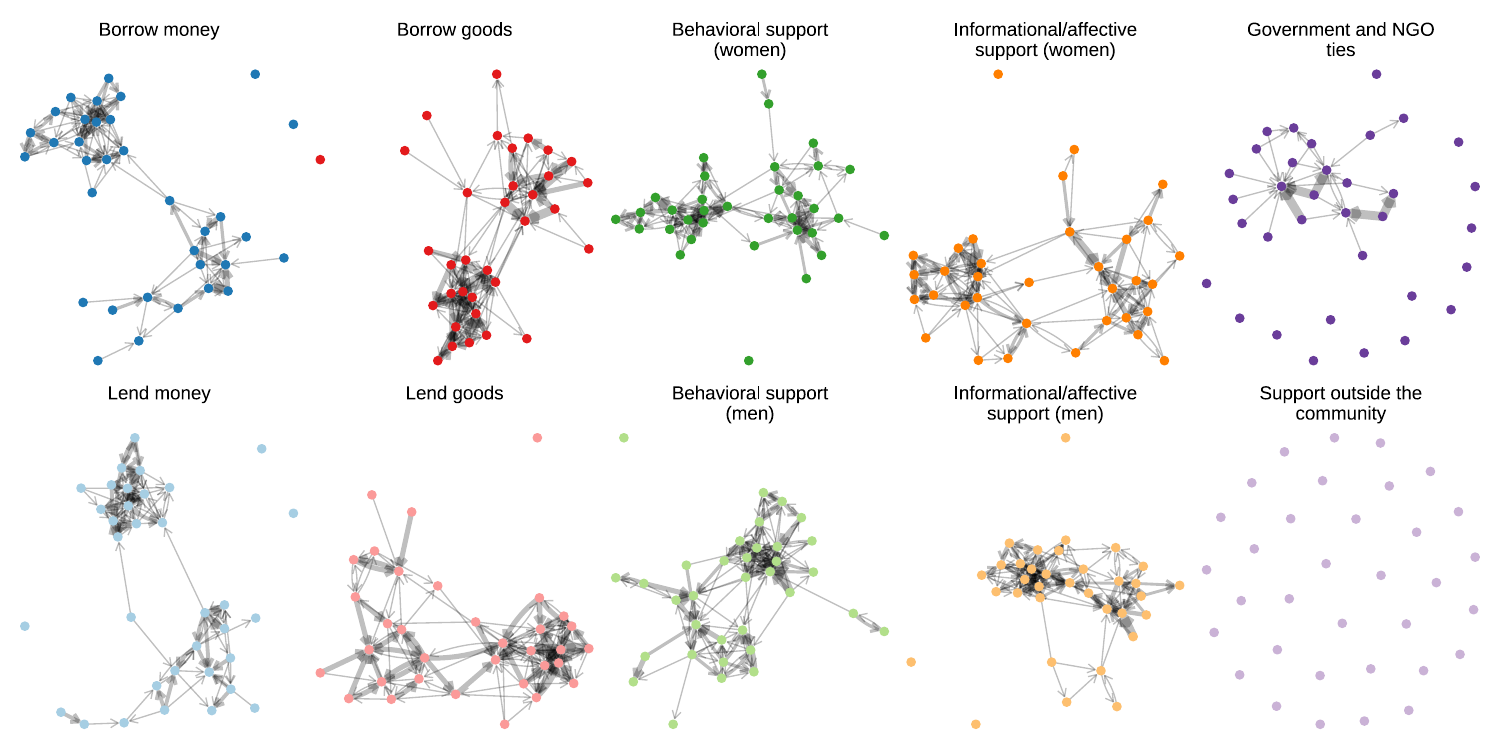}
\end{center}
\caption{Sharing unit-level networks for community SN}
\label{fig:networks_sn}
\end{figure}

\begin{figure}
\begin{center}
\includegraphics[width = \textwidth]{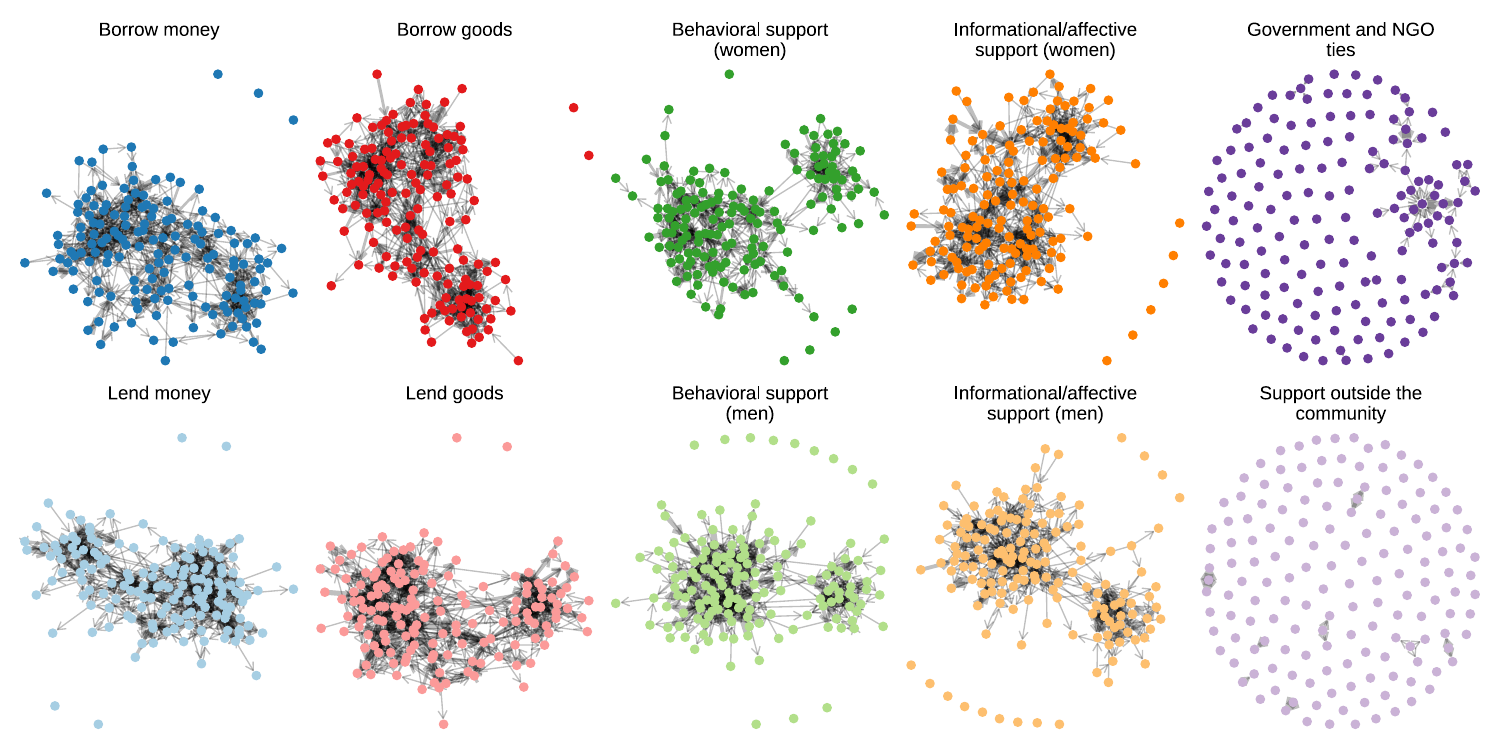}
\end{center}
\caption{Sharing unit-level networks for community TE}
\label{fig:networks_te}
\end{figure}

\begin{figure}
\begin{center}
\includegraphics[width = \textwidth]{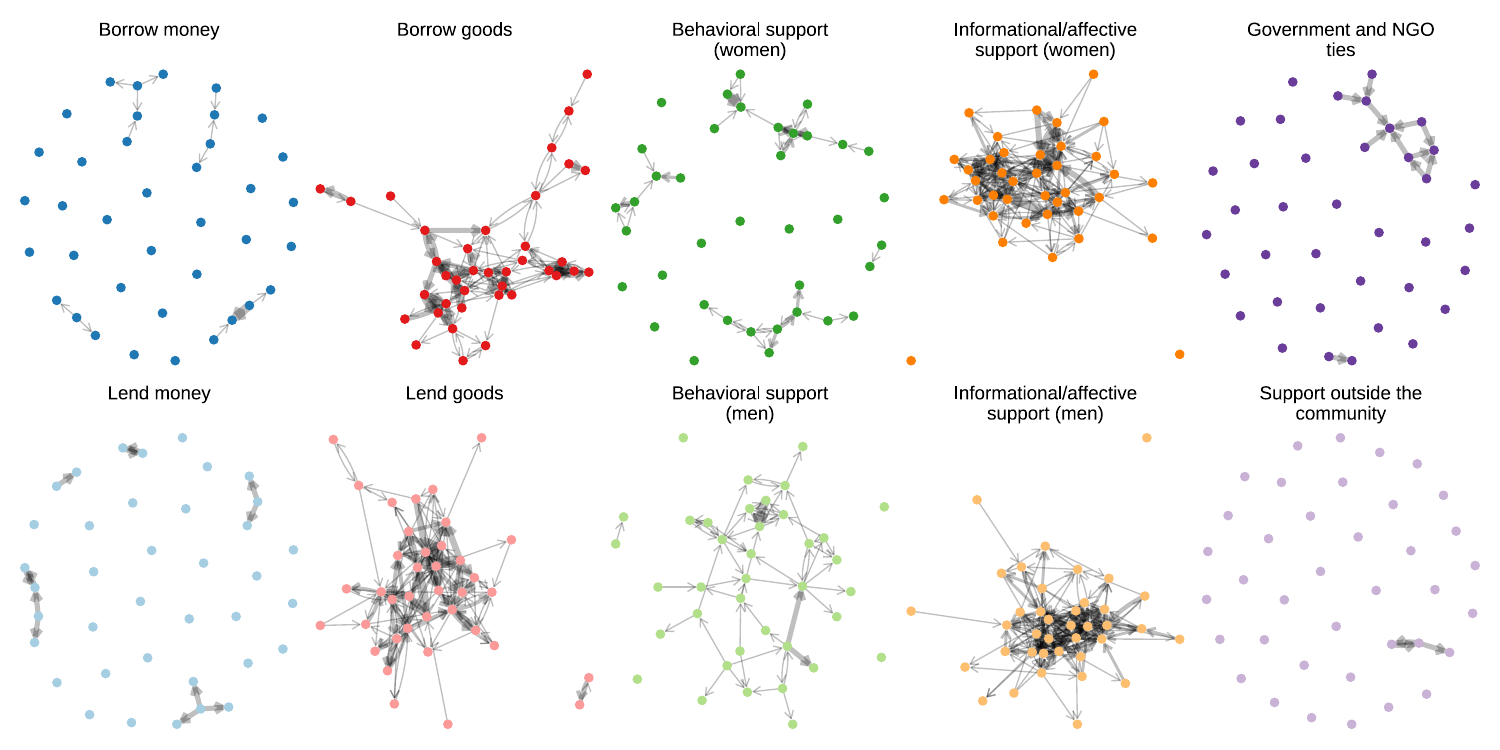}
\end{center}
\caption{Sharing unit-level networks for community TI}
\label{fig:networks_ti}
\end{figure}

\begin{figure}
\begin{center}
\includegraphics[width = \textwidth]{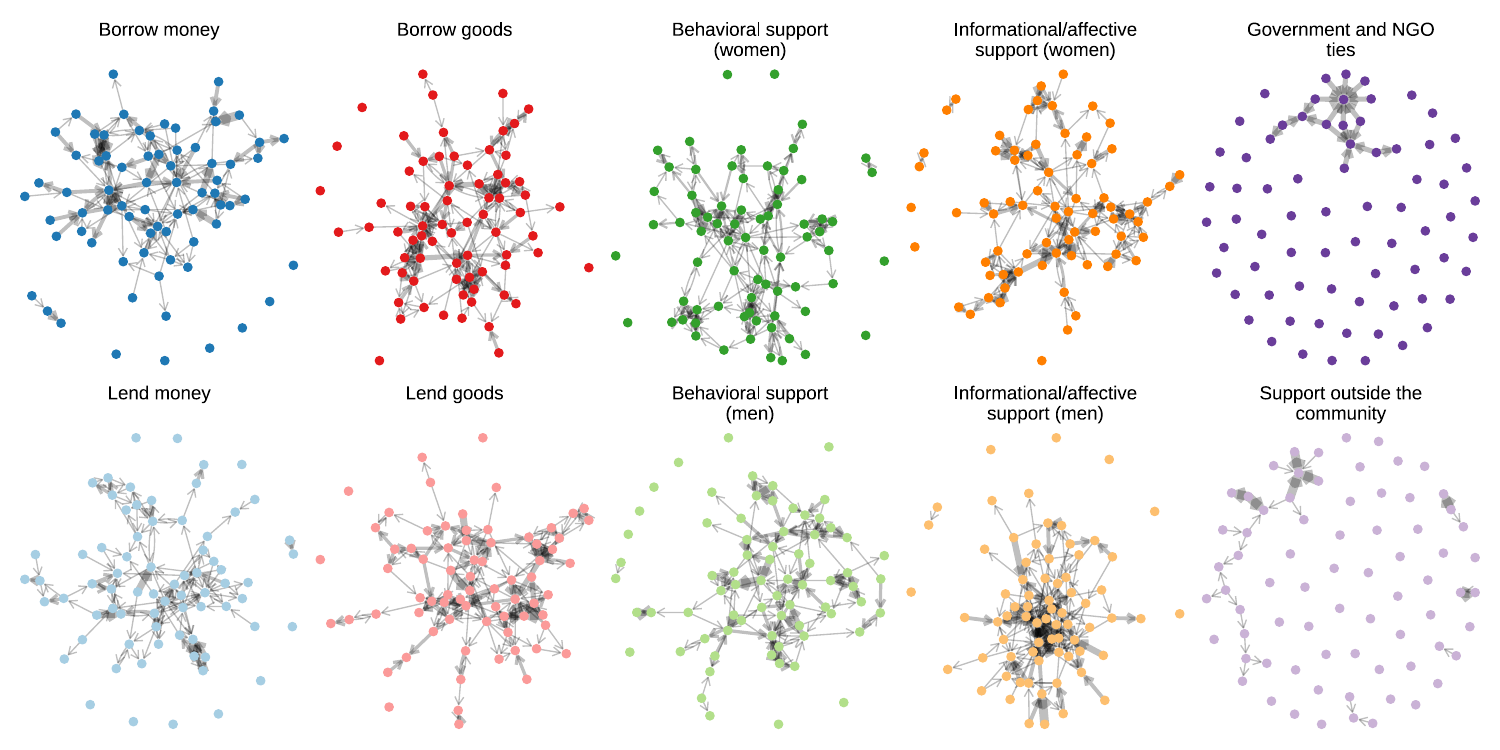}
\end{center}
\caption{Sharing unit-level networks for community TM}
\label{fig:networks_tm}
\end{figure}

\begin{figure}
\begin{center}
\includegraphics[width = \textwidth]{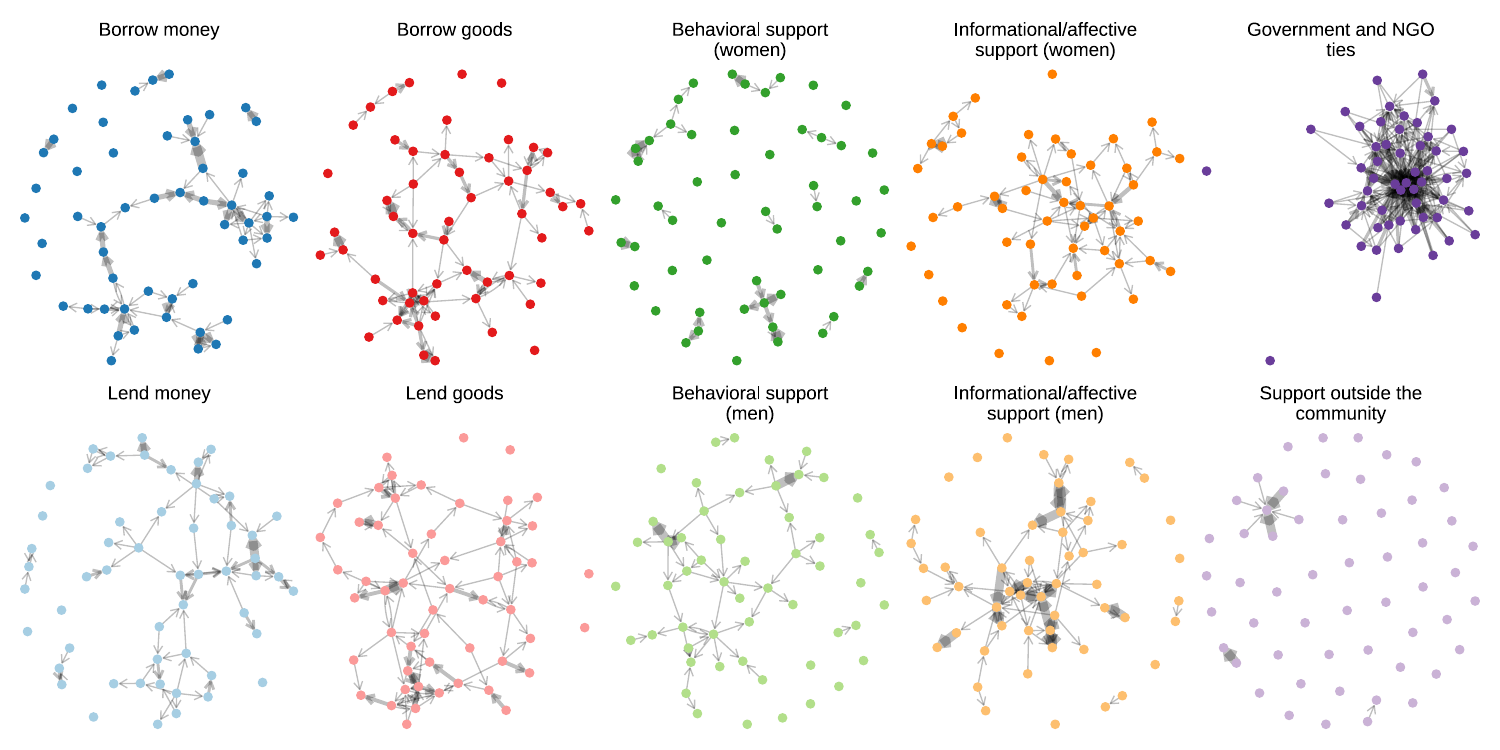}
\end{center}
\caption{Sharing unit-level networks for community TP}
\label{fig:networks_tp}
\end{figure}

\begin{figure}
\begin{center}
\includegraphics[width = \textwidth]{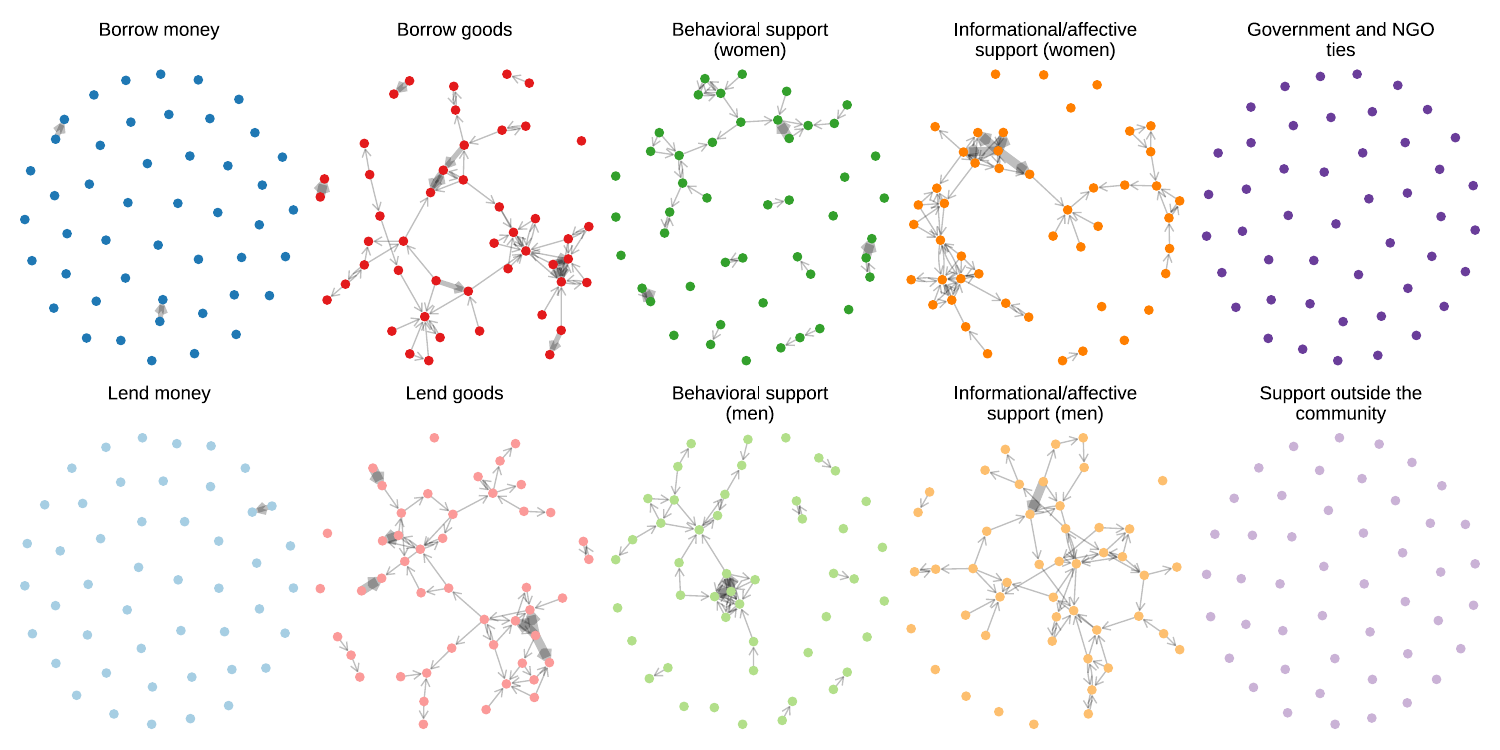}
\end{center}
\caption{Sharing unit-level networks for community TS}
\label{fig:networks_ts}
\end{figure}

\begin{figure}
\begin{center}
\includegraphics[width = \textwidth]{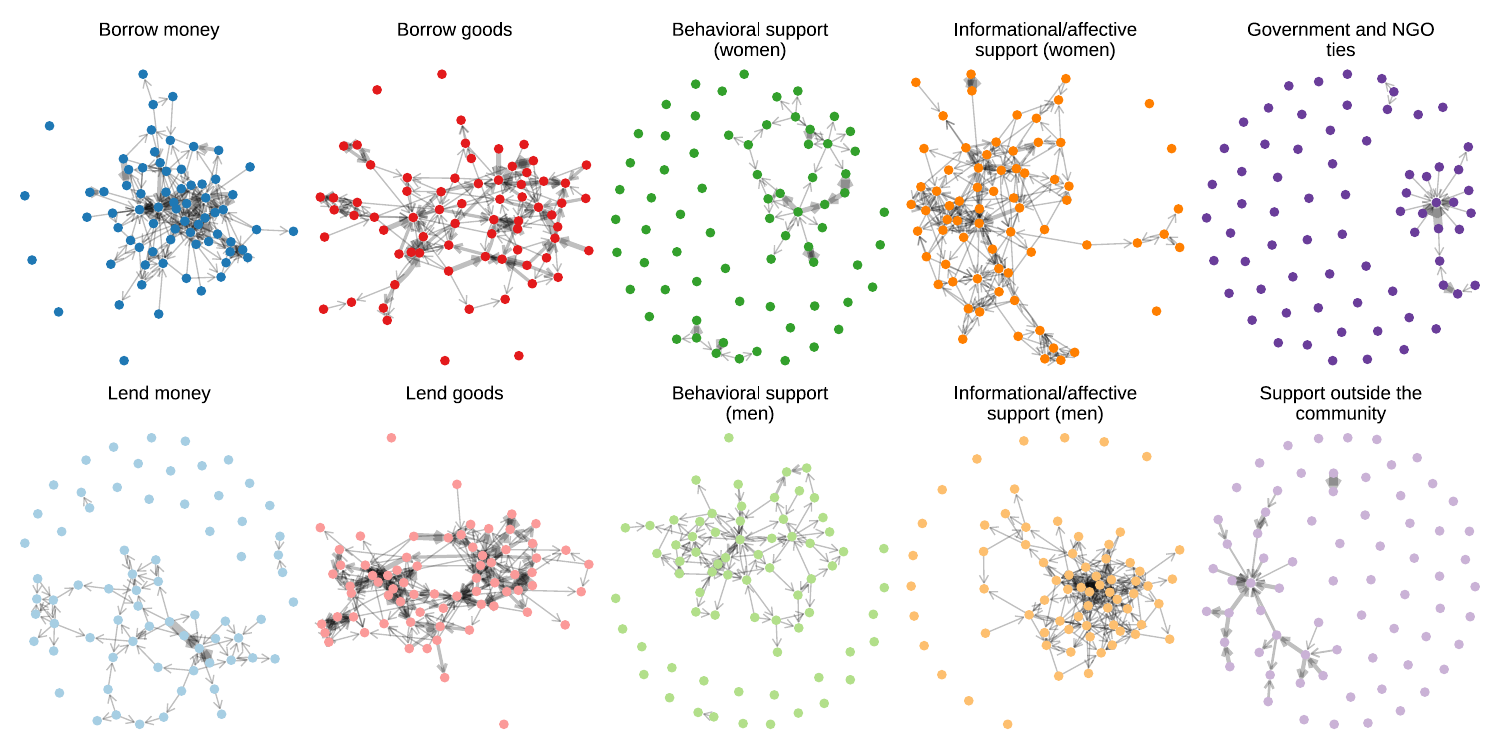}
\end{center}
\caption{Sharing unit-level networks for community TZ}
\label{fig:networks_tz}
\end{figure}

\begin{figure}
\begin{center}
\includegraphics[width = \textwidth]{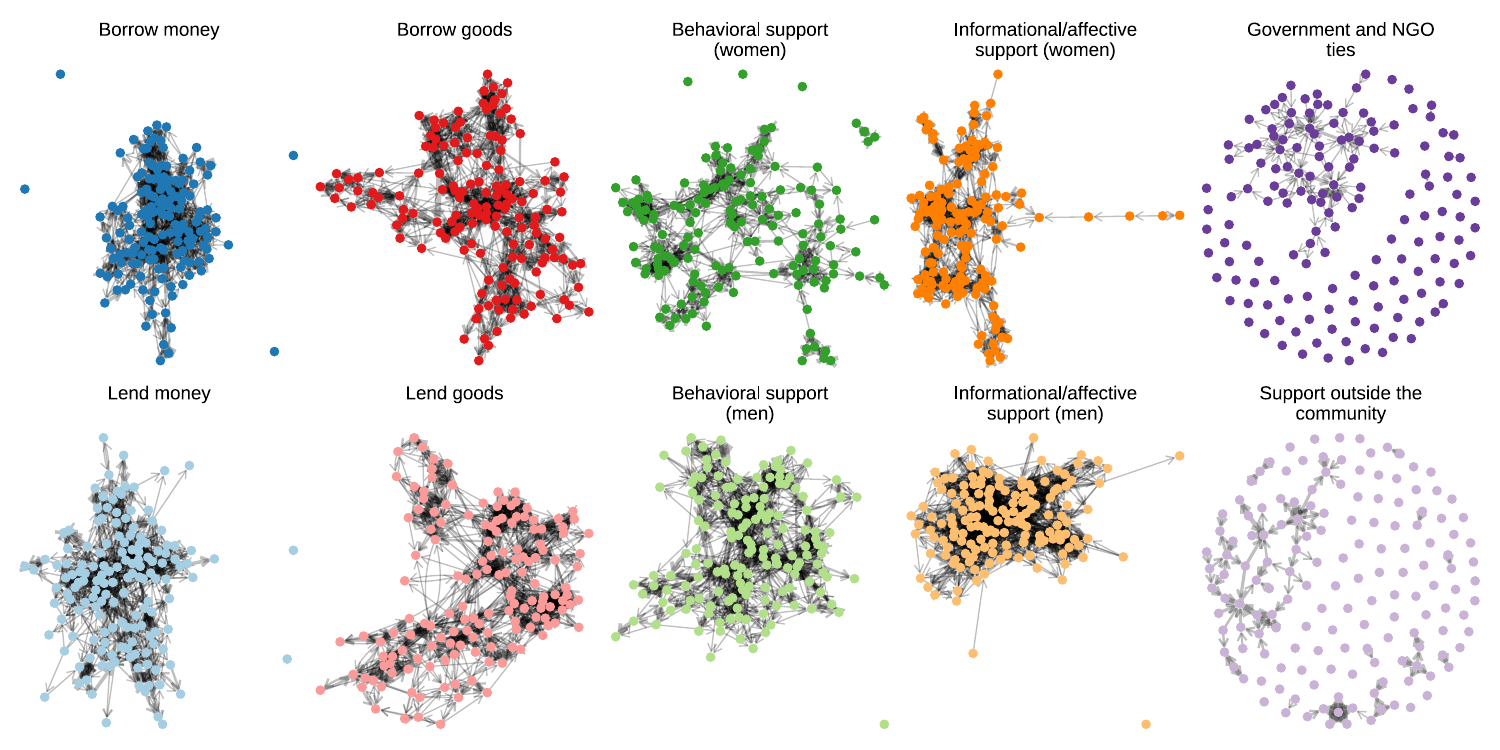}
\end{center}
\caption{Sharing unit-level networks for community UP}
\label{fig:networks_up}
\end{figure}

\begin{figure}
\begin{center}
\includegraphics[width = \textwidth]{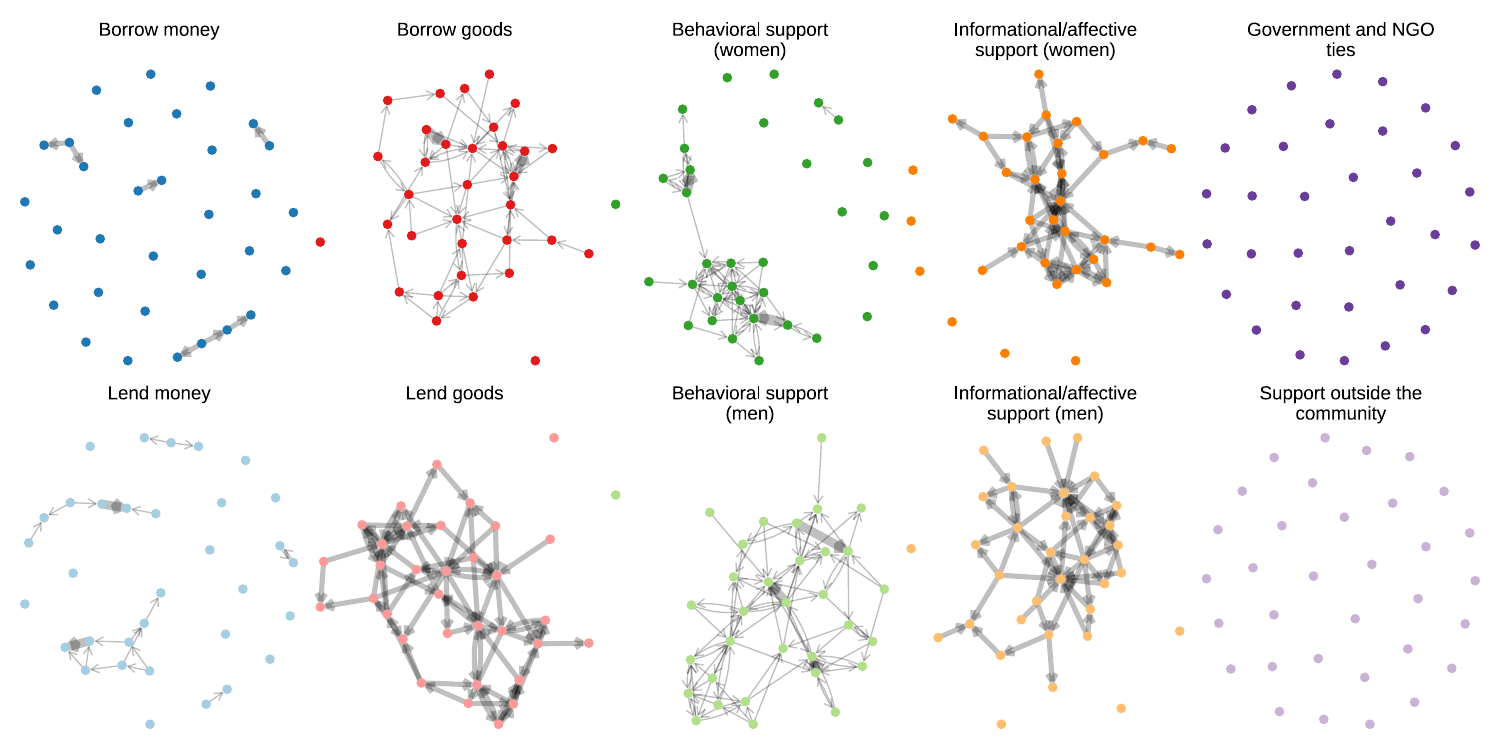}
\end{center}
\caption{Sharing unit-level networks for community VT}
\label{fig:networks_vt}
\end{figure}

\begin{figure}
\begin{center}
\includegraphics[width = \textwidth]{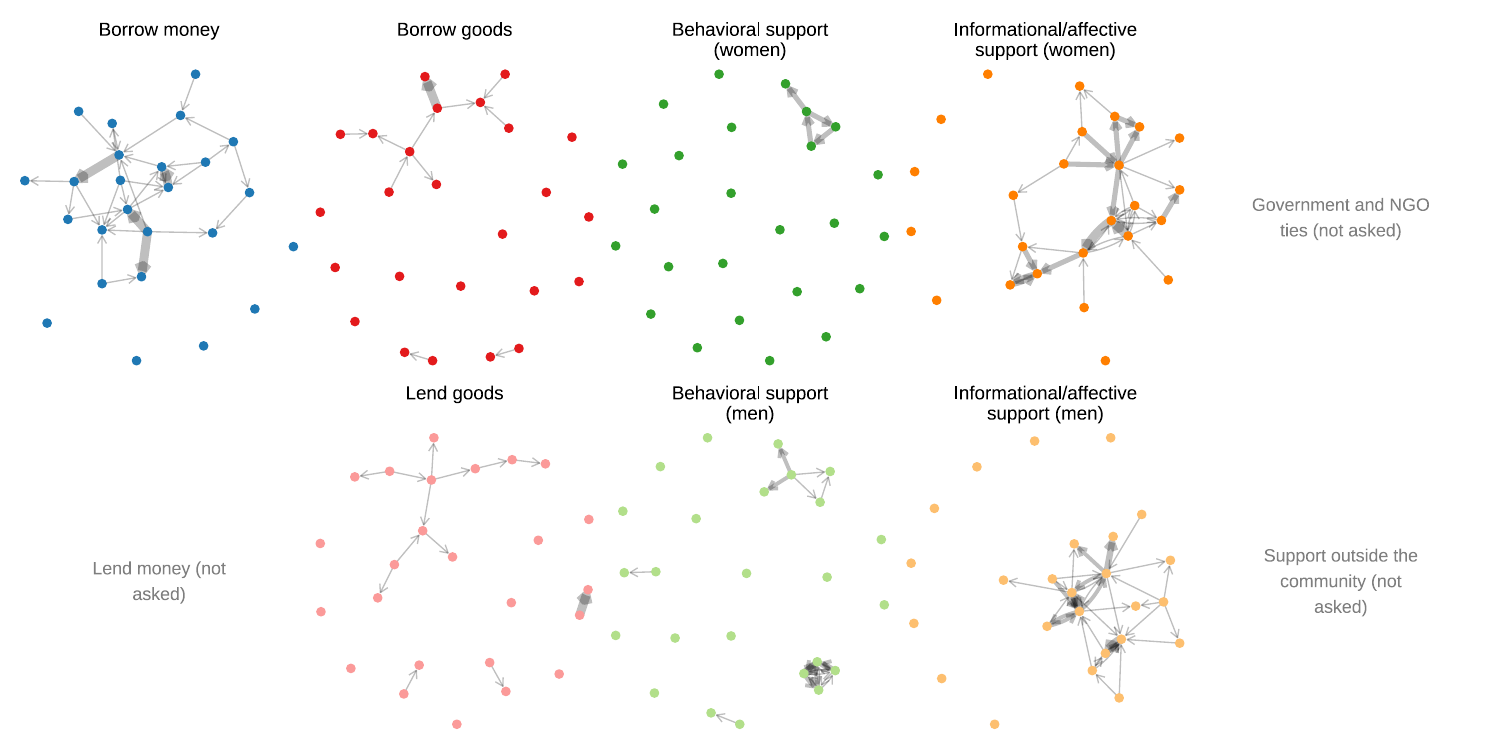}
\end{center}
\caption{Sharing unit-level networks for community WH}
\label{fig:networks_wh}
\end{figure}

\begin{figure}
\begin{center}
\includegraphics[width = \textwidth]{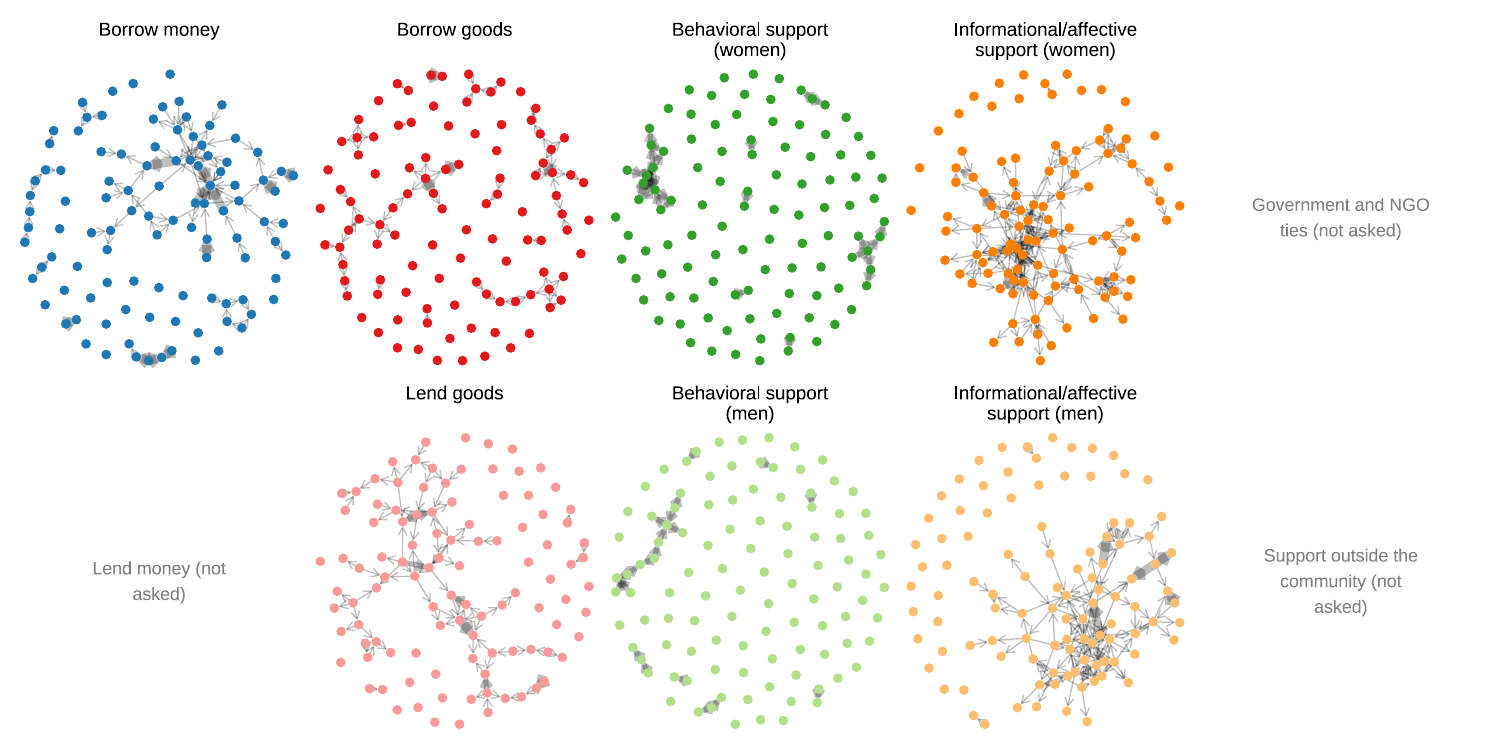}
\end{center}
\caption{Sharing unit-level networks for community WL}
\label{fig:networks_wl}
\end{figure}

\begin{figure}
\begin{center}
\includegraphics[width = \textwidth]{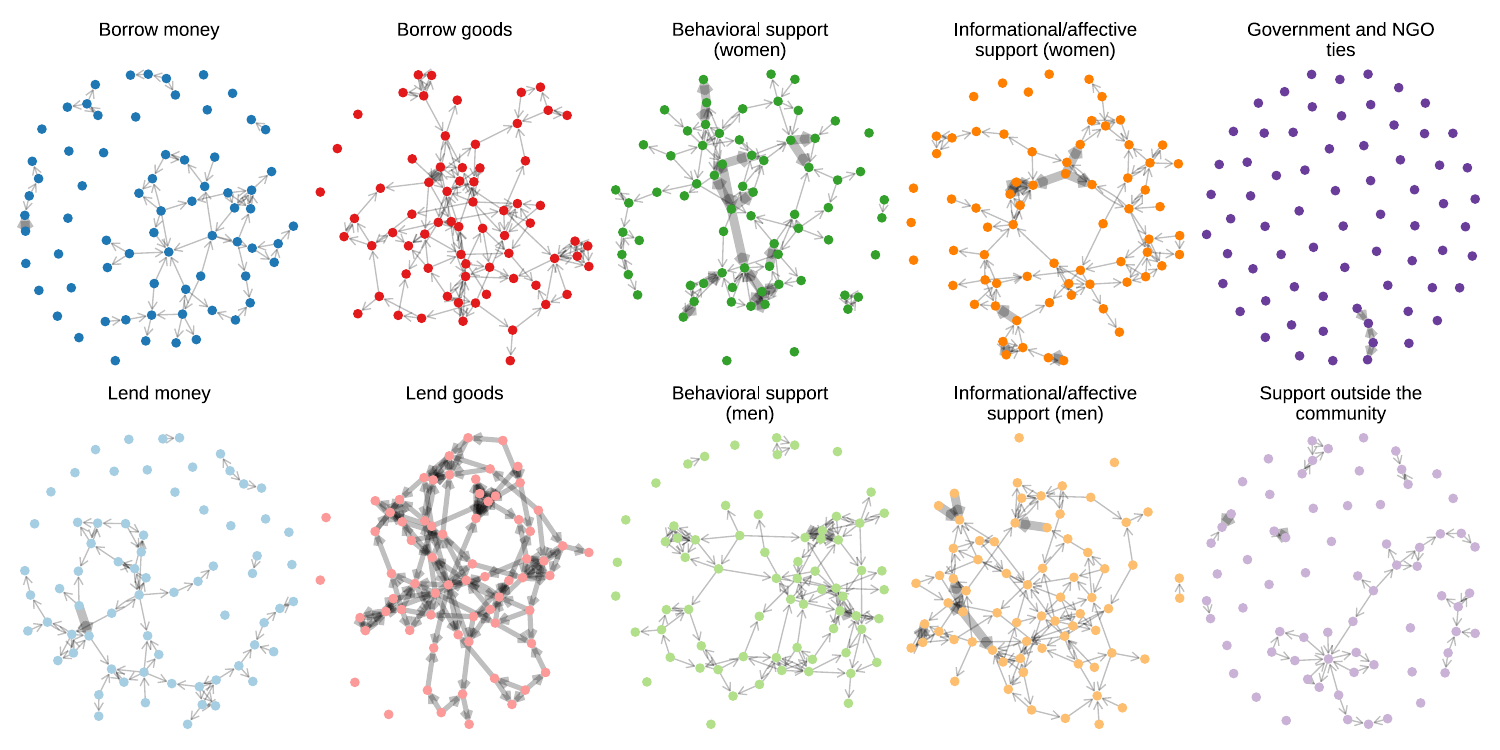}
\end{center}
\caption{Sharing unit-level networks for community YI}
\label{fig:networks_yi}
\end{figure}

\begin{figure}
\begin{center}
\includegraphics[width = \textwidth]{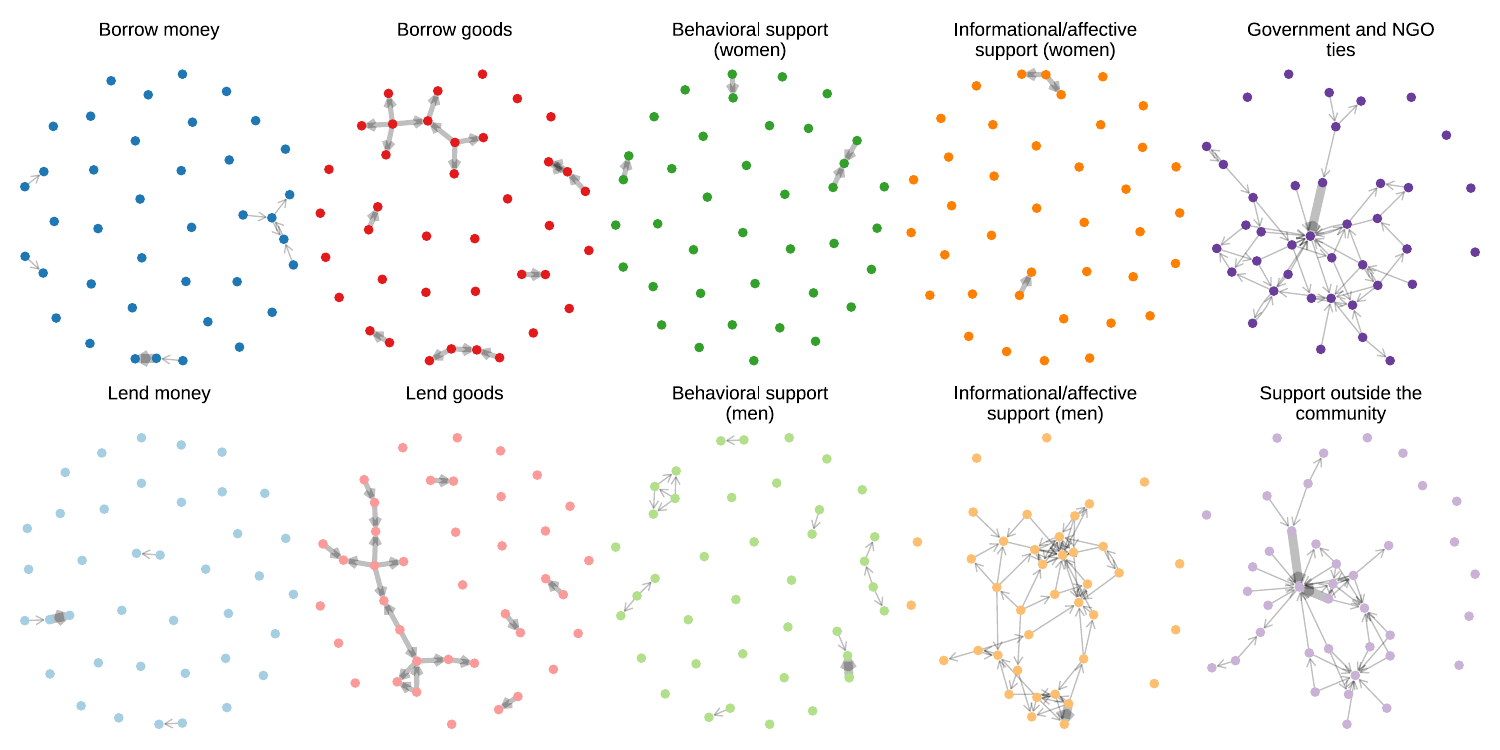}
\end{center}
\caption{Sharing unit-level networks for community YN}
\label{fig:networks_yn}
\end{figure}

In Figures~\ref{fig:networks_ah}-\ref{fig:networks_yn} we show the sharing unit-level social support networks derived from each of the standard network prompts for each community. In some cases, some prompts were not asked, as is noted in each figure.

\clearpage
\subsubsection{Aggregate Network}\label{si:network_plots_agg}

\begin{figure}
\begin{center}
\includegraphics[page = 1, width = \textwidth]{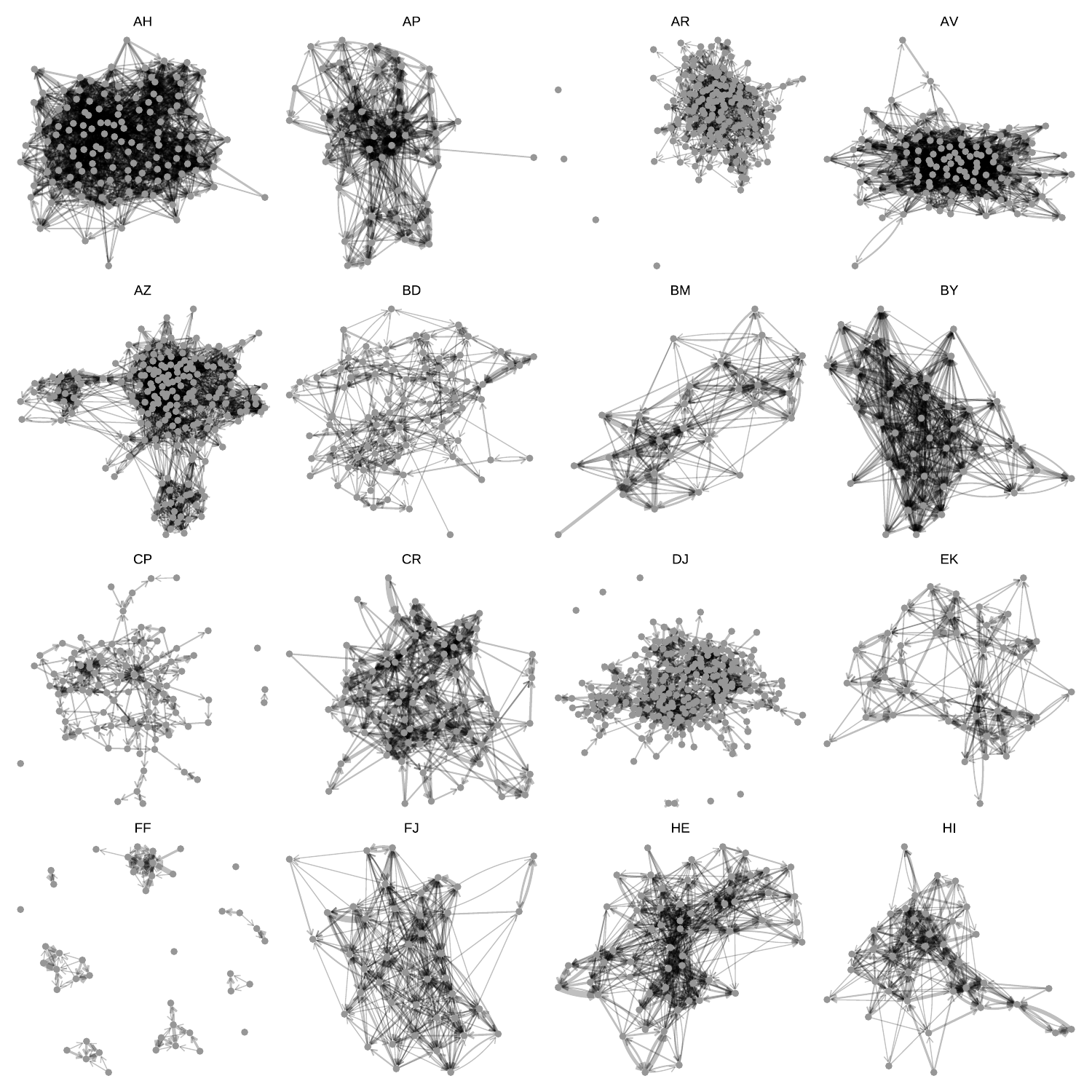}
\end{center}
\caption{Aggregate network for each community}
\label{fig:networks_agg1}
\end{figure}

\begin{figure}
\begin{center}
\includegraphics[page = 2, width = \textwidth]{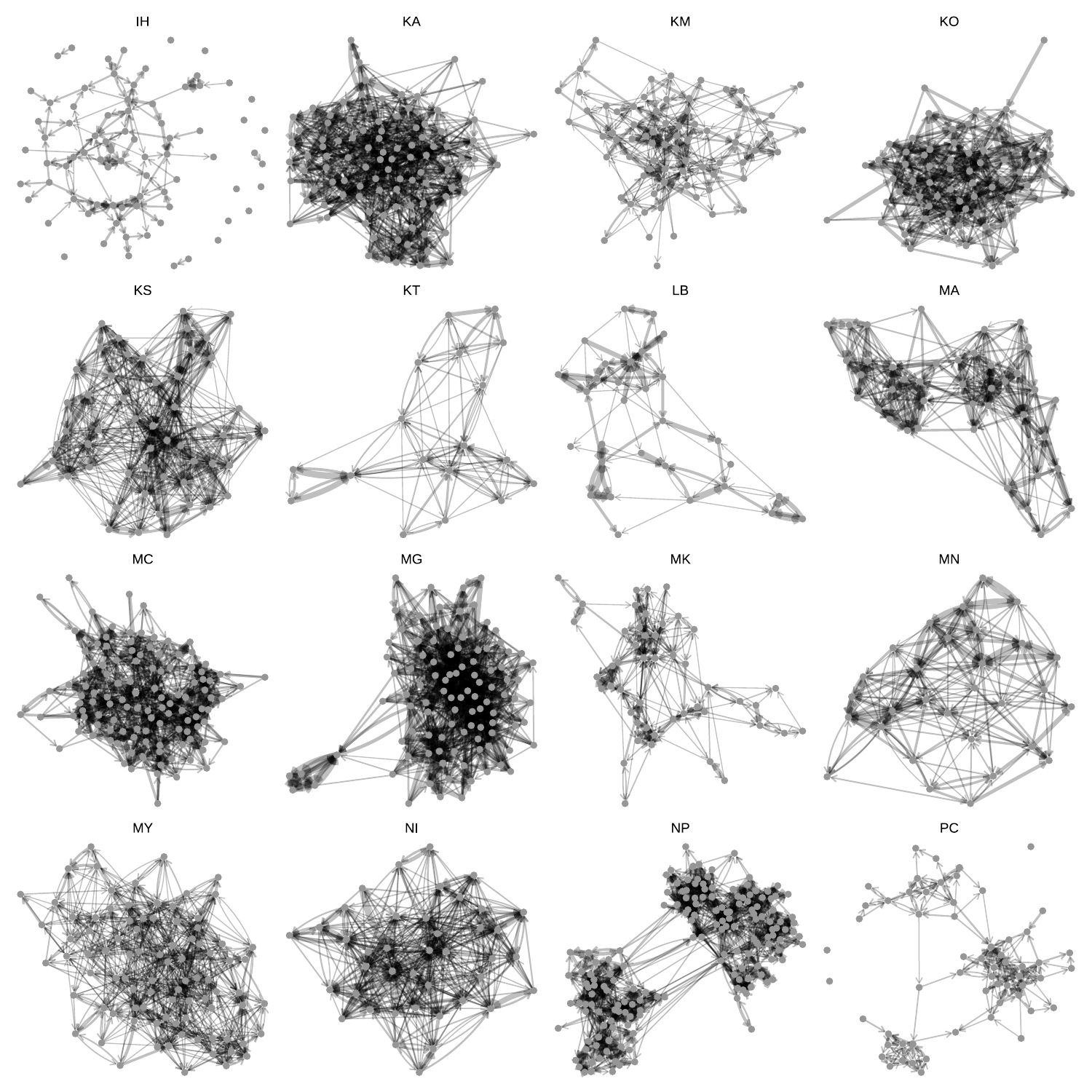}
\end{center}
\caption{Aggregate network for each community}
\label{fig:networks_agg2}
\end{figure}

\begin{figure}
\begin{center}
\includegraphics[page = 3, width = \textwidth]{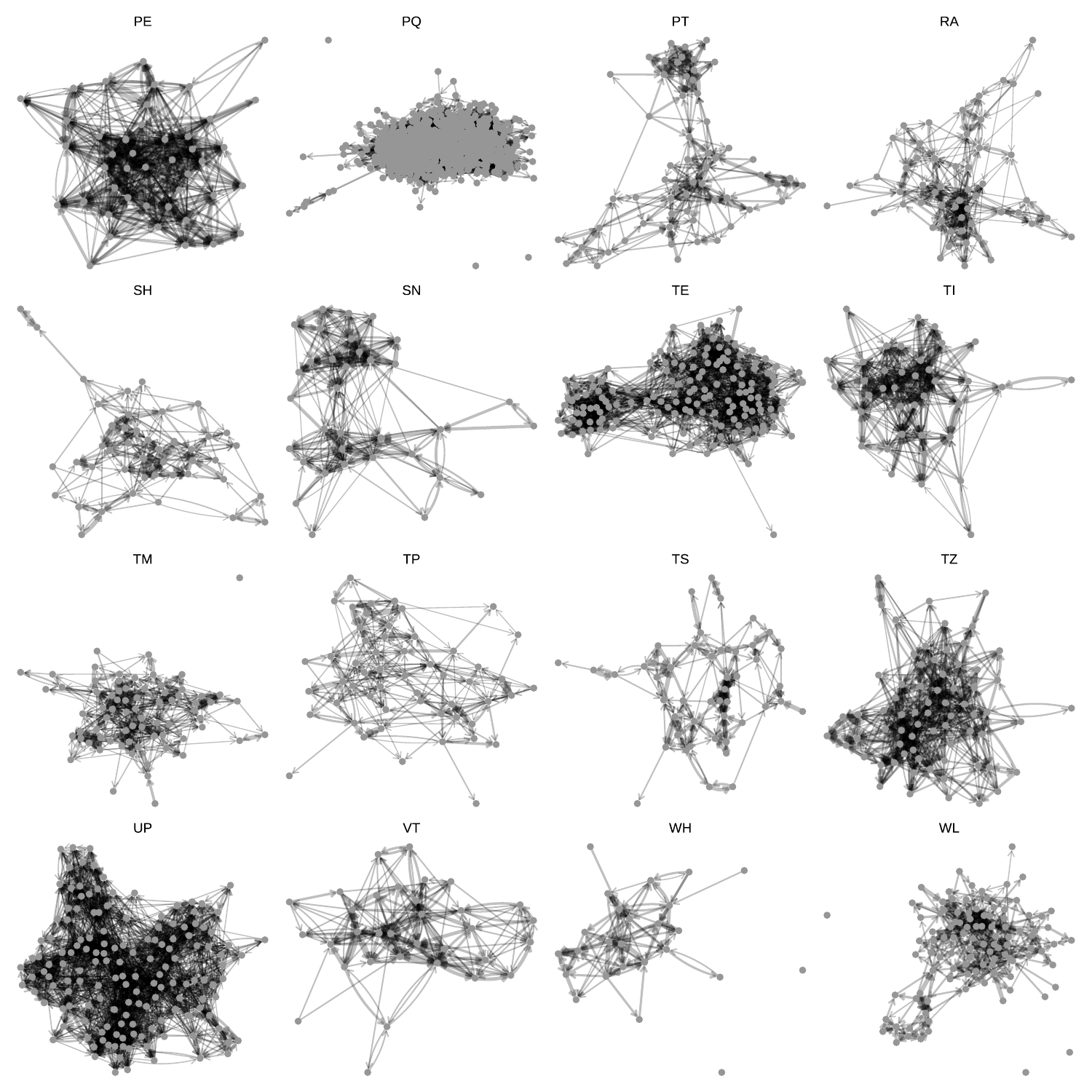}
\end{center}
\caption{Aggregate network for each community}
\label{fig:networks_agg3}
\end{figure}

\begin{figure}
\begin{center}
\includegraphics[page = 4, width = \textwidth]{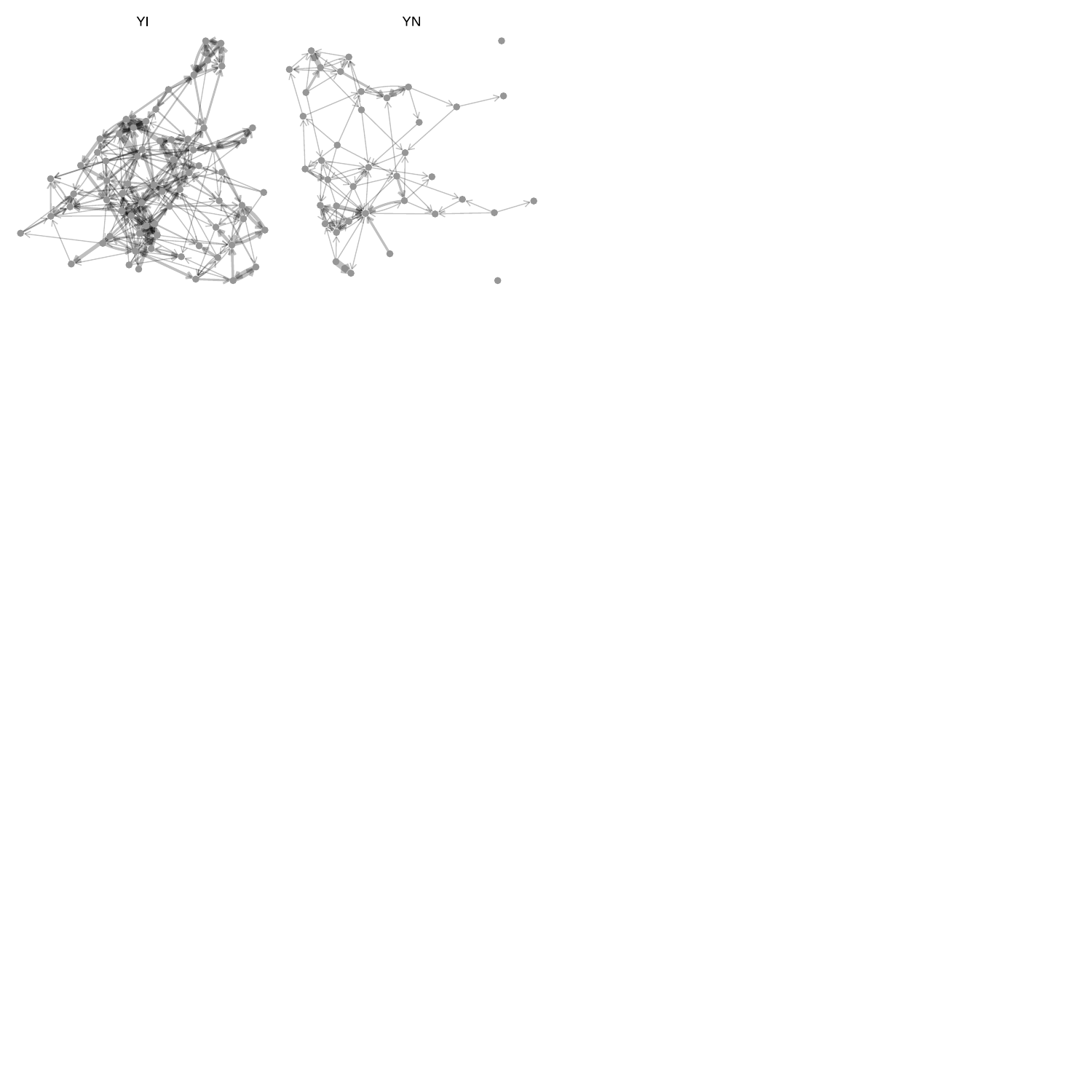}
\end{center}
\caption{Aggregate network for each community}
\label{fig:networks_agg4}
\end{figure}

In Figures~\ref{fig:networks_agg1}-\ref{fig:networks_agg4} we show the aggregate network for each community.

\clearpage
\subsection{Kinship}\label{si:kin}

\begin{figure}
\begin{center}
\includegraphics[page = 1, width = \textwidth]{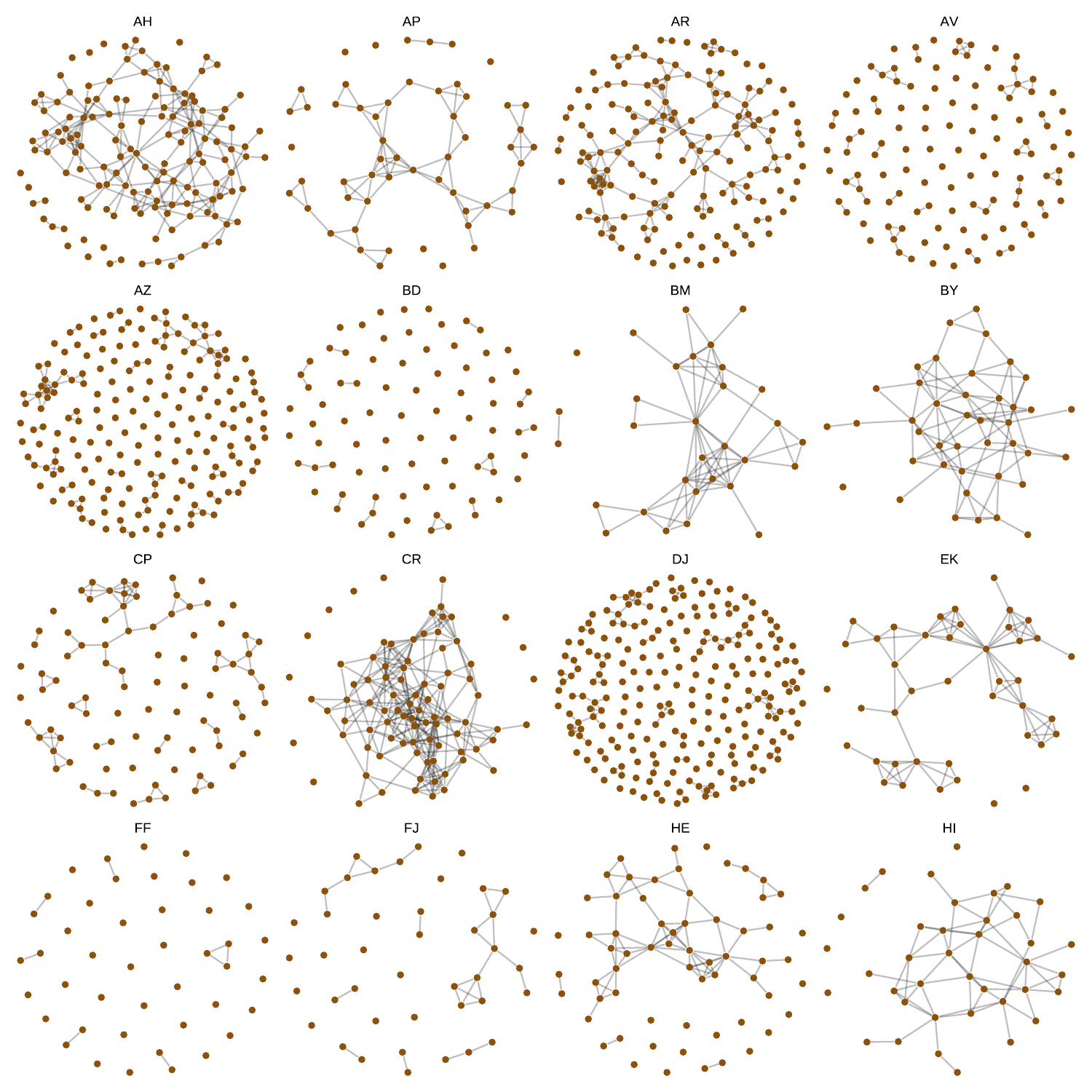}
\end{center}
\caption{Close kin (i.e., parent-child or full sibling) relationships between sharing units in each community.}
\label{fig:networks_kin1}
\end{figure}

\begin{figure}
\begin{center}
\includegraphics[page = 2, width = \textwidth]{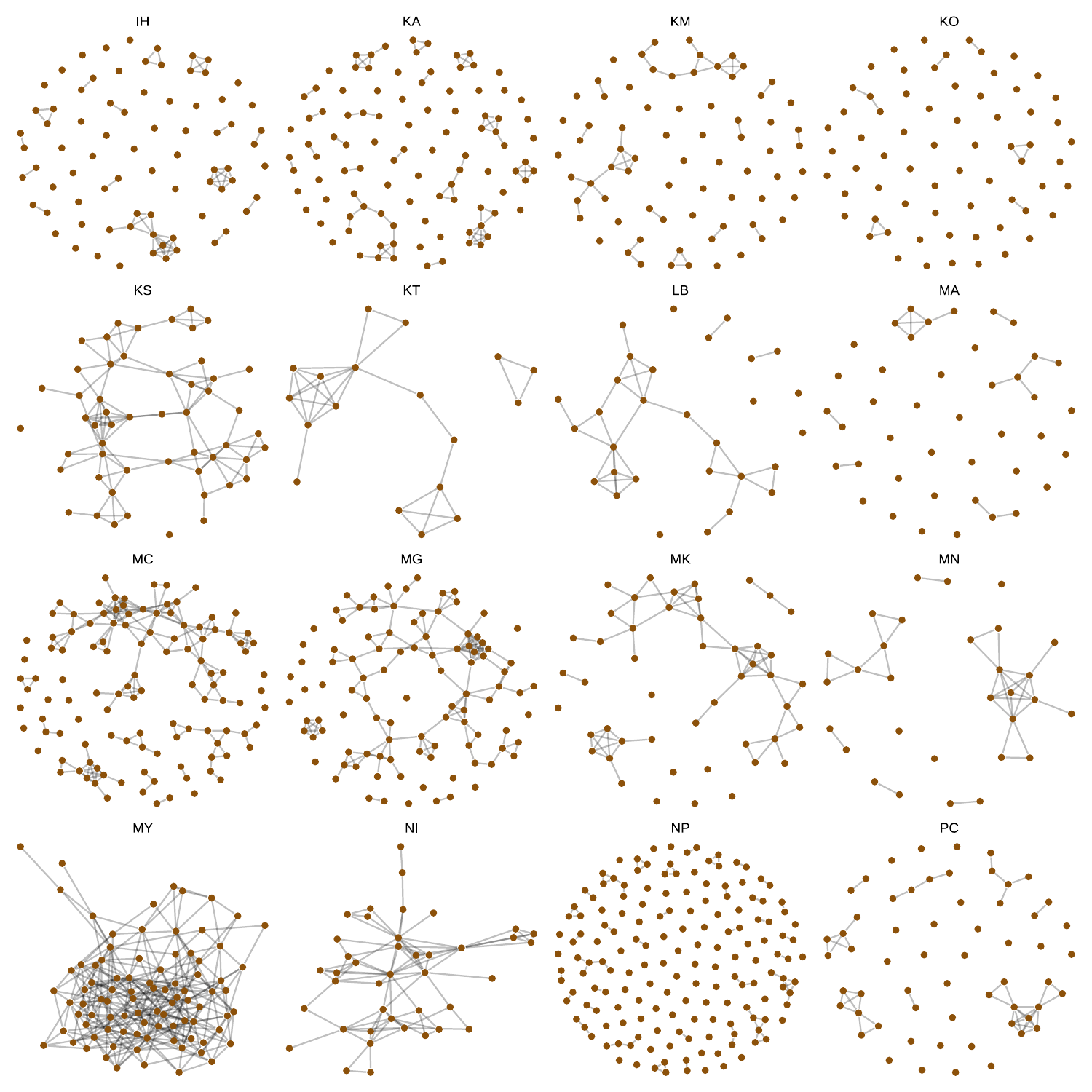}
\end{center}
\caption{Close kin (i.e., parent-child or full sibling) relationships between sharing units in each community.}
\label{fig:networks_kin2}
\end{figure}

\begin{figure}
\begin{center}
\includegraphics[page = 3, width = \textwidth]{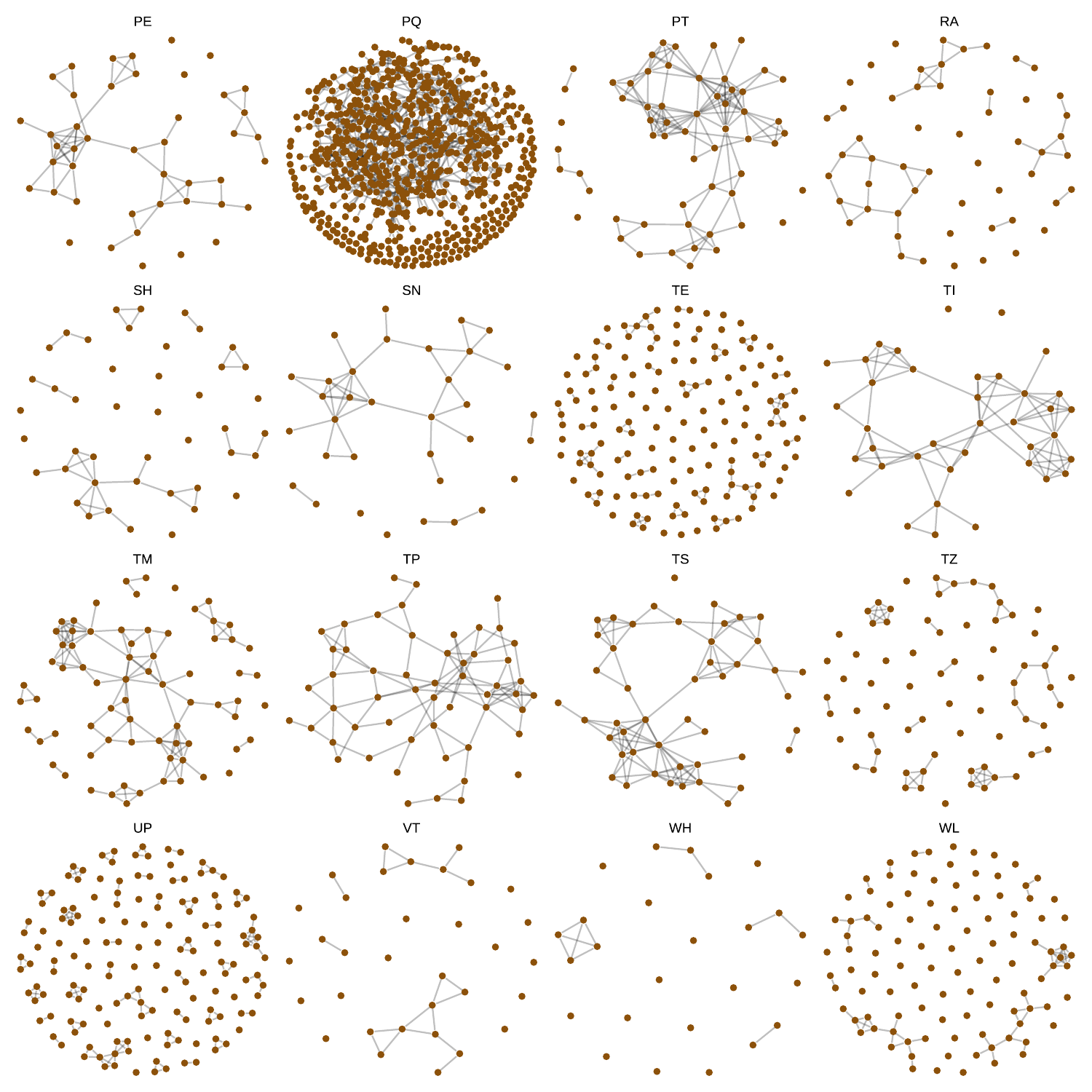}
\end{center}
\caption{Close kin (i.e., parent-child or full sibling) relationships between sharing units in each community.}
\label{fig:networks_kin3}
\end{figure}

\begin{figure}
\begin{center}
\includegraphics[page = 4, width = \textwidth]{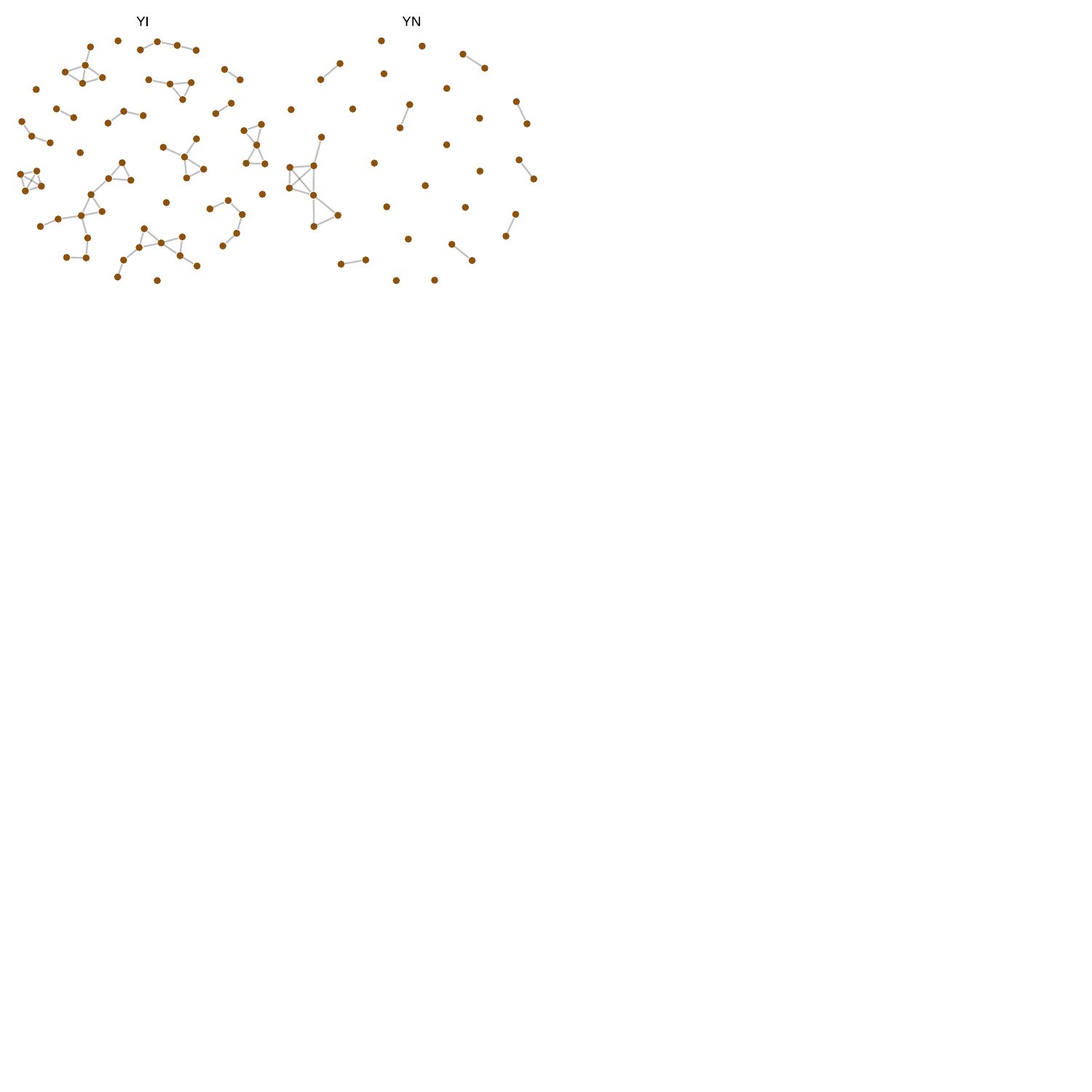}
\end{center}
\caption{Close kin (i.e., parent-child or full sibling) relationships between sharing units in each community.}
\label{fig:networks_kin4}
\end{figure}

The close kin networks, plotted in Figures~\ref{fig:networks_kin1} to \ref{fig:networks_kin4}, with summary statistics shown in Table~\ref{tab:net-summary-stats-primary_kin}, are derived not from network prompts, but from the records of relatedness reported by sharing units. While we attempted to record the spouses, parents and grandparents of all community residents (allowing us to calculate relatedness to at least cousins), our attention here stays focused on close, immediate kin, i.e., parent-child and full sibling relationships. Self-loops, which would show that there are close kin relations in virtually all sharing units, are not included.

\begin{landscape}
\begingroup\fontsize{8}{10}\selectfont

\begin{longtable}[t]{lrrrrrrrrrrrrrr}
\caption{\label{tab:net-summary-stats-primary_kin}Network summary statistics by community for the close kin network. `Mean Deg.' shows the mean degree. `Deg. Centr.' shows the degree centralization. `Deg. Assort.' shows the degree assortativity. `APL' shows the average path length. `Diam.' shows the diameter. `N Isol.' shows the number of isolates, and `Prop. Isol.' their proportion of all nodes. `N Comp.' shows the number of weakly-connected components. `Largest Comp.' shows the proportion of nodes in the largest component. `Global Clust.' shows the global clustering coefficient. `Av. Local Clust.' shows the average local clustering coefficient.}\\
\toprule
Site & Nodes & Edges & Density & \shortstack{Mean\\ Deg.} & \shortstack{Deg.\\ Centr.} & \shortstack{Deg.\\ Assort.} & APL & Diam. & N Isol. & \shortstack{Prop.\\ Isol.} & N Comp. & \shortstack{Largest\\ Comp.} & \shortstack{Global\\ Clust.} & \shortstack{Av. Loc.\\ Clust.}\\
\midrule
\endfirsthead
\caption[]{Network summary statistics by community for the close kin network. `Mean Deg.' shows the mean degree. `Deg. Centr.' shows the degree centralization. `Deg. Assort.' shows the degree assortativity. `APL' shows the average path length. `Diam.' shows the diameter. `N Isol.' shows the number of isolates, and `Prop. Isol.' their proportion of all nodes. `N Comp.' shows the number of weakly-connected components. `Largest Comp.' shows the proportion of nodes in the largest component. `Global Clust.' shows the global clustering coefficient. `Av. Local Clust.' shows the average local clustering coefficient. \textit{(continued)}}\\
\toprule
Site & Nodes & Edges & Density & \shortstack{Mean\\ Deg.} & \shortstack{Deg.\\ Centr.} & \shortstack{Deg.\\ Assort.} & APL & Diam. & N Isol. & \shortstack{Prop.\\ Isol.} & N Comp. & \shortstack{Largest\\ Comp.} & \shortstack{Global\\ Clust.} & \shortstack{Av. Loc.\\ Clust.}\\
\midrule
\endhead

\endfoot
\bottomrule
\endlastfoot
AH & 137 & 257 & 0.03 & 3.75 & 0.05 & 0.22 & 4.77 & 12 & 10 & 0.07 & 14 & 0.88 & 0.49 & 0.58\\
AP & 55 & 84 & 0.06 & 3.05 & 0.09 & 0.39 & 4.65 & 12 & 6 & 0.11 & 9 & 0.78 & 0.57 & 0.71\\
AR & 158 & 214 & 0.02 & 2.71 & 0.05 & 0.42 & 5.68 & 13 & 25 & 0.16 & 34 & 0.70 & 0.54 & 0.54\\
AV & 116 & 52 & 0.01 & 0.90 & 0.04 & 0.31 & 1.44 & 3 & 52 & 0.45 & 74 & 0.06 & 0.61 & 0.71\\
AZ & 190 & 126 & 0.01 & 1.33 & 0.04 & 0.59 & 2.51 & 6 & 71 & 0.37 & 102 & 0.11 & 0.61 & 0.63\\
BD & 79 & 19 & 0.01 & 0.48 & 0.02 & 0.37 & 1.14 & 2 & 50 & 0.63 & 62 & 0.04 & 0.67 & 0.67\\
BM & 31 & 74 & 0.16 & 4.77 & 0.31 & 0.15 & 2.37 & 5 & 1 & 0.03 & 3 & 0.90 & 0.62 & 0.74\\
BY & 41 & 88 & 0.11 & 4.29 & 0.12 & 0.02 & 2.77 & 6 & 1 & 0.02 & 2 & 0.98 & 0.32 & 0.42\\
CP & 90 & 80 & 0.02 & 1.78 & 0.07 & 0.52 & 3.38 & 8 & 21 & 0.23 & 35 & 0.27 & 0.62 & 0.66\\
CR & 87 & 277 & 0.07 & 6.37 & 0.10 & 0.13 & 2.77 & 6 & 10 & 0.11 & 11 & 0.89 & 0.49 & 0.55\\
DJ & 231 & 168 & 0.01 & 1.45 & 0.02 & 0.43 & 2.33 & 7 & 76 & 0.33 & 111 & 0.06 & 0.60 & 0.61\\
EK & 42 & 85 & 0.10 & 4.05 & 0.27 & -0.05 & 3.58 & 8 & 3 & 0.07 & 4 & 0.93 & 0.60 & 0.79\\
FF & 44 & 8 & 0.01 & 0.36 & 0.04 & 1.00 & 1.00 & 1 & 31 & 0.70 & 37 & 0.07 & 1.00 & 1.00\\
FJ & 42 & 28 & 0.03 & 1.33 & 0.07 & 0.50 & 2.27 & 5 & 13 & 0.31 & 20 & 0.26 & 0.51 & 0.54\\
HE & 63 & 85 & 0.04 & 2.70 & 0.12 & 0.35 & 3.20 & 7 & 15 & 0.24 & 20 & 0.59 & 0.54 & 0.60\\
HI & 39 & 64 & 0.09 & 3.28 & 0.12 & 0.00 & 2.84 & 6 & 4 & 0.10 & 6 & 0.85 & 0.27 & 0.30\\
IH & 77 & 54 & 0.02 & 1.40 & 0.09 & 0.68 & 1.36 & 3 & 32 & 0.42 & 47 & 0.13 & 0.87 & 0.96\\
KA & 99 & 71 & 0.01 & 1.43 & 0.05 & 0.51 & 2.02 & 7 & 34 & 0.34 & 52 & 0.11 & 0.77 & 0.76\\
KM & 72 & 39 & 0.02 & 1.08 & 0.06 & 0.47 & 2.42 & 6 & 29 & 0.40 & 41 & 0.15 & 0.53 & 0.53\\
KO & 67 & 11 & 0.00 & 0.33 & 0.03 & 0.61 & 1.08 & 2 & 52 & 0.78 & 58 & 0.04 & 0.86 & 0.86\\
KS & 53 & 108 & 0.08 & 4.08 & 0.11 & 0.13 & 3.68 & 9 & 2 & 0.04 & 3 & 0.96 & 0.54 & 0.65\\
KT & 18 & 31 & 0.20 & 3.44 & 0.27 & 0.22 & 2.77 & 6 & 0 & 0.00 & 2 & 0.83 & 0.74 & 0.80\\
LB & 30 & 35 & 0.08 & 2.33 & 0.16 & 0.19 & 3.47 & 8 & 5 & 0.17 & 8 & 0.70 & 0.54 & 0.60\\
MA & 42 & 16 & 0.02 & 0.76 & 0.08 & 0.37 & 1.46 & 3 & 23 & 0.55 & 29 & 0.12 & 0.60 & 0.50\\
MC & 115 & 184 & 0.03 & 3.20 & 0.08 & 0.47 & 3.69 & 9 & 13 & 0.11 & 22 & 0.52 & 0.59 & 0.67\\
MG & 101 & 153 & 0.03 & 3.03 & 0.07 & 0.55 & 5.28 & 12 & 16 & 0.16 & 20 & 0.75 & 0.62 & 0.64\\
MK & 48 & 66 & 0.06 & 2.75 & 0.09 & 0.21 & 3.63 & 9 & 7 & 0.15 & 11 & 0.60 & 0.61 & 0.64\\
MN & 30 & 36 & 0.08 & 2.40 & 0.16 & 0.43 & 1.79 & 3 & 3 & 0.10 & 9 & 0.40 & 0.66 & 0.83\\
MY & 89 & 334 & 0.09 & 7.51 & 0.13 & 0.07 & 2.66 & 6 & 0 & 0.00 & 1 & 1.00 & 0.38 & 0.44\\
NI & 41 & 97 & 0.12 & 4.73 & 0.26 & 0.01 & 2.84 & 7 & 0 & 0.00 & 1 & 1.00 & 0.50 & 0.70\\
NP & 161 & 69 & 0.01 & 0.86 & 0.02 & 0.47 & 1.40 & 3 & 69 & 0.43 & 103 & 0.04 & 0.60 & 0.69\\
PC & 59 & 46 & 0.03 & 1.56 & 0.09 & 0.45 & 1.56 & 3 & 23 & 0.39 & 31 & 0.17 & 0.70 & 0.73\\
PE & 44 & 65 & 0.07 & 2.95 & 0.12 & 0.52 & 3.66 & 8 & 7 & 0.16 & 9 & 0.70 & 0.63 & 0.66\\
PQ & 730 & 1192 & 0.00 & 3.27 & 0.01 & 0.22 & 6.27 & 15 & 94 & 0.13 & 116 & 0.80 & 0.39 & 0.47\\
PT & 59 & 148 & 0.09 & 5.02 & 0.21 & 0.28 & 3.10 & 7 & 4 & 0.07 & 7 & 0.83 & 0.61 & 0.69\\
RA & 54 & 42 & 0.03 & 1.56 & 0.06 & 0.18 & 2.57 & 7 & 14 & 0.26 & 22 & 0.28 & 0.38 & 0.43\\
SH & 44 & 34 & 0.04 & 1.55 & 0.13 & 0.29 & 2.34 & 5 & 12 & 0.27 & 19 & 0.32 & 0.52 & 0.65\\
SN & 34 & 43 & 0.08 & 2.53 & 0.17 & 0.27 & 3.02 & 6 & 3 & 0.09 & 7 & 0.71 & 0.52 & 0.58\\
TE & 152 & 88 & 0.01 & 1.16 & 0.03 & 0.38 & 1.65 & 5 & 58 & 0.38 & 86 & 0.05 & 0.65 & 0.75\\
TI & 38 & 84 & 0.12 & 4.42 & 0.15 & -0.03 & 2.92 & 6 & 2 & 0.05 & 3 & 0.95 & 0.59 & 0.73\\
TM & 72 & 125 & 0.05 & 3.47 & 0.09 & 0.32 & 3.75 & 9 & 3 & 0.04 & 11 & 0.65 & 0.58 & 0.67\\
TP & 54 & 104 & 0.07 & 3.85 & 0.10 & 0.33 & 3.77 & 9 & 1 & 0.02 & 2 & 0.98 & 0.39 & 0.40\\
TS & 47 & 108 & 0.10 & 4.60 & 0.23 & 0.15 & 3.45 & 7 & 1 & 0.02 & 3 & 0.94 & 0.60 & 0.74\\
TZ & 70 & 53 & 0.02 & 1.51 & 0.05 & 0.69 & 2.17 & 7 & 23 & 0.33 & 34 & 0.13 & 0.77 & 0.59\\
UP & 151 & 126 & 0.01 & 1.67 & 0.03 & 0.63 & 1.32 & 3 & 30 & 0.20 & 69 & 0.06 & 0.83 & 0.91\\
VT & 33 & 19 & 0.04 & 1.15 & 0.09 & 0.24 & 2.06 & 4 & 14 & 0.42 & 18 & 0.27 & 0.43 & 0.61\\
WH & 25 & 11 & 0.04 & 0.88 & 0.09 & 0.76 & 1.15 & 2 & 13 & 0.52 & 17 & 0.16 & 0.86 & 0.67\\
WL & 104 & 55 & 0.01 & 1.06 & 0.05 & 0.47 & 2.43 & 6 & 55 & 0.53 & 64 & 0.12 & 0.61 & 0.47\\
YI & 69 & 62 & 0.03 & 1.80 & 0.03 & 0.01 & 2.11 & 6 & 6 & 0.09 & 20 & 0.16 & 0.49 & 0.55\\
YN & 39 & 18 & 0.02 & 0.92 & 0.11 & 0.52 & 1.45 & 3 & 16 & 0.41 & 25 & 0.18 & 0.62 & 0.82\\*
\end{longtable}
\endgroup{}

\end{landscape}

\begin{figure}[h]
\begin{center}
\includegraphics[width=\textwidth]{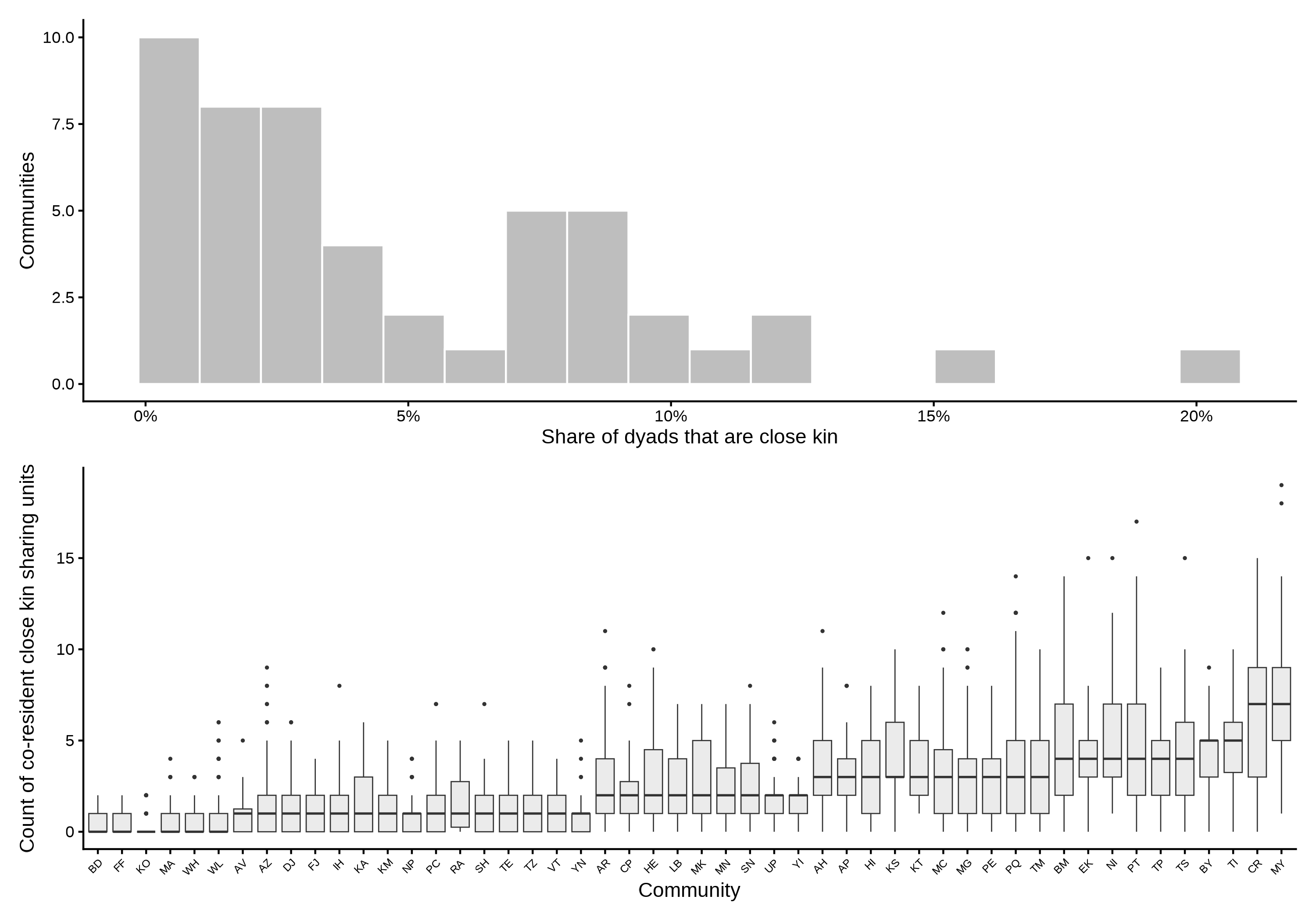}
\end{center}
\caption{\textbf{Availability of co-residing close kin, across communities}. (Top) Histogram showing the distribution of the share of all possible dyads in a community that are close kin. (Bottom) The distribution within each community of the number of co-residing close kin sharing units that each sharing unit has available to them.}
\label{fig:kin_availability}
\end{figure}

Communities vary substantially in the extent to which they are comprised of kin or non-kin.
Ample research has established the importance of kin, especially \emph{close} kin, in provisioning support \cite{wellman_brothers_1989}, meaning that these neighbours may be particularly important for sharing units.
Across all ENDOW communities, sharing units have a median of two other sharing units with close kin, but this obscures substantial within- and cross-site variation.
In Figure~\ref{fig:kin_availability} we show this variation more directly.

\begin{figure}[h]
\begin{center}
\includegraphics[width=\textwidth]{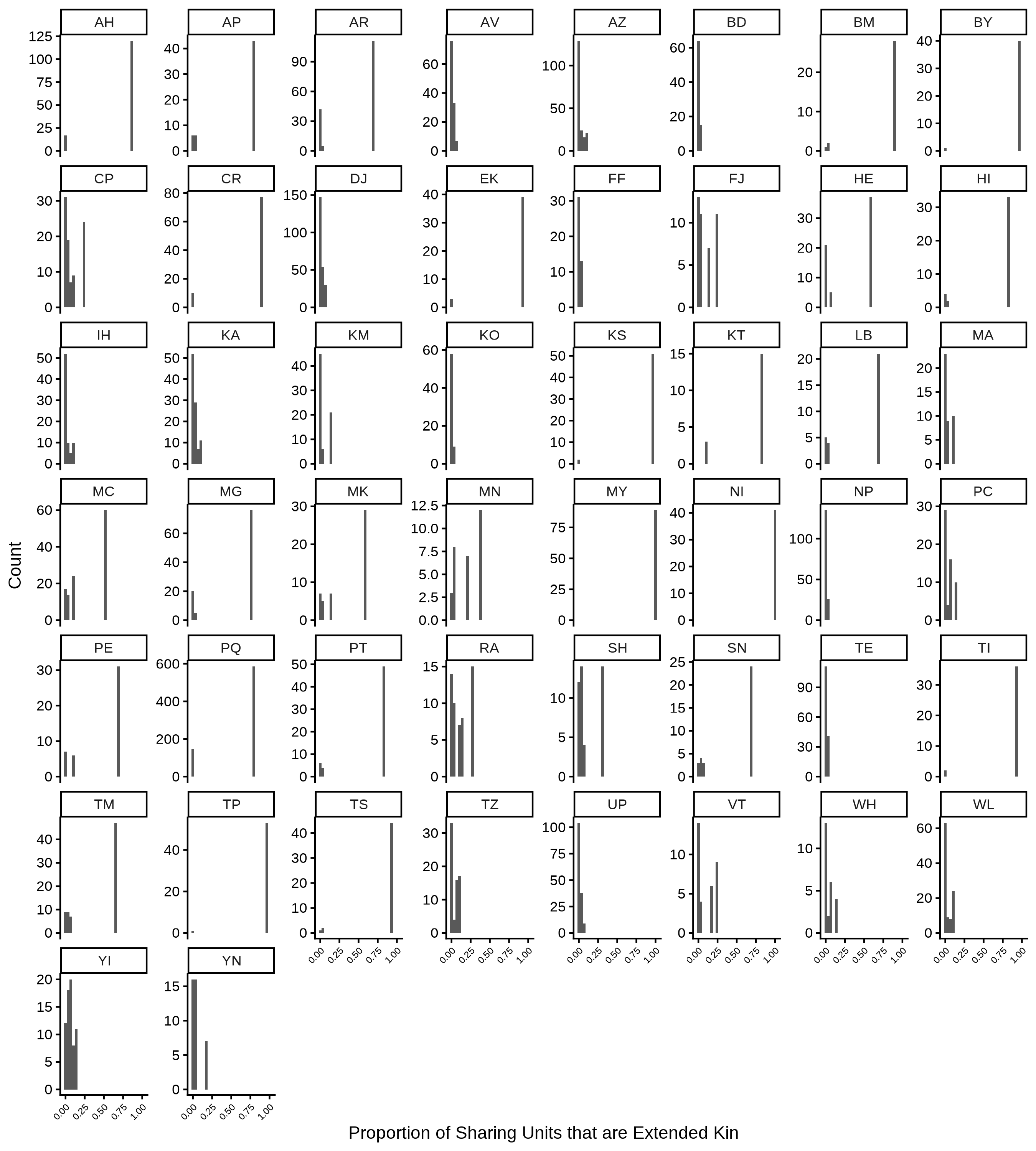}
\end{center}
\caption{\textbf{Histograms of a sharing unit's access to extended kin.} For each sharing unit in each community, we calculate the proportion of other sharing units that are in the same component as it in the close kin network. }
\label{fig:prop_extended_kin}
\end{figure}

The values for mean degree and the proportion of isolates in Table~\ref{tab:net-summary-stats-primary_kin} are particularly informative of how many potential kin-based support alters sharing units have in their communities. While in MY (a Mayan farming community in Mexico) sharing units have, on average, eight other sharing units with close kin as neighbours, in FF (a pastoralist community in Namibia), $70\%$ of the households have none and the others have only one other sharing unit with close kin.
In Figure~\ref{fig:prop_extended_kin} we use the close kin network to get a sense of the possible access to more extended kin, as close kin ties indirectly link sharing units together, implying more distant relatedness. Note, however, that this is not necessarily exhaustive of extended kin ties in these communities: only extended kin relations that are mediated by other sharing units would be connected in this way, so this may considerably underestimate extended kin relations in some communities. Still, we can see that there are a few distinct regimes across communities:
in some communities, almost all sharing units are indirectly connected together through kin relations to form one large extended kin group (\textit{e.g.}, AH, BM, BY, CR, EK, HI, KT, MY, NI, PT, TI, TP, TS),
in others, sharing units are rather mixed with some clusters of kin groups (\textit{e.g.}, FJ, MN, RA),
and in others sharing units are largely lacking local kin (\textit{e.g.}, AV, BD, FF, KM, KO, MA, NP).

While co-resident kin \emph{may} potentially provide support, this need not be the case. Certainly, Figure~\ref{fig:kin_supp_violin} suggests that there is consistent reliance on kin, but we can investigate further.
Table~\ref{tab:global-kin-support} shows the mean, median, and standard deviation across communities of measures showing the extent to which sharing units in each community rely upon co-residing close kin (and non-kin), for each network layer and the aggregate network. It is again evident that close kin are important support alters in every network layer, and that overall there is a strong likelihood that co-resident kin will be support alters. One notable pattern is that the women-specific networks show a higher reliance on kin than the men-specific networks (consistent with what has been observed in other settings, \textit{e.g.}, \cite{moore_structural_1990, dunbar_social_1995}). 

Table~\ref{tab:sitewise-kin-support} illustrates this on a site-by-site basis with a focus on the aggregate network: first pooling together all sharing units across all communities and then individually for each community, we show the typical number of co-residing kin and the percent of sharing units with no co-residing close kin (equivalent to the proportion of isolates shown above in Table~\ref{tab:net-summary-stats-primary_kin}), before showing the average proportion of support alters in the aggregate network who are close kin and the average proportion of close kin who are support alters (equivalent to the summaries in Figure~\ref{fig:kin_supp_violin}), and finally the average proportion of sharing units that are \emph{not} close kin who are named as support alters in the aggregate network. Again, while there is variation in the availability of close kin, there is consistent reliance on them across communities.
To further emphasise this, in Figure~\ref{fig:kin_availability_vs_reliance} we contrast (left) the availability of co-residing close kin with the extent to which they are named as support alters in the aggregate network, showing that reliance on close kin is substantially higher than what their availability alone would suggest (indicated by the dotted line, which would imply that kin are alters proportional to their availability). Second, we contrast (right) the probability that sharing units in each community have a non-kin sharing unit as a alter in the aggregate network to the equivalent probability for close kin sharing unit, demonstrating again that close kin are much more likely to provide support than non-kin (with the dotted line showing an expectation of no difference between the two).

\begin{landscape}

\begingroup\fontsize{8}{10}\selectfont

\begin{longtable}[t]{lccccccccc}
\caption{\label{tab:global-kin-support}Median (above) and mean and standard deviation (below) of kin availability and tie probability measures for each network layer. `Prop Partners Kin' is the proportion of connected dyads (i.e., partners) that are kin. `P(Tie $\mid$ Kin)' is the probability that a kin dyad is connected in the layer. `P(Tie $\mid$ Non-kin)' is the probability that a non-kin dyad is connected in the layer. `Risk Ratio' is the ratio of P(Tie $\mid$ Kin) to P(Tie $\mid$ Non-kin). `Odds Ratio' shows the odds of a tie existing between kin, relative to the odds of an tie existing between non-kin.}\\
\toprule
 & Borrow money & Lend money & Borrow goods & Lend goods & Women behav. & Men behav. & Women info. & Men info. & Aggregate\\
\midrule
\endfirsthead
\caption[]{Median (above) and mean and standard deviation (below) of kin availability and tie probability measures for each network layer. `Prop Partners Kin' is the proportion of connected dyads (i.e., partners) that are kin. `P(Tie $\mid$ Kin)' is the probability that a kin dyad is connected in the layer. `P(Tie $\mid$ Non-kin)' is the probability that a non-kin dyad is connected in the layer. `Risk Ratio' is the ratio of P(Tie $\mid$ Kin) to P(Tie $\mid$ Non-kin). `Odds Ratio' shows the odds of a tie existing between kin, relative to the odds of an tie existing between non-kin. \textit{(continued)}}\\
\toprule
 & Borrow money & Lend money & Borrow goods & Lend goods & Women behav. & Men behav. & Women info. & Men info. & Aggregate\\
\midrule
\endhead

\endfoot
\bottomrule
\endlastfoot
Prop Partners Kin & 0.32 & 0.37 & 0.31 & 0.29 & 0.35 & 0.25 & 0.26 & 0.19 & 0.22\\
 & 0.35 (0.22) & 0.38 (0.25) & 0.36 (0.21) & 0.33 (0.2) & 0.41 (0.24) & 0.32 (0.21) & 0.3 (0.2) & 0.23 (0.16) & 0.24 (0.15)\\
P(Tie $\mid$ Kin) & 0.21 & 0.25 & 0.29 & 0.29 & 0.21 & 0.2 & 0.26 & 0.19 & 0.62\\
 & 0.24 (0.15) & 0.26 (0.15) & 0.33 (0.16) & 0.32 (0.17) & 0.24 (0.17) & 0.22 (0.16) & 0.29 (0.15) & 0.21 (0.12) & 0.61 (0.16)\\
P(Tie $\mid$ Non-kin) & 0.02 & 0.02 & 0.02 & 0.02 & 0.01 & 0.02 & 0.03 & 0.02 & 0.08\\
 & 0.02 (0.02) & 0.02 (0.02) & 0.03 (0.03) & 0.03 (0.03) & 0.02 (0.02) & 0.02 (0.02) & 0.04 (0.03) & 0.04 (0.03) & 0.1 (0.07)\\
Risk Ratio & 16.5 & 14.82 & 15.58 & 12.72 & 19.47 & 11.59 & 9.53 & 6.44 & 8.09\\
 & 21.28 (20.26) & 20.91 (20.15) & 20.77 (24.29) & 17.04 (16.02) & 25.61 (23.64) & 16.52 (14.68) & 16.08 (17.51) & 9.95 (10.6) & 10.58 (10.32)\\
Odds Ratio & 21.97 & 20.96 & 22.73 & 17.85 & 27.44 & 17.44 & 13.3 & 7.68 & 21.64\\
 & 29.9 (27.8) & 28.98 (27.64) & 30.01 (28.4) & 24.66 (19.27) & 34.76 (31.63) & 21.98 (18.1) & 22.3 (22.14) & 12.66 (12.69) & 26.2 (20.46)\\*
\end{longtable}
\endgroup{}

\begin{figure}[h]
\begin{center}
\includegraphics[width=0.9\textwidth]{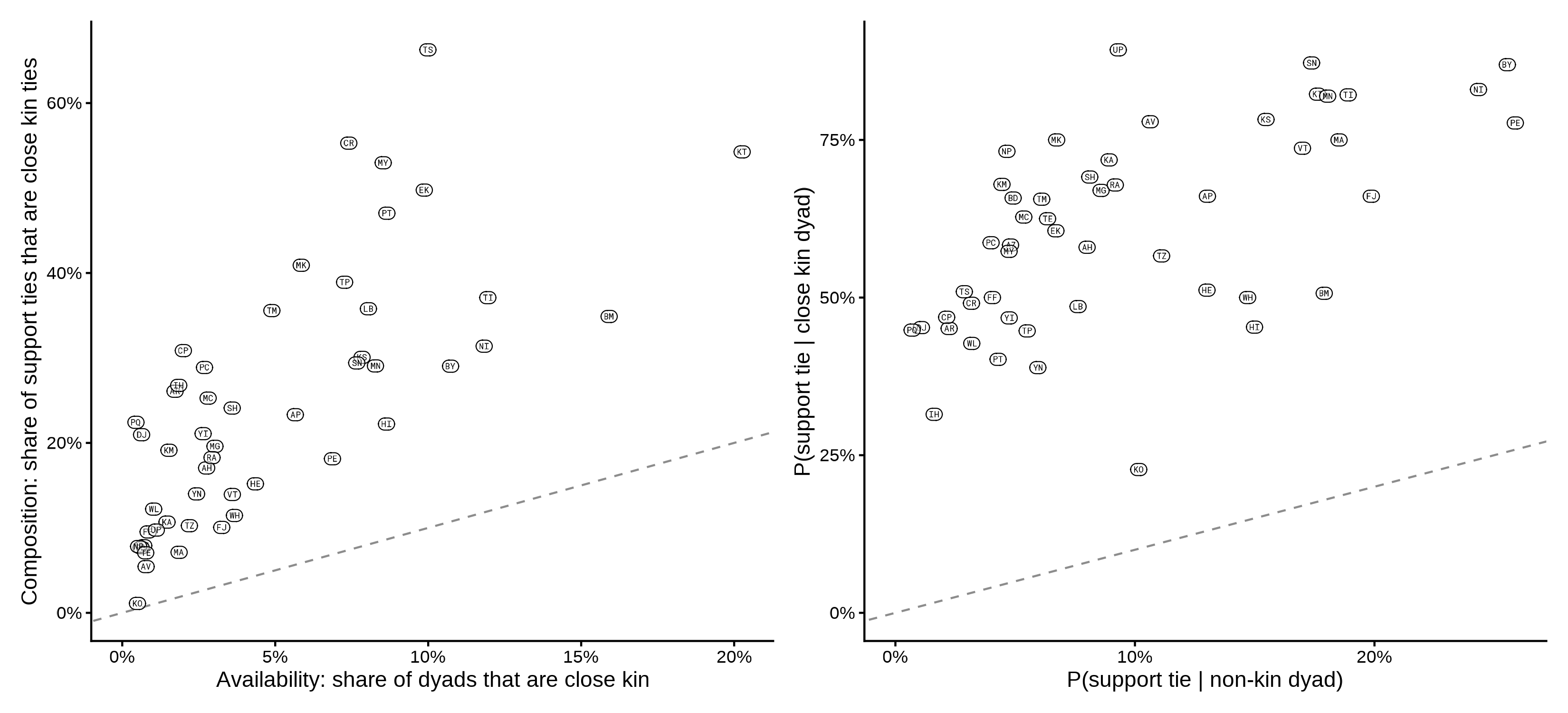}
\end{center}
\caption{\textbf{Close kin availability versus reliance on close kin, in each community}. (Left) The proportion of all dyads that are close kin versus the proportion of connected dyads that are close kin; the dotted line represents what we would observe if kin were alters proportional to their availability. (Right) The probability that a non-kin dyad is connected in the aggregate network versus the probability that a close kin dyad is connected in the aggregate network; the dotted line represents equivalent probabilities.}
\label{fig:kin_availability_vs_reliance}
\end{figure}

\end{landscape}

\begingroup\fontsize{8}{10}\selectfont

\begin{longtable}[t]{lcccccc}
\caption{\label{tab:sitewise-kin-support}Availability of close kin and ties to them in the aggregate network, across all communities (`Pooled') and separately by community. `N' is the number of sharing units. `Co-Res. Kin' is the mean and median (in brackets) of number of co-residing sharing units with close kin that each sharing unit has. `No Kin' is the percent of sharing units with no co-residing close kin. `Partners Kin' is the mean proportion of partners in the aggregate network who are close kin. `Named Kin' is the mean proportion of co-resident kin sharing units that are partners in the aggregate network. `Named Non-Kin' is equivalently the proportion of non-kin sharing units that are named as partners in the aggregate network.}\\
\toprule
Community & N & Co-Res. Kin & No Kin & Partners Kin & Named Kin & Named Non-Kin\\
\midrule
\endfirsthead
\caption[]{Availability of close kin and ties to them in the aggregate network, across all communities (`Pooled') and separately by community. `N' is the number of sharing units. `Co-Res. Kin' is the mean and median (in brackets) of number of co-residing sharing units with close kin that each sharing unit has. `No Kin' is the percent of sharing units with no co-residing close kin. `Partners Kin' is the mean proportion of partners in the aggregate network who are close kin. `Named Kin' is the mean proportion of co-resident kin sharing units that are partners in the aggregate network. `Named Non-Kin' is equivalently the proportion of non-kin sharing units that are named as partners in the aggregate network. \textit{(continued)}}\\
\toprule
Community & N & Co-Res. Kin & No Kin & Partners Kin & Named Kin & Named Non-Kin\\
\midrule
\endhead

\endfoot
\bottomrule
\endlastfoot
Pooled & 4366 & 2.49 (2) & 24.12 & 0.24 & 0.61 & 0.07\\
AH & 137 & 3.75 (3) & 7.30 & 0.19 & 0.59 & 0.08\\
AP & 55 & 3.05 (3) & 10.91 & 0.23 & 0.69 & 0.13\\
AR & 158 & 2.71 (2) & 15.82 & 0.32 & 0.54 & 0.02\\
AV & 116 & 0.90 (1) & 44.83 & 0.07 & 0.79 & 0.11\\
AZ & 190 & 1.33 (1) & 37.37 & 0.09 & 0.65 & 0.05\\
BD & 79 & 0.48 (0) & 63.29 & 0.08 & 0.72 & 0.05\\
BM & 31 & 4.77 (4) & 3.23 & 0.34 & 0.60 & 0.18\\
BY & 41 & 4.29 (5) & 2.44 & 0.34 & 0.88 & 0.26\\
CP & 90 & 1.78 (2) & 23.33 & 0.36 & 0.49 & 0.02\\
CR & 87 & 6.37 (7) & 11.49 & 0.58 & 0.53 & 0.03\\
DJ & 231 & 1.45 (1) & 32.90 & 0.21 & 0.48 & 0.01\\
EK & 42 & 4.05 (4) & 7.14 & 0.52 & 0.65 & 0.07\\
FF & 44 & 0.36 (0) & 70.45 & 0.12 & 0.62 & 0.04\\
FJ & 42 & 1.33 (1) & 30.95 & 0.13 & 0.74 & 0.20\\
HE & 63 & 2.70 (2) & 23.81 & 0.17 & 0.57 & 0.13\\
HI & 39 & 3.28 (3) & 10.26 & 0.23 & 0.45 & 0.15\\
IH & 77 & 1.40 (1) & 41.56 & 0.29 & 0.36 & 0.02\\
KA & 99 & 1.43 (1) & 34.34 & 0.10 & 0.70 & 0.09\\
KM & 72 & 1.08 (1) & 40.28 & 0.19 & 0.62 & 0.04\\
KO & 67 & 0.33 (0) & 77.61 & 0.01 & 0.30 & 0.10\\
KS & 53 & 4.08 (3) & 3.77 & 0.33 & 0.79 & 0.15\\
KT & 18 & 3.44 (3) & 0.00 & 0.59 & 0.84 & 0.18\\
LB & 30 & 2.33 (2) & 16.67 & 0.43 & 0.54 & 0.07\\
MA & 42 & 0.76 (0) & 54.76 & 0.07 & 0.75 & 0.18\\
MC & 115 & 3.20 (3) & 11.30 & 0.26 & 0.65 & 0.05\\
MG & 101 & 3.03 (3) & 15.84 & 0.22 & 0.73 & 0.09\\
MK & 48 & 2.75 (2) & 14.58 & 0.47 & 0.82 & 0.07\\
MN & 30 & 2.40 (2) & 10.00 & 0.27 & 0.76 & 0.18\\
MY & 89 & 7.51 (7) & 0.00 & 0.56 & 0.62 & 0.05\\
NI & 41 & 4.73 (4) & 0.00 & 0.31 & 0.87 & 0.24\\
NP & 161 & 0.86 (1) & 42.86 & 0.09 & 0.77 & 0.05\\
PC & 59 & 1.56 (1) & 38.98 & 0.30 & 0.59 & 0.04\\
PE & 44 & 2.95 (3) & 15.91 & 0.21 & 0.77 & 0.26\\
PQ & 730 & 3.27 (3) & 12.88 & 0.26 & 0.49 & 0.01\\
PT & 59 & 5.02 (4) & 6.78 & 0.50 & 0.50 & 0.04\\
RA & 54 & 1.56 (1) & 25.93 & 0.20 & 0.74 & 0.09\\
SH & 44 & 1.55 (1) & 27.27 & 0.29 & 0.77 & 0.08\\
SN & 34 & 2.53 (2) & 8.82 & 0.30 & 0.90 & 0.17\\
TE & 152 & 1.16 (1) & 38.16 & 0.08 & 0.64 & 0.06\\
TI & 38 & 4.42 (5) & 5.26 & 0.43 & 0.81 & 0.19\\
TM & 72 & 3.47 (3) & 4.17 & 0.44 & 0.68 & 0.06\\
TP & 54 & 3.85 (4) & 1.85 & 0.42 & 0.49 & 0.05\\
TS & 47 & 4.60 (4) & 2.13 & 0.67 & 0.53 & 0.03\\
TZ & 70 & 1.51 (1) & 32.86 & 0.11 & 0.53 & 0.11\\
UP & 151 & 1.67 (2) & 19.87 & 0.12 & 0.90 & 0.09\\
VT & 33 & 1.15 (1) & 42.42 & 0.13 & 0.66 & 0.17\\
WH & 25 & 0.88 (0) & 52.00 & 0.14 & 0.61 & 0.15\\
WL & 104 & 1.06 (0) & 52.88 & 0.13 & 0.52 & 0.03\\
YI & 69 & 1.80 (2) & 8.70 & 0.24 & 0.48 & 0.05\\
YN & 39 & 0.92 (1) & 41.03 & 0.14 & 0.40 & 0.06\\*
\end{longtable}
\endgroup{}

\begin{figure}[h]
\begin{center}
\includegraphics[width=\textwidth]{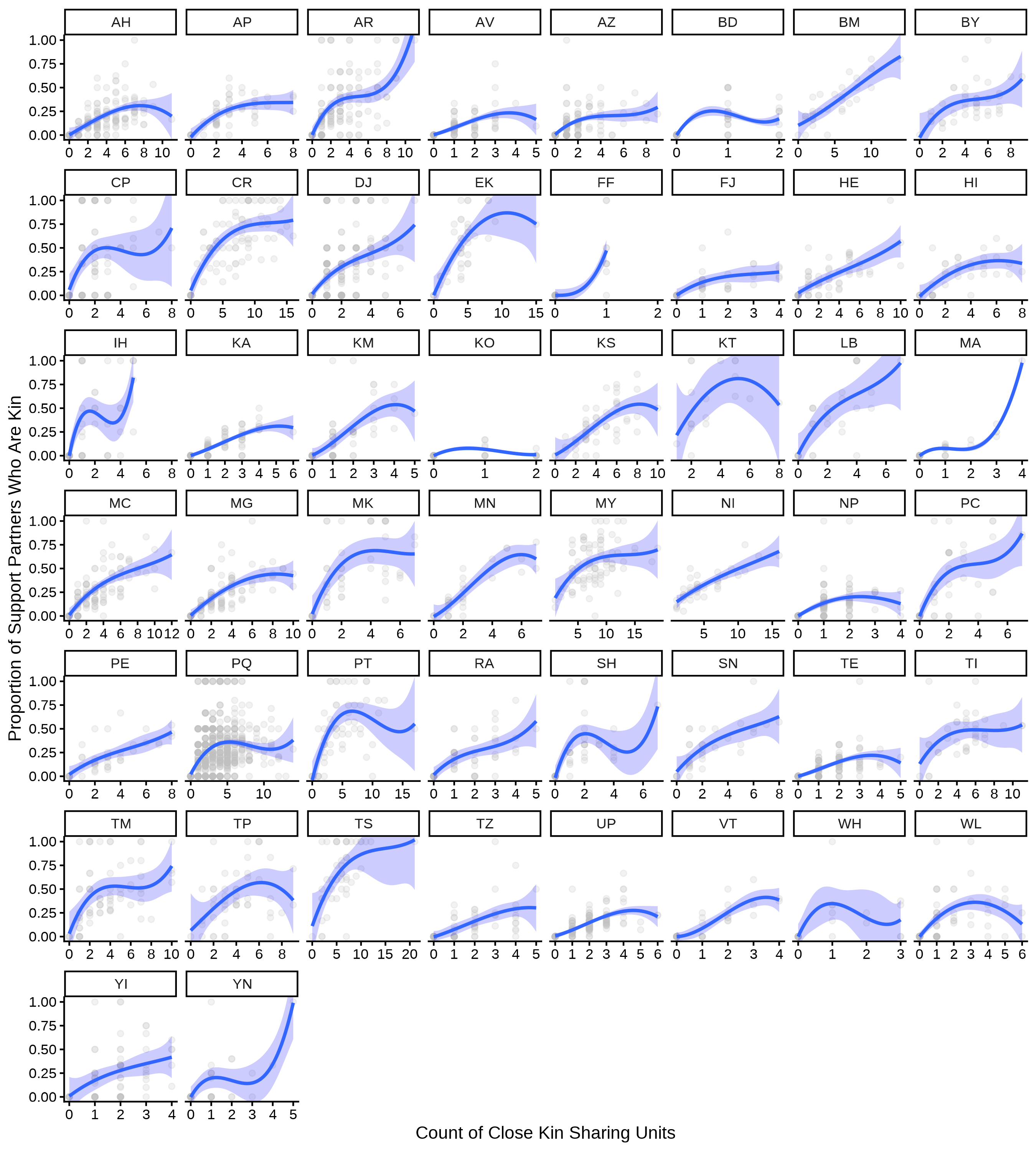}
\end{center}
\caption{\textbf{Relationship between a sharing unit's co-resident kin and their reliance on those kin.} Based on support ties in the aggregate network. Curves show cubic polynomial fits to the data. Note that axes vary across plots.}
\label{fig:prop_kin_vs_cores_kin}
\end{figure}

\begin{figure}[h]
\begin{center}
\includegraphics[width=\textwidth]{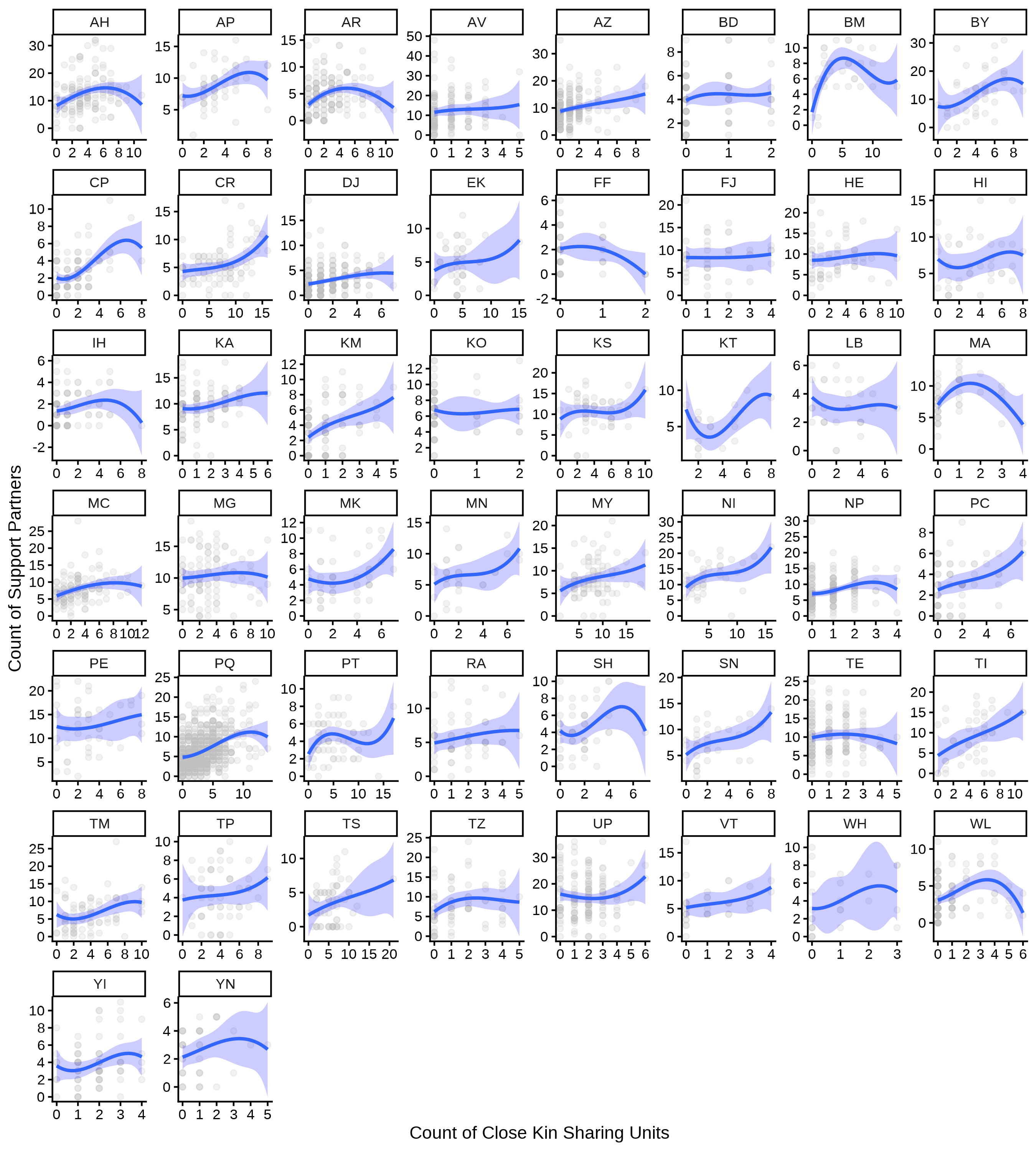}
\end{center}
\caption{\textbf{Relationship between a sharing unit's co-resident kin and their support alters.} Based on support ties in the aggregate network. Curves show cubic polynomial fits to the data. Note that axes vary across plots.}
\label{fig:supp_vs_cores_kin}
\end{figure}

Figure~\ref{fig:prop_kin_vs_cores_kin} illustrates whether sharing units with a greater number of co-resident kin are more or less reliant on those kin. We observe a broadly positive association that tempers at higher numbers across the vast majority of communities, suggesting that sharing units with more co-resident kin are more reliant on and entangled with those kin, but that this need to rely on kin may ``saturate'' at some point.
In Figure~\ref{fig:supp_vs_cores_kin} we see whether sharing units with a greater number of co-resident kin also have greater overall access to support within their community. Having more sharing units that might be expected or obligated to assist may make it easier for sharing units to form supportive relationships within the community. However, there is little suggestion of this association across communities. Not having (more) co-resident kin does not seem to be associated with fewer overall support alters. While kin may often be important sources of support, there appear to be other ways for sharing units to forge supportive relationships.

\begin{figure}[h]
\begin{center}
\includegraphics[width=\textwidth]{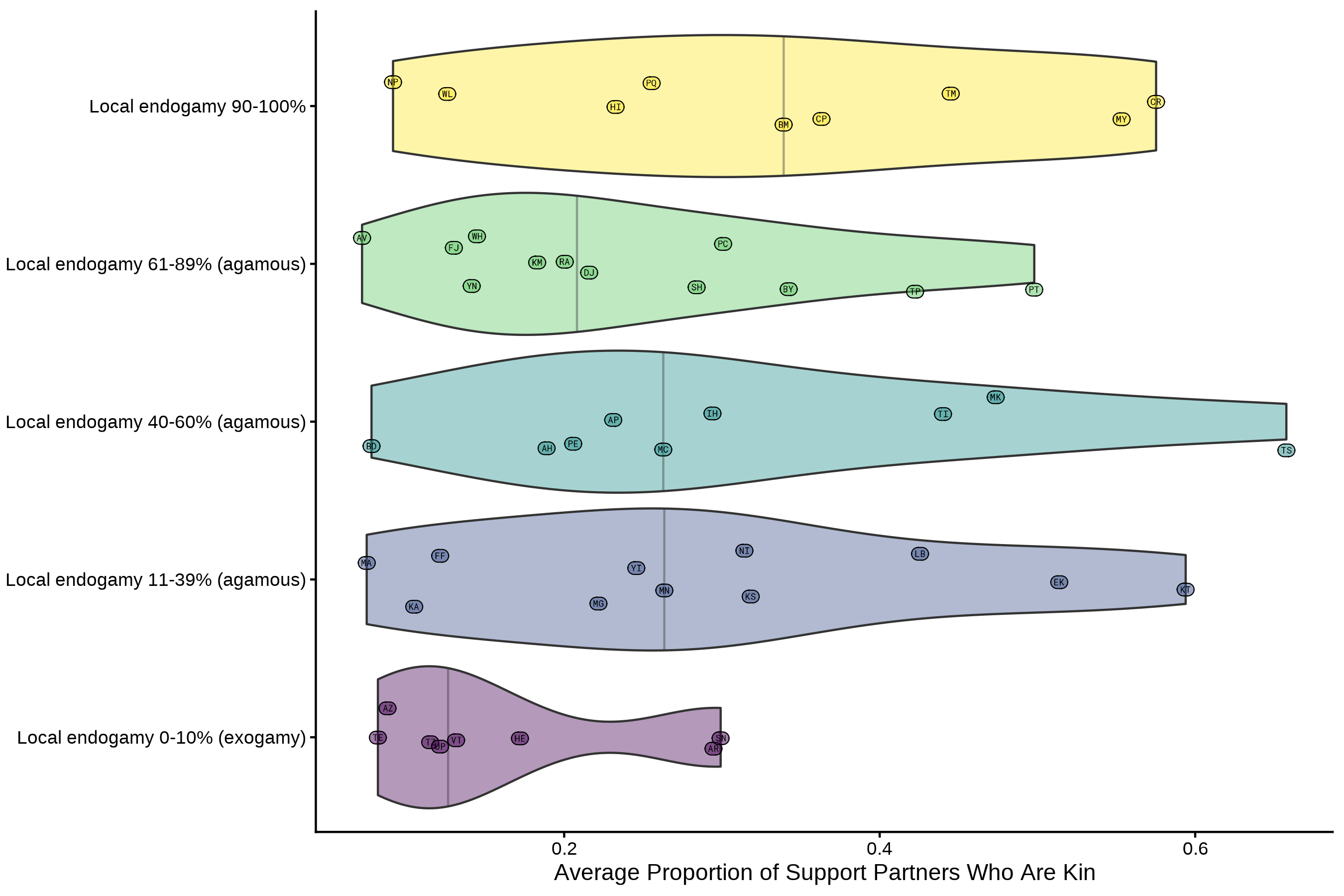}
\end{center}
\caption{\textbf{The average proportion of co-resident kin by community endogamy.} Rates of endogamy as reported by the ethnographers working in each community.}
\label{fig:endogamy}
\end{figure}

Marriage endogamy (whether people are expected to marry within a community, however defined), residence patterns, and sharing unit composition/definition will all shape the extent to which kin are present and called upon. In Figure~\ref{fig:endogamy} we show each community's average proportion of support alters who are close kin, grouped by the ethnographers' reports of local endogamy patterns. Communities that are exogamous (where people are generally expected to marry outside their local community) are ones where sharing units have fewer kin among their support alters, and communities with high local endogamy are ones where sharing units have a high proportion of kin in their support alters, though the differences are slight and only suggestive.

\end{document}